\documentclass[12pt,narrowdisplay,onecolumn]{elsart3p}
\usepackage{graphicx,amssymb,feynmp}
\def\be{\begin{eqnarray}}
\def\ee{\end{eqnarray}}
\def\bc{\begin{center}}
\def\ec{\end{center}}
\def\om{\omega}
\def\dsp{\displaystyle}
\def\prt{\partial}
\def\lsim{\lesssim}
\def\gsim{\gtrsim}

\def\rmd{{\rm d}}
\journal{ArXiv
}
\let\sectiontmp\section
\def\section{\setcounter{equation}{0}\sectiontmp}

\begin{document}
\begin{frontmatter}

\title{Time  delays and  advances in classical and quantum
systems
}
\author[UMB]{E.E.~Kolomeitsev}
\author[MEPHI]{D.N.~Voskresensky}
\address[UMB]{Matej Bel  University, SK-97401 Banska Bystrica, Slovakia}
\address[MEPHI]{National Research Nuclear University "MEPhI", Kashirskoe sh. 31, Moscow 115409, Russia}
\begin{abstract}
The paper reviews positive and negative time delays in various
processes of classical and quantum physics. In the beginning, we demonstrate
how a time-shifted response of a system to an external
perturbation appears in classical mechanics and classical
electrodynamics. Then we quantify durations of various quantum mechanical
processes. The duration of the quantum tunneling is studied. An
interpretation of the Hartman paradox is suggested. Time delays and
advances appearing in the three-dimensional scattering problem on a central potential are considered. Then we discuss delays and advances appearing in quantum field theory and
after that we focus on the issue of time delays and advancements
in quantum kinetics. We discuss problems of the
application of generalized kinetic equations in simulations of
the system relaxation towards equilibrium and analyze the
kinetic entropy flow. Possible measurements of time delays and
advancements in experiments similar to the recent OPERA neutrino
experiment are also discussed.
\end{abstract}


\setcounter{secnumdepth}{5}
\setcounter{tocdepth}{5}

\end{frontmatter}

\tableofcontents

%


\section{Introduction}

Many definitions of time, as a measure of a duration of  a
process, are possible in classical mechanics because for the
measuring of the time duration any process is suitable, which
occurs at a constant pace. Naively thinking, a response of a
system to an external perturbation should be delayed in accordance
with the causality principle. However, it is not always the case.
There may arise  both delays and advancements (negative time
delays) in system responses without contradiction with causality.

Time delays and possible time advancements in quantum mechanical
phenomena have been extensively discussed in the literature, see
Refs.~\cite{Wigner,WuOm,GW,Nusszv-book,Bosanas,Carvalho,Razavi,Winful,Shvartsburg,Muga-I,Muga-II,Campo}
and references therein. In spite of that many questions still
remain not quite understood. Worth mentioning is the Hartman
effect~\cite{Hartman}, that the transition time of a quantum
particle through a one-dimensional barrier is seemingly
independent of the barrier length for broad barriers. This causes
apparent superluminal phenomena in the quantum mechanical
tunneling. Many, at first glance, supporting experiments with
single photons, classical light waves and microwaves have been
performed, see Refs.~\cite{Garrett,Faxvog,Garrison98,Stenner,Boyd}
and references therein. Different definitions of time delays, such
as the group  transmission time delay $\delta t_{\rm T}$, the
group reflection  time delay $\delta t_{\rm R}$, the interference
time delay $\delta t_{\rm i}$, the dwell time $t_{\rm d}$, the
sojourn time $t_{\rm soj}$, and some other quantities have been
introduced to treat the problem. All these time scales suffer from
the Hartmann effect and are in odd with the natural expectation
that the tunneling time should be proportional to the length of
the barrier. A re-interpretation consistent with special
relativity suggested  in \cite{Winful} is that these times should
be treated as the live times of the corresponding wave packets
rather than the traveling time. If so, the so far performed
experiments measured an energy dissipation at the edges of the
barrier rather than a particle traveling time.

Additionally to the mentioned time delays other relevant time
quantities  were introduced, and the differences between the
averaged scattering time delay $\delta t_{\rm s}$ and the Wigner
scattering time delay $\delta t_{\rm W}$ were discussed in
Ref.~\cite{GW,Nussenzweig,DP}, see also
Refs.~\cite{Nusszv-book,Bosanas,Carvalho} and references therein.
Based on these analyses authors of Ref.~\cite{DP} argued that
kinetic simulations describing relaxation of a system, first,
towards the local equilibrium and, then, towards  the global one
must account for delays in scattering events consistently with
mean fields acting on particles, in order to model consistently
thermodynamic properties of the system. For practical simulations,
as the relevant relaxation time they suggested to use the
scattering time delay $\delta t_{\rm s}$, as it follows from the
phase shift analysis, rather than the collision time $t_{\rm
col}$, as it appears in the original Boltzmann equation. A number
of BUU simulations of heavy ion collision reactions were performed
using this argumentation, see Ref. \cite{Mosel} and references
therein.

The appropriate frame for the description of non-equilibrium
many-body processes is the real-time formalism of quantum filed
theory developed by Schwinger, Kadanoff, Baym and
Keldysh~\cite{Schwinger,Kad,Kad62,Keldysh}. A generalized kinetic
description of off-mass-shell (virtual) particles has been
developed  based on the quasiclassical treatment of the Dyson
equations for non-equilibrium systems, see Refs.~\cite{Kad62,Daniel84,BezzDuB72,Seren-Rainer83,Chou-Su85,Rammer86,Berges05}. This
treatment assumes the validity of the first-order gradient
approximation  to the Wigner-transformed Dyson equations. As it is
ordinary sought, the gradient approximation is valid, if the
typical time-space scales are much larger than the microscopic
scales, such as $1/E_{\rm F}$ and $1/p_{\rm F}$ for a slightly
excited Fermi systems, where $E_{\rm F}$ is the Fermi energy and
$p_{\rm F}$ is the Fermi momentum. As the result, a quantum
kinetic equation  is derived for off-mass shell particles, for
which the energy and momentum are not connected by any dispersion
relation. We call this generalized kinetic equation the
Kaddanoff-Baym (KB) equation. Among other terms, this equation
contains the Poisson-bracket term, which origin has not been quite
understood during a long time. Botermans and Malfliet in
Ref.~\cite{Bot90} suggested to replace the production rate in that
Poisson-bracket term by its approximate quasi-equilibrium value.
This allowed to simplify the KB equation for near equilibrium
configurations. The resulting form of the kinetic equation is
called the Botermans-Malfliet (BM) form. It is argued that the BM
replacement does not spoil the validity of the first order
gradient approximation. The so-called $\Phi$-derivable
self-consistent approximations in the quantum field theory were
introduced by Baym in Ref.~\cite{Baym} for quasi-equilibrium
systems. His derivation was then generalized to an arbitrary
Schwinger-Keldysh contour in Ref.~\cite{IKV}.
Reference~\cite{IKV2} developed the self-consistent treatment of
the quantum kinetics. References~\cite{KIV01,IKV3} demonstrated
that the KB kinetic equation is compatible with the exact
conservation of the Noether 4-current and the Noether
energy-momentum, whereas the Noether 4-current and the Noether
energy-momentum related to the BM form of equation are conserved
only approximately, up to zeroth gradients. Fulfillment of the
conservation laws is important  in practical simulations of
dynamical processes. For example, in kinetic simulations of heavy
ion collisions the gradient approximation may not work at least on
an initial stage of the expansion of the fireball. In this case
the KB form of the kinetic equation should be preferable compared
to the BM one due to inherent exact conservation laws for the
Noether quantities in the former case. However, up to now the
simulation scheme, the so called test particle method, has been
realized in applications to heavy-ion collisions only for the BM
form of the kinetic equation, see
Refs.~\cite{Cassing,leupold,Mosel}. The relaxation time arising in
the kinetic equation presented in the BM form is the scattering
time delay, $\delta t_{\rm s}$, rather than the average collision
time $t_{\rm col}$, as it appears in the original KB equation.
Since $\delta t_{\rm s}$ can be naturally interpreted in terms of
the virial expansion \cite{DP}, this was considered as an argument
in favor of the BM form of the kinetic equation.

Recent work~\cite{IV09} suggested a non-local form of the quantum
kinetic equation, which up to second gradients coincides with the
KB equation and up to first gradients, with the BM equation. Thus,
the non-local form keeps the Noether 4-current and Noether
energy-momentum conserved at least up to first gradients. Second
advantage of the non-local form is that it allows interpretation
of mentioned difference in the Poisson-bracket terms in the KB and
BM equations, as associated with space-time and energy-momentum
delays and advancements. Also the non-local form of the kinetic
equation permits, in principle, to develop a test particle method,
similar to that is used for the BM form of the kinetic equation.

In this paper we study problems related to determination of time
delays and advancements in various phenomena. In Sect. \ref{Mech}
we discuss how time delays and lesser time advancements arise in
the description of oscillations in classical mechanics and in
classical field theory of radiation. In Sect. \ref{sec:quant}
 we consider time delays and advancements in one-dimensional quantum mechanical tunneling and in
scattering of particles above the barrier. Problem of an apparent
superluminality in the tunneling (the Hartman effect)  is
considered and a solution of the paradox is suggested.  In Sect.
\ref{scattering}  we consider time delays and advancements in the
three-dimensional scattering problem. Then in Sect. \ref{sec:qft}
we introduce the non-equilibrium Green's function formalism and
show that not only space-time delays but also advancements appear
in Feynmann diagrammatic description of quantum processes within
the quantum field theory. In Sect. \ref{sec:kin} we focus on the
quasiclassical description of non-equilibrium many-body phenomena.
We introduce gradient expansion scheme and arrive at a set of
equations for the kinetic quantities, which should be solved
simultaneously. The kinetic equation for the Wigner density is
presented in three different forms, the KB, the BM and the
non-local form. We discuss time delays and advancements, as they
appear in the non-local form of the kinetic equation (and in the
KB equation equivalent to it up to the second-order gradient
terms) and consider their relation to those quantities, which
arise in the  quantum mechanical one-dimensional tunneling,  in
motion above the barrier and in 3-dimensional scattering. To
demonstrate that all three forms of the kinetic equation are not
fully equivalent  in the region of a formal applicability of the
first order gradient expansion we calculate the kinetic entropy
flow in all three cases and explicate their differences. Then  we
find some solutions for all three forms of the kinetic equation,
rising the question about applicability of  the gradient expansion
in the  description of the  relaxation of a slightly
non-equilibrium system towards equilibrium. Basing on this
discussion we put in question applicability of the BM kinetic
equation for simulations of violent heavy-ion collisions.
 A possibility for appearance of instabilities for superluminal virtual particles is also discussed.
In Sect. \ref{sec:measurement} we discuss  measurements of time
delays and advancements. The origin of an apparent
superluminality, as  might be seen   in experiments similar to
those performed by the OPERA and MINOS neutrino collaborations
\cite{OPERA,MINOS} is discussed.
 In  Appendix \ref{app:virial} we present formulation of the
virial theorem in classical mechanics in terms of the scattering
time delay. Appendix \ref{app:schroed} demonstrates derivation of
some helpful relations between wave functions.
  In Appendix \ref{minentr} we discuss the $H$ theorem and demonstrate the
minimum of the entropy production at the system relaxation towards
the equilibrium.


Starting from Sect. \ref{sec:qft} we use units $\hbar =c=1$. Where
necessary we recover $c$ and $\hbar$.
\section{Time shifts in classical mechanics and in classical field theory}
\label{Mech}


In this section we introduce a number of time characteristics of
the dynamics of physical processes. We demonstrate how a time
shifted response of a system to an external perturbation appears
in classical mechanics and classical electrodynamics. We show that
there may arise as delays as advancements in the system response.

\subsection{Time shifts in classical mechanics}

Let us introduce some  definitions of time, as a measure of
duration  of processes in classical mechanics, which will further
appear in quantum mechanical description.

For measuring of a time duration any process is suitable, which
occurs at constant pace. For example to measure time of motion one
can use a camel moving straightforwardly with  constant velocity
$\vec{v}$, then $t=l/v$, where $l\simeq N\,l_0$ is the distance
passed by the camel, $N$ is number of its steps, $l_0$ is the step
size. Such a simple measurement of time (in camel's steps) is
certainly inconvenient, because a distance between initial and
final camel's positions can be very large for large times. To
overcome the problem one may use a 'mechanical camel' moving
around a circle with a constant angular velocity or linear speed.
Our hand watches are constructed namely in such a manner, where
the  clock arrow takes the role of the camel. More generally, for
a time measurement one may use any periodic process describing by
an ideal oscillator (e.g. one may use for that the atomic clock).
Then the time is measured in a number of half-periods $P/2$ of the
oscillator motion.

Another way to measure time is to exploit the particle
conservation law.  One of the oldest time-measuring devices
constructed in such a manner is a clepsydra or a water clock. Its
usage is based on the principle of the conservation of an amount
of water. Water can be of course replaced by any substance, which
local density $\rho(\vec{r},t)$ obeys the continuity equation
$\prt \rho/\prt t+{\rm div}\vec{j}=0$, where $\vec{j}=\rho\,
\vec{v}$ is a 3D-flux density dependent of a local velocity
$\vec{v}(\vec{r}, t)$ of an element of the substance. Now, if we
take a large container of volume $V$ with a hole of area $S$, the
time passed can be defined, as the ratio of the amount of
substance inside the container to the flux draining out of the
container through the hole:
\be
 t_{\rm d}^{\rm (cl)} =\intop_{V}{\rho \,\rmd^3 r }\Bigg/{\Big|\intop_{S}\vec{j}(\rho)\,\rmd \vec{s}\,\Big|}\,.
 \label{dwell3d}
\ee
We will call this quantity \emph{a dwell time} since similar
definition of a time interval is used in  quantum mechanics in
stationary problems.

In one dimensional case the time  particles dwell in some segment
of the $z$ axis   open at the  ends $z_1$ and $z_2$, through which
particles flow outside the segment, can be found as
\be\label{dwell1d}
 t_{\rm d}^{\rm (1,cl)}= \frac{\int_{z_1}^{z_2}\rho \, dz}{|j(z_1)+j(z_2)|},
\ee
where $\rho (z)$ is the particle density and $j(z)=v(z)\rho(z)$ is
a 1D flux density.  Obviously, for a   particle flux from a hole
at $z=z_2$ (at $j (z_1)=0$)  with constant density $\rho$ and
constant velocity ${v}$ we then have $t_{\rm d}^{\rm (cl)} =l/v$
with $l=z_2-z_1$. If $\rho$ depends on $t$, the definitions (\ref
{dwell3d}), (\ref{dwell1d}) become inconvenient, since $t_{\rm
d}^{\rm (cl)}$ is then a non-linear function of $t$.

Another relevant time-quantity reflecting a temporal extent of a
physical process can be defined as follows. Consider the  motion
of a classical particle in an arbitrary time-dependent
one-dimensional potential $U(z,t)$. The particle trajectory is
described by the function $z(t)\in\mathcal{C}$, where
$\mathcal{C}$ is the space region allowed for classical motion.
Let the particle moves for a time $\tau$, then a part of this
time, which particle spends within an interval $[z_1 ,z_2 ]\in
\mathcal{C}$, is given  by the integral
\be
 t^{\rm (cl)}_{\rm soj}(z_1 ,z_2 ,\tau)=\intop_0^\tau \rmd t \,\theta\big(z(t)-z_1 \big)\, \theta\big(z_2 -z(t)\big)
  =\intop_0^\tau \rmd t \intop_{z_1}^{z_2}\rmd s\, \delta\big(s-z(t)\big)\,.
   \label{sojt-class}
\ee
Such a temporal quantity can be called \emph{a classical sojourn time}. What is notable is that exactly this
time has a well defined counterpart in quantum mechanics.

Now consider particle motion in a stationary field $U(z)$. Using
the equation of motion $\rmd z/\rmd t=v(z;E)$, where
$v(z;E)=\sqrt{\frac{2}{m}\, (E-U(z))}$ is the particle velocity
and $E$, the energy, for an infinite motion we can recast the
sojourn time (\ref{sojt-class}) as
 \be
t^{\rm (cl)}_{\rm soj}(z_1 ,z_2,\tau)=\intop_{z(0)}^{z(\tau)}
\frac{\rmd z}{v(z;E)} \intop_{z_1}^{z_2}\rmd s\,
\delta\big(s-z\big) = \intop_{\max\{z_1 ,z(0)\}}^{\min\{z_2
,z(\tau)\}} \frac{\rmd z}{v(z;E)}
 \label{soj}
  \ee
provided the interval $[z_1 ,z_2 ]$ overlaps with the interval
$[z(0),z(\tau)]$\,. If the particle motion is infinite one can put
$\tau\to\infty$\,. For finite motion the integral would diverge in
this limit and $\tau$ must be kept finite. It is convenient to
restrict $\tau$ by the half of period $\tau\le P/2$, which depends
on the energy of the system and is given by \cite{LLmekh}
\begin{eqnarray}P(E)= 2\, \intop_{z_1(E)}^{z_2(E)}\frac{\rmd z}{v(z;E)}\,,
 \label{class-P}
 \ee
where now $z_{1,2}(E)$ are the turning points, given by equation
$U(z_{1,2})=E$\,. For $\tau> P/2$ the sojourn time contains a
trivial part, which is a multiple of the half-period, $t^{\rm
(cl)}_{\rm soj}(z_1 ,z_2 ,\tau)=n\, P/2 +t^{\rm (cl)}_{\rm
soj}(z_1 ,z_2 ,\tau-n\,P/2)$, where $n$ is an integer part of the
ratio $2\,\tau/P$.

Following (\ref{soj}), the classical sojourn time $t^{\rm
(cl)}_{\rm soj}(z_1,z,\tau(z_1,z))$ can be rewritten through the
derivative of the shortened action
 \be
&&t^{\rm (cl)}_{\rm soj}(z_1,z,\tau(z_1,z))=\frac{\prt S_{\rm
sh}(z_1, z, E;U)}{\prt E},\\
  &&S_{\rm sh}(z_1, z,
E;U)=\intop_{z_1}^{z} p\, \rmd z =\intop_{z_1}^{z} \sqrt{2\, m\,
\big(E-U(z)\big)}\,\rmd z\,.\nonumber
 \ee
Taking $z=z_2$ we get
 \be
t^{\rm (cl)}_{\rm soj}(z_1,z_2,P/2)=P/2\,,
 \ee
provided  $z_{1,2}$ are the turning points.

For an infinite motion with $E > \max U(z)$, following (\ref{soj})
we can define a classical sojourn time delay/advance for the
particle traversing the region of the potential compared to a free
motion as
\begin{eqnarray}\delta t^{\rm cl}_{\rm soj}
 = t^{\rm (cl)}_{\rm soj}(-\infty,\infty,\infty;U)- t^{\rm (cl)}_{\rm soj}(-\infty,\infty,\infty;U=0)
 =\sqrt{\frac{m}{2}} \intop_{-\infty}^{+\infty} \Big(\frac{1}{\sqrt{E-U(z)}}-\frac{1}{\sqrt{E}}\Big)\rmd z \,.
 \label{class-delay-1D}
  \ee
Calculating $t^{\rm (cl)}_{\rm soj}(-\infty,\infty,\infty)$ we
extended the lower limit in the time integration in
(\ref{sojt-class}) to $-\infty$\,. The classical sojourn time
delay/advance (\ref{class-delay-1D}) for infinite motion can be
then rewritten as
\be
 \delta t^{\rm (cl)}_{\rm soj}=\frac{\prt \big(S_{\rm sh}(E;U)-S_{\rm sh}(E;0)\big)}{\prt E} \,,\,
 \label{trQMC}\ee
 where
 $S_{\rm sh}(E;U)=\intop_{-\infty}^{+\infty} p\, \rmd z$.

The definition~(\ref{trQMC}) of the time delay is similar to the
definition of the group time delay $\delta t_{\rm gr}$ appearing
in consideration of waves in classical and quantum mechanics. In
the later case the $\Psi$-function of quasi-classical stationary
motion is expressed as $\Psi\propto e^{iS_{\rm sh}(z_1, z,
E;U)/\hbar}$. With the help of a classical analog of the phase
shift,
 \be
\hbar\delta^{\rm (cl)}(z_1,z,E;U)\equiv S_{\rm sh}(z_1,z,E;U)\,,
\label{cl1dtrav1}
  \ee
  we now introduce the group time
 \be\label{grTime}
 t^{\rm (cl, 1D)}_{\rm gr}(z_1,z,E;U)\equiv \hbar \frac{\partial\delta^{\rm
 (cl)}(z_1,z,E;U)}{\partial E}.
\ee
 Thus,
 \begin{eqnarray}t^{\rm (cl, 1D)}_{\rm gr}(z_1,z_2,E;U)=\hbar \frac{\partial\delta^{\rm (cl)}(z_1,z_2
,E;U)}{\partial E}=P/2\,,
 \ee
 provided $z_{1,2}$ are turning points.

For one-dimensional infinite motion, introducing  $\delta^{\rm
(cl)}=S_{\rm sh}(-\infty,\infty, E;U)/\hbar \equiv S_{\rm
sh}(E;U)/\hbar$ and $\delta^{\rm (cl)}_{\rm free}= S_{\rm
sh}(E;0)/\hbar$, we can write the group time delay respectively
the free motion
 as
\be
 \delta t^{\rm (cl, 1D)}_{\rm gr}= \hbar \frac{\partial (\delta^{\rm (cl)}- \delta^{\rm (cl)}_{\rm free})}{\partial E}
  = \delta t^{\rm (cl, 1D)}_{\rm soj}\, .
\label{cl1dtrav}
  \ee
Moreover, one may introduce another temporal scale --- \emph{a
phase time delay}
 \be
 \delta t^{\rm (cl)}_{\rm ph}=\hbar\,\delta^{\rm cl}/E\,.
 \label{clph}
 \ee
Also, from Eq.~(\ref{class-delay-1D}) we immediately conclude that \emph{in 1D the time shift is negative
(advance), $\delta t^{\rm cl}_{\rm soj}<0$, for an attractive potential $U<0$ and it is positive (delay) for a
repulsive potential $U>0$.}

 Extensions of the definitions of the full classical sojourn time
and classical sojourn time delay/advance concepts to the
three-dimensional (3D) motion are straightforward. In analogy to
Eq.~(\ref{sojt-class}) the time a particle spends within  a 3D
volume  $\Omega$ during the time $\tau$ can be defined as
\be
 t^{\rm (cl)}_{\rm soj}( \Omega ,\tau) = \intop_0^\tau \rmd t \intop_{\vec{r}\in \Omega}\rmd
\vec{r}\, \delta\big(\vec{r}-\vec{r}(t)\big)\,.
 \label{sojt-class3}
 \ee

Consider now {\em{ a radial motion of a particle  in a central
stationary field}} decreasing sufficiently rapidly with the
distance from the center. Using the symmetry of the motion towards
the center and away from it, we can choose the moment $t=0$, as
corresponding to the position of the closest approach to the
center. Then for times $t\to\pm \infty$ the particle moves freely
and its speed is $v_{\infty}$. We can define a classical time
delay by which the free particle motion differs from the motion in
the potential as
\begin{eqnarray}\delta t^{\rm (cl)}_{\rm W }=2\lim_{t\to \infty}\big(t(r,U)
-r(t,U=0)/v_{\infty}\big)\,,
 \label{class-Wigner-t}
  \ee
where $r(t,U=0)$ is the particle's radial coordinate for free
motion. Factor $2$ counts forward and backward motions in radial
direction. We will call this time delay, \emph{the Wigner time
delay}. One can see that this time is equivalent to a classical
sojourn time delay, $\delta t^{\rm (cl)}_{\rm W }=\delta t^{\rm
(cl)}_{\rm soj}$, defined similarly to Eq.~(\ref{class-delay-1D}).
Using the virial theorem for classical scattering on a central
potential $U(r)$~\cite{Demkov}, one may show that (see
Appendix~\ref{app:virial})
\be
 \delta t^{\rm (cl)}_{\rm soj}= \delta t^{\rm (cl)}_{\rm W }=
 \frac{1}{ E} \intop_{0}^{\infty} \Big(2\, U(r(t)) +r(t)\, U'(r(t))\Big)\rmd t\,,
 \label{class-delay-3D}
  \ee
where the integration goes along the particle trajectory $r(t)$.
The result holds for potentials decreasing faster than $1/r$. We
see  that \emph{ in $3D$-case there is no direct correspondence
between the signs of the potential and the time shift $\delta
t^{\rm (cl)}_{\rm soj}$.} For a power-law potential
$U=a/r^\alpha$, $\alpha
>0$, we have a delay, $\delta t^{\rm (cl)}_{\rm W}>0$, for $a(2-\alpha)>0$, and we have an advance, $\delta
t^{\rm (cl)}_{\rm W}<0$, for $a(2-\alpha)<0$. For $\alpha =2$
there is no any time shift compared to the free motion.

Now, using that in a central field~\cite{LLmekh}
 \be
t(r)=\int_{r_0}^r \frac{\rmd r}{v_r},\quad v_r =\sqrt{v_{\infty}^2 -\frac{2\,U(r)}{m}-\frac{M^2}{m^2 r^2}}\,,
 \ee
where $r_0 =r(v_r =0)$ is the turning point, \footnote{If there is
no turning point, one puts $r_0 =0$.} and $M$ is the angular
momentum, we can rewrite the limit in Eq.~(\ref{class-Wigner-t})
as
\be
 \lim_{r\to\infty}\big(t(r)-r/v_{\infty}\big)=\lim_{r\to\infty}\Big(\intop_{r_0}^r \frac{\rmd
r}{v_r}-\frac{r}{v_{\infty}}\Big)\,.
 \label{lim}
  \ee
For a central potential the shortened action is $S_{\rm
sh}(r_0,r,E,U)=\int_{r_0}^r p_r \rmd r$, $S_{\rm
sh}(E,U)=\int_{r_0}^{\infty} p_r \rmd r$, and the classical analog
of the phase shift is given by
 \be
\hbar\,\delta^{\rm cl}(v_{\infty},M) -\hbar\,\delta^{\rm cl}(v_{\infty},M,U=0)
=\lim_{r\to\infty}\left[\int_{r_0}^r p_r \rmd r
-\int_{r_0}^r p_r (U=0) \rmd r\right]\,,\quad p_r =m\,v_r.
\label{class-delta-delta}
 \ee
Then, similarly to Eq.~(\ref{trQMC}) we can define the group time
delay, as the energy derivative of the phase acquired during the
whole period of motion (forward and backward), and from comparison
with Eq.~(\ref{lim}) we have
 \be
\delta t^{\rm (cl,3D)}_{\rm gr}\equiv 2\,\hbar \frac{\partial (
\delta^{\rm (cl)}- \delta^{\rm (cl)}_{\rm free})}{\partial E}
=\delta t^{\rm (cl)}_{\rm W}\,.
 \label{clW}
  \ee
As we see, {\em compared to the one-dimensional case
(\ref{cl1dtrav}) (where integration limits in expression for
$S_{\rm sh}$ are from $-\infty$ to $\infty$), in the
three-dimensional case (\ref{clW}) for the delay in the radial
motion there appears extra factor 2.} In sect. \ref{sec:quant} we
shall see that such a delay undergo divergent waves, whereas
scattered waves are characterized by twice less delay, as it is in
one dimensional classical motion. Also, in three-dimensional case
one may introduce a phase time scale given by the same expression
(\ref{clph}), as in one-dimensional case.

Moreover, for systems  under the action of external time dependent
forces there appear extra time-scales characterizing dynamics.
Above we considered undamped mechanical motion. Below we study
damped motion. We consider several examples of such a kind, when
mechanical trajectories can be explicitly found. We introduce
typical time scales and demonstrate possibility, as time delays of
the processes, as time-advancements.

\subsubsection{Anharmonic damped 1D-oscillator under the action of an external force. General
solution}\label{AnharmonicGeneral}

  Consider a particle with a mass $m$ performing a
one-dimensional motion along $z$ axis in a slightly anharmonic
potential under the action of an external time-dependent force
$F(t)$ and some non-conservative force (friction) leading to a
dissipation. The equation of motion of the particle is
\be
\ddot{z}(t) + E_{\rm R}^2\, z(t) +\Gamma\, \dot{z}(t) +\Lambda\,
z^2(t)=\frac{1}{m}\,F(t)\,, \label{class-eom} \ee
where $E_{\rm R}$ is  the oscillator frequency  and  $\Gamma>0$ is
the energy dissipation parameter. The anharmonicity of the
oscillator is controlled by the parameter $\Lambda$. Within the
Hamilton or Lagrange formalism, Eq.~(\ref{class-eom}) can be
derived, e.g., with the help of introduction of an artificial
doubling of the number of degrees of freedom, as in
Ref.~\cite{Bateman31,Majima11}, or if one assumes that the
oscillator is coupled to the environment ("a viscous medium"), as
in Ref.~\cite{Caldeira83}. To establish a closer link to the
formalism of the quantum field theory, which we will pursue in
Sect. ~\ref{sec:qft}, we introduce  the dynamical variable
("field") $\phi (t)=m\,z(t)$ obeying the equation
\begin{eqnarray}
-\hat{S}_t\, \phi (t)=J(t), \quad -\hat{S}_t=\frac{\rmd^2}{\rmd
t^2}+E_{\rm R}^2+\Gamma \frac{\rmd }{\rmd t}\,, \quad J(t)=F(t)-
\frac{1}{m}\,\Lambda\, \phi^2(t) \,,
%
%
\label{sourcecl}
\end{eqnarray}
with the differential operator $\hat{S}_t$ and the source term
$J$, which depends non-linearly on $\phi$ and on the external
force $F(t)$.

In  absence of  anharmonicity, $\Lambda = 0$, solution of Eq.~(\ref{class-eom}) can be written as
\be
z(t;\Lambda=0)=z_0(t) - \intop_{-\infty}^{+\infty}\rmd t' G_0(t-t')w(t')\,,
\quad
w(t')=\frac{1}{m}\,F(t')\,,
\label{freesol}
\ee
where the Green's function $G_0(t-t')$ satisfies the equation
\be
\hat{S}_t G_0(t-t')=\delta(t-t')\,.
\label{class-eqG0}
\ee
The quantity $z_0(t)$ in Eq.~(\ref{freesol}) stands for the solution of the homogeneous  equation $\hat{S}_t\,
z(t)=0$ with initial conditions of the oscillator, namely, its position $z_0(0)$ and velocity $\dot{z}_0(0)$
(both are encoded in the oscillation amplitude $a_0$ and the phase $\alpha_0$):
\begin{eqnarray}z_0(t)=a_0 \exp\Big(-\half \Gamma t\Big)\,\cos\Big(\om_{\rm R}
\, t+\alpha_0\Big)\,,
 \label{class-z0}
  \ee
where $\om_{\rm R} =\sqrt{E_{\rm R}^2-\quart\Gamma^2}$. Two
time-scales characterize this solution: the time of the amplitude
quenching - \emph{the decay time}
\be
 t_{\rm dec}^{\rm (cl)}= 2/\Gamma\,
  \label{tdec}
  \ee
and  the period of oscillations $P=\frac{2\, \pi}{\om_{\rm R}}$,
see Eq.~(\ref{class-P}). The value $t_{\rm dec}^{\rm (cl)}$
describes  decay  of the field ($\phi=mz$ variable). The $\phi^2$
quantity is damping on two times shorter scale. Note that in
quantum mechanics we ordinary consider  damping of the density
variable, $|\Psi|^2$.
 The definition of the sojourn time
(\ref{soj}) provides a relation for the period $t^{\rm (cl)}_{\rm
soj}(z_0(P/2),z_0(0),\tau=P/2)=P/2$. The phase time shift $\delta
t_{\rm ph}=\alpha_0/\om_{\rm R}$  can be eliminated by the choice
of the initial time moment.

In the Fourier representation Eq.~(\ref{freesol}) acquires simple form
\be
z(\om;\Lambda=0)=z_0(\om)-G_0 (\om)w(\om)\,,
 \ee
where $w(\om)$ is the Fourier transform of the external
acceleration $w(t)$,
\be
w(\om)= \intop_{-\infty}^{+\infty} \rmd t e^{+i\,\om\, t}\,
\frac{F(t)}{m}\,.
\ee
The Fourier transform of Eq.~(\ref{class-eqG0}) yields the Green's function
\begin{eqnarray}G_0(\om)= \intop_{-\infty}^{+\infty}  e^{i\om t} G_0(t)  \rmd
t = \frac{1}{\om^2\,-E_{\rm R}^2+i\,\Gamma\, \om}\,.
 \label{GF0-om} \ee
This Green's function has the retarded property having poles in
the lower complex semi-plane at $\om=\pm\om_{\rm R}
-{\textstyle\frac{i}{2}}\Gamma$. As a function of time, it equals
to
\be
G_0(t)=\frac{e^{-\half\Gamma t}}{\om_{\rm R}}\sin\Big(\om_{\rm
R}\,t+\pi\Big)\theta(t)\,, \quad
\theta(t)=\left\{\begin{array}{ccc} 0 &,& t<0\\ 1 &,& t\ge
0\end{array}\right. \,. \label{GF0-class} \ee
For $\Gamma < 2\,E_{\rm R}$ the particle oscillates in  response
to the external force while for $\Gamma \ge 2E_{\rm R}$ the
oscillations become over-damped. In further to be specific we
always assume that $\Gamma < 2E_{\rm R}$.

Note that the Green's function $G_0 (\om)$ satisfies exact
sum-rule
 \be\label{sumrulemech}
 \int_{-\infty}^{\infty}A\,2\om\,
\frac{d\om}{2\pi}=1\,,\quad A=-2\Im G_0\,.
 \ee
 This sum-rule is actually a
general property of the retarded Green's function for the
stationary system of relativistic bosons, see \cite{Martin68} and
our further considerations in Sect. \ref{sec:kin}.

 The
solution (\ref{class-z0}) of the homogeneous equation can be also
represented through the Green's function convoluted with the
source term $w_0(t)$ expressed through the $\delta$-function and
its derivative
\be
z_0(t)&=&-\intop_0^{t}\rmd t' G_0(t-t')\, w_0(t')\,,
 \quad
w_0(t)=a_0 \,E_R\,\sin(\beta-\alpha_0)\,\delta(t-0)- a_0\, \cos\alpha_0\, \delta'(t-0)\,,
 \label{Class-j0}\\
&& \beta ={\rm arctan}\left(\frac{\Gamma}{2\, \om_{\rm
R}}\right)\,.
   \label{Class-beta}
\ee
In Fourier representation we have $z_0(\om)=-G_0(\om)\, w_0(\om)$,
where $w_0(\om)=a_0\,\big(E_{\rm R}\,\sin(\beta-\alpha_0)+i\,\om\,
\cos\alpha_0\big)$\,.

Now we are at the position to include effects of anharmonicity, $\Lambda\neq 0$. In the leading order with
respect to a small parameter $\Lambda$ the Fourier transform of the solution $z(\om)$ of the equation of
motion acquires the form
\begin{eqnarray}z(\om, \Lambda)&=& -G_0(\om)\, \widetilde{w}(\om)
+ \Lambda G_0(\om)\intop_{-\infty}^{+\infty}\frac{\rmd \om'}{2\pi} \frac{\rmd \om''}{2\pi} (2\pi)\,
\delta(\om-\om'-\om'') \big[ G_0(\om')\,\widetilde{w}(\om')\big]\,\big[G_0(\om'')\,\widetilde{w}(\om'') \big]\,,
 \label{class-zom} \ee
where $\widetilde{w}(\om)=w_0(\om)+w(\om)$.  Eq. (\ref{class-zom}) has a straightforward diagrammatic interpretation
\be
z(\om)=\parbox{8cm}{\includegraphics[width=8cm]{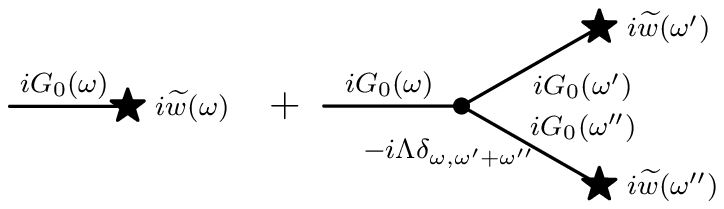}}\,,
\ee
where the thin line stands for the free Green's function
$iG_0(\om)$, the cross depicts the source $i\,\widetilde{w}(\om)$,
and the dot represents the coupling constant $-i\Lambda$. The
integration is to be performed over the source frequencies with
the $\delta$-function responsible for the proper frequency
addition. The diagrammatic representation can, of course, be
extended further to higher orders of $\Lambda$. The full solution
$z(\om)$ is presented  by the thick line with the cross
\be
z(\om)=\parbox{1.5cm}{\includegraphics[width=1.5cm]{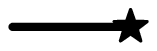}}\,,
\ee
 where the thick line stands for the full Green's function $iG(\om)$ satisfying
the Dyson equation shown in Fig.~\ref{fig:class-dyson}.

\begin{figure}
\centerline{
\parbox{7cm}{\includegraphics[width=7cm]{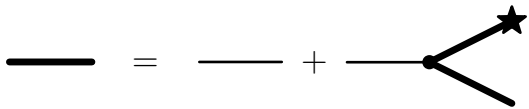}}}
\caption{ Dyson equation for the full Green's function
of the anharmonic  oscillator described by the equation of motion
(\ref{class-eom}). } \label{fig:class-dyson}
\end{figure}

Let us consider another aspect of the problem. For simplicity
consider a linear oscillator ($\Lambda =0$). Assume that in vacuum
oscillations are determined by equation
 \be
\ddot{z}(t) + E_{\rm 0}^2\, z(t) =0\,, \label{class-eom1}
 \ee
 The Fourier transform of the retarded Green's function describing these oscillations is as
 follows
 \be
G_0^0 (\om)=\frac{1}{\om^2 -\om_0^2+i0\om}\,.
 \ee
Being placed in an absorbing medium the oscillator changes its
frequency and acquires the width, which can be absorbed in the
quantity $\Re \Sigma =E_{\rm R}^2 -E_0^2$, $\Im \Sigma
=-\Gamma\om$ heaving a meaning of a retarded self-energy. Then we
rewrite (\ref{GF0-om}) as
 \be
G_0 (\om)=\frac{1}{\om^2
-\om_0^2-\Sigma}=\frac{1}{(G_0^0)^{-1}-\Sigma}\,,
 \ee
and we arrive at equation
 \be\label{Dysonmech}
G_0 =G_0^0 +G_0^0 \Sigma G_0
 \ee
 known in quantum field theory,  as  the
Dyson equation for the retarded Green's functions.

\subsubsection{Anharmonic damped oscillator under the action of an external force. Specific solutions}

Now we illustrate the above general formula at hand of  examples. To be specific we assume that the oscillator
was at rest initially, and we start with the case $\Lambda =0$.


\emph{Example 1.} Consider a response of the system to a sudden change of an external constant force
\be
F(t)\equiv F_1(t)=F_0\,\theta(-t)\,.
\label{Class-const}
\ee
The solution of Eq.~(\ref{class-eom}) for $\Lambda=0$ is
\be
z(t)\equiv z_1(t) &=& -\intop_{-\infty}^{+\infty} \frac{\rmd\om}{2\pi\, i}
e^{-i\om t} G_0(\om) \frac{F_0/m}{\om+i\epsilon}
=
\frac{F_0/m}{E_{\rm R}\om_{\rm R}} e^{-\half\Gamma t}
\cos\big(\om_{\rm R} t-\beta \big)\theta(t) + \frac{F_0}{mE_{\rm
R}^2} \theta (-t)\,, \label{Class-sol-const} \ee
here $\beta$ is defined as in Eq.~(\ref{Class-beta}). The solution is purely causal, meaning that there are no
oscillations for $t<0$ and that they start exactly at the moment when the force ceases. This naturally follows
from the retarded properties of the Green's function (\ref{GF0-class}), which has the $\theta$-function
cutting off any response for negative times. The latter occurs because  both poles of the Green's function are
located in the lower complex semi-plane and the parameter $\Gamma$ is positive corresponding to the
dissipation of the energy in the system.

Solution (\ref{Class-sol-const}) is characterized by three time
scales. Two time scales,  the period of oscillations $P=\frac{2\,
\pi}{\om_{\rm R}}$, cf.
~(\ref{class-P}), and the time of the  amplitude quenching, i.e.
the decay time $t_{\rm dec}^{\rm (cl)}=2/\Gamma$, cf.
~(\ref{tdec}),
appear already in the free solution (\ref{class-z0}).
Another time scale appears as the phase time delay in the response
of the system on the perturbation occurred at the time moment
$t=0$ (cf. Eq.~(\ref{clph})),
\be
\delta t_{\rm ph}^{\rm (cl)} =\beta / \om_{\rm R}\,>0. \ee
The solution (\ref{Class-sol-const}) is depicted on the left panel
of Fig.~\ref{fig:class-1} for three values of $\Gamma$. Arrows
demonstrate that for $\Gamma \neq 0$ the response of the
oscillator on the action of the external perturbation is purely
causal. {\em{ The larger $\Gamma$ is the smaller is $t_{\rm
dec}^{\rm (cl)}$ and the larger is $\delta t_{\rm ph}^{\rm (cl)}$,
i.e. the larger is the time shift of the oscillations.}} For
$\Gamma \to 2E_{\rm R}$ the oscillation period $P \to 0$ and the
phase shift $\delta t_{\rm ph}^{\rm (cl)}$ becomes infinite, but
the ratio $\delta t_{\rm ph}^{\rm (cl)}/P$ remains finite, $\delta
t_{\rm ph}^{\rm (cl)}/P=\beta/2\pi\to 1/4$.

\begin{figure}
\centerline{
\includegraphics[width=5cm]{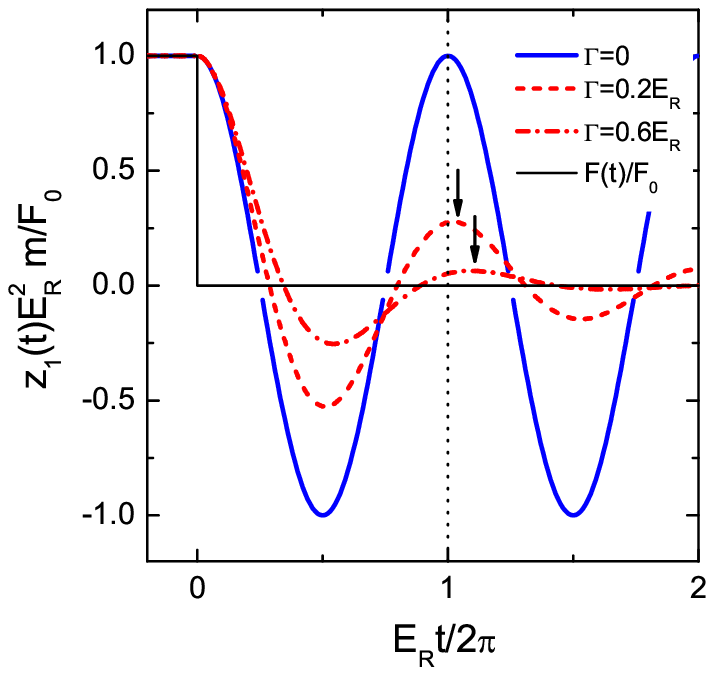}\quad\quad
\includegraphics[width=5cm]{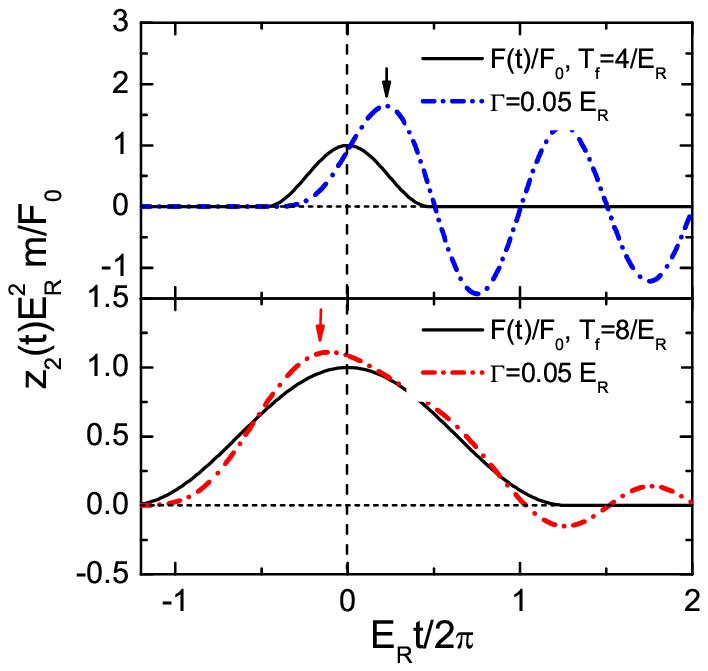}
}
\caption{ Response of the oscillator to the external force. Left
panel -- Example 1: the external force is given by
(\ref{Class-const}). Solution (\ref{Class-sol-const}) is shown for
different values of $\Gamma$. Right panel -- Example 2:  the
external force (\ref{class-force-cos2})  is shown by the solid
line. Dash-dotted lines depict solutions (\ref{class-cos2}).
Values of $\Gamma$ and $T_f$ are shown in legends. }
\label{fig:class-1}
\end{figure}


\emph{Example 2.} Interestingly, the same oscillating system,
being placed in another external field, can exhibit apparently
acausal reaction. To demonstrate this possibility consider the
driving force acting within a finite time interval $[-T_{\rm
f},+T_{\rm f}]$ and having a well defined peak occurring at $t=0$:
\be
F(t)\equiv F_2(t)=F_0\cos^2\left(\frac{\pi\, t}{2\, T_{\rm
f}}\right)\,\theta\big(T_{\rm f}-|t|\big)\,.
\label{class-force-cos2} \ee
The oscillator response to this pulse-force is given by
\be
z(t)\equiv z_2(t)&=& -\frac{F_0}{m} \intop_{-\infty}^{+\infty}
\frac{\rmd \om}{2\, \pi}\, e^{-i\, \om\, t}\,G_0(\om)
\frac{\sin(\om\, T_{\rm f})}{\om+i\epsilon}\frac{\pi^2/T_{\rm
f}^2}{(\om+i\epsilon)^2-\pi^2/T_{\rm f}^2}\,. \ee
After some manipulations the solution acquires the form
\be
z_2(t)&=&\frac{F_0}{m\, E_{\rm R}^2} \big[\zeta(t+T_{\rm
f})\,\theta(t+T_{\rm f})-\zeta(t-T_{\rm f})\, \theta(t-T_{\rm
f})\big]\,, \nonumber\\ \zeta(t) &=& \frac12\Big[ 1 - \frac{E_{\rm
R}^2}{r_+r_-}\cos\big(\frac{\pi}{T_{\rm f}}\, t -
\beta_-+\beta_+\big) + \frac{E_{\rm R}}{\om_{\rm
R}}\frac{(\pi^2/T^2_{\rm f})}{r_+\, r_-}
      e^{-\half\Gamma t} \cos\big(\om_{\rm R}\, t-\beta-\beta_--\beta_+\big)
\Big]\,, \nonumber\\ r_\pm &=&\sqrt{(\om_{\rm R}\pm \pi/T_{\rm
f})^2+\quart \Gamma^2}\,,\quad \beta_\pm=\arctan
\big(\half\Gamma/[\om_{\rm R}\pm \pi/T_{\rm f}]\big) \,,
\label{class-cos2} \ee
and the phase shift $\beta$ here is given by
Eq.~(\ref{Class-beta}). The first two terms in $\zeta(t)$ are
operative only for $-T_{\rm f}\le t\le T_{\rm f}$ and cancel out
exactly for $t>T_{\rm f}$\,. If the interval of the action of the
force is very short, i.e. $T_{\rm f}\, E_{\rm R}\ll 1$, then for
$t>T_{\rm f}$ the oscillator moves like after a single momentary
kick similarly to that in Example~1, and up to the terms $\sim
O(E_{\rm R}^2\,T_{\rm f}^2/\pi^2)$ the solution (\ref{class-cos2})
yields $z_2(t)\approx z_1(t+T_{\rm f})$\,. In the opposite case,
i.e. for $T_{\rm f}\, E_{\rm R}\gg 1$ and $t\in [-T_{\rm f},T_{\rm
f}]$, the solution $z_2(t)$ oscillates around the profile of the
driving force (\ref{class-force-cos2}) with a small amplitude
$\sim (F_0 /mE_{\rm R}^2) O(\pi^2/E_{\rm R}^2)T_{\rm f}^2$,
\begin{eqnarray}
(m\, E_{\rm R}^2/F_0)z_2(t) &=& \frac{1}{F_0} F_2(t-\Gamma/E_{\rm
R}^2)+ \frac{\pi^2}{2\,T_{\rm f}^2\, E_{\rm R}^2}
\Big\{\Big(1-\half\frac{\Gamma^2}{E_{\rm
R}^2}\Big)\,\cos\Big(\frac{\pi}{T_{\rm
f}}\Big[t-\frac{\Gamma}{E_{\rm R}^2}\Big]\Big) \nonumber\\ &+&
e^{-\half\Gamma(t+T_{\rm f})}\frac{E_{\rm R}}{\om_{\rm R}\, }\,
\cos\big(\om_{\rm R}\, (t+T_{\rm f})-3\, \beta\big) \Big\}\,. \ee

In the given example besides $P$ and $t_{\rm dec}^{\rm (cl)}$ the
system is characterized by  the initial pulse-time
\be
 t_{\rm pulse}=2T_{\rm f}
\ee
and by two phase time scales
\be
 \delta t_{\rm ph}^{(1)}=T_{\rm f}(\beta_{-} - \beta_{+})/\pi
 \quad \mbox{and}\quad
 \delta t_{\rm ph}^{(2)}=(\beta +\beta_{-} +\beta_{+})/\om_{\rm R}\,.
 \ee

The solution (\ref{class-cos2}) is shown in
Fig.~\ref{fig:class-1}, right panel. As we see from the lower
panel, for some values of $T_{\rm f}$ and $\Gamma$ the maximum of
the oscillator response may occur \emph{before} the maximum of the
driving force. Therefore, {\em{ if for the identification of a
signal we would use a detector with the threshold close to the
pulse peak, such a detector would register a peak of the response
of the system before the input's peak.}} In Ref.~\cite{Garrison98} a
similar mathematical model was used to simulate and analyze "a
causal loop paradox", when a signal from the ``future'' switches
off the input signal. The system with such a bizarre property has
been realized experimentally~\cite{Mitchell97}.


\emph{Example 3.} The temporal response of the system depends on characteristic frequencies of the
driving force variation. For a monochromatic driving force
\be
F(t)\equiv F_3(t)=F_0\,\cos(E_p\, t)
\label{class-force-mono}
\ee
the solution of the equation of motion for $t>0$  is
\be
z(t)=z_3(t)=\frac{F_0}{m}\, |G_0(E_p)|\,\cos(E_p\,t - \delta(E_p))
= \frac{(F_0/m)\,\cos(E_p\,t - \delta(E_p))}{\sqrt{(E_{\rm R}^2-
E_p^2)^2+\Gamma^2\,E_p^2}}\,, \label{class-mono} \ee
where the phase shift of the oscillations compared to the
oscillations of the driving force, $\delta(E_p)$, is determined by
the argument of the Green's function
\be
\delta(E_p)=\pi+\arg G_0(E_p)= \frac{i}{2}\left( \log\big[(E_{\rm
R}^2-E_p^2)/(E_p\, \Gamma) - i\big] - \log\big[(E_{\rm
R}^2-E_p^2)/(E_p\, \Gamma) + i\big] \right)\,. \label{class-phase}
\ee
The phase shift $\delta$ is determined such that
$\delta(E_p=0)=0$. In Eq.~(\ref{class-phase}) the logarithm is
continued to the complex plane as $\log(\pm i)=\pm \pi$ so that
the function $\delta(E_p)$ is continuous at $E_p=E_{\rm R}$, see
Fig.~\ref{fig:class}a, and in other points
\be\label{tange} \tan\delta(E_p)=-E_p\Gamma/(E_p^2-E_{\rm R}^2)\,.
\ee
The amplitude of the solution (\ref{class-mono}) has a resonance
shape peaking at $E_p=E_{\rm R}$ with a width determined by the
parameter $\Gamma$\,.  In contrast to Examples 1 and 2 solution
(\ref{class-mono}) does not contain the time-scale $t_{\rm
dec}^{\rm (cl)}$, since the external force does not cease with
time and continuously pumps-in the energy in the system. So, two
time scales, the period $P=2\pi/E_p$, and the phase time
\begin{eqnarray}
\label{class-tphase}
\delta t_{\rm ph}^{(1)}=\delta (E_p)/E_p
\end{eqnarray}
fully control the dynamics.
Note that in difference with (\ref{clph}), here $E_p$ is the
frequency rather than the particle energy.

We have seen in Example~2 that for some choices of the external
force restricted in time the oscillating system can provide an
apparently advanced response. The anharmonicity can produce a
similar effect. For the case of small anharmonicity, $\Lambda\neq
0$, the solution (\ref{class-mono}) acquires a new term (an
overtone)
\be
z_{3\Lambda}(t)=z_3(t) -\frac{(F_0/m)^2\, \Lambda/(2E_{\rm
R}^2)}{\big[(E_{\rm R}^2- E_p^2)^2+\Gamma^2\,E_p^2\big]} \Big[1+
E_{\rm R}^2\,\frac{\cos\big(2\,[E_p\,t-\delta(E_p)] -
\delta(2E_p)\big)}{\sqrt{(E_{\rm R}^2-4 E_p^2)^2 + 4\Gamma^2
E_p^2}} \Big]\,, \label{class-mon-L} \ee
which oscillates on the double frequency $2E_p$ and the phase is
shifted with respect to the solution (\ref{class-mono}) by
$\delta(2E_p)$\,. The Fourier transform of this solution is given
by Eq.~(\ref{class-zom}) provided $w_0$ is put  zero.
Respectively, there appears an additional phase time scale
\be\delta t_{\rm ph}^{(2)}=(\delta (E_p)+\frac{1}{2}\delta
(2E_p))/E_p\ee characterizing dynamics of the overtone.

In Fig.~\ref{fig:class}b we show the solution (\ref{class-mon-L})
for several frequencies $E_p$. If we watch for maxima in the
system response $z(t)$ (shown by arrows) and compare how their
occurrence is shifted in time with respect to maxima of the
driving force, we observe that for most values of $E_p$ the
overtone in (\ref{class-mon-L}) induces a small variation of the
phase shift with time. However {\em{ for $E_p\sim \half E_{\rm R}$
the overtone can produce an additional maximum in $z(t)$, which
would appear as occurring before the actual action of the force.}}
So the system would seem to ``react'' in advance.

\begin{figure}
\centerline{
\parbox{5cm}{\includegraphics[width=5cm]{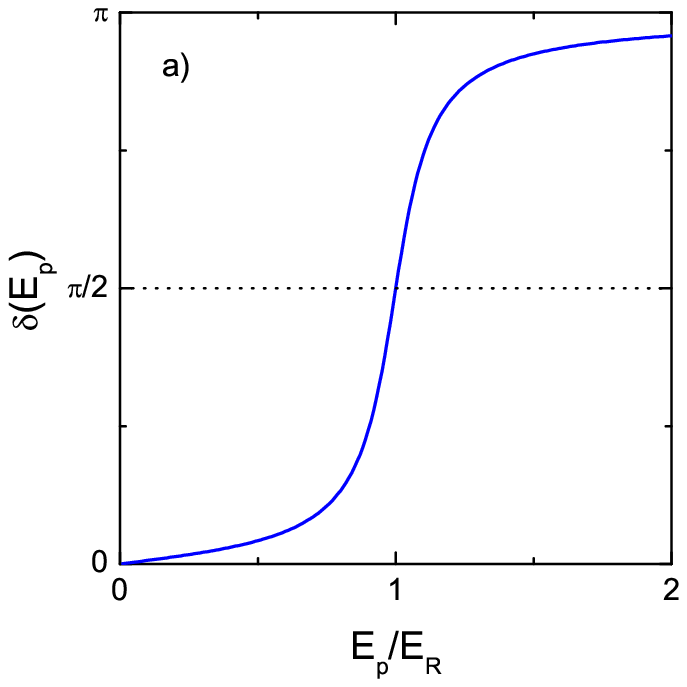}}
\parbox{5cm}{\includegraphics[width=5cm]{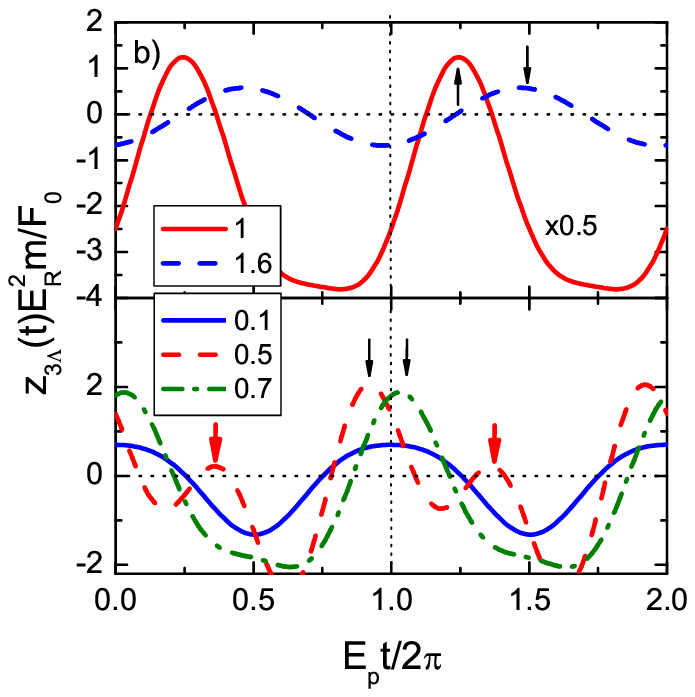}}
} \caption{ Panel a): Phase shift $\delta(E_p)$
given by Eq.~(\ref{class-phase}). Panel b): Response of the damped
anharmonic oscillator to a harmonic external force
(\ref{class-force-mono}) for different values of the force
frequency $E_p$ shown by line labels in units of $E_{\rm R}$ for
$\Gamma=0.2\, E_{\rm R}$ and $\Lambda=0.3\, E_{\rm R}^4 m/F_0$\,.
Arrows show response maxima. Vertical dotted line shows maximum of
the driving force.)} \label{fig:class}
\end{figure}


{\em Example 4.} In realistic cases the driving force can rarely
be purely monochromatic, but is usually a superposition of  modes
grouped around a frequency $E_p$:
\be
F(t)\equiv F_4(t)=F_0\intop_{-\infty}^{+\infty} \rmd E \,
g(E-E_p;\gamma)\,\cos(E\, t)\,, \label{ft-envelop} \ee
where an envelope function $g(\epsilon;\gamma)$, $\epsilon
=E-E_p$, is a symmetrical function of frequency deviation picked
around $\epsilon=0$ with a width $\gamma$ and normalized as
$\intop_{-\infty}^{+\infty} \rmd \epsilon \,
g(\epsilon;\gamma)=1$\,. The integral (\ref{ft-envelop}) can be
rewritten as
\be
F_4(t) &=&F_0\,\cos(E_p t) \intop_{-\infty}^{+\infty} \rmd
\epsilon\, g(\epsilon;\gamma)\, \cos(\epsilon\, t) = \mathcal{A}_F
(\gamma t) \cos(E_p t)\,, \label{decomp-envel} \ee
that allows us to identify $E_p$ as the carrier frequency and
$\mathcal{A}_F(\gamma t)$, as the amplitude modulation depending
on dimension-less variable $\gamma t$.

For $\Lambda=0$, the particle motion is described by the function
\be
z_4(t) &=& -\intop_{-\infty}^{+\infty}\frac{\rmd \om}{2\, \pi}
e^{-i\, \om\,t} G_0(\om)\, \frac{1}{m}\,F(\om) =-\frac{F_0}{m}
\intop_{-\infty}^{+\infty}\frac{\rmd \om}{2\, \pi} e^{-i\, \om\,t}
G_0(\om)\, \pi\big[g(\om+E_p;\gamma)+g(\om-E_p;\gamma)\big]
\nonumber\\ &=&-\frac{F_0}{m} \Re\intop_{-\infty}^{+\infty}\rmd
\epsilon e^{-i\, (E_p+\epsilon)\,t} G_0(E_p+\epsilon)\,
g(\epsilon;\gamma) \, . \label{class-z4} \ee
The last integral can be formally written as
\be
m\,z_4(t)&=&  |G_0(E_p)| \Re e^{-i(E_p\, t-\delta(E_p))} e^{-\half
\prt^2_E \log G_0(E_p)\, \prt_t^2 + O(\prt_t^3)}
\,\mathcal{A}_F\big(\gamma(t+i\, \prt_E \log G_0(E_p))\big).
\label{class-z4-exact} \ee
Here $ O(\prt^3_t)$ represents time derivatives of the third order
and higher. We used the relation $\log G_0(E)=\log|G_0(E)|+i\,
\delta(E)-i\,\pi$, where $\delta (E)$ is defined as in
Eq.~(\ref{class-phase}), but now as function of $E$ rather than
$E_p$. The first-order derivatives generate the shift of the
argument of the amplitude modulation via the relation
$\exp(a\,\prt_t)\, \mathcal{A}_F(t)= \mathcal{A}_F(t+a)$\,. Note
that the time shift of $\mathcal{A}_F(t)$ involves formally the
"imaginary time". As we will see later in Sect.~\ref{sec:quant},
the same concept appears also in  quantum mechanics.

To proceed further with Eq.~(\ref{class-z4-exact}) one may assume
that the function $\mathcal{A}_F(t)$ varies weakly with time so
that the second and higher time derivatives can be neglected. In
terms of the envelop function, this means that $g(\epsilon)$ is a
very sharp function falling rapidly off for $\epsilon\gsim\gamma$
while $\gamma\ll\Gamma$. A typical time, on which the function
$\mathcal{A}_F(t)$ fades away, can be estimated as
\begin{eqnarray}
 \label{tgammacl}
t^{\gamma,(\rm cl)}_{\rm dec}=1/\gamma,
\end{eqnarray}
If, additionally, the oscillator system has a high quality factor,
i.e., $\Gamma\ll E_{\rm R}$ and  $|\partial_E \log |G_0(E_p)||\ll \delta'(E_p)$, that is correct for $E_p$ very
near $E_{\rm R}$, we arrive at the expression
\be
m\,z_4(t)=\mathcal{A}_F\left(t-\delta'(E_p))\right)\,|G_0(E_p)|\,\cos(E_p\,
t-\delta(E_p))\,.
\label{class-z4-shift}
\ee
We see that in this approximation there are five time scales
determining the response of the system.  The oscillations are
characterized by the period $P= 2\pi/E_p$ and damping time $t_{\rm
dec}^{\rm (cl)}=2/\Gamma$. Moreover, the envelope function is
damping on the time scale $t^{\gamma,(\rm cl)}_{\rm dec}$.
Additionally, there are two delay time scales: \emph{ Oscillations
of the carrier wave are delayed by the phase time}
$\delta t_{\rm ph}^{\rm (cl)}(E_p)={\delta(E_p)}/{E_p}\,,$ see
(\ref{class-tphase}),
\emph{whereas the amplitude modulation $\mathcal{A}_F$ is delayed by the group time}
\be
t_{\rm gr}^{\rm (cl)}(E_p)=\frac{\partial \delta(E_p)}{\partial
E_p} =-\frac{E_R^2+E_p^2}{E_p}\, \Im G_0(E_p)\,.
\label{class-tgroup}
\ee
This time shift appears because the system
responses slightly differently to various frequency modes
contributing to the force envelop (\ref{ft-envelop}).

\begin{figure}
\centerline{
\parbox{5cm}{\includegraphics[width=5cm]{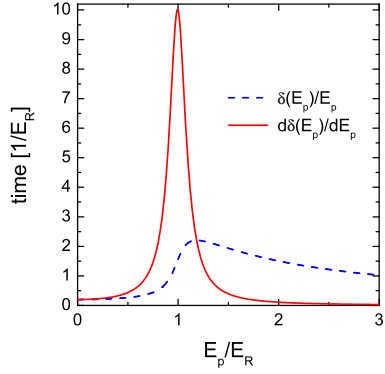}}
} \caption{The phase and group time delays given by
Eqs.~(\ref{class-tphase}) and (\ref{class-tgroup}), respectively,
calculated for $\Gamma=0.2E_{\rm R}$.  The horizontal line shows
the decay width $t_{\rm dec}^{(cl)}$.} \label{fig:class-group}
\end{figure}

The group and phase times are shown in Fig.~\ref{fig:class-group}.
The group time is much more rapidly varying function of the
external frequency $E_p$ and is strongly peaked at $E_p\sim E_{\rm
R}$\,. Close to the resonance the group time  can be written as
\be
\label{A1}
t_{\rm gr}^{(1)}(E_p)\approx \frac{\Gamma/2}{(E_p
-E_{\rm R})^2+\quart\Gamma^2}=\half A_1(E_p)>0\,.
\ee
For $\Lambda\neq 0$ there also appear another resonances in the
system response, see Eq.~(\ref{class-mon-L}). In the linear in
$\Lambda$ approximation the resonance with $E_p\simeq \half E_{\rm
R}$ is excited. Close to this resonance the group time  is
\be\label{A2}  t^{(2)}_{\rm gr}(E_p)\approx  \frac{\Gamma/4}{(E_p
-E_{\rm R}/2)^2+\quart(\Gamma/2)^4} =\half A_2(E_p)>0\, \ee
with a maximum at $E_p=E_{\rm R}/2$. The width of the peak is
$\Gamma/4$. Note that for  both modes the functions $A_{1,2}(E_p)$
satisfy the sum-rule
\be
\int_{-\infty}^{\infty} A_{1,2}(E) \rmd E/(2\pi)=1\,,\quad
\mbox{or}\quad \int_{-\infty}^{\infty}  t_{\rm gr}^{(1,2)}(E) \rmd
E/\pi=1\,. \label{sumrulecl} \ee
The energy-time sum-rules demonstrate relation of the group times
to the density of states, i.e. re-grouping  of the number of
degrees of freedom.

The time-difference
 \be
 \delta t_{\rm f}^\gamma =t^{(i)}_{\rm gr}-t^{\gamma,(\rm cl)}_{\rm
 dec},
\label{forwgamma}
 \ee
we call it  {\em{ forward time delay/advance}}, demonstrates are
the groups of waves delayed on the scale of degrading of the
envelop function.
As is seen from Fig. \ref{fig:class-group},
in the near resonance region $\delta t_{\rm f}^\gamma >0$, whereas
in the off-resonance region $\delta t_{\rm f}^\gamma <0$. As we
shall see in Sect. \ref{sec:quant}, an important case is when
$\gamma \sim \Gamma$.

To study  corrections to Eq.~(\ref{class-z4-shift}) due to the
second-order derivatives in Eq.~(\ref{class-z4-exact}) we turn
back to the case $\Lambda =0$ and  take the Gaussian envelope
function $g_{\rm Gauss}(\epsilon)$ and the corresponding amplitude
modulation $\mathcal{A}_{F,{\rm Gauss}}(\gamma t)$,  such that
\be
g_{\rm Gauss}(\epsilon;\gamma)=\frac{\exp(-\epsilon^2/2\gamma^2)}{\sqrt{2\, \pi\, \gamma^2}} \,, \quad
\mathcal{A}_{F,{\rm Gauss}}(t;\gamma)=F_0\exp(-\gamma^2 t^2/2)\,.
\label{class-gauss-F} \ee
Then, using the identity
\be
e^{a\, \prt^2_t}\, e^{-\gamma^2 t^2/2}
\equiv \sum_{n=0}^{\infty} \frac{a^n}{n!}\, \prt_t^{2\, n} e^{-\gamma^2 t^2/2}
= \frac{e^{-\gamma^2 t^2/2(1+2\, a\, \gamma^2)}}{\sqrt{1+2\,a\,\gamma^2 }}\,,
\ee
we obtain the response of the system to the Gaussian force in the form
\be
z_{4{\rm Gauss}}(t) &=&
\frac{F_0}{m} |G_0(E_p)|\Re \,
\frac{e^{-i\,(E_p\,t - \delta(E_p))}}{\sqrt{1 -\gamma^2\prt^2_E\log G_0(E_p) }}
\exp\left[
-\gamma^{2}\frac{\big(t+i\prt_E \log G_0(E_p)\big)^2}
{2\big(1-\gamma^{2}\prt^2_E\log G_0(E_p) \big)}
\right]\,.
\label{class-Gauss}
\ee
The derivatives of the Green's function can be conveniently expressed through the Green's function as
\begin{eqnarray}
&& \prt_E \log G_0(E) = i\,\delta'(E)+ \prt_E \log |G_0(E)|=-(2\, E + i\, \Gamma)\, G_0(E)\,,
\nonumber\\
&&\prt^2_E \log G_0(E)= i\,\delta''(E)+ \prt_E^2 \log |G_0(E)|
=2\, G_0(E)+ (4\, E_{\rm R}^2 - \Gamma^2)\, G_0^2(E)\,.
\end{eqnarray}
After some algebra we can cast this expression in the form similar
to Eq.~(\ref{class-z4-shift}) with the amplitude modulation
(\ref{class-gauss-F}):
\begin{eqnarray}
m\, z_{4{\rm Gauss}}(t)= \mathcal{A}_{F,{\rm
Gauss}}\big(t-\tilde{t}_{\rm gr}^{\rm (cl)};\tilde{\gamma}\big)\,C_\gamma\,|G_0(E_p)|\,\cos(\widetilde{E}_p(t)\,
 t-\tilde{\delta}(E_p))\,,
\label{class-Gauss-reduced}
\end{eqnarray}
where, however, we have  to redefine parameters of both the
carrier wave and the amplitude modulation function. The width of
the Gaussian packet is determined from expression
\begin{eqnarray}
\tilde{\gamma}^2=\gamma^2\frac{1-\gamma^2\, \prt_E^2\log|G_0(E_p)|}{|1-\gamma^2\prt_E^2\log
G_0(E_p)|^2}\,,
\label{class-g-Gauss-cor-gamma}
\end{eqnarray}
and  the amplitude modulation is delayed by the group time
\be
\tilde{t}_{\rm gr}^{\rm (cl)}=\delta'(E_p) +
\prt_E\log|G_0(E_p)|\, \frac{\gamma^2
\delta''(E_p)}{1-\gamma^2\prt_E^2\log|G_0(E_p)|}\,.
\label{class-g-Gauss-cor-dt} \ee
An interesting effect is that {\em the frequency of the carrier
wave is changed and even becomes time dependent,}
\be
\widetilde{E}_p(t)= E_p + \tilde{\gamma}^2\,
\Big[\prt_E\log|G_0(E_p)|+ (\half t-\delta'(E_p))\frac{\gamma^2
\delta''(E_p)}{1-\gamma^2\prt_E^2\log|G_0(E_p)|} \Big] \,,
\label{class-g-Gauss-cor-Ep} \ee
and the phase shift is given by
\begin{eqnarray}
\tilde{\delta}&=&\delta(E_p) +\half \arctan\Big(\frac{\gamma^2\,
\delta''(E_p)}{1-\gamma^2\prt^2_E\log|G_0(E_p)|}\Big) \nonumber\\
&+&\tilde{\gamma}^2\, \Big[\delta'(E_p)\, \prt_E\log|G_0(E_p)| +
\half\big(
\big[\prt_E\log|G_0(E_p)\big]^2-\big[\delta'(E_p)\big]^2\big)
\frac{\gamma^2 \delta''(E_p)}{1-\gamma^2\prt_E^2\log|G_0(E_p)|}
\Big] \,.
\label{class-g-Gauss-cor-delta}
\end{eqnarray}
The amplitude of the system response is modulated by the factor
\be
C_\gamma
=\exp\Big[\half\gamma^2\frac{\big[\prt_E\log|G_0(E_p)|\big]^2}{(1-\gamma^2\prt_E^2\log
|G_0(E_p)|)^2}\Big] \bigg/{|1-\gamma^2\prt_E^2\log G_0(E_p)|}\,.
\label{class-g-Gauss-cor-A}
\ee

\begin{figure}
\centerline{
\parbox{4.85cm}{\includegraphics[width=4.85cm]{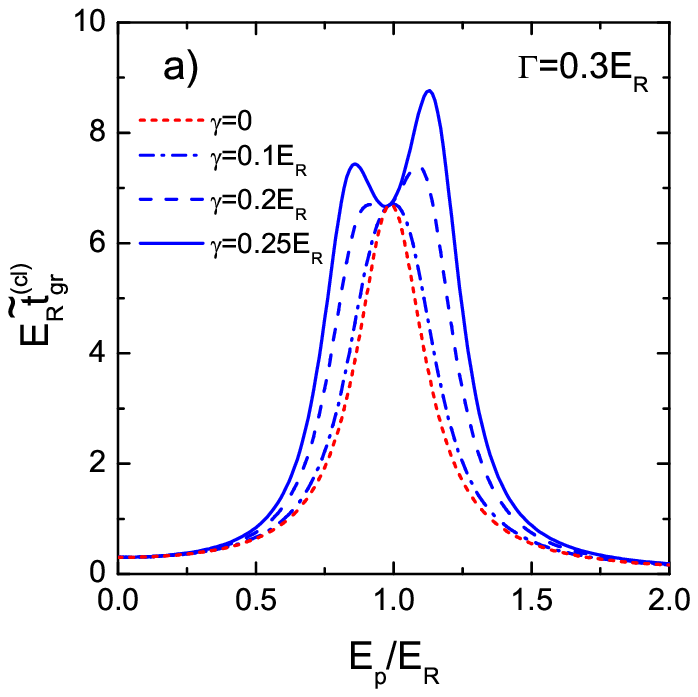}}
\parbox{5cm}{\includegraphics[width=5cm]{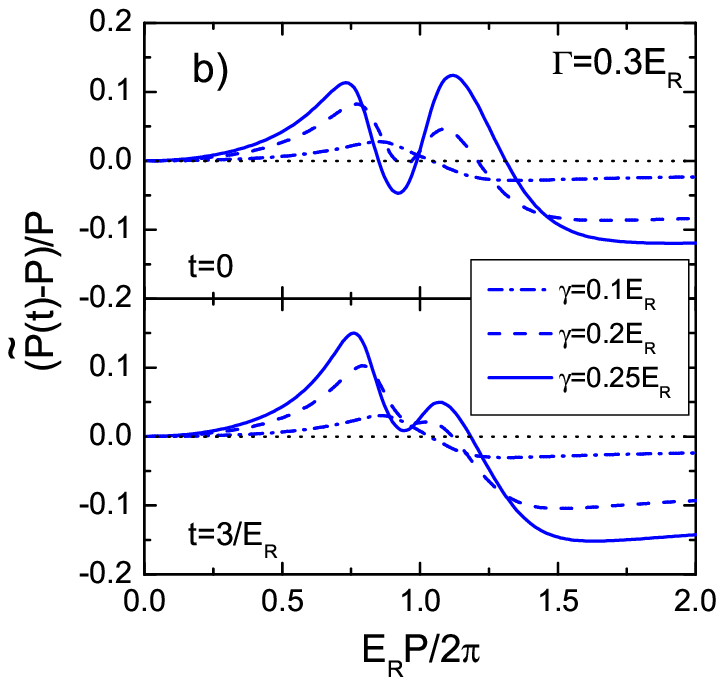}}
\parbox{5cm}{\includegraphics[width=5cm]{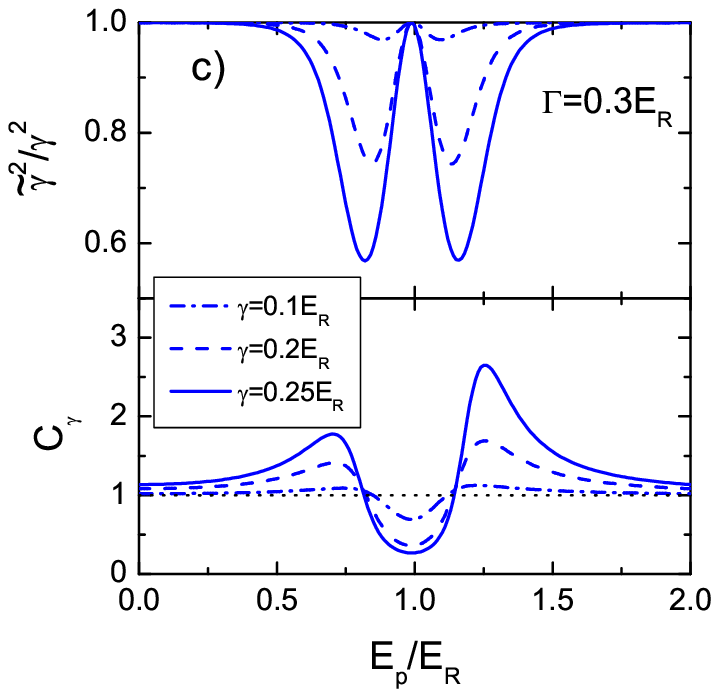}}
} \caption{Parameters of the system response (\ref{class-Gauss})
to the modulated periodic external force (\ref{decomp-envel}) with
the Gaussian envelop (\ref{class-gauss-F}) calculated including
the second-order derivatives for several values of the envelop
width $\gamma$ as functions of the force oscillation frequency
$E_p$. The damping parameter of the system is $\Gamma=0.3\, E_{\rm
R}$\,. Panel a): Time shift of the amplitude modulation
(\ref{class-g-Gauss-cor-dt}) for various values of $\gamma$. The
dotted line shows the time shift (\ref{class-tgroup}) entering in
the leading-order expression (\ref{class-z4-shift}) involving the
first-order derivatives only. Panel b): Relative deviation of the
oscillation quasi-period
$\widetilde{P}(t)=2\pi/\widetilde{E}_p(t)$, see Eq.
(\ref{class-g-Gauss-cor-Ep}), from the force oscillation period
$P=2\pi/E_p$ for two moments of time $t=0$ and $t=3/ E_{\rm R}$\,.
Panel c): Modification of the packet width
$(\tilde{\gamma}/\gamma)^2$ given by
Eq.~(\ref{class-g-Gauss-cor-gamma}) and the amplitude scaling
factor $C_{\gamma}$ given by Eq.~(\ref{class-g-Gauss-cor-A})\,.
 }
\label{fig:class-Gauss-corr}
\end{figure}
Keeping terms quadratic in $\gamma$ we find the corrected group and phase times
\be\label{class-g-Gauss-cor-dtA}
&&\tilde{t}_{\rm gr}^{\rm (cl)}\simeq  \delta'(E_p) + \gamma^2
\,\prt_E\log|G_0(E_p)|\, \delta''(E_p)+O(\gamma^4)\,,
\\
&&E_p\delta\tilde{t\,}_{\rm ph}^{\rm (cl)}\simeq
\delta(E_p)+\gamma^2\,\Big(\half  \delta''(E_p)+
\prt_E\log|G_0(E_p)|\,\big(\delta'(E_p)\, -
\delta(E_p)/E_p\big)\Big) +O(\gamma^4)\,.
\label{class-g-Gauss-cor-deltaA}
\ee

The importance of various correction terms  depends on how close
the carrier frequency $E_p$ is to the resonance frequency $E_{\rm
R}$. Assuming that the oscillator has a high quality factor
$\Gamma\ll E_{\rm R}$,  we can distinguish three different
regimes: $(i)$ very near to the resonance, $|E_p-E_{\rm
R}|\lsim\Gamma^2/E_{\rm R}$, $(ii)$ an intermediate regime,
$\Gamma^2/E_{\rm R}\ll|E_p-E_{\rm R}|\lsim\Gamma$, and $(iii)$ far
from the resonance $\Gamma\ll|E_p-E_{\rm R}|$\,. In the regime
$(i)$ corrections in (\ref{class-g-Gauss-cor-dtA}),
(\ref{class-g-Gauss-cor-deltaA}) are respectively of the order of
$O(\gamma^2/E_{\rm R}^2)$ and $O(\gamma^2/E_{\rm R}\Gamma)$. In
the regime $(ii)$ correction terms are of the order of
$O(\gamma^2/\Gamma^2)$. In the regime $(iii)$ corrections are
respectively of the order of $O(\gamma^2/E_{\rm R}^2)$ and
$O(\gamma^2\Gamma/E_{\rm R}^3)$ at most.

To illustrate the applicability range of the leading-order
expression (\ref{class-z4-shift}) and the size of the corrections
in Eq.~(\ref{class-Gauss-reduced}) we plot in
Fig.~\ref{fig:class-Gauss-corr} the
quantities~(\ref{class-g-Gauss-cor-gamma}),
(\ref{class-g-Gauss-cor-dt}), (\ref{class-g-Gauss-cor-Ep}),
(\ref{class-g-Gauss-cor-A}) versus the force oscillation frequency
$E_p$ for various values of the envelop width $\gamma$ and
$\Gamma=0.3\,E_{\rm R}$\,. We see that, as argued before, the
corrections are small for $E_p$ far from the resonance frequency
$E_{\rm R}$ and right at the resonance. {\em The corrections are
maximal for $E_p\sim E_{\rm R}\pm \Gamma$\,. Remarkably, at these
frequencies the system response could become significantly broader
(i.e. it lingers longer in time) than the driving force,
$\tilde{\gamma}<\gamma$.} Figure~\ref{fig:class-Gauss-corr} shows
also that Eq.~(\ref{class-z4-shift}) can be used only for
$\gamma/\Gamma\ll 0.3$. The expression (\ref{class-Gauss-reduced})
is applicable for $\gamma/\Gamma\lsim 0.5$ and $\Gamma/E_{\rm
R}\lsim 0.3$ at the 30\% accuracy level. For higher values of
$\gamma$ the corrections become too large and further terms in
expansion (\ref{class-z4-exact}) have to be taken into account.

In Fig.~\ref{fig:Class-Gauss-exact} we depict the response of the
system to the force (\ref{decomp-envel}) with the 'broad' Gaussian
envelop (\ref{class-gauss-F}), $\gamma=\Gamma$, as it follows from
  numerical evaluation of the integral
(\ref{class-z4}). We clearly see that {\em{ when $E_p$ approaches
the interval $E_{\rm R}\pm \Gamma$ not only the amplitude of the
system response grows, but also the response lasts much longer
than the force acts.}} Thus we demonstrated peculiarities of the
effect of a smearing of the wave packet in classical mechanics.

\begin{figure}
\includegraphics[width=\textwidth]{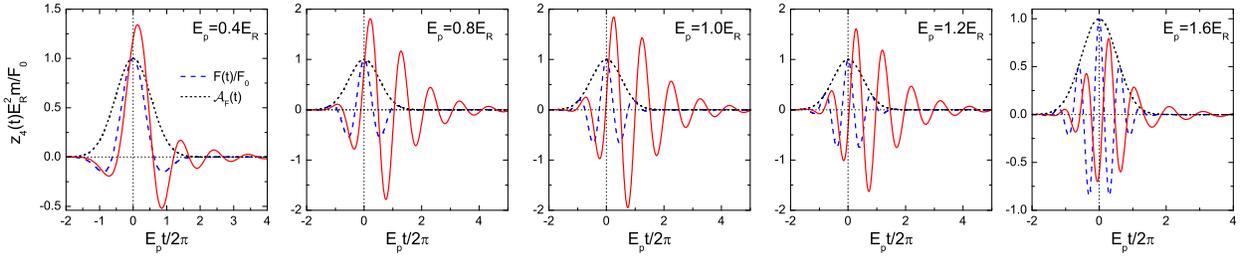}
\caption{Solid lines: response (\ref{class-z4}) of the oscillatory
system to the modulated periodic external force
(\ref{decomp-envel}) with the Gaussian envelop
(\ref{class-gauss-F}) calculated for $\gamma=\Gamma=0.3E_{\rm R}$
and various values of the force frequency $E_p$. Dashed lines show
the external force (\ref{decomp-envel}) and dotted lines depict
the envelop function (\ref{class-gauss-F}).}
\label{fig:Class-Gauss-exact}
\end{figure}

\subsubsection{Simple 3D-example. Scattering of particles on hard spheres}

Consider a simplest case when a beam of (point-like) particles
falls onto a hard sphere of a radius $R$, cf. \cite{Nusszv-book}.
The particles scatter at different angles $\theta$, $\sin
(\theta/2)=\sqrt{1-b^2/R^2}$\, depending on the impact parameter
$b$. The arrival time of the scattered particle to the detector
decreases with an increase of the size of the sphere. The time
advance for the particle scattered on the sphere surface compared
to that it would scatter on the center in the same $\theta$
direction  is
\begin{eqnarray}
\delta t_{\rm W}^{\rm (cl)} = -2\frac{\Delta l}{v} =
-\frac{2\,R}{v}\sin\frac{\theta}{2}\geq -\frac{2\,R}{v}\,,
\label{hsWigner}
\end{eqnarray}
see Fig.~\ref{fig:class-hardsph}. In the given example $\delta
t_{\rm soj}^{\rm (cl)} =\delta t_{\rm W}^{\rm (cl)}$, as they were
introduced above, see Eqs. (\ref{sojt-class3}),
(\ref{class-delay-3D}). As is seen from (\ref{class-delay-3D}),
for the repulsive potential $V=a/(r-b)^{\alpha}$, $r>r_0=r(v_r
=0)$,  $a>0$, $\alpha >0$, as in case of the scattering on the
hard sphere, there appears a time advancement,  provided $r_0$
being very close to $b$. However the value of the Wigner time
advancement is  limited.

As we shall see below,  the relevant quantity related to the
advance/delay of the scattered wave, the scattering advance/delay
time, is the half of the Wigner advance/delay time. In the given
hard sphere example thus introduced quantity,
\begin{eqnarray}
\delta t_{\rm s}^{\rm (cl)}  = \half\delta t_{\rm W}^{\rm (cl)}\,,
\label{hsScat}
\end{eqnarray}
is the difference of time, when the particle touches the sphere
surface, and the time, when the particle freely reaches the center
of the sphere. The advance $\delta t_{\rm s}^{\rm (cl)}$ is
limited by the value $-R/v$.
\begin{figure}
\centerline{\includegraphics[width=5cm]{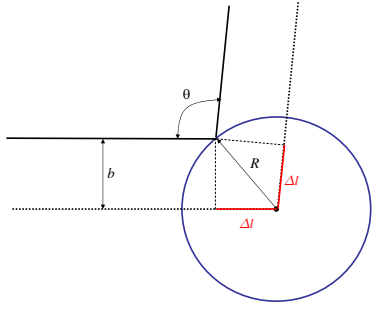}}
\caption{Particle scattering at angle $\theta$ off the hard sphere
of radius $R$} \label{fig:class-hardsph}
\end{figure}

Note that the averaged  advance time for all scattered particles
incident on the sphere at various impact parameters $0\le b\le R$
is
\begin{eqnarray}
\langle\delta t_{\rm W}^{\rm (cl)}\rangle = \intop_0^{R} \delta
t_{\rm W}^{\rm (cl)}\,\frac{2\pi\,b\rmd b}{\pi R^2} =
-\frac{4}{v\, R^2}\intop_0^{R}\sqrt{R^2-b^2}\,b\rmd b
=-\frac43\frac{R}{v}\,.
\end{eqnarray}
 From the above analysis we are able also to conclude that
  the collision term in the kinetic equation describing behavior of a non-equilibrium
gas of hard spheres should incorporate mentioned non-local time
advancement effects.

\subsection{Time shifts in classical electrodynamics}

\subsubsection{Dipole radiation of charged oscillator}\label{electro}

Let us consider the same damped oscillatory system,  as in the
previous subsections, assuming now that the particle is charged
and oscillates in the $z$ direction near the point $z=0$ under the
action of an incident electromagnetic wave propagating in the $x$
direction with the electric field polarized along the $z$ axis:
\begin{eqnarray}
\vec{\mathcal{E}}_{\rm in}(t,\vec{r}\,)
=\mathcal{E}_{0}\,\vec{e}_z\, \cos\big(\vec{p}\,\vec{r} -
E_p\,t\big)\,, \quad \vec{\mathcal{H}}_{\rm in}(t,\vec{r}\,)
=[\vec{e}_x\times\vec{\mathcal{E}}_{\rm in}]\,, \quad \vec{p}=p\,
\vec{e}_x\,.
\label{in-field}
\end{eqnarray}
Here $\vec{e}_{x(yz)}$ denotes the unit vector along $x(yz)$
direction, $E_p =c\,p$ and  $c$ is the speed of  light. We assume
that the field weakly changes over the range of particle
oscillations.
Then the force acting on the charge is $\vec{F}(t)\approx
\mathcal{E}_{0}\,\vec{e}_z \cos(E_p \,t)$ and the oscillations are
described by Eq.~(\ref{class-mono}) of the previous section. The
electric dipole moment induced in the system by the incident wave
is given by
\be
\vec{d}(t)=ez(t)=\frac{e^2\,\mathcal{E}_0}{m}
\frac{\cos\big(E_p\,t -\delta(E_p)\big)} {\sqrt{(E_{\rm R}^2-
E_p^2)^2+\Gamma_{\rm tot}^2\,E_p^2}}\,\vec{e}_z\,, \quad
\Gamma_{\rm tot}=\Gamma+\Gamma_{\rm rad}\,, \label{class-dipole}
\ee
$\Gamma_{\rm tot}$ is the total width of the oscillator. The
oscillating dipole emits electromagnetic waves. Therefore, there
is a dissipative process due to the radiation friction force,
$\Gamma_{\rm rad}= 2e^2\,E_p^2 /(3\,m\,c^3)$\, for $E_p\lsim
E_{\rm R}$, which we consider. The additional damping effects
included in $\Gamma$ are, e.g., due to atomic collisions, provided
the charged particle oscillates in medium. The formulated model is
the well-known Lorentz model for vibrations of an electron in an
atom. An ensemble of such oscillators resembles a dispersive
medium.

Far from the dipole in the so-called wave zone $|d_{\rm max}/e|\ll
\lambda\ll r$, where $\lambda=2\pi/p$ is the radiation wave-length
and $|d_{\rm max}/e|$ is the amplitude of the oscillations, the
outgoing waves of electric and magnetic fields are given
by~\cite{LLField}
\be
\vec{\mathcal{E}}_{\rm out}(t,\vec{r})
=\frac{1}{r\, c^2} [\vec{n}_r\times[\vec{n}_r\times \ddot{\vec{d}}(t-r/c)]]\,,
\quad
\vec{\mathcal{H}}_{\rm out}(t,\vec{r})=
[\vec{n}_r\times\vec{\mathcal{E}}_{\rm out}(t,\vec{r})]=\frac{1}{r\, c^2}
[\ddot{\vec{d}}(t-r/c)\times \vec{n}_r]
\label{out-field}
\ee
with $\vec{n}_r=\vec{r}/|\vec{r}\,|$\,. The time shift $t-r/c$ arises due to finiteness of the
speed of light. The scattered electric field is polarized along a meridian,
$\vec{\mathcal{E}}_{\rm out}\parallel \vec{e}_\theta$\,.

The differential cross section for the scattering process can be
defined as the ratio of the time-averaged intensity of the induced
radiation $\rmd \bar{I}$, passing through a sufficiently large
sphere of radius $R_0$, to the time-averaged energy flux of the
incident wave falling on the oscillator, see \S 78
in~\cite{LLField},
\be
d\sigma =\frac{1}{\overline{\vec{S}_{\rm in}}} \rmd \bar{I}\,,
\quad \rmd I= \vec{S}_{\rm out} \rmd \vec{s}_r\,. \quad
\vec{S}_{\rm in(out)}=\frac{c}{4\pi}\big[\vec{E}_{\rm
in(out)}\times \vec{H}_{\rm in(out)}\big]\,. \nonumber \ee
Here the line over a symbol means a time average over the
oscillation period and $\rmd \vec{s}_r$ is element of the surface
oriented in the direction $\vec{n}_r$\,. With the help of
Eqs.~(\ref{in-field}) and (\ref{out-field}) we find
\be
\vec{S}_{\rm in}=\frac{c}{4\pi}\, \mathcal{E}^2_0\, \cos^2(E_p\,
t)\, \vec{e}_x\,, \quad \rmd I= (\vec{S}_{\rm out}\vec{n}_r) R_0^2
\rmd \Omega
 = \frac{1}{4\pi\, c^3}\big[\ddot{\vec{d}}(t-R_0/c)\times
\vec{n}_r\big]^2\,\rmd\Omega \,.
\ee
Using Eq.~(\ref{class-dipole}) and performing the averaging over
the time we obtain~\cite{LLField}
\be
\frac{\rmd \sigma}{\rmd \Omega}=\Big(\frac{e^2}{m\, c^2}\Big)^2
\frac{E_p^4}{(E_{\rm R}^2- E_p^2)^2+\Gamma_{\rm tot}^2\,E_p^2}
[\vec{e}_z\times \vec{n}_r]^2\,. \label{sigmaField} \ee
We chose  the spherical coordinate system so that the polar angle
corresponds to the scattering angle $\theta$ --- the angle between
the propagation directions of incoming and outgoing waves,
$\cos\theta=(\vec{n}_r\,\vec{p})/p$. Then the vector product in
(\ref{sigmaField}) can be written as $[\vec{e}_z\times
\vec{n}_r]^2 = \cos^2\theta+\sin^2\theta\,\sin^2\phi$\,. Thus the
cross section depends on the azimuthal angle that corresponds to a
scattering of photons with different magnetic quantum numbers:
$[\vec{e}_z\times
\vec{n}_r]^2=\frac{4\pi}{3}|Y_{1,0}(\theta,\phi)+(Y_{1,+1}(\theta,\phi)+Y_{1,-1}(\theta,\phi))/\sqrt{2}|^2$,
where  $Y_{l,m}(\theta,\phi)$ are the spherical functions. The
magnetic number dependence appears because we have confined the
oscillator motion to one dimension. For a spherically symmetric
scattering this  dependence would be averaged out.

The differential cross section can now be written as
\be
\frac{\rmd \sigma}{\rmd \Omega} 
&=&\frac{\pi}{p^2}\frac{3\,\, \Gamma_{\rm tot}^2 E_p^2} {(E_{\rm
R}^2- E_p^2)^2+\Gamma_{\rm tot}^2\,E_p^2} B_{\rm rad}^2
\Big|Y_{10}(\theta,\phi)+\frac{1}{\sqrt
2}\big(Y_{1-1}(\theta,\phi)+Y_{1-1}(\theta,\phi)\big)\Big|^2 \,,
\label{sigmaField-2} \ee
where we introduced the branching ratio
$B_{\rm rad} = \Gamma_{\rm rad}/\Gamma_{\rm tot}$.
One can introduce the scattering amplitude as
\be
\frac{\rmd \sigma}{\rmd \Omega} &=& \big|f(\theta,\phi)\big|^2\,,\
\quad f(\theta,\phi)=
\sum_{l=0}^\infty\sqrt{2\,l+1}\, \sqrt{4\pi} \sum_{m=-l}^{l}
{f}_{l,m}\,Y_{l,m}(\theta,\phi) \,. \label{EM-scattampl} \ee
For the spherically symmetrical scattering the amplitude would be
\be\label{spham} f(\theta)=
\sum_{l=0}^\infty
(2\,l+1) {f}_l\, P_l(\cos\theta)\,. \ee
Here $P_l(\cos\theta)$ are Legendre polinomials normalized as
\cite{LL}: $\int_{-1}^{1} P_l^2 (x)dx=2/(2l+1)$.

In our case the scattering amplitude has only terms with $l=1$ and
$\sqrt{2}\,{f}_{1,\pm1}= {f}_{1,0}\equiv {f}_1$ with
\be
&& 2p{f}_1 = \frac{B_{\rm rad}\,\Gamma_{\rm tot}\, E_p}{E_{\rm
R}^2-E_p^2-i\,\Gamma_{\rm tot}\,E_p} =  \frac{B_{\rm
rad}}{\cot\delta (E_p) -i} =  B_{\rm rad}\sin\delta (E)e^{i\delta
(E_p)}. \ee
The phase of the scattered waves $\delta (E_p)$ is defined as in
Eqs.~(\ref{class-phase}), and (\ref{tange}) but now with
$\Gamma_{\rm tot}$ instead of $\Gamma$, i.e. $\tan\delta
(E_p)=-\Gamma_{\rm tot}E_p/(E_p^2-E_{\rm R}^2) $\,.

After integration over the scattering angle the total cross section can be cast in the standard
spin-averaged Breit-Wigner resonance form (see page~374 in Ref.~\cite{PDG-book})
\be
\sigma
&=& {2(2l+1)} {4\pi} |{f}_l|^2 = \frac{3}{2}\, \frac{4\pi}{p^2}
\frac{\Gamma_{\rm tot}^2\,E_p^2} {(E_{\rm R}^2-
E_p^2)^2+\Gamma_{\rm tot}^2\,E_p^2} B_{\rm rad}^2\,. \ee
Here the statistical factors correspond to the  angular momentum,
$l=1$ in our case.

From the structure of Eq.~(\ref{class-dipole}) we see that the
concepts of the phase and group time delays (\ref{class-tphase})
and (\ref{class-tgroup}) are also applicable to electromagnetic
waves, if we deal with not a monochromatic wave but a wave packet
instead. If the incoming wave were like $|\vec{\mathcal{E}}_{\rm
in}|=\mathcal{E}_0\, f_{\mathcal{E}}(\vec{p}\vec{r}-E_p\,t)
\cos\big(\vec{p}\vec{r} - E_p\, t\big)$ with some function
$f_{\mathcal{E}}(x)$ integrable in the interval
$(-\infty,+\infty)$, then the outgoing wave would be
$|\vec{\mathcal{E}}_{\rm out}|\propto \mathcal{E}_0\,
f_{\mathcal{E}}(t-\delta t_{\rm s}^{\rm (cl)}-r/c) \cos\big(p\,r -
E_p\,[t-\delta (E_p)]\big)$\,. The propagation of the scattered
wave packet is delayed by the group time (\ref{class-tgroup}), see
also (\ref{A1}), (\ref{A2}),
\begin{eqnarray}
 t_{\rm s}^{\rm (cl)} = \frac{\partial \delta}{\partial E_p}
\approx \frac{A}{2}=\frac{\Gamma_{\rm tot} /2}{(E_p-E_{\rm R})^2
+\Gamma^2_{\rm tot}/4}>0, \label{trad}
\end{eqnarray}
which here in three dimensional case has meaning of {\em{ the
scattering delay time, being twice as small compared to the Wigner
delay time}} introduced above, see Eq. (\ref{clW}). Here we
performed expansion in frequencies close to the resonance $E_p\sim
E_{\rm R}$\,. With $t_{\rm s}^{\rm (cl)}$ from Eq.~(\ref{trad}),
the scattered wave appears with a delay compared to the condition
$t-r/c\geq 0$. Thus causality requires that the scattered wave
arises for $t- t_{\rm s}^{\rm (cl)}-r/c\geq 0$.

\subsubsection{Scattering of light on hard spheres}

For the scattering of light on a hard sphere of radius $R$, the
causality condition can be formulated as~\cite{Nusszv-book,WuOm}:
\emph{if the incident wave propagating along $z$ direction
vanishes for $t<z/c$, the scattered wave in the direction $\theta$
must vanish for $t<\big(r-2\, R\sin(\theta/2)\big)/c$. } The
quantity $2R\sin(\theta/2)$ is the difference in the paths of the
light scattered at angle $\theta$ on the sphere surface and on the
sphere center (\ref{hsWigner}).

 The scattering process (when the
beam just touches the sphere) proceeds with twice shorter advance
compared the time $R/c$ which the light would pass to the center
of the sphere, cf. (\ref{hsWigner}). Correspondingly,  the
advancement in the scattering time, $\delta t_{\rm s}^{\rm (cl)}$,
proves to be  twice as small compared to the advancement in the
Wigner time, $\delta t_{\rm W}^{\rm (cl)}$.


\section{Time shifts in non-relativistic quantum mechanics: 1D-scattering}\label{sec:quant}

The problem of how to quantify a duration of quantum mechanical
processes has a long and vivid history. It started with a
statement of Wolfgang Pauli~\cite{Pauli} that in the framework of
traditional non-relativistic quantum mechanics it is impossible to
introduce a hermitian (self-adjoint) linear operator of time,
which is canonically conjugate to the Hamiltonian. The reason for
this is that for most of the systems of physical interest the
Hamiltonian is bounded from below.~\footnote{Nowadays there
continue  attempts to introduce a formal quantum observable for
time, e.g., see ~\cite{Olkhov}.} Later on a variety of 'time-like'
observables were introduced tailored for each particular system.
For a comprehensive review of the history of this question we
address the reader to the Introduction in Ref.~\cite{Muga-I}.
Various inter-related definitions of time appeared, for instance,
in considerations of the following questions: How long does the
quantum transition last (quantum jump
duration)~\cite{Hegerfeldt,Schulman}? What are interpretations of
time-energy uncertainty relations~\cite{Tam-Tamm}? How one can
quantify a time of flight or a time of arrival of a particle to a
given point~\cite{Aharonov-Vaidman,Allcock}? How long does it take
for a particle to tunnel through a
barrier~\cite{MacColl,Hartman,Hauge-Stoevneng,Landauer-Martin,Collins,Winful}?
What is a life time of a
resonance~\cite{GW,Baz66,Fonda78,Bosanas,DP}? What is the duration
of particle collision~\cite{Wigner,Smith,GW,WuOm}?

Without any pretense to address all these issues, in this section
we would like  to introduce the basic concepts related to the
temporal characteristics of typical quantum mechanical processes,
such as tunneling, scattering, and decay.


\subsection{Stationary problem}\label{sssec:stationary}

We begin with a one-dimensional quantum-mechanical system,
described by the Hamiltonian $\hat{H}=\hat{H}_0+\hat{U}$
consisting of the free motion Hamiltonian
$\hat{H}_0=-\frac{\hbar^2}{2\, m}\frac{\prt^2}{\prt z^2}$ for a
particle with mass $m$ and of an arbitrary potential
$\hat{U}=U(z)\geq 0$, which is assumed to be localized within the
interval $-L/2<z<L/2$ and vanishing elsewhere outside. This
Hamiltonian has a continuous spectrum $0<E<+\infty$ and the
complete set of eigenfunctions $\psi(z;E)$ obeying the equation
$\hat{H}\psi(z;E)= E\, \psi(z;E)$.
We will consider the wave functions satisfying the asymptotic conditions for the standard
scattering problem~\footnote{Instead of the basis wave-functions for unilateral incidence one could use the symmetrical and anti-symmetrical wave functions $\psi_s=\psi_1+\psi_2$ and
$\psi_a=\psi_1-\psi_2$ corresponding to bilateral incidence~\cite{Nussenzveig00}. }
\begin{eqnarray}
\psi_{1}(z;E) &=& \left\{
\begin{array}{lll}
e^{i\, k\, z/\hbar}+R_1(E)\, e^{-i\, k\,z/\hbar} &,  &  z< -\half L,\\
\psi_{U,1}(z;E) &,  & -\half L\le z \le \half L\\
T_1(E)\, e^{+i\, k\,z/\hbar} &,  & \half L < z,
\end{array}\right. \,,
\label{psi-f-1}\\
\nonumber\\
\psi_{2}(z;E) &=&
\left\{\begin{array}{lll}
T_2(E)\, e^{-i\, k\,z/\hbar} &, & z <
-\half L,\\ \psi_{U,2}(z;E) &, & -\half L\le z \le \half L,\\
R_2(E)\,e^{+i\, k\, z/\hbar}+ e^{-i\, k\,z/\hbar} &, & \half L < z,
\end{array}\right.
\label{psi-f-2}
\end{eqnarray}
with $k=\sqrt{2\, m\, E}>0$\,. The wave functions $\psi_1$ and $\psi_2$ describe the physical situation when a particle beam from the left or from the right, respectively, incident on the potential becomes split into a reflected part with the amplitude $R_{1,2}$ and a transmitted part with the amplitude $T_{1,2}$. The wave functions are normalized to the unit incident amplitude. Then the quantities $|R_{1,2}|^2$ and $|T_{1,2}|^2$ have the meaning of the reflection and transmission probabilities, respectively, and $|R_{1,2}|^2+|T_{1,2}|^2=1$\,. For any given wave function $\Psi$ the current is calculated standardly
\be
\mathcal{J}[\Psi]
=\frac{i\hbar}{2\,m}\left(\Psi\nabla_z \Psi^*-\Psi^*\nabla_z \Psi\right)
\,.
\label{current}
\ee
Thus, for the wave function $\psi_1$ we can define three currents:
the incident current $j_{\rm I}=\mathcal{J}[\exp(i\, k\,
z/\hbar)]=\frac{k}{m}$, the transmitted current $j_{\rm
T}=\mathcal{J}[\psi_1(z>\half L)]=|T_1(E)|^2 j_{\rm I}$ and the
reflected current $j_{\rm R}=\mathcal{J}[\psi_1(z<-\half
L)]-j_{\rm I}=-|R_1(E)|^2 j_{\rm I}$. The current conservation is
fulfilled and $j_{\rm I}=j_{\rm T}-j_{\rm R}$\,. Here, it is
important to notice that in the region of the potential there
exists an "internal" current $j_{\rm
int}=\mathcal{J}[\psi_{U,1}(z)]$. In case of the classically
allowed motion above the barrier $j_{\rm int}=j_{\rm T}$ is
determined by the sum of the currents of the forward-going wave
and of the backward-going wave, whereas in the region under the
barrier $j_{\rm int}$ is determined  by the contribution of
interference of waves, since the coordinate dependence of the
stationary wave function is given then by real functions. Namely
the latter circumstance is the reason of the so called Hartman
paradox of apparent superluminality of the under-the-barrier
motion surviving in case of infinitely narrow in energy space wave
packets (stationary state limit), which we will consider below.

The time-reversal invariance of the Schr\"odinger equation implies
that $T_1(E)=T_2(E)$\,. In general case of asymmetric potential
$R_1\neq R_2$. The functions $R_1(E)$, $R_2(E)$ and
$T(E)=T_1(E)=T_2(E)$ form the S-matrix of the one-dimensional
scattering problem~\cite{Faddeev64}.
The unitarity of the S-matrix implies the relation $T^*(E)\,
R_2(E)=- R_1^*(E)\, T(E)$\,.

To simplify further consideration we will assume that the potential $U$ is symmetric, $U(-z)=U(z)$.
Then there is a symmetry between the reflected amplitudes
$R_1(E)=R_2(E)=R(E)$, and the 'internal' parts of the
wave-functions $\psi_{U,1}(z;E)=\psi_{U,2}(z;E)=\psi_{U}(z;E)$ in
Eqs.~(\ref{psi-f-1}), (\ref{psi-f-2}) can be written as
superpositions of symmetric and anti-symmetric wave-functions
$\chi_+(z;E)$ and $\chi_-(z;E)$, respectively,
\begin{eqnarray}
\psi_{U}(z;E)= C_+\, \chi_+(z;E)+ C_-\, \chi_-(z;E)\,.
\label{internal-wf}
\end{eqnarray}
The functions are chosen such that $\chi_\pm(0;E) =
L\,\chi'_\mp(0;E)/2 = (1\pm 1)/2$, where the prime means the coordinate derivative. The coefficients $C_\pm$ in Eq.~(\ref{internal-wf}) can be expressed through the scattering
amplitudes as follows
\begin{eqnarray}
C_\pm=\frac{(T\pm R)\,e^{i\,k\,L/2\hbar}\pm e^{-i\,k\,L/2\hbar}}{2\,\chi_\pm(L/2;E)}\,.
\label{C-coeff}
\end{eqnarray}
The transmitted and reflected amplitudes are then expressed
through the logarithmic derivatives of these functions
\begin{eqnarray}
d_\pm(E)=\frac{L}{2}\frac{\prt}{\prt z}\ln\chi_\pm(z;E)\Big|_{z=L/2}\,,
\label{logd}
\end{eqnarray}
which can be chosen real. The amplitudes
\begin{eqnarray}
&&R(E)=-\half e^{-i\,k\,L/\hbar} \left[D_+(E) + D_-(E) \right]\,,
\quad
T(E)=-\half e^{-i\,k\,L/\hbar} \left[D_+(E) - D_-(E)
\right]\,
\label{RTtrhoughD}
\end{eqnarray}
are expressed through the functions
\begin{eqnarray}
D_\pm(E) &=& \frac{d_\pm(E) + i\, k\, L/2\hbar}{d_\pm(E) - i\, k\, L/2\hbar}=e^{i\,2\delta_{\pm}(E)}
\,,
\quad
\delta_{\pm}(E) = {\rm arctan} \Big(\frac{k\, L}{2\, \hbar\, d_\pm(E)}\Big)
\,,
\label{Dfun}
\end{eqnarray}
which have simple poles. The reflected and transmitted amplitudes can be now written as
\begin{eqnarray}
R(E) &=& e^{i\,\phi_{\rm R}}\,\cos\big(\delta_+(E) - \delta_-(E)\big)\,,
\quad
\phi_{\rm R}(E)=\pi-\frac{k\, L}{\hbar}+\delta_{\rm s} (E) \,,
\nonumber\\
T(E) &=& e^{i\,\phi_{\rm T}}\,\sin\big(\delta_+(E) - \delta_-(E)\big)\,,
\quad
\phi_{\rm T}(E) = {\textstyle \frac32}\pi-\frac{k\, L}{\hbar}+\delta_{\rm s}(E) \,,
\label{ampl-phase}
\end{eqnarray}
where we introduced an ordinary 1D-scattering phase shift \cite{Faddeev64}
\begin{eqnarray}
\delta_{\rm s}(E)=\delta_+(E) + \delta_-(E)\,.
\label{delta-s}
\end{eqnarray}
For sum and differences of the phases $\delta_+$ and $\delta_-$ one can use the following
relation
\begin{eqnarray}
\tan\big(\delta_+(E)\pm \delta_-(E)\big)=
\frac{k\, L}{2\, \hbar} \frac{d_-\pm d_+}{d_-\,d_+\mp \frac{k^2\, L^2}{4\,\hbar^2}}\,.
\end{eqnarray}

The coefficients of the internal wave function (\ref{internal-wf}) can be expressed with the help of Eqs.~(\ref{RTtrhoughD}) and (\ref{Dfun}) through the logarithmic derivatives as follows
\begin{eqnarray}
2\, \chi_{\pm}(L/2;E) C_{\pm}=\mp \frac{i k\, L}{\hbar}\frac{e^{-ik\,L/2\hbar}}{d_\pm - i k\, L/2\hbar}\,.
\label{C-coeff-d}
\end{eqnarray}

Substituting a scattering wave function $\psi_{1}$ or $\psi_2$ in
Eq.~(\ref{app:Int-rel}) of Appendix \ref{app:schroed} we find the
relation between the integral of the internal part of the wave
function, $\psi_{U,i}$, and the scattering amplitudes and phase
derivatives
\be
\intop_{-L/2}^{+L/2}\rmd z |\psi_{U}(z;E)|^2 = L
+\hbar\,\frac{k}{m}\,|T(E)|^2\, \phi_{\rm T}'(E) +
\hbar\,\frac{k}{m}\,|R(E)|^2 \,\phi_{\rm R}'(E)
+\frac{\hbar}{k}\,\Im\big(R(E) e^{+i\, k\,L/\hbar}\big),
\label{phase-deriv-integ} \ee
here the prime stands for the derivative with respect to the
energy. The last term appeared due to interference of the
reflected and incident waves.

Note that all derived expressions are valid for description of the
scattering on an arbitrary (symmetric) finite-range potential.
Thus we are able to consider on equal footing the particle
tunneling, scattering above the barrier, as well as the scattering
on quasistationary levels, provided in the latter case the
potential has a hole $U_{\rm min}<U<U_{\rm max}$ in some interval
$-L/2<-a<z<a<L/2$ and $U_{\rm min}<E<U_{\rm max}$.

The above expressions can be also applied for the situation, when
only a half of the coordinate space is available for the particle
motion.
Such a situation is discussed in Sect.~\ref{quasistationary},
where we describe the decay of quasistationary states. Then we can
use the wave function $\psi_1$, see Eq.~(\ref{psi-f-1}), with the
condition $\psi_1 (0)=0$, if the particle motion is allowed in the
left half-space ($z<0$), or we can use the wave function $\psi_2$
[Eq.~(\ref{psi-f-1})] with the condition $\psi_2 (0)=0$, if
particles move in the right half-space ($z>0$). The presence of
the wall at $z=0$ requires that only anti-symmetric wave function
survives in (\ref{internal-wf}) and the internal wave function
becomes equal to
\begin{eqnarray}
\psi_{U}(z;E)= \widetilde{C}_-\, \chi_-(z;E)\,,\quad
\widetilde{C}_-= 2 C_-\,.
\label{internal-wf-hs}
\end{eqnarray}

 This is easily taken into account
in the above general expressions by  the replacement $d_+\to d_-$.
After this, the transmitted wave disappears, $T=0$, and the
reflected wave amplitude reduces to a pure phase multiplier,
$R=e^{i\,\phi_{\rm s}(E)}$, with
\begin{eqnarray}
\phi_{\rm s}(E)=\pi-\frac{k\, L}{\hbar}+2\,\delta_-(E)\,.
\end{eqnarray}
Note that similarly is described the wave function of the radial
motion in a three-dimensional scattering problem, where $\delta_-$
($\delta_- = \delta_+$ for symmetric potential) plays a role of
the scattering phase, see Sect. \ref{scattering} below.



\begin{figure}
\centerline{\includegraphics[width=14cm]{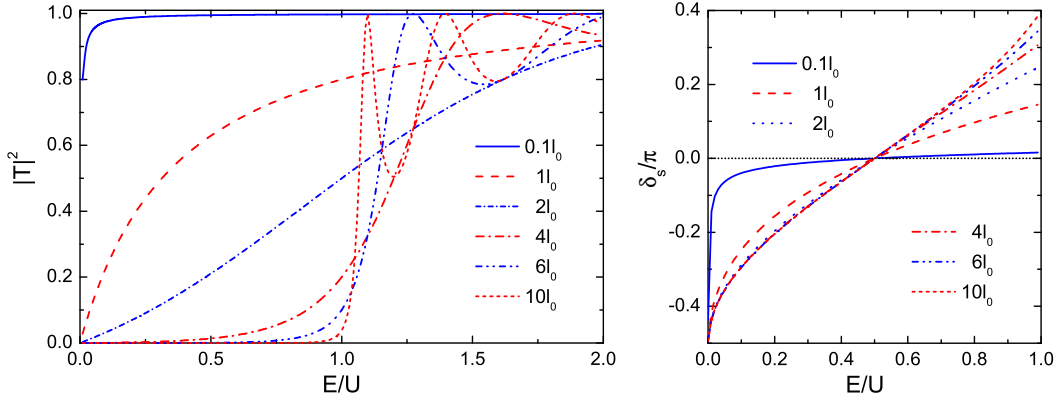}}
\caption{Amplitude and phase of the transmission wave for the
rectangular barrier of hight $U$ and length $L$ calculated
according to Eqs.~ (\ref{rectang-delta-s}) and
(\ref{rectang-R-T}). Various curves correspond to barriers of
different lengths $L$, shown by labels in units of
$l_0=\hbar/\sqrt{2\,m\,U}$.  } \label{fig:Tsq-rect}
\end{figure}

\emph{Example: scattering on a rectangular barrier}

Consider a rectangular potential barrier of length $L$:
$U(z)=U=const>0$ for $-L/2< z <L/2$. We assume first that $E<U$.
Then we deal with  a tunneling problem. The wave function
$\psi_{U}$ in internal region, see Eq.~(\ref{internal-wf}),  is
decomposed into the following even and odd functions:
\begin{eqnarray}
&&
\chi_+(z;E)=\cosh(\varkappa\, z/\hbar)\,,\quad
\chi_-(z;E)=\sinh(\varkappa \, z/\hbar)\,,\quad
\varkappa=\sqrt{2\,m\, (U-E)}>0\,.
\label{rectang-chis}
\end{eqnarray}
The logarithmic derivatives follow then as
\begin{eqnarray}
d_+ &=&\frac{\varkappa^2\,L^2}{4\hbar^2\, d_-} =\frac{\varkappa\,L}{2\hbar}\tanh(\varkappa\, L/2\hbar)\,.
\label{rectang-ds}
\end{eqnarray}
The phases of transmitted and reflected amplitudes in
(\ref{ampl-phase}) can now be written through the scattering
phase:
\begin{eqnarray}
\delta_{\rm s}(E)= -
\arctan\left(\frac{\varkappa^2-k^2}{2\,k\,\varkappa}\tanh(\varkappa\,
L/\hbar)\right)\,. \label{rectang-delta-s}
\end{eqnarray}
We used here the relation $\tan\big({\pi}/{2}+{\rm
arctan}(1/x)\big)=-x$\,. The squared amplitudes are given by
\begin{eqnarray}
|R|^2&=&\frac{(\varkappa^2+k^2)^2}{(\varkappa^2-k^2)^2}
\sin^2\delta_{\rm s} =\frac{(\varkappa^2+k^2)^2\,
\sinh^2(\varkappa\, L/\hbar)} {(\varkappa^2 +
k^2)^2\sinh^2(\varkappa\, L/\hbar) + 4\,k^2\, \varkappa^2}\,,
\nonumber\\ |T|^2 &=&1- |R|^2=\frac{\cos^2\delta_{\rm
s}}{\cosh^2(\varkappa\, L/\hbar)} = \frac{4\,k^2\, \varkappa^2}
{(\varkappa^2+k^2)^2\,\sinh^2(\varkappa\, L/\hbar) +4\,k^2\,
\varkappa^2}\,.
 \label{rectang-R-T}
\end{eqnarray}
The coefficients $C_\pm$ in~(\ref{C-coeff}) can be expressed now as follows
\begin{eqnarray}
C_+ =\frac{- i e^{-i k\, L/2\hbar}}
{\frac{\varkappa}{k}\,\sinh\big(\frac{\varkappa L}{2\,\hbar}\big)-i\, \cosh\big(\frac{\varkappa L}{2\,\hbar}\big)}\,, \quad
C_- =\frac{ i e^{-i k\, L/2\hbar}}
{\frac{\varkappa}{k}\,\cosh\big(\frac{\varkappa L}{2\,\hbar}\big)-i\, \sinh\big(\frac{\varkappa L}{2\,\hbar}\big)}.
\label{rectan-C^2}
\end{eqnarray}

The amplitudes $R$ and $T$ can be written as functions of two
dimensionless variables characterizing the energy of the incident
particle, $E/U$, and the width of the potential, $L/l_0$, where
$l_0=\hbar/\sqrt{2\,m\,U}$. These functions are illustrated in
Fig.~\ref{fig:Tsq-rect}. For a thin barrier, $L\lsim l_0$, the
transmission probability is close to unity and the scattering
phase is small accept for very small energies. For $E<U$ the
transmission probability decreases gradually with an increase of
$L$ until $L\simeq 2 l_0 $ and than falls off exponentially for
larger $L$. The scattering phase is a monotonously growing
function of the energy.

If we replace $\varkappa\to i|\varkappa|$,
Eqs.~(\ref{rectang-delta-s}) and (\ref{rectang-R-T}) also  can be
used for $E>U$ (scattering above the barrier). For $E>U$, the
transmission probability is finite approaching unity for $E\gg U$
unsteadily exhibiting peaks at $E/U=1-\pi^2\,n^2\, l_0^2/4\,L^2$
with integer $n$. At the peaks $|T|=1$, see
Fig.~\ref{fig:Tsq-rect}.

\subsection{Characteristics of time in stationary scattering problem}
\label{sssec:time-stat}

Within a stationary scattering problem formulated above there is
no notion of time {\it per se}, since the only $t$ dependent
overall factor $e^{-i\, E\, t/\hbar}$ does not enter physical
quantities. However, we have at our disposal quantities, which can
be used to construct a measure with the dimensionality of time.
Such a quantity describing the transmitted waves (at $z>L/2$)
arises, for example, if we divide the integral of the squared wave
function $\intop_a^b |\psi(z;E)|^2\rmd z$ by a current. The flux
density outside the barrier
does not depend on the coordinate. So we can use expression for
the transmitted flux density $j_{\rm T}=|T(E)|^2\, k/m$. Then for
any interval with $a,b>\half L$ the quantity
\be
\frac{1}{j_{\rm T}} \intop_a^b |\psi(z;E)|^2\rmd z =\frac{b-a}{v}
\ee
is just a passage time of the segment $[a,b]$ by a particle with
the velocity $v=k/m$\,. An application of this quantity to the
left from the barrier for $a,b<-\half L$ could be meaningless,
since, e.g., in the case of the full wave reflection from an
infinite barrier the total flux vanishes $|j_{\rm I}|-|j_{\rm
R}|=j_{\rm T}=0$. On the other hand, the reflected current cannot
be used also since it vanishes for the free particle motion. Thus,
in order to construct a relevant time-quantity for a particle
moving in the segment $[a,b]$ with $a,b<-\half L$ we divide the
squared wave function by the incident current
\begin{eqnarray}
\frac{1}{j_{\rm I}} \intop_a^b |\psi(z;E)|^2\rmd z =\frac{b-a}{v}\, \big( 1+ |R|^2\big) +
\frac{\hbar}{k\,v}\,|R|\,\sin\big(2\,k\, z /\hbar-\phi_{\rm R}\big)\Big|^b_a\,.
\label{dwell-reflect}
\end{eqnarray}
The first term represents the passage time of the incident wave in
the forward direction through the segment $[a,b]$ (the unity in
the brackets) and the passage time of the reflected wave in the
backward direction ($|R^2|$ in the brackets). For a fully opaque
barrier $|R|=1$ we, obviously, get $2(b-a)/v$. The second term
appeared due to interference of the incident and reflected waves.
It can be neglected only in the short de Broglie wave-length limit
$\hbar/k\ll (b-a)$.

Another approach to the definition of time is to introduce an
explicit "clock" -- a microscopic device characterized by a simple
time variation with a constant well defined period -- which is
weakly coupled to a quantum system under investigation. From a
change of the clock's "pointer" one can then read off a duration
of the process in the quantum system measured in terms of the
clock's period. Such a procedure was proposed by Salecker and
Wigner in~\cite{Salecker58} for measurements of space-time
distances. Peres in~\cite{Peres80} extended this concept to
several quantum mechanical problems including a time-of-flight
measurement of the velocity of a free non-relativistic particle.

Back in 1966, Baz'~\cite{Baz66} proposed the use of the Larmor
precession, as a measure of a scattering time in  quantum
mechanics. He ascribed spin $\half$ and a magnetic moment $\mu$ to
the scattered particle and assumed presence of a weak magnetic
field $B$ within the finite space region of interest, e.g. within
a range of potential. The difference in the spin polarization
before and after the region proportional to $-\half
\hbar\om_L\,t_{\rm L}$, where $\om_L=\mu\, B/\hbar$ is the Larmor
frequency, gives the time the particle takes to traverse the
region. For a one dimensional case this approach was adopted by
Rybachenko in Ref.~\cite{Rybachenko}. In the framework of the
time-dependent formalism the spin-clock method was analyzed in
Ref.~\cite{Verhaar78}.

In Ref.~\cite{Buttiker83} B\"uttiker showed that  for a
one-dimensional scattering problem the Larmor precession time
introduced in~\cite{Baz66,Rybachenko} is equivalent to the
\emph{dwell time}
\begin{eqnarray}
t_{\rm d}(a,b,E)=\frac{1}{j_{\rm I}} \intop_a^b |\psi(z;E)|^2\rmd
z\,, \label{dwell}
\end{eqnarray}
which  tells how long the incident current $j_{\rm I}$ must be
turned on to produce the necessary particle storage within the
segment $[a,b]$, see (\ref{dwell-reflect}). This time is a
quantum-mechanical counter part of the classical 1D dwell time
(\ref{dwell1d}). Indeed, as follows from the Schr\"odinger
equation, the probability density given by the square of the
wave-function satisfies the continuity equation, as for water in a
clepsydra.

The value
\begin{eqnarray}
 \delta t_{\rm d}(a,b,E)= t_{\rm d}(a,b,E)-(b-a)/v
\end{eqnarray}
shows difference of the time, which particle spends in the segment $[a,b]$ of the potential and the time, if the potential in this region  were switched off.

For the case  $E>\max U$ the classical motion is allowed for any
$z$ and the time a particle needs to move from $-L/2$ to $+L/2$
--- \emph{the classical traversal time} --- is
\begin{eqnarray}
t_{\rm trav}^{\rm (cl)}(E)= \intop_{-L/2}^{+L/2}\frac{\rmd
z}{\sqrt{2 (E-U(z))/m}}\,,
\label{trav-cl}
\end{eqnarray}
cf. with the definition of the classical sojourn time (\ref{soj}).
However, when the energy is smaller than a potential maximum,
there appears an imaginary contribution to this quantity from the
integration between turning points $z_1(E)$ and $z_2(E)$, which
are solutions of the equation $U(z_{1,2})=E$. The imaginary time
pattern  is used in the so-called imaginary-time formalism, being
successfully applied  in the problems of quantum tunneling through
varying in time barriers, see the review~\cite{Popov05}.
Nevertheless the imaginary time can be hardly used as the typical
time for passing of the barrier.

Ref. \cite{Popov71} considering electron-positron pair production
within imaginary time formalism estimated the traversal time of
the barrier as its length divided by the velocity of light $c$
(for relativistic particles). The inverse quantity $\omega_{\rm
tun}\sim |\nabla U|/mc$ separates then two regimes of particle
production in rapidly varying potentials (for $\omega>\omega_{\rm
tun}$) and that in static fields (for $\omega\ll\omega_{\rm
tun}$). Similarly, B\"uttiker and Landauer Ref.~\cite{Buttiker}
argued to use the quantity
\begin{eqnarray}
t_{\rm trav}^{\rm (BL)}(E) = \intop_{-L/2}^{+L/2}\frac{\rmd
z}{\sqrt{2 |E-U(z)|/m}} = \int_{-L/2}^{+L/2}\frac{m \rmd
z}{\varkappa(z,E)}\,
\label{trav}
\end{eqnarray}
for description of the tunneling time  trough rapidly varying
barriers at a non-relativistic particle motion.
Also, they conjectured to use this value to estimate the traversal
time of the tunneling through stationary potential barriers.  Ref.
\cite{OlkhovskyPhRep04} has shown that this time arises, as a
standard  dispersion of the tunneling time distribution. A support
for the usage of (\ref{trav}) to estimate time of particle passage
through barriers comes from analysis of the radiation spectral
density for charged particles traversing the barrier, which  is
determined by the ordinary classical formula \cite{Dyakonov}:
$(\partial E_{\omega}/\partial \omega)_t \propto e^2 \omega^2 k^2
(t^{\rm BL}_{\rm trav})^2/m^2$, where $t_{\rm trav}^{\rm BL}$
enters as the time of passing the barrier region.

Also, one can formally construct an analogue of the phase time, as
in Eq.~(\ref{class-tphase}) in Sect.~\ref{Mech}, e.g., $\delta
t_{\rm ph,R}(E)=\hbar\big(\phi_{\rm R}(E)-\pi\big)/E$ and $\delta
t_{\rm ph,T}(E)=\hbar\big(\phi_{\rm T}(E)-\pi\big)/E$, as  time
shifts between incident and reflected and transmitted waves, but
these time shifts are not associated with observables.

Relevant quantities are the group times $\hbar \rmd\phi_{\rm
R}(E)/\rmd E$ and $\hbar \rmd\phi_{\rm T}(E)/\rmd E$, cf.
Eq.~(\ref{grTime}), similar to those we introduced in Sect.
\ref{Mech}. These quantities will be discussed in a more detail
below.


Another part of stationary problems relates to description of
bound states arising in case of  attractive stationary potentials.
Inside the potential well, i.e. for $z\in [z_1,z_2]$, where $z_1$
and $z_2$ are turning points, the semiclassical wave-function can
be written in two ways~\cite{Migd}:
\begin{eqnarray}
\psi^{\rm (scl)}(z;E)&=&\frac{C_1}{\sqrt{|\varkappa(z,E)|}}
\cos\Big(\intop^{z}_{z_1}|\varkappa(z',E)|\rmd z'/\hbar
-\phi_{<}\Big)\,,
\end{eqnarray}
or
\begin{eqnarray}
\psi^{\rm (scl)}(z;E)&=&\frac{C_2}{\sqrt{|\varkappa(z,E)|}}
\cos\Big(\intop_{z}^{z_2}|\varkappa(z',E)|\rmd z'/\hbar -\phi_{>}\Big) \,,
\label{scl-wf-bound}
\end{eqnarray}
where $\phi_{<} =\phi_{>}=\pi/4$, provided  potential is a smooth
function of $z$ near  turning points. Note that in purely quantum
case  the phase shifts of in-going and out-going waves for the
bound states may depend on $E$. Condition of coincidence of these
solutions yields $C_1=C_2 (-1)^n$ and we get the Bohr-Sommerfeld
quantization rule
\begin{eqnarray}
\int_{z_1}^{z_2} |\varkappa(z,E)|dz =\hbar (\pi n+ \phi_{<}+\phi_{>})\,,\quad n=0,\pm 1,...
\end{eqnarray}
From this rule for the  passage time of the potential well one
gets
\begin{eqnarray}
t_{\rm trav}^{\rm (scl)}(z_1,z_2,E)=\half P=\hbar\pi \frac{\rmd
n}{\rmd E}+t_{\rm gr}^{<}+t_{\rm gr}^{>} ,
\label{t-trav-scl}\end{eqnarray}
where $P$ is the period of motion, and $\frac{\rmd n}{\rmd E}$ is
the number of states per unit energy, $t_{\rm gr}^{<}=\hbar \rmd
\phi_{<}/\rmd E$ and $t_{\rm gr}^{>}=\hbar \rmd\phi_{<}/\rmd E$,
and for the semiclassical motion $t_{\rm gr}^{<}+t_{\rm
gr}^{>}=0$. Replacing (\ref{scl-wf-bound})  with  appropriate
normalization in (\ref{dwell}) we get $t_{\rm trav}^{\rm
(scl)}(z_1,z_2,E)\simeq t_{\rm d}(z_1,z_2,E)$.


\emph{Example: dwell time for a rectangular barrier}

\noindent We apply the dwell time definition (\ref{dwell}) to the
wave function ~(\ref{internal-wf}), (\ref{rectang-chis}) and
calculate the dwell time of the particle under the barrier. The
incident current is $j_{\rm I}=k/m$ and
\begin{eqnarray}
t_{\rm d}(-L/2,L/2,E) &=& \frac{m}{k}  \intop_{-L/2}^{+L/2}
\Big\{
 |C_+|^2\, \cosh^2(\varkappa\, z/\hbar)
+|C_-|^2\, \sinh^2(\varkappa\, z/\hbar)\Big\}\rmd z
\nonumber\\
&=& \frac{m\, L}{2\,k} \Big(| C_+|^2 -|C_-|^2 \Big)
+\frac{m}{2\,k} \frac{\hbar}{\varkappa}\sinh(\varkappa L/\hbar)\Big(| C_+|^2 +|C_-|^2 \Big)
\,.
\label{tdwell-1}
\end{eqnarray}
We see that the dwell time contains two time scales: the one is
the free  traversal time $m\,L/k$ and the other is a purely
quantum scale $m\hbar/k|\varkappa|$.  Namely, the former quantity
determines the traveling time for classically allowed motion with
$E\gg U$. The internal wave function given by
Eqs.~(\ref{internal-wf}) and (\ref{rectang-chis}), $\psi_{U}$, is
expressed in terms of evanescent and growing functions
$\psi_U^{\rm (evan)}= \half(C_+ - C_-) e^{-\varkappa z/\hbar}$ and
$\psi_U^{\rm (grow)}=\half(C_+ + C_-) e^{+\varkappa z/\hbar}$.
Using Eqs. (\ref{rectang-delta-s}) and ~(\ref{rectan-C^2}), after
some algebra we obtain from Eq.~(\ref{tdwell-1}):
\begin{eqnarray}
t_{\rm d}^{\rm (evan)} &=& \frac{m}{k}\intop_{-L/2}^{+L/2}\quart
|C_+ - C_-|^2 e^{-2\varkappa z/\hbar}\rmd z
=\frac{m\,\hbar}{k\varkappa }\cos^2\delta_{\rm
s}\frac{k^2+\varkappa^2}{2\,\varkappa^2} \tanh(\varkappa L/\hbar)
\frac{e^{\varkappa L/\hbar}}{2\cosh(\varkappa
L/\hbar)}\,,\nonumber \label{td-evan}\\ t_{\rm d}^{\rm (grow)} &=&
\frac{m}{k}\intop_{-L/2}^{+L/2}\quart |C_+ + C_-|^2 e^{2\varkappa
z/\hbar}\rmd z=\frac{m\,\hbar}{k\varkappa }\cos^2\delta_{\rm
s}\frac{k^2+\varkappa^2}{2\,\varkappa^2} \tanh(\varkappa L/\hbar)
\frac{e^{-\varkappa L/\hbar}}{2\cosh(\varkappa
L/\hbar)}\,,\nonumber \label{td-grow}\\ t_{\rm d}^{\rm (evan)} &+&
t_{\rm d}^{\rm (grow)} =\frac{m\,\hbar}{k\varkappa
}\cos^2\delta_{\rm s}\frac{k^2+\varkappa^2}{2\,\varkappa^2}
\tanh(\varkappa L/\hbar)\,, \label{tdwell-evan-grow}
\end{eqnarray}
and the correlation term
\begin{eqnarray}
t_{\rm d}^{\rm(cor)} = \half \Re\{(C_+ - C_-)(C_+ + C_-)^*\}\,L=
\frac{m\,L\,}{k} \cos^2\delta_{\rm s} \frac{(\varkappa^2-k^2)}{2\varkappa^2\cosh^2(\varkappa L/\hbar)}\,.
\label{tdwell-cor}
\end{eqnarray}
Interestingly, the  traversal time scale $\propto L$ appears in an
interference term $t_{\rm d}^{\rm(cor)}\propto U$ between
evanescent and growing waves, whereas the quantum term appears in
a sum of the dwell times constructed from the pure  evanescent and
growing waves, $t_{\rm d}^{\rm (evan)}$ and $t_{\rm d}^{\rm
(grow)}$.

For tunneling, $E<U$, through a thick barrier, $\varkappa L/\hbar \gg 1$, we have
$$
t_{\rm d}
\simeq t_{\rm d}^{\rm (evan)},
$$
since $t_{\rm d}^{\rm (cor) }/t_{\rm d}^{\rm (evan)}\sim t_{\rm d}^{\rm (grow) }/t_{\rm d}^{\rm (evan)}\sim
e^{-2\varkappa L/\hbar}\ll 1$. The integral in $t_{\rm d}^{\rm
(evan)}$ is determined by the region near $z=-L/2$ (provided
particles flow on the barrier from the left). Thus in this case
{\em{ the dwell time of particles  under the barrier is determined
 by the inflow near the left edge of the barrier and does not describe particle transmission.}}
This observation a bit corrects  statement \cite{Winful} p. 7,
that the dwell time of particles under the barrier "...does not
distinguish transmitted particles from reflected particles" and
"tells us the dwell or sojourn time in the barrier regardless of
whether the particle is transmitted or reflected at the end of its
stay".

Combining Eqs. (\ref{tdwell-evan-grow}) and (\ref{tdwell-cor}) we
obtain
\begin{eqnarray}
&t_{\rm d}(-L/2,L/2,E) =\frac{m\,L\,\cos^2\delta_{\rm s}}{2\,k}
\left[\frac{k^2+\varkappa^2}{\varkappa^2} \frac{\tanh(\varkappa
L/\hbar)}{\varkappa L/\hbar}-\frac{k^2-\varkappa^2}{\varkappa^2}
\frac{1}{\cosh^2(\varkappa L/\hbar)}\right]\,. \label{tdwell-rect}
\end{eqnarray}
Behavior of this value is illustrated in
Fig.~\ref{fig:Tdwell-rect}, left. We see that for $E>U$ expression
(\ref{tdwell-rect}) exhibits peaks, when the system gets stuck
above the barrier in resonance states, for which the barrier
becomes effectively absolutely transparent (maxima of $|T|^2$ in
Fig.~\ref{fig:Tsq-rect}). The resonance energy is determined by
the condition  $|\varkappa| L/\hbar=\pi\,n$ for an integer $n>0$.
The peak heights increase with the barrier thickness as
$(m\,L/k)[1 + 2 L^2/(l_0^2\,n^2\,\pi^2)]$\,. The time
(\ref{trav-cl}) of the traversal of the distance $L$ at resonance
energies can be related to the density of the resonance states:
\begin{eqnarray}
t^{\rm (scl)}_{\rm trav}=\frac{L m}{|\varkappa|}=\pi \hbar\frac{\rmd
n}{\rmd E}\,.
\end{eqnarray}
For $E\gg U$  we get
\begin{equation}
t_{\rm d}(-L/2,L/2,E)=\frac{m L}{k}\Big[ 1+\frac{U}{2\, E}
-\sin(2\varkappa
L/\hbar)\frac{l_0^2}{4L^2}\Big(\frac{U}{E}\Big)^{3/2}
+O\Big(\frac{U^{2}}{E^{2}}\Big)\Big]\,.
\end{equation}
As we see from Fig.~\ref{fig:Tdwell-rect}, for $E>U$  the dwell
time oscillates around the classical traversal time and approaches
it for $E\gg U$ as
\begin{eqnarray}
\big|t_{\rm trav}^{\rm (cl)}(-L/2,L/2,E)-t_{\rm d}(-L/2,L/2,E)\big|
<\frac{m L}{k}\Big[\frac{l_0^2}{4L^2}\Big(\frac{U}{E}\Big)^{3/2} + O\Big(\frac{U^2}{E^2}\Big)\Big]\,.
\end{eqnarray}

For a broad barrier in the limit $|\varkappa|L/\hbar\gg 1$ and for
$E\geq U$  we have for the dwell time
\begin{eqnarray}
 t_{\rm d}(-L/2,L/2,E_p)\simeq t_{\rm d}^{\rm (cor)}
\simeq \frac{m\,L}{k} \frac{2\, k^2\,(k^2+|\varkappa|^2)}{
\left[(|\varkappa|^2-k^2)^2\sin^2( |\varkappa| L/\hbar)+4|\varkappa|^2 k^2\right]}\,.
 \label{uptrav}
\end{eqnarray}
As follows from this expression the dwell time  exceeds the
classical free traversal time $t_{\rm trav}^{\rm free}=m L/k$.

For $E=U$ and arbitrary $L$ the dwell time is
\begin{eqnarray}
 t_{\rm d}(-L/2,L/2,E=U)
 = \frac{4}{3}\frac{m\,L}{k} \frac{1+3 l_0^2/L^2}{1+4l_0^2/L^2}\,.
 \end{eqnarray}
In the tunneling regime $E<U$ the dwell time starts from zero at
$E=0$, increases with increase of $E$ and reaches the free
traversal time $m L/k$ at
\begin{equation}
E_1=U\frac{1+3\,b\,l_0^2/L^2 }{1+4\,b\,l_0^2/L^2}\,,\quad b\approx 2.5484\,.
\end{equation}
It is interesting to note that {\em{the dwell time is
\emph{always} smaller than the classical traversal time for
energies of the scattered particle $E<U$ for a thick barrier and
for $E<{\textstyle\frac{3}{4}}U$\, for a thin barrier.}}

 Since in the tunneling regime the dwell time decreases with increase of the barrier depth,
and $t_{\rm d}(E\to 0)\to 0$, the dwell time cannot be appropriate
measure of the time passage through the barrier.

\begin{figure}
\centerline{\includegraphics[width=6.7cm]{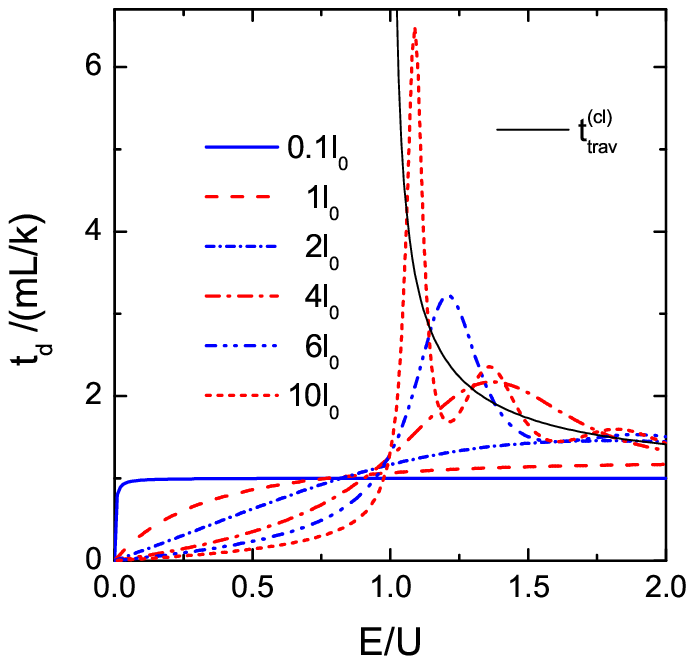}
\quad
\includegraphics[width=7cm]{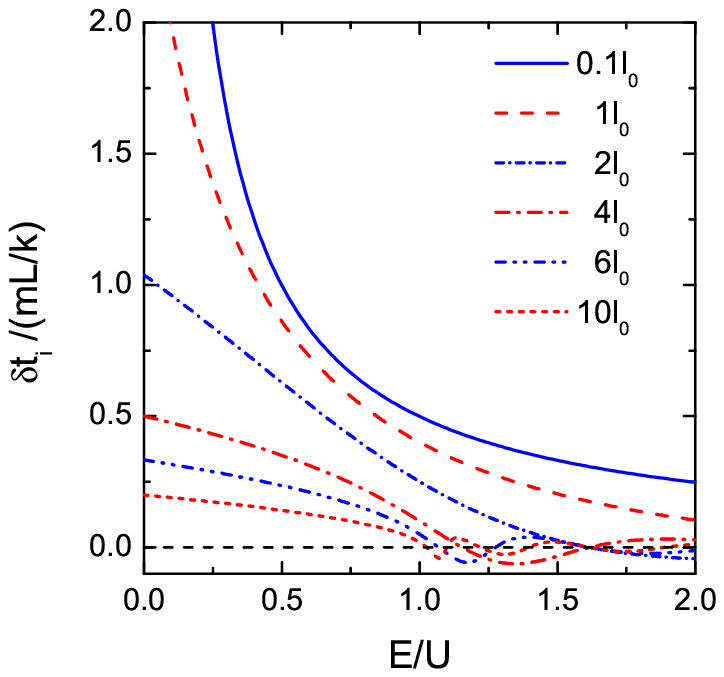}}
\caption{
The dwell time (\ref{tdwell-rect}) (left panel) and the interference time (\ref{tinter-rect}) (right panel) for the rectangular barrier  depicted as a function of the energy for various barrier lengths measured in units $l_0=\hbar/\sqrt{2\,m\,U}$.
The thin solid curve on the left panel shows the classical traversal time (\ref{trav-cl}) for the rectangular barrier. }
\label{fig:Tdwell-rect}
\end{figure}


\subsection{Non-stationary problem: scattering of a wave packet}

The evolution of a quantum-mechanical system from the time moment
$t_0$ until the time moment $t$ is determined by the Hamilton
operator: $\Psi(z,t)=\exp(-i\hat{H}\, (t-t_0)) \Psi(z,t_0)$. A
non-stationary quantum state, i.e. a state for which physical
observables change with time, thus, cannot be an eigenstate of the
Hamiltonian. Otherwise the time variations  reduce to a phase
factor $\exp(-i\,E\,(t-t_0))$, which does not enter  observables.
Hence, in order to describe the passage time of some spatial
interval by a quantum particle we need to deal with  {\it a wave
packet} describing by a superposition of stationary states with
various energies $E$, $\psi(z;E)$,
\begin{eqnarray}
\Psi (z,t)=\intop_{0}^{\infty}\frac{\rmd E}{2\pi\hbar} \Phi(E) \psi(z;E)\, e^{-i\,E\, t/\hbar} \,,
\label{wp}
\end{eqnarray}
with some $\Phi(E)$ as the energy envelop function. Such a packet
would necessarily have some spatial extension, which is the larger
the smaller is the energy spread of the states collected in the
packet. As we discuss in this section, mentioned delocalization
makes  determination of the passage time of a spatial interval by
a quantum particle to be a delicate problem.

As the stationary wave-function $\psi (z;E)$ we can take
wave-function~(\ref{psi-f-1}), $\psi(z;E)=C\,\psi_1(z,E)$.
Normalization constant $C$ can be determined from the relation
\begin{eqnarray}
\intop_{-\infty}^\infty \rmd z  \psi^*(z;E) \psi(z;E')
= 2\pi\hbar \sqrt{\frac{2E}{m}}\delta(E-E')
= 2\,\pi\hbar\,\delta(k-k')\,,
\label{moment-norm}
\end{eqnarray}
where $k=\sqrt{2mE}$ and $k'=\sqrt{2mE'}$. The wave function of
the wave packet (\ref{wp}) can  be normalized as
\begin{eqnarray}
\intop_{-\infty}^\infty \rmd z\, \big|\Psi (z,t)\big|^2
=
\intop_{0}^{\infty}\frac{\rmd E}{2\pi\hbar}\,
\sqrt{\frac{2E}{m}}\big| \Phi(E)\big|^2=1\,.
\label{wp-norm}
\end{eqnarray}
Then the quantity
\begin{eqnarray}
\rmd W_E
=\sqrt{\frac{2E}{m}}\big| \Phi(E)\big|^2 \frac{\rmd E}{2\,\pi\,\hbar}
\label{dWE}
\end{eqnarray}
is interpreted as the probability for the particle described by
the wave packet to have the energy within the segment $[E, E+dE]$.
The average energy of the state, $\overline{E}$, is given by
\begin{eqnarray}
\overline{E}=\intop_{-\infty}^\infty \rmd x  \Psi^*(x,t)\hat{H}
\Psi(x,t)=\intop_{0}^{\infty}\frac{\rmd E}{2\pi\hbar} E\,
\sqrt{\frac{2E}{m}} \big| \Phi(E)\big|^2\,. \label{aver-E}
\end{eqnarray}
Similarly, the energy dispersion of the wave packet is given by
\begin{eqnarray}
\gamma^2= \intop_{-\infty}^\infty \rmd x \,
\Psi^*(x,t)\,\big(\hat{H}^2-\overline{E}^2\big)\, \Psi(x,t)
=
\intop_{0}^{\infty}\frac{\rmd E}{2\pi\hbar}
\big(E^2-\overline{E}^2\big)\,\sqrt{\frac{2E}{m}}
 \big|\Phi(E)\big|^2\,.
\label{disp-E}
\end{eqnarray}
Formally, we can change an integration variable from $E$ to
$k=\sqrt{2\,m\,E}$ and rewrite the distribution (\ref{dWE}) as
\begin{eqnarray}
 \rmd W_E=\big| \varphi(k)\big|^2\frac{\rmd k}{2\,\pi\,\hbar}\,,\quad
 \varphi(k)=\frac{k}{m}\Phi(k^2/2m),
 \label{dWp}
\end{eqnarray}
and the wave packet (\ref{wp}), as
\begin{eqnarray}
\Psi(z,t)=\intop_{0}^{\infty}\frac{\rmd k}{2\pi\hbar}\,
\varphi(k)\, \psi(z;k^2/2m)\, e^{-i\,k^2 t/2m\hbar} \,.
\label{wpp}
\end{eqnarray}
We emphasize that the quantity  $\big| \varphi(k)\big|^2$ cannot
be identified with a momentum distribution of the state, since in
general the wave function $\psi(z;E=k^2/2m)$ is not an eigen
function of the momentum operator. However, in the remote past,
i.e., for large and negative $t$, when the peak of the packet is
at large and negative $z\ll -L$, we deal with a free wave packet.
Then  only one term of the wave function (\ref{psi-f-1})
contributes to the integral (\ref{wpp}). Indeed, only in the term
proportional to $e^{i\,k\,z}$ for $z\ll -L$ the exponents under
the integral in Eq.~(\ref{wpp}) can cancel each other for $z\sim
t\,k /m$. Thus, in the past the maximum of the packet located far
to the left from the barrier,\footnote{Had we taken the wave
function (\ref{psi-f-2}) in Eq.~(\ref{wpp}) we would get that for
large negative $t$ the maximum of the packet is located far to the
right from the barrier and the packet proceeds to the left.} ---
\emph{an incident wave packet}
\begin{eqnarray}
\Psi (z, t)\approx \Psi_{\rm I} (z,t)
=\intop_{0}^{\infty}\frac{\rmd k}{2\pi\hbar}\, \varphi(k)\,
e^{ik\,z/\hbar} e^{-i\,k^2 t/2m\hbar} \,, \label{wp-free}
\end{eqnarray}
moves to the right. In this limit the quantity $\big|
\varphi(k)\big|^2$ defines the asymptotic momentum distribution in
the packet. Note that there is always a small but finite
probability  for the particle to be in any point of the $z$ axes.

The momentum average and variance are then given by
\begin{eqnarray}
p = \langle k\rangle_k\,,
\quad
\gamma_{p}^2 = \langle (k^2-p^2)\rangle_k \,.
\label{p0-gammap}
\end{eqnarray}
Here we use the notation
\begin{eqnarray}
\big\langle \dots \big\rangle_k = \intop_0^\infty \frac{\rmd
k}{2\,\pi\,\hbar}\, \big(\dots\big)\, \big| \varphi(k)\big|^2
\label{p-aver}
\end{eqnarray}
for the average over the momentum distribution. The average energy
and momentum are related as $\overline{E}=(p^2+\gamma_p^2)/2m$\,.
For evaluation of the $k$-averages (\ref{p-aver}) of a function
$f$ dependent on $k$ we can use the relation
\begin{eqnarray}
\langle f(k) \rangle_k\approx f(p) + \frac{1}{2} \gamma_p^2
\frac{\rmd^2 f(p) }{\rmd p^2} =f(E_p) + \frac{\gamma_p^2}{2\, m}
\frac{\rmd f(E_p)}{\rmd E_p} + \frac{\gamma_p^2\, p^2}{2\, m^2}
\frac{\rmd^2 f(E_p)}{\rmd E_p^2}\,,\label{f-aver}
\end{eqnarray}
provided $\gamma_p$ is small. We used  Eq.~(\ref{p0-gammap}) and
in the second equality we changed  variables from  $p$ to
$E_p=p^2/2m$.

The momentum profile function $\varphi(k)$ is complex,
$\varphi(k)=|\varphi(k)|\,e^{i\xi(k)}$. The derivative of its
phase with respect to momentum,  $\xi'(k)$, determines the average
coordinate of the incident packet~\cite{HFF87}
\begin{eqnarray}
\bar{z}_{\rm I}(t)&=&\intop^{-L/2}_{-\infty} \rmd z\, z |\Psi_{\rm
I}(z,t)|^2 /\intop_{-\infty}^{-L/2} \rmd z\,  |\Psi_{\rm
I}(z,t)|^2\approx \bar{z}_{\rm I}^{\rm
(as)}(t)=\intop_{-\infty}^{+\infty} \rmd z\, z |\Psi_{\rm
I}(z,t)|^2
\nonumber\\
&=& \big\langle
-\hbar\xi'(k)+\frac{k}{m}\,t \big\rangle_k= \langle -\hbar\xi'(k)
\rangle_k + v_{\rm I}\,t \,,\quad v_{\rm I}=\frac{p}{m}\,.
\label{xI-aver}
\end{eqnarray}
Note that the second approximate equality in the first line is
valid, if at time $t$ the packet is located almost entirely to the
left from the barrier. It is valid for  $\gamma_p (-z_0 +L)\gg \hbar$ for large negative $t$.
The derivation of this relation is given in Appendix~\ref{app:I-centroid}.
Let us fix the phase $\xi(k)$ so that in the remote past at $t_0$ the packet center was at $z_{\rm 0}=v_{\rm I}\,t_0$, then $-\hbar\xi'(k)=z_0-k\, t_0/m$ and therefore
$\langle -\hbar\xi'(k) \rangle_k =0.$

The evolution of the packet width in the coordinate space is
determined by the function $\varphi(k)$ as
\begin{eqnarray}
\overline{z_{\rm I}^2}(t)-\bar{z}_{\rm I}^2(t)
&\approx&\frac{\hbar}{4\pi}  |\varphi(0)|'|\varphi(0)| +
\hbar^2\Big\langle\Big(\frac{|\varphi(k)|'}{|\varphi(k)|}\Big)^2  \Big\rangle_k
+\Big\langle \Big(\hbar\xi'(k)-\frac{k}{m}\,t \Big)^2\Big\rangle_k
-
\Big\langle \Big(\hbar\xi'(k)-\frac{k}{m}\,t \Big)\Big\rangle^2_k\,.
\label{packet-width-x}
\end{eqnarray}

For description of a remote solitary incident packet moving to the
right with an initial average energy $\simeq \overline{E}$ the
envelop function $\Phi(E)$ must be sharply peaked at
$E=\overline{E}$, or equivalently for the description of the
packet moving with the average momentum $p$ the function
$\varphi(k)$ must be sharply peaked at $k=p$. If the widths of the
peaks of the functions $\Phi(E)$ and $\varphi(k)$ are sufficiently
small, i.e. $\gamma_{p}\ll p$  and  $\gamma\ll \overline{E}$, the
lower limit in all momentum and energy integrations can be
extended to $-\infty$. Often, the normalized momentum profile
function is chosen in the Gaussian form
\begin{eqnarray}
\varphi(k)=\Big(\frac{2\,\pi\, \hbar^2}{\gamma_{p}^2}\Big)^{1/4}\,
\exp\Big(-\frac{(k-p)^2}{4\, \gamma_{p}^2}+i\, \xi(k)\Big)\,.
\label{Gauss-phi-prof}
\end{eqnarray}
Then, using that $-\hbar\xi'(k)=z_0-k\, t_0/m$ we find from Eq.~(\ref{packet-width-x})
\begin{eqnarray}
\overline{z_{\rm I}^2}(t)-\bar{z}_{\rm I}^2(t)=
\frac{\hbar^2}{4\,\gamma_{p}^2}+ \frac{\gamma_{p}^2}{m^2}\, (t-t_0)^2\,,
\label{gauss-width-x}
\end{eqnarray}
where we used that for a narrow packet $|\varphi(0)|\to 0$\,. We
recover the well-known result that the width of a free packet
increases with time. For the typical time of the smearing of the
packet we immediately get $t-t_0\sim t_{\rm sm}$, where
\begin{eqnarray}
t_{\rm sm}= \hbar\,  m/\gamma_p^2\,. \label{tsmear}
\end{eqnarray}

\subsection{Characteristics of time for scattering of a wave packet with
negligibly small momentum uncertainty}\label{ssc:group-time}

Consider  a wave packet (\ref{wpp}) prepared far away from the
potential region, so that the packet could be made sufficiently
broad to assure a small momentum uncertainty and at the same time
it would take a long time for the packet to reach the potential
barrier. After the wave packet has reached the barrier it is split
into the reflected wave packet and two forward going (evanescent
and growing) waves propagating under the barrier, which outside of
the barrier  transform to a transmitted wave packet.

\emph{The transmitted packet}  is determined by~\cite{Bohm}
\begin{eqnarray}
\Psi_{\rm T} (z,t)=\intop_0^\infty \frac{\rmd k}{2\pi \hbar}
\varphi (k) |T(E)| e^{i\phi_{\rm T}(E)} e^{ik z/\hbar-i E
t/\hbar}\,. \label{twp}
\end{eqnarray}
 \emph{The reflected packet}  moving backwards  is
\begin{eqnarray}
\Psi_{\rm R} (z,t)=\intop_0^\infty \frac{\rmd k}{2\pi \hbar}
\varphi(k) |R(E)| e^{i\phi_{\rm R}(E)} e^{-ik z/\hbar-i E
t/\hbar}\,, \label{rwp}
\end{eqnarray}
where, as before, $E=k^2/2m$.

Also, one can introduce  two measures of time that could
characterize the wave propagation within the potential region.
Consider the difference of the time, when the maximum of the
incident packet (\ref{wp-free}) is at the coordinate $z=-L/2$, and
the time, when the maximum of the transmitted packet (\ref{twp})
is at $z=+L/2$, and the difference of the time, when the maximum
of the incident packet  and the maximum of the reflected packet
(\ref{rwp}) are at the same spatial point  $z=-L/2$. We call these
time intervals  \emph{the transmission and reflection group
times}, $t_{\rm T}$ and $t_{\rm R}$. The construction of the delay
times  goes back to pioneering works by
Eisenbud~\cite{Eisenbud48}, Wigner~\cite{Wigner} and
Bohm~\cite{Bohm}. According to the method of stationary phase, the
position of the maximum of an oscillatory integral, as those in
Eqs.~(\ref{wp-free}), (\ref{twp}), and (\ref{rwp}), is determined
by the stationarity of the complex phase of the integrand. For
sufficiently narrow initial momentum distribution, $\gamma_p\ll
p$, we can write
\begin{eqnarray}
t_{\rm T}(E_p)= \Big(\hbar\xi'(p)+\frac{L}{2\,v_{\rm I}
}+\frac{\hbar}{v_{\rm I}}\frac{\rmd}{\rmd p}\phi_{\rm T}(E_p)\Big)
-
\Big(\hbar\xi'(p) - \frac{L}{2\,v_{\rm I}}\Big)
=
\frac{L}{v_{\rm I}} + \hbar\frac{\rmd  \phi_{\rm T}(E_p)}{\rmd
E_p}\,, \label{gt}
\end{eqnarray}
and
\begin{eqnarray}
t_{\rm R}(E_p)= \Big(\hbar\xi'(p) + \frac{L}{2\,v_{\rm
I}}+\frac{\hbar}{v_{\rm I}}\frac{\rmd}{\rmd p} \phi_{\rm
R}(E_p)\Big)
-
\Big(\hbar\xi'(p) - \frac{L}{2\,v_{\rm I}}\Big) =
\frac{L}{v_{\rm I}} + \hbar \frac{\rmd  \phi_{\rm R}(E_p)}{\rmd E_p}\,,
\label{gr}
\end{eqnarray}
here and below $E_p=p^2/2m$. For $E_p\gg U$, $\frac{\rmd \phi_{\rm
T,R}(E_p)}{\rmd E_p}\to 0$ and $t_{\rm T}(E_p)\simeq t_{\rm
R}(E_p)\simeq\frac{L}{v_{\rm I}}$ reduce to the passage time of
the distance $L$. However,  interpretation of these times for
$E<U$ needs a special care. Recall that  in case of the tunneling
the transmission and reflection group times $t_{\rm T}(E_p)$ and
$t_{\rm R}(E_p)$ are asymptotic quantities since they count time
steps for events happened at $z=-L/2$ and $z=L/2$ rather than at
the turning points. Moreover, as we shall see, for the tunneling
through thick barriers the dependence of these times on $L$
ceases.

One can introduce conditional
transmission and reflection group times by multiplying the times
$t_{\rm T}$ and $t_{\rm R}$ with the transmission and reflection
probabilities, respectively. Summing them up we define \emph{a
bidirectional scattering time}, as the sum of the weighted average
of transmitted and reflected group delays \cite{Winful}
\begin{eqnarray}
{t}_{\rm bs}(E_p)=|T(E_p)|^2\, t_{\rm T}(E_p)+|R(E_p)|^2\,   t_{\rm R}(E_p)
\,.
\label{gtil}
\end{eqnarray}
This time can be also expressed through the induced,
 transmitted and reflected currents defined for the stationary problem, see  Eq.~(\ref{current}),
\begin{eqnarray}
{t}_{\rm bs}=\frac{j_{\rm T}}{j_{\rm I}}\, t_{\rm T}+\frac{|j_{\rm R}|}{j_{\rm I}}\,  t_{\rm R}.
\label{gtil-2}
\end{eqnarray}
For symmetrical barrier with the help of Eqs.~(\ref{ampl-phase})
and (\ref{delta-s}) we get
\begin{eqnarray}
t_{\rm bs}(E_p)=t_{\rm T}(E_p)=t_{\rm R}(E_p)=\hbar\frac{\rmd
\delta_{\rm s}(E_p)}{\rmd E_p}\,. \label{t_RT-sym}
\end{eqnarray}
We should emphasize a direct correspondence of the
transmission and reflection group times defined here to the
classical group times defined in Eqs.~(\ref{clW}) and
(\ref{class-tgroup}). For $E_p\gg U$, $\hbar\frac{\rmd \delta_{\rm
s}(E_p)}{\rmd E_p}\to L/v_{\rm I}$.

There is a relation~\cite{GarciaRubio89,Winful03} between the
bidirectional scattering time $t_{\rm bs}$, Eq. (\ref{gtil}), and
the dwell time $t_{\rm d}$, Eq.~(\ref{dwell}), which follows from
the general relation for the stationary wave function
(\ref{phase-deriv-integ}) and definitions (\ref{gt}), (\ref{gr}):
\begin{eqnarray}
t_{\rm d}(-L/2,L/2,E_p)=t_{\rm bs}(E_p)-\delta t_{\rm i}(E_p)\,.
\label{tD-tR-tT}
\end{eqnarray}
The last term here is the interference time delay. It arises due
to the interference of the incident part of the wave function (the
incident packet) with its reflected part. This term is of the same
origin as the last term on the left-hand side of
Eq.~(\ref{dwell-reflect}),
\begin{eqnarray}
\delta t_{\rm i}(E_p)=-\frac{\hbar}{pv_{\rm I}}\,\Im\big(R(E_p)
e^{+i\, p\,L/\hbar}\big) =-\frac{\hbar\, }{pv_{\rm I}}
|R(E_p)|\sin\big(\phi_{\rm R}(E_p)+p\,L/\hbar\big)\,.
\label{t-inter}
\end{eqnarray}
The interference time can be as positive as negative, so it
represents delay or advance of the incident packet. This term
$\delta t_{\rm i}$ is especially important for low energies (small
momenta), when the packet approaches the barrier very slowly.
Taking into account that $|T|^2+|R|^2=1$ we can rewrite
Eq.~(\ref{tD-tR-tT}) in the form
\begin{eqnarray}
t_{\rm d}=|T|^2\, (t_{\rm T}-\delta t_{\rm i})+|R|^2\,  (t_{\rm
R}-\delta t_{\rm i}). \label{tD-tR-tT-2}
\end{eqnarray}
 The times $(t_{\rm
T}-\delta t_{\rm i})$ and $(t_{\rm R}-\delta t_{\rm i})$ coincide
with the Larmor times introduced by Baz' and
Rybachenko~\cite{Baz66,Rybachenko} in general case of asymmetric
potentials. Naively \cite{Hauge-Stoevneng}, one interprets result
(\ref{tD-tR-tT-2}), as the time spend by the particle under the
barrier ($t_{\rm d}$) is the sum of the tunneling traversal time
in transmission $t_{\rm T}-\delta t_{\rm i}$  times probability of
transmission and the tunneling traversal time in the reflection
$t_{\rm R}-\delta t_{\rm i}$   times probability of reflection.
Such an interpretation is actually false, since in quantum
mechanics one should sum amplitudes rather than probabilities
\cite{Winful}. Moreover, as we mentioned, for thick barriers
$t_{\rm d}$ is almost entirely determined by the behavior of the
wave function on the left edge of the barrier and thereby does not
relate to the transmission process. Else, $t_{\rm T}$ and $t_{\rm
R}$ are determined when the peaks of packets are at $z=\pm L/2$
rather than at the turning points and thereby they cannot control
only the tunneling.

Some authors, see \cite{Ricco,Moura,Splitter}, introduce
tunneling transit times by dividing the probability stored within
the potential region by the local transmitted flux and the
reflected flux
\begin{eqnarray}
\widetilde{t}_{\rm T}= \frac{1}{j_{\rm T}}
\int_{-L/2}^{L/2} \rmd z |\psi(z,E_p)|^2 = \frac{t_{\rm d}}{|T|^2}
,\quad
\widetilde{t}_{\rm R}= \frac{1}{|j_{\rm R}|}
\int_{-L/2}^{L/2} \rmd z |\psi(z,E_p)|^2 = \frac{t_{\rm d}}{|R|^2}\,,
\label{dr}
\end{eqnarray}
from where we get $t_{\rm d}^{-1}=\widetilde{t}_{\rm
T}^{-1}+\widetilde{t}_{\rm R}^{-1}$.  It follows from analogy with
fluid mechanics: the local velocity $v(z)$ is related to the local
density $\rho =|\psi (z)|^2$ through $j=\rho v$, see
(\ref{dwell1d}).
 Since $|T|^2$ is exponentially
small for a broad barrier, $\widetilde{t}_{\rm T}$ is
exponentially large in this case. It is perfectly luminal and does
not saturate with barrier length \cite{Winful03}.
Ref.~\cite{Winful} argues that the quantities (\ref{dr})
characterize net-delays of transmitted and reflected fluxes rather
than tunneling times. Indeed the time $\widetilde{t}_{\rm T}$ is a
property of entire wave function made up of forward and backward
going components and thereby cannot  be considered as traversal
time of transmitted particles only \cite{Winful}. Performing
minimization of $\widetilde{t}_{\rm T}$, Ref. ~\cite{Riza} finds a
variationally determined  tunneling time $\widetilde{t}_{\rm
T}^{\rm min}\propto 1/|T|$. Both $\widetilde{t}_{\rm T}$ and
$\widetilde{t}_{\rm T}^{\rm min}\to \infty$ for $|T|\to 0$.
 Note that the typical time after passing of
which we are able to observe the particle with probability of the
order of one to the right from the barrier, if it initially were
to the left from the barrier is indeed proportional to $1/|T|^2$.
But the time $\propto 1/|T|^2$ does not correspond to our
expectations for the quantity characterizing traversal time of the
given particle  from $a$ to $b$. It is associated with the
life-time of metastable states, being in this case the tunneling
particles treated as quasiparticles decaying from a state on one
side of the barrier into another state on other side of the
barrier \cite{Collins}. This time represents a mean time, in which
a certain likelihood of a tunneling event may take place. After
passage of this time it becomes probable that approximately a half
of the original particle density has managed to tunnel away. This
does not reflect actual time of the tunneling.


\emph{Example~1: group times for a rectangular barrier}\\ The
scattering phase for a rectangular barrier is given in
Eq.~(\ref{rectang-delta-s}). Substituting this expression in
Eq.~(\ref{t_RT-sym}) we find
\begin{eqnarray}
t_{\rm bs}=t_{\rm T}=t_{\rm R} =\frac{L\cos^2\delta_{\rm
s}}{2\,v_{\rm I}}
\left[\frac{(p^2+\varkappa^2)^2}{p^2\varkappa^2}\frac{\tanh(\varkappa
L/\hbar)}{\varkappa
L/\hbar}-\frac{p^2-\varkappa^2}{\varkappa^2}\frac{1}{\cosh^2(\varkappa
L/\hbar)}\right]\,. \label{tgr-rect}
\end{eqnarray}
The interference time (\ref{t-inter}) can be written  as
\begin{equation}
\delta t_{\rm i}=\frac{L}{v_{\rm I}} \cos^2 \delta_{\rm s}
\frac{p^2+\varkappa^2}{2\,p^2} \frac{\tanh(\varkappa
L/\hbar)}{\varkappa L/\hbar}\,. \label{tinter-rect}
\end{equation}
For $E_p\ll U$, performing expansion in $E_p/U$   we have
\begin{eqnarray}
 \delta t_{\rm i}\approx 2\frac{ l_0}{v_{\rm I}}\coth(L/l_0) -
 \frac{
 l_0}{v_{\rm I}}\Big(\coth(L/l_0)+(8\coth(L/l_0)-L/l_0)/\sinh^2(L/l_0)\Big)\frac{E_p}{U}\,.
\end{eqnarray}

The interference time is shown in Fig.~\ref{fig:Tdwell-rect},
right, as a function of $E_p/U$ for various barrier lengths. As we
see, the interference time is especially significant  for small
energies when the incident packet approaches the barrier slowly.
For $E_p <U$, $\delta t_{\rm i}>0$, for $E_p >U$, at some energies
$\delta t_{\rm i}$ becomes negative. For $E_p \gg U$, $\delta
t_{\rm i}\to 0$ and  $t_{\rm bs}\simeq t_{\rm d}\simeq L/v_{\rm
I}$.

\emph{Example~2: group times in the semi-classical
approximation}\\ The wave function of the stationary scattering
problem, which enters the wave packet (\ref{wp}), can be written
in the semiclassical approximation as follows~\cite{LL}
\begin{eqnarray}
\psi^{\rm(scl)}(z;E_p)&=&\left\{
\begin{array}{lll}
  \sqrt{\frac{m}{|\varkappa(z,E_p)|}}\left[ e^{\frac{i}{\hbar} \int_{z_1}^z |\varkappa(z',E_p)|\rmd z' +i\phi_0}
  +e^{-\frac{i}{\hbar} \intop_{z_1}^z |\varkappa(z',E)|\rmd z' -i\phi_0}\right] &, & z<z_1 ,\\
  \sqrt{\frac{D\,m}{\varkappa(z,E_p)}}\, e^{\intop^{z_2}_z \varkappa(z',E_p)\rmd z'/\hbar} &,
  & z_1\le z\le z_2 ,\\
  \sqrt{\frac{D\,m}{|\varkappa(z,E_p)|}}\,
   e^{\frac{i}{\hbar} \intop_{z_2}^z |\varkappa(z',E_p)|\rmd z' +i\phi_0} &, & z_2\le
   z\,,
\end{array}
\right. \nonumber\\
&&D=\exp\Big(-\frac{2}{\hbar}\int_{z_1}^{z_2}\varkappa(z',E_p)\rmd
z'\Big)\,,
  \label{scl-wf-scatt}
\end{eqnarray}
where $z_1$ and $z_2$ are the left and right turning points
($z_1<z_2$) and the phase $\phi_0=\pi/4$ for a smooth scattering
potential, cf. Eq.~(\ref{scl-wf-bound}). Note that in the
framework of the semiclassical approximation  ~\cite{LL} it is
legitimate to take into account only evanescent wave inside the
barrier. Being derived with the same accuracy, the reflection
coefficient equals unity. Respectively, the incident current is
then totally compensated by the reflected one and the current
inside the barrier is absent, whereas it is present outside the
barrier for $z>z_2$. This current non-conservation is
inconvenient, when we study particle propagation inside the
barrier. To recover the current conservation  one should include
the contribution of the growing wave inside the barrier, despite
this procedure  is beyond the scope of the formal applicability
of the semiclassical approximation, see \cite{Razavi}. Similarly,
in non-equilibrium quantum field  description one introduces so
called self-consistent approximations to keep the conservation
laws on exact level, see \cite{Baym,IKV,IKV2,KIV01,IKV3} and
discussion in Sect. \ref{sec:kin}.

Repeating the procedure that leads to Eqs.~(\ref{gt}) and
(\ref{gr}) from (\ref{scl-wf-scatt}) we obtain
\begin{eqnarray}
t^{\rm (scl)}_{\rm T} &=& \Big(\hbar\xi'(p)+\frac{1}{v_{\rm
I}}\frac{\rmd }{\rmd p}  \intop_{z_2}^z |\varkappa(z',E_p)|\rmd z'
\Big)\Bigg|_{z=z_2}- \Big(\hbar\xi'(p)+\frac{1}{v_{\rm
I}}\frac{\rmd }{\rmd p}   \int_{z_1}^z |\varkappa(z',E_p)|\rmd z'
\Big)\Bigg|_{z=z_1}= 0\,, \nonumber\\ t^{\rm (scl)}_{\rm R} &=&
\Big(\hbar\xi'(p)+\frac{1}{v_{\rm I}}\frac{\rmd }{\rmd p}
\intop_{z_1}^z |\varkappa(z',E_p)|\rmd z' \Big)\Bigg|_{z=z_1}-
\Big(\hbar\xi'(p)-\frac{1}{v_{\rm I}}\frac{\rmd }{\rmd p}
\int_{z_1}^z |\varkappa(z',E_p)|\rmd z' \Big)\Bigg|_{z=z_1}= 0\,.
\end{eqnarray}
We see that  {\em{ within  semiclassical approximation  $t^{\rm
(scl)}_{\rm T}=t^{\rm (scl)}_{\rm R}=t^{\rm (scl)}_{\rm bs}=0$,}}
if we compare the moments of time, when the maxima of the  packets
are at the turning points.  This  was first announced in
\cite{Campi} but basing on this fact concluded that the tunneling
time in semiclassical approximation is zero. In our opinion, being
zero, the quantity $t_{\rm T}^{\rm (scl)}$, as well as $t_{\rm
T}$, can hardly be considered as appropriate characteristic of the
time passage of the barrier. The values $t^{\rm (scl)}_{\rm
T}=t^{\rm (scl)}_{\rm R}=0$ just show that the delay of wave
packets within the region of finite potential appears due to
purely quantum effects, being vanishing in semiclassical
approximation. It also demonstrates that in case of the tunneling
the group delays are accumulated in the region near the turning
points where semiclassical approximation is not applicable.

Finally,  we repeat  that in general case the reflection group
time shows nothing else that a time delay between formation of the
peak of the reflected wave at $z=-L/2$ compared to the moment,
when the incident wave peak reached $z=-L/2$. The transmission
group time demonstrates difference of time moments, when the peak
of the transmission wave starts its propagation at $z=L/2$ and the
incident wave peak reaches $z=-L/2$. In semiclassical
approximation these time delays are absent.

\subsection{Sojourn time for  scattering of an arbitrary wave-packet}

So far we have considered the time-like quantities, which are
precise only to the extend that the packet has a small momentum
uncertainty, as the group times [Eqs.~(\ref{gt}), (\ref{gr}), and
(\ref{t_RT-sym})]~\cite{Hauge-Stoevneng}, and  the dwell time
[Eq.~(\ref{dwell})], originated  within stationary problem.
Nevertheless it is possible to introduce another  time-like
quantity, which measures how long the system stays within a
certain coordinate region. In classical mechanics the time, which
a system committing  1D motion spends  within the segment $[a,b]$,
is determined by the integral (\ref{sojt-class}). In quantum
mechanics the $\delta$-function over the classical trajectory is
to be replaced with the quantum probability density
$|\Psi(z,t)|^2$, see \cite{Baskin-Sokolovskii}. Now, if we
consider a wave packet starting from the left  at large negative
$z$ for large negative  $t$ and proceeding to $z=+\infty$, then
the time it spends within the segment $[a,b]$ is given by
\emph{the quantum mechanical sojourn time} defined as
\begin{eqnarray}
t_{\rm soj}(a,b)=\intop_{-\infty}^{+\infty}\rmd t\intop_a^b\rmd z |\Psi(z,t)|^2\,.
\label{soj-quant}
\end{eqnarray}
The packet wave function is normalized as (\ref{wp-norm}). Between
the dwell time and the sojourn time there is a
relation~\cite{LeavensAers}, see derivation in
Appendix~\ref{app:I-centroid1},
\begin{eqnarray}
&&t_{\rm soj}(a,b)
=\intop_{-\infty}^{+\infty} \frac{\rmd k}{2\pi\hbar} |\varphi(k)|^2 t_{\rm d}(a,b,k^2/2m)
=\big\langle t_{\rm d}(a,b,k^2/2m)\big\rangle_k\,.
\label{soj-dwell}
\end{eqnarray}
Using that the wave function obeying the Schr\"odinger equation
satisfies the continuity equation
\begin{eqnarray}
\frac{\rmd }{\rmd t}|\Psi(z,t)|^2=-\frac{\rmd}{\rmd z} j(z,t)\,,
\label{cont-eq}
\end{eqnarray}
where $j(z,t)=\mathcal{J}[\Psi(z,t)]$ with the current
$\mathcal{J}$  defined in (\ref{current}), we can rewrite  the
sojourn time through the currents on the borders of the interval
\begin{eqnarray}
t_{\rm soj}(a,b)= -\intop_{-\infty}^{+\infty}\rmd t
\intop_{-\infty}^{t} \rmd t' \big[ j(b,t') - j(a,t') \big]\,.
\label{tsoj-q}
\end{eqnarray}
From these relations we see that the sojourn time, has the same
deficiencies, as the dwell time. Namely, for a broad barrier both
quantities  demonstrate how long it takes for the particles to
enter the barrier from the left end, but they do not describe
particle transmission to the right end.

We now apply the relation (\ref{tsoj-q}) and the sojourn time
definition to the wave function (\ref{psi-f-1}). The total time,
which the packet spends in the barrier region, $-L/2\le z\le L/2$,
is $t_{\rm soj}(-L/2,L/2)=\big\langle t_{\rm
d}(-L/2,L/2,k^2/2m)\big\rangle_k$. As we show in
Appendix~\ref{app:I-centroid1}
the integration of
currents in Eq.~(\ref{tsoj-q}) gives~\cite{HFF87}
\begin{eqnarray}
\intop_{-\infty}^{+\infty}\rmd t\!\!\intop_{-\infty}^{t}\!\! \rmd
t' \Big(j(L/2,t')-j(-L/2,t')\Big)= -\big\langle|T(E)|^2\,t_{\rm
T}(E)\big\rangle_k -\big\langle|R(E)|^2\,t_{\rm R}(E)\big\rangle_k
+\big\langle \delta t_{\rm i}(E)\big\rangle_k\,.
\label{int-cur-soj}
\end{eqnarray}
Thus,
 in case of an arbitrary momentum distribution we obtain  generalization of Eqs.~(\ref{tD-tR-tT}) and
 (\ref{t-inter}):
\begin{eqnarray}
t_{\rm soj}(-L/2,L/2) = \big\langle t_{\rm d}(-L/2,L/2,E_k)\big\rangle_k
= \big\langle|T(E)|^2\,t_{\rm T}(E)\big\rangle_k
+ \big\langle|R(E)|^2\,t_{\rm R}(E)\big\rangle_k
-\big\langle \delta t_{\rm i}(E)\big\rangle_k\,.
\label{tD-tR-tT-soj}
\end{eqnarray}
Thereby, {\em{from definition of the sojourn time we extract the
same information as from definition of the dwell time but averaged
over energies of the packet.}} We stress that both quantities do
not describe time of the particle passage of the barrier.

\subsection{The Hartman effect}\label{sssec:hartman}

For energies above the barrier the proper time for the  particle
to pass the region of the potential is  the traversal time $t_{\rm
trav}^{\rm(cl)}$. Other times $t_{\rm d}(-L/2,L,E_p)$, $t_{\rm
soj}(-L/2,L,E_p)$, $t_{\rm bs}(E_p)$ introduced above also
appropriately characterize the particle motion. For energies well
above the barrier, $E\gg U$, we find $t_{\rm trav}^{\rm(cl)}\simeq
t_{\rm bs}(E_p)\simeq t_{\rm d}(-L/2,L,E_p)\simeq L/v_{\rm I}$.
However there appear problems with interpretation of all these
times, as a characteristic of a particle's passage under the
barrier (for $E<U$).

In numerous works the dwell time was interpreted as a mean time
the particle spends under the barrier regardless of whether it is
ultimately transmitted or reflected, see discussion in
Ref.~\cite{Winful}. The link (\ref{tD-tR-tT-2}) between the dwell
time and the group times suggested a naive interpretation of the
times $t_{\rm T}-\delta t_{\rm i}$ and $t_{\rm R}-\delta t_{\rm
i}$ as mean times the transmitted and reflected particles spend
under the barrier. As we mentioned, such an interpretation is
based on a classical counting of possible outcomes of a scattering
process in 1D, when an incident particle can be either reflected
or transmitted, cf. Sec.~IIIB in Ref.~\cite{Hauge-Stoevneng}.
Accepting such an interpretation of the group and dwell times one
encounters a paradox. In order to understand it  more clearly
consider tunneling ($E<U$) through a  thick rectangular barrier
$\varkappa L /\hbar=\sqrt{1-E_p/U}L/l_0\gg 1$. From
Eqs.~(\ref{rectang-delta-s}), (\ref{t_RT-sym}) and
(\ref{tD-tR-tT-2}) we find
\begin{eqnarray}
 &&t_{\rm
d}(-L/2,L/2,E_p) = t_{\rm T}(E_p)-\delta t_{\rm i}(E_p) =t_{\rm
R}(E_p)-\delta t_{\rm i}  (E_p)= \frac{p^2}{p^2+
\varkappa^2}\hbar\frac{\rmd \delta_{\rm s}(E_p)}{\rmd E_p},
 \nonumber\\
&&t_{\rm bs}(E_p)= t_{\rm T}(E_p) =t_{\rm R}(E_p) \simeq \hbar
\frac{\rmd \delta_{\rm s}(E_p)}{\rmd E_p}=2\frac{\hbar}{v_{\rm I}
\varkappa} \,. \label{Hart-time}
\end{eqnarray}
The characteristic length entering these expressions is  the quantum depth of particle
penetration inside the barrier region $\sim \hbar/\varkappa$
rather than the length of the barrier $L$. Therefore all these
{\em{ characteristic times are reduced to the quantum time $t_{\rm
quant}\sim \hbar/v_{\rm I}\varkappa$ not proportional to the
barrier length $L$,}} as one could expect for a proper passage
time of the distance $L$ with a constant velocity. This would mean
that, being evaluated with the help of these times, the average
velocity of the particle passage of the barrier would exceed the
speed of light for sufficiently large $L$. Such a phenomenon first
described in Ref.~\cite{Hartman} was then called the Hartmann
effect. The effect survives independently of the specific form of
the potential.  The same effect arises, if one uses the
relativistic Dirac and Klein-Gordon equations instead of the
Schr\"odinger equation \cite{Win}. As Winful writes \cite{Winful}:
"Because of this apparent superluminality,  there are some who
dismiss it as a relevant time scale for the tunneling process.
This is part of the motivation for the ongoing search of other
tunneling times."

The Hartman effect has not yet been observed for matter waves.
However, one has used the identity of the form of the Helmoltz
equation for wave propagation in a bulk inhomogeneous medium and
the time-dependent Schr\"odinger equation, and studied the
tunneling of electromagnetic waves through a barrier.
Reference~\cite{photo-tunnel}  reported that a superluminal
tunneling of light was observed, that caused a vivid discussion in
the literature, see for example reviews~\cite{Muga-I,Muga-II}. The
group velocity may become superluminal and even negative without
any contradiction with causality
\cite{Landauer-Martin,Carvalho,Chiao}. General
arguments~\cite{Segev2000} based on unitarity and causality show
that the peak of the transmitted pulse is constructed mainly from
leading edge of the incident one. Namely, pulse reshaping leading
to apparent causality was found in absorbing or amplifying media
whose relaxation times are long compared with pulse duration. For
thorough analysis of the Hartman effect and  re-interpretations of
the experiments free of problems with causality we refer the
reader to the review of Winful ~\cite{Winful}. The saturation of
the group delay with the barrier length is explained by the
saturation of the stored energy. The Winful's argument to avoid
the Hartman effect is that "the transmitted pulse is not the same
entity as the incident pulse." However such an interpretation does
not answer the question whether it is possible to get an
appropriate time for the passage of the barrier, which is
proportional to its length. This problem has not yet been solved.

Let us first formulate arguments why the group times and the dwell
time are not appropriate quantities to  measure the tunneling
time. First of all  the group times $t_{\rm T}$, $t_{\rm R}$ and
${t}_{\rm bs}$ ought to be understood as asymptotic quantities
(cf., Refs.\cite{Hauge-Stoevneng,LeavensAers,Landauer-Martin}),
which apply to events with distinct wave packets measured, in
reality, far from the barrier. Approaching the barrier, the
incident wave packet interferes with the reflected part of itself.
One can extrapolate the time to that the freely propagating
incident wave packet would need to arrive at the left border of
the potential region ($z=-L/2$) in absence of the reflection.
Similarly, the transmitted wave packet can be extrapolated to the
right border of the potential ($z=L/2$). One can of course
extrapolate further into the potential regions until the turning
points $z_1$ and $z_2$ determined by the equation
$U(z_{1,2})=E_p$. All that one can deduce from such extrapolations
is that, if the incident wave packet, being extrapolated from the
past, reaches the point $z=-L/2$ at $t=t_{-}$, then the peak of
the remote transmitted wave packet, being extrapolated backwards
from the future, occurs at the coordinate $z=L/2$ at the time
$t=t_{-}+t_{\rm T}$. One cannot say, where the transmitted wave
packet peak was at $t<t_{-}+t_{\rm T}$ and, thus, the group times
do not  measure the traveling time from input to output.
Reference~\cite{Winful} goes even further considering the incident
and transmitted wave packets as different entities arguing that
there is no obvious causal relation between the measurement of the
incident packet somewhere to the left from the barrier and the
measurement of the transmitted packet to the right from the
barrier. The problem is even more subtle, if considering
transmission one uses  centroids $\overline{z}_{\rm T}(t)$ and
$\overline{z}_{\rm R}(t)$ related to the transmitted, incident and
reflected wave packets, see below Sect. \ref{Centroids}.

Another argument is against the usage of Eq.~(\ref{tD-tR-tT-2})
for the interpretation of the group times $t_{\rm T}$ and $t_{\rm
R}$, as the transmission and reflection times~\cite{Winful}. The
counting of possible outcomes for the scattering of the packet on
the barrier, as being either transmitted or reflected, is not
valid for a quantum system: a wave packet can be both transmitted
and reflected. In quantum mechanics one sums complex amplitudes
rather than probabilities~\cite{Landauer-Martin}. As we argued on
example of the thick  barrier, values $t_{\rm bs}$ and $t_{\rm d}$
show time delays of the wave packet on the barrier edge $z=-L/2$,
not a life time of a stored energy within the whole barrier region
escaping from both sides of the barrier. For example, as follows
from Eqs.~(\ref{tdwell-1}), (\ref{tdwell-evan-grow}) and
(\ref{tdwell-cor}) the dwell time can be written as a
superposition of the dwell times constructed separately from the
evanescent wave and the growing wave, $t_{\rm d}^{\rm (evan)}$ and
$t_{\rm d}^{\rm (grow)}$, and their interference, $t_{\rm d}^{\rm
(\rm cor)}$. For tunneling through a thick barrier, $E<U$ and
$\varkappa L/\hbar \gg 1$, we have $t_{\rm d}^{\rm (grow) }/t_{\rm
d}^{\rm (evan)}\ll1$ and $t_{\rm d}^{\rm (cor) }/t_{\rm d}^{\rm
(evan)}\ll1$, therefore $t_{\rm d}\simeq t_{\rm d}^{\rm (evan)}$.
Thus, in this particular limit, knowing the value $t_{\rm d}$, one
may conclude only about a delay of reflection but one cannot say
about the delay of transmission. These comments also concern the
quantities (\ref{dr}), which, as argued in
Refs.~\cite{Winful,Winful03,Riza}, could characterize net-delays
of transmitted and reflected fluxes.

Concluding, as Winful \cite{Winful} writes: "The Hartman effect is
at the heart of the tunneling time conundrum. Its origin has been
a mistery for decades \cite{Bonifacio,Muga-I,Razavi}. Its
resolution would be of fundamental importance as it would lead to
conclusive answers regarding superluminality and the nature of
barrier tunneling." Our contribution to the resolution of the
Hartmann paradox is presented below. Since the origin of the
problem is that the semiclassical local current is absent for the
under barrier motion (there is no propagating packet  peak in
under the barrier motion), the solution is based on that the
particle transit time through the barrier should be associated
with the time variation of the amplitude of the waves under the
barrier, rather than with a particle flux there, information on
which one tries to extract considering motion of the peaks of the
transmitted and incident wave packets.

\subsection{Centroid transmission and reflection time delays and asymptotic motion of packets}\label{Centroids}

The quantities $t_{\rm T}$ and $t_{\rm R}$,  Eqs.~(\ref{gt}),(\ref{gr}) were found with the help of the stationary phase approach. These times characterize the time delays
within the segment $[-L/2,L/2]$  in transmission and reflection processes.
Defining them we used the assumption that the position of a particle can be identified with the position of the maximum of the wave packet. However, it is not so easy to experimentally
distinguish the peak position of a spatially broad packet. Moreover information not only about the spatial distribution in the packet but also about its average width is lost.

As a simple alternative, Hauge {\it et al.} proposed in Ref.~\cite{HFF87} to operate with the average coordinates of the packets to specify the position of the particle and to study the
motion of the centroids of the incident, transmitted, and reflected wave packets. The result depends on the packet width but only through average quantities $ \overline{z}_{\rm T,R,I}(t)$. A price for simplicity is a loss of an information about specific energy distribution within the packet that results in a loss of an information about specific spatial distribution on a spatial scale $\hbar/\gamma_p$. This is the minimum length characterizing the
centroid (compare  size  of a thick guy).

The average coordinate of the incident wave packet [Eq.~(\ref{wp-free})], i.e. {\it the incident centroid, $\overline{z}_{\rm I}(t)$},  is given by Eq.~(\ref{xI-aver}). After the collision  we deal with reflected and transmitted packets (\ref{twp}) and (\ref{rwp}) to the right ($z>L/2$) and to the left ($z<-L/2$) from the potential region, respectively. The average coordinates of these packets --- {\it the transmitted and reflected centroids} --- are defined as follows
\begin{eqnarray}
\overline{z}_{\rm T}(t) &=& \frac{\int_{L/2}^{+\infty} \rmd z\,
z\, |\Psi_{\rm T}(z,t)|^2} {\int_{L/2}^{+\infty} \rmd z |\Psi_{\rm T}(z,t)|^2}
\,,\qquad
\overline{z}_{\rm R}(t) = \frac{\int^{-L/2}_{-\infty} \rmd z \,z\,|\Psi_{\rm R}(z,t)|^2} {\int^{-L/2}_{-\infty} \rmd z |\Psi_{\rm R}(z,t)|^2}\,.
\label{centroids}
\end{eqnarray}
Note that in  relations (\ref{xI-aver}) and (\ref{centroids}) the centroid motions invoke the same time parameter $t$. The moments when  the particle enters the segment $[-L/2,L/2]$ or leaves it, being reflected or transmitted, can be specified  by the
requirement that the centroids are at some chosen positions nearby
or at the borders of the potential region. Comparing  these
moments of time one can determine the delay times of the particle
within the segment $[-L/2,L/2]$ during the reflection and
transmission processes. In this way it is possible to study the
corrections to the group times (\ref{gt}) and (\ref{gr}) induced
by the packet finite width and the change of the packet shape in
the process of tunneling and reflection.

The integrals in Eq.~(\ref{centroids}) is difficult to calculate
in general, as it requires a solution of the time-dependent Schr\"odinger equation. Nevertheless one can derive some rigorous results for the asymptotic time dependencies of the
centroids~\cite{HFF87}. If we consider sufficiently large times, for which $\overline{z}_{\rm T}(t)\gg L/2$ and $\overline{z}_{\rm I}(t),\overline{z}_{\rm R}(t)\ll -L/2$, then the transmitted packet is located almost entirely to the right from the region of
non-zero potential  and the reflected packet is to the left from it. In this case we can extend the integration limits in (\ref{centroids}) to $\pm\infty$ and define the asymptotic
centroids of transmitted and reflected packets
\begin{eqnarray}
 \overline{z}_{\rm T}^{\rm (as)}(t) = \frac{
\int_{-\infty}^{+\infty} \rmd z\,z\, |\Psi_{\rm T}(z,t)|^2}
{\int_{-\infty}^{+\infty} \rmd z |\Psi_{\rm T}(z,t)|^2} \,, \quad
 \overline{z}_{\rm
R}^{\rm (as)}(t) = \frac{\int^{+\infty}_{-\infty} \rmd z\,z\,
|\Psi_{\rm R}(z,t)|^2} {\int^{+\infty}_{-\infty} \rmd z |\Psi_{\rm
R}(z,t)|^2}\, \label{centroid-2}
\end{eqnarray}
in analogy to the asymptotic centroid of the incident packet
(\ref{xI-aver}). From the definition (\ref{centroids}) follows
that the centroid of the transmitted packet $\overline{z}_{\rm T}$
is an increasing function of the lower integration limit $L/2$,
indeed
\begin{eqnarray}
\frac{\prt}{\prt L} \overline{z}_{\rm T}(t;L) &=&
\frac12\, |\Psi(L/2,t)|^2 \frac{\int_{L/2}^{+\infty} \rmd z\,
(z-L/2)\, |\Psi(z,t)|^2} {\left(\int_{L/2}^{+\infty} \rmd z |\Psi(z,t)|^2\right)^2}
>0\,,
\end{eqnarray}
as an integral of two positive non-zero functions cannot be equal
zero. Similarly, we obtain that the centroids of the incident and
reflected packets are decreasing functions of the upper
integration limits, $-L/2$ in this case. From these inequalities
immediately follows that
\begin{eqnarray}
\overline{z}_{\rm T}^{\rm (as)}(t)\equiv \overline{z}_{\rm
T}(t;L=-\infty)<\overline{z}_{\rm T}(t;L/2) \,,\quad
\overline{z}_{\rm R(I)}^{\rm (as)}(t)\equiv \overline{z}_{\rm
R(I)}(t;L=-\infty)>\overline{z}_{\rm R(I)}(t;L/2)\,.
\end{eqnarray}
The difference between  Eq.~(\ref{centroids}) and
Eq.~(\ref{centroid-2}) increases, when $\overline{z}_{\rm T}^{\rm
(as)}(t)$ approaches $L/2$  and when $\overline{z}_{\rm R(I)}^{\rm
(as)}(t)$ approaches $-L/2$, and it becomes of the order of the
spatial packet width, i.e. $\sim \hbar/\gamma_{p}$, for the
incident packet and $\sim \hbar/\gamma_{p,\rm R(T)}$ for the
packets after scattering: the  widths of the transmitted and
reflected packets can deviate from the width of the incident
packet after interaction, as we shall see in Sect.~\ref{Wpexample}. The width of the momentum distribution for transmitted and reflected packets can be calculated as inverse
spatial width of the packet [in analogy to Eq.~(\ref{packet-width-x})]
\begin{eqnarray}
\frac{\hbar^2}{4\,\gamma_{p,\rm T}^2}= \overline{[z^{\rm (as)}_{\rm
T}(t)]^2}-[\overline{z}_{\rm T}^{\rm (as)}(t)]^2 \,,\quad
\frac{\hbar^2}{4\,\gamma_{p,\rm R}^2}= \overline{[z^{\rm (as)}_{\rm
R}(t)]^2}-[\overline{z}_{\rm R}^{\rm (as)}(t)]^2\,.
%
\label{RT-width-def}
\end{eqnarray}
Taking the packet widths into account, we can parameterize the centroid
motion as
\begin{eqnarray}
\overline{z}_{\rm T}(t)&=&\overline{z}_{\rm T}^{\rm (as)}(t) +\varsigma_{\rm T}\frac{\hbar}{\gamma_{p,\rm T}}
f_c\big(\gamma_{p,\rm T}\,|L/2-\overline{z}^{\rm (as)}_{\rm T}(t)|/\hbar\big)\,,\quad t>0\,,\quad \overline{z}_{\rm T}(t)>L/2
 \,,
\nonumber\\ \overline{z}_{\rm R}(t)&=&\overline{z}_{\rm R}^{\rm
(as)}(t) - \varsigma_{\rm R}\frac{\hbar}{\gamma_{p,\rm R}}
f_c\big(\gamma_{p,\rm R}\,|L/2+\overline{z}^{\rm (as)}_{\rm
R}(t)|/\hbar\big)\,, \quad t>0 \,,\quad \overline{z}_{\rm
R}(t)<-L/2\,, \nonumber\\ \overline{z}_{\rm
I}(t)&=&\overline{z}_{\rm I}^{\rm(as)}(t) - \varsigma_{\rm
I}\frac{\hbar}{\gamma_p} f_c\big(\gamma_p\,|L/2+\overline{z}^{\rm
(as)}_{\rm I}(t)|/\hbar\big)\,,\qquad \quad t<0,\quad
\overline{z}_{\rm I}(t)<-L/2\,, \label{z-aver-cor}
\end{eqnarray}
where $\varsigma_{\rm I(R,T)}$ are some positive constants of the
order of unity, and the transition to the  asymptotic motion is
controlled by the function $f_{c}(\zeta)$: $f_{c}(\zeta\lsim
1)\sim 1$ for $0\le\zeta\lsim 1$ and $f_{c}$ vanishes for
$\zeta\gg 1$. For the Gaussian momentum distribution
(\ref{Gauss-phi-prof}), which leads to the Gaussian spatial form
of the wave packets, we find $\varsigma_{\rm I}=\varsigma_{\rm
T}=\varsigma_{\rm R}=1/\sqrt{2\pi}$ and
$f_c(\zeta)=\exp(-2\zeta^2)/\mathop{\mathrm{erfc}}(-\sqrt{2}\zeta)$,
where $\mathrm{erfc}$ stands for the complementary error function.

We can proceed further with the evaluation of the integrals
(\ref{centroid-2}), see Appendix~\ref{app:I-centroid}, and obtain
the results for arbitrary momentum distribution
$\varphi(k)$~\cite{HFF87}:
\begin{eqnarray}
\overline{z}_{\rm T}^{\rm (as)}(t) &=&
\big\langle\big( -\hbar \xi'(k) -\hbar\phi_{\rm T}'(k)
+ t\,k/m\big)\big\rangle_{k,\rm T}\,,
\qquad
\overline{z}_{\rm R}^{\rm (as)}(t) =
\big\langle\big( \hbar \xi'(k) + \hbar\phi_{\rm R}'(k)
- t\,k/m\big)\big\rangle_{k,\rm R} \,.
\label{zT-zR}
\end{eqnarray}
Here we introduce the averaging over the momentum weighted with the transmission or reflection probability
\begin{eqnarray}
\langle (\dots ) \rangle_{k,\rm T} =\frac{\langle |T(E)|^2(\dots ) \rangle_{k}}
{\langle |T(E)|^2\rangle_{k}}\,,
\quad
\langle (\dots ) \rangle_{k,\rm R} =\frac{\langle |R(E)|^2(\dots ) \rangle_{k}}
{\langle |R(E)|^2\rangle_{k}}\,.
\label{RT-aver}
\end{eqnarray}
Recall that the momentum averaging $\langle\dots\rangle_k$ is defined in Eq.~(\ref{p-aver}).
The evaluation of the $T$-weighted $k$-averages (\ref{RT-aver}) for small $\gamma_p$ can be done with the help of the relations
\begin{eqnarray}
\langle f(k)\rangle_{k,\rm T}\approx f(p) +\half \gamma_p^2 \frac{\rmd^2f(p)}{\rmd p^2}
+ \gamma_p^2 \frac{\rmd f(p)}{\rmd p} \frac{\rmd}{\rmd p}\log|T(p)|^2\,.
\label{fT-aver}
\end{eqnarray}
Here the primes mean derivatives with respect to the momentum. The
analogous relation can  be written for the $R$-weighted
$k$-average. We can also use the relation, which holds up to the
order $\gamma_p^2$:
\begin{eqnarray}
\langle f(k)\, g(k)\rangle_{k,\rm T(R)}\approx \langle
f(k)\rangle_{k,\rm T(R)}\langle g(k)\rangle_{k,\rm T(R)}+
\gamma_p^2 f'(p)\, g'(p)\,, \label{fg-Taver}
\end{eqnarray}
where $f(k), g(k)$ are arbitrary functions.  For the packet widths
defined in Eq.~(\ref{RT-width-def}) we obtain now, using
Eqs.~(\ref{app:z2T-1}) and (\ref{app:z2R-1}) of
Appendix~\ref{app:I-centroid},
\begin{eqnarray}
\frac{\hbar^2}{\gamma_{p,\rm T}^2}&=& \left\langle
\hbar^2\,\Big[\frac{\rmd}{\rmd k}\log(|\varphi(k)||T(k)|)\Big]^2\right\rangle_{k,\rm T}
+\left\langle\Big(\hbar\xi'(k)+ \hbar\phi_{\rm T}'(k)
-\frac{k}{m}t\Big)^2 \right\rangle_{k,\rm T}
\nonumber\\
&-&\left\langle\hbar\xi'(k)+ \hbar\phi_{\rm T}'(k) -\frac{k}{m}t \right\rangle^2_{k,\rm T}\,,
\nonumber\\
\frac{\hbar^2}{\gamma_{p,\rm R}^2}&=& \left\langle
\hbar^2\,\Big[\frac{\rmd}{\rmd k}\log(|\varphi(k)||R(k)|)\Big]^2\right\rangle_{k,\rm R}
+\left\langle\Big(\hbar\xi'(k)+ \hbar\phi_{\rm R}'(k) -\frac{k}{m}t\Big)^2 \right\rangle_{k,\rm R}
\nonumber\\
&-&\left\langle\hbar\xi'(k)+ \hbar\phi_{\rm R}'(k) -\frac{k}{m}t
\right\rangle^2_{k,\rm R}\,.
\label{centroid-width}
\end{eqnarray}

First of all, from these expressions  we see that asymptotically
reflected and transmitted centroids move with velocities, which
differ from the velocity of the incident packet. From
(\ref{zT-zR}) we get
\begin{eqnarray}
v_{\rm R}=\frac{\rmd \overline{z}^{\rm (as)}_{\rm R}(t)}{\rmd t}=
-\Big\langle\frac{k}{m}\Big\rangle_{k,\rm R}=\frac{p_{\rm R}}{m}\,,
\quad
v_{\rm T}=\frac{\rmd \overline{z}^{\rm (as)}_{\rm T}(t)}{\rmd t}=
\Big\langle\frac{k}{m}\Big\rangle_{k,\rm T}=\frac{p_{\rm T}}{m}\,.
\label{vRvT-def}
\end{eqnarray}
For sufficiently narrow initial momentum distribution $\varphi(k)$
peaked around $p$ with the dispersion $\gamma_p$, see
Eqs.~(\ref{p0-gammap}), (\ref{f-aver}),  with the help of
Eq.~(\ref{f-aver}) we find
\begin{eqnarray}
v_{\rm R}\approx -v_{\rm I}\Big(1+
2\frac{\gamma_p^2}{m} \frac{\rmd }{\rmd E_p} \log |R(E_p)| \Big)
\,,\quad
v_{\rm T}\approx v_{\rm I}\Big(1+
2\frac{\gamma_p^2}{m} \frac{\rmd }{\rmd E_p} \log |T(E_p)| \Big)\,.
\label{vRvT-approx}
\end{eqnarray}
In case of the tunneling through a thick barrier the  expansion
(\ref{vRvT-approx}) holds for $\gamma_p \ll \sqrt{\hbar
|\varkappa|/L}$.

From Fig.~\ref{fig:Tsq-rect} we see that for $E_p <U$ (tunneling
regime)\footnote{For $U- E_p \lsim \gamma$ one cannot distinguish
between a tunneling regime and a particle motion above the
barrier. To deal with the pure tunneling one should assume that
$U- E_p \gg\gamma$.} the transmission amplitude $|T|$ is an
increasing function of energy, therefore the reflection amplitude
$|R|$ decreases with an energy increase. Hence in the tunneling
for the transmitted packet  $v_{\rm T}>v_{\rm I}$, while for the
reflected packet  $|v_{\rm R}|< v_{\rm I}$. The reason for this
phenomenon is obvious, the barrier acts as a filter letting  with
higher probability penetration for the modes with higher energies.
This serves as an  argument against a direct comparison of
characteristics of the transmitted and incident packets without a
normalization to the characteristics of the corresponding
stationary problem.

Using Eqs.~(\ref{fT-aver}) and (\ref{fg-Taver}) we can rewrite
expressions for the asymptotic centroid of the transmitted packet
(\ref{zT-zR}) as follows:
\begin{eqnarray}
\overline{z}^{\rm (as)}_{\rm T}(t)\approx \langle(-\hbar\xi'(k)) \rangle_{k,\rm T}
 +\frac{v_{\rm T}}{|v_{\rm T}|} L + v_{\rm T}\, \big(t-\overline{t}_{\rm T}\big) + v_{\rm I}\hbar \frac{\gamma_p^2}{m} \frac{\rmd^2\delta(E_p)}{\rmd E_p^2}\,,\quad
\overline{t}_{\rm T}=\left\langle\hbar \frac{\rmd\delta_{\rm s}(E)}{\rmd E} \right\rangle_{k,\rm T}\,.
\label{zT-g2}
\end{eqnarray}
We see that the centroid motion is delayed by the time
$\overline{t}_{\rm T}$, which is the averaged group time
(\ref{t_RT-sym}). From Eq.~(\ref{centroid-width}) the momentum
width of the transmitted packet including $\gamma_p^2$ corrections
is given by
\begin{eqnarray}
\frac{1}{\gamma_{p,\rm T}^2}\simeq
\frac{1}{\gamma_p^2}-2\frac{\rmd^2}{\rmd p^2} \log |T(k)|\,.
\label{width-g2}
\end{eqnarray}
The corresponding expressions for the reflected centroid differ only in the subindex ``R''.
Note that the second term in Eq.~(\ref{zT-g2}) will change the sign if we replace $v_{\rm T}$ with $v_{\rm R}$.

The {\em{ centroid transmission and reflection time delays}} can be defined as
\begin{eqnarray}
t^{\rm (cen)}_{\rm R} &=&  \tau^{(\rm R)}_- - \tau^{(\rm I)}_-\,,
\quad
t^{\rm (cen)}_{\rm T} =  \tau^{\rm (T)}_+ - \tau^{(\rm I)}_- \,.
\label{tauC-def}
\end{eqnarray}
where $\tau^{(\rm T)}_+$  is the time, when the transmitted packet
emerges to the right from the potential region, $\tau^{(\rm I)}_-$ is
the time, when the incident packet enters the potential region,
 and $\tau^{(\rm R)}_-$ is the time, when the reflected packet emerges
to the left from the potential region. Quantification of the
emergence moments requires some care. In
Refs.~\cite{HFF87,LeavensAers} the authors used the asymptotic
expressions (\ref{zT-zR}) and (\ref{xI-aver}) and extrapolated
them right up to the  borders of the region of non-zero potential
$z=\pm L/2$, going thereby beyond their application domain, since
the correction terms in Eqs.~(\ref{z-aver-cor}) cannot be
neglected for those $z$. Moreover, from the very definitions of
the centroids [Eq.~(\ref{centroids})] {\em one can easily see that
$\overline{z}_{\rm T}(t)$ can never reach the point $z=+L/2$ and
$\overline{z}_{\rm R}(t)$, the point $z=-L/2$.} To avoid this
problem we are forced to step away from the borders of the region
of non-zero potential by a quantity $\sim \hbar/\gamma_p$ and
define $\tau^{(\rm I)}_-$, $\tau^{(\rm R)}_-$, and $\tau^{(\rm
T)}_+$ from relations
\begin{eqnarray}
\overline{z}_{\rm I}^{\rm (as)}(\tau^{\rm (I)}_-)=-\frac{L}{2}- \tilde\varsigma_{\rm I}\frac{\hbar}{\gamma_p}\,,
\quad
\overline{z}_{\rm R}^{\rm (as)}(\tau^{\rm (R)}_-)=-\frac{L}{2}- \tilde\varsigma_{\rm R}\frac{\hbar}{\gamma_{p,\rm R}}\,,
\quad
\overline{z}_{\rm T}^{\rm (as)}(\tau^{\rm (T)}_+)=+\frac{L}{2}+ \tilde\varsigma_{\rm T}\frac{\hbar}{\gamma_{p,\rm T}}\,.
\label{t_pm-defin}
\end{eqnarray}
The constants $\tilde\varsigma_{\rm I,R,T}$ are positive and
$\tilde\varsigma_{\rm I,R,T}\sim \varsigma_{\rm I,R,T}\sim 1$.
Note that for the Gaussian packets at the time moments defined by these
conditions with $\tilde\varsigma_{\rm I,R,T}= \varsigma_{\rm I,R,T}$,
the maxima of the packets are located  exactly at the barrier borders $z=\pm L/2$.

Let us use such initial wave packet distributions  that correspond
to $\hbar\xi'(k)=z_0-k\, t_0/m$, see
Eqs.~(\ref{xI-aver}). Then the solutions of Eqs.~(\ref{t_pm-defin}) are
\begin{eqnarray}
\tau^{(\rm I)}_- &=&  - \frac{1}{v_{\rm I}}\,\Big( \frac{L}{2}+
\frac{\tilde\varsigma_{\rm I}\hbar}{\gamma_p}\Big)\,,
\nonumber\\
\tau^{(\rm R)}_- &=& -\frac{1}{v_{\rm R}} \Big( \frac{L}{2}+
\frac{\tilde\varsigma_{\rm R}\hbar}{\gamma_{p,\rm R}}\Big) -
\frac{1}{v_{\rm R}} \left\langle \frac{k}{m}\hbar\frac{\rmd}{\rmd
E}\phi_{\rm R}(E)\right\rangle_{k,\rm R} \,,
\nonumber\\
\tau^{\rm (T)}_+ &=&\frac{1}{v_{\rm  T}} \Big( \frac{L}{2}+
 \frac{\tilde\varsigma_{\rm T}\hbar}{\gamma_{p,\rm T}}\Big) +\frac{1}{v_{\rm T}} \left\langle
 \frac{k}{m}\hbar\frac{\rmd}{\rmd E}\phi_{\rm T}(E)\right\rangle_{k,\rm T}
 \,.
\label{t^RT}
\end{eqnarray}
Substituting Eq.~(\ref{ampl-phase}) into Eqs.~(\ref{t^RT}) for the
centroid reflection and transmission time delays (\ref{tauC-def})
we find
\begin{eqnarray}
t^{\rm (cen)}_{\rm R} &=& t_{\rm form,R} +\frac{L}{2}\frac{v_{\rm
I}-|v_{\rm R}|}{v_{\rm I}\,|v_{\rm R}|} +\frac{1}{|v_{\rm R}|}
\left\langle \frac{k}{m} \hbar\frac{\rmd \delta_s(E)}{\rmd E}\right\rangle_{k,\rm R} \,,
\nonumber\\
t^{\rm (cen)}_{\rm T} &=& t_{\rm
form,T} +\frac{L}{2}\frac{v_{\rm T} -v_{\rm I}}{v_{\rm I}v_{\rm
T}} + \frac{1}{v_{\rm T}} \left\langle \frac{k}{m} \hbar\frac{\rmd
\delta_s(E)}{\rmd E}\right\rangle_{k,\rm T} \,, \label{tauC}
\end{eqnarray}
where we introduced new quantities
\begin{eqnarray}
t_{\rm form, R}=t_{\rm form, I}
+\frac{\tilde\varsigma_{\rm R}\hbar}{\gamma_{p,\rm R}\,|v_{\rm R}|}
\,,\quad
t_{\rm form, T}=t_{\rm form, I}
+\frac{\tilde\varsigma_{\rm T}\hbar}{\gamma_{p,\rm T}\,v_{\rm T}}
\,,\quad
t_{\rm form, I}=\frac{\tilde\varsigma_{\rm I}\hbar}{\gamma_p\, v_{\rm I}}
\,.
\label{tquant-form}
\end{eqnarray}
which can be called {\em{the wave packet formation times.}} These
quantities characterize the time needed to the packets to
'complete the scattering event', i.e., enter the potential zone
and emerge from it. We note that $t^{\rm (cen)}_{\rm R}\neq t^{\rm (cen)}_{\rm T}$
even for symmetrical barrier in contrast to the group  times
(\ref{t_RT-sym}). These times show averaged passage times by
particles of the typical  spatial packet length $\hbar/\gamma_p$.

Due to performed averaging, dealing with centroids one loses an
information about specific form of spatial distribution in the
packet on a scale $\lsim\hbar/\gamma_p$, which could be extracted,
if one worked with not averaged spatial distributions. Mentioned
uncertainty is small provided formation times are shorter than
other quantities in (\ref{tquant-form}), for $\gamma_p \gg
|\varkappa|$, i.e. when the incident wave packet is very narrow in
space and broad in momentum. Then $t_{\rm form, R(T)}\ll t_{\rm
quant}\sim \hbar/v_{\rm I}|\varkappa|$  and the formation times in
Eq.~(\ref{tauC}) can be neglected. In this case the wave packet is
well localized spatially and the centroids can serve as
appropriate characteristics of the particle position. However,
unfortunately, for the case of large $\gamma_p$ we cannot anymore
speak about tunneling, since the large part of the wave packet
propagates above the barrier.

Contrary, for a very narrow momentum distribution ($\gamma_p \ll |\varkappa|$)  we would expect to recover previously obtained results for the group times (\ref{t_RT-sym}) and (\ref{gtil}). The latter quantities determined with the help of the packet peaks (by method of the stationary phase) do not depend on the widths of the packets. Therefore, to make both approaches
compatible we have to subtract from $\tau_{\rm R,T}$ formation times, being divergent for $\gamma_p\to 0$. This reflects the fact that the spatially broad packet needs a very long time to complete the scattering, in a line with uncertainty relation. In the limit
$\gamma_p\to 0$ from Eq.~(\ref{tauC}) we obtain
\begin{eqnarray}
v_{\rm T}\approx |v_{\rm R}|\approx v_{\rm I} \,,\quad t^{\rm (cen)}_{\rm
T}-t_{\rm form,T}\approx t^{\rm (cen)}_{\rm R}-t_{\rm form,T}\approx t_{\rm
bs}= \hbar\frac{\rmd \delta_s(E_p)}{\rmd E_p} \,.
\end{eqnarray}

Now, let us apply  results (\ref{vRvT-approx}) and (\ref{tauC}) to
the case of a narrow momentum distribution (small $\gamma_p$) and  a very thick rectangular barrier $\varkappa L/\hbar\gg 1$.
The transmission and reflection amplitudes (\ref{rectang-R-T}) and
their log-derivatives can be approximated as
\begin{eqnarray}
&&|T(E_p)|\approx \frac{4\,\varkappa\, p}{\varkappa^2+p^2} \,
e^{-\varkappa L/\hbar}\,, \quad |R|\approx 1 \,, \nonumber\\ &&
\frac{\rmd}{\rmd E_p}\log|T(E_p)|= \frac{m\, L}{\hbar\,\varkappa}
+\frac{m}{p^2}-\frac{m}{\varkappa^2}\,, \quad \frac{\rmd}{\rmd
E}\log|R(E_p)|=0\,, \label{broad-rect}
\end{eqnarray}
and
 \be
v_{\rm T}\simeq v_{\rm I}\left[1+2\gamma_p^2\left(\frac{L}{\hbar
\varkappa}+\frac{1}{p^2}-\frac{1}{\varkappa^2}\right)\right],\quad
v_{\rm R}\simeq v_{\rm I}\,.\label{broad-rectVel}
 \ee
 Recall that $p=\sqrt{2\, m\, E_p}$ and $\varkappa=\sqrt{2\,
m\, (U-E_p)}$\,. Since in case of a thick barrier the reflection
probability is close to unity, we have $v_{\rm R}\approx -v_{\rm
I}$.  Using Eq.~(\ref{f-aver}), for the centroid reflection time
delay we find
\begin{eqnarray}
t^{\rm (cen)}_{\rm R }-t_{\rm form,R} &\approx& \frac{1}{v_{\rm I}}
\left\langle\frac{k}{m}\hbar\frac{\rmd \delta_s(E)}{\rmd E} \right\rangle_k
\approx \hbar \frac{\rmd \delta_s(E_p)}{\rmd E_p}+
\frac{3\gamma_p^2}{2m}\hbar\frac{\rmd^2 \delta_s(E_p)}{\rmd E_p^2}
+ \frac{\gamma_p^2 p^2}{2\,m^2}\hbar\frac{\rmd^3 \delta_s(E_p)}{\rmd E_p^3}
\nonumber\\
&=& 2\,\frac{ \hbar}{v_{\rm I} \,\varkappa} \left( 1+\gamma_p^2\frac{\varkappa^2+3\,
p^2}{2\varkappa^4} \right)\,.
\end{eqnarray}
For the Gaussian wave packet the reflection packet formation time
coincides with the incident packet formation time  $t_{\rm
form,I}\simeq t_{\rm form,R}\simeq (\hbar\sqrt{2/\pi})/(\gamma_p\,
v_{\rm I})$.

Expression for the centroid transmission time delay is more
cumbersome. Using approximate relations (\ref{vRvT-approx}) and
the expansion
\begin{eqnarray}
\frac{1}{v_{\rm T}}
\left\langle \frac{k}{m}\hbar\frac{\rmd \delta_s(E)}{\rmd E}\right\rangle_{k,\rm T}
=\frac{\rmd \delta_s(E_p)}{\rmd E_p}
\Bigg[ 1 +2\gamma_p^2\,
\Bigg(\frac{\delta''_s(p)}{\delta'_s(p)}-\frac{1}{p}\Bigg)
\frac{\rmd }{\rmd p}\log |T(p)|
 + \gamma_p^2\frac{\delta'''_s(p)}{2\delta'_s(p)}
\Bigg],
\end{eqnarray}
we finally find
\begin{eqnarray}
t^{\rm (cen)}_{\rm T}-t_{\rm form,T}\simeq  \frac{2\,\hbar}{v_{\rm I}\,
\varkappa} \Bigg[1+ \gamma_p^2\Bigg(\frac{L}{\hbar}\, \left(
\frac{L}{\hbar}+\frac{\varkappa^2-p^2}{\varkappa\, p^2}
\right)+\frac{2L\,(p^2-\varkappa^2)}{\hbar\varkappa^3} +
\frac{9\,\varkappa^2\,p^2-4\varkappa^4-p^4}{\varkappa^4\, p^2}
 \Bigg) \Bigg]\,.
\label{tformT}
\end{eqnarray}
For the Gaussian wave packet the transmission quantum formation
time becomes
\begin{eqnarray}
t_{\rm form,T}\simeq \frac{\hbar\sqrt{2}}{\sqrt{\pi}\gamma_p\,
v_{\rm I}}
\left[1-\frac{\gamma_p^2}{2} \left(\frac{L}{\hbar\varkappa}\Big(1-\frac{p^2}{\varkappa^2}\Big)
+\frac{3\varkappa^4-\varkappa^2\,p^2+2\, p^4}{\varkappa^4\,p^2}\right)\right]\,.
\label{tformT1}
\end{eqnarray}
Expansions in (\ref{tformT}), (\ref{tformT1}) hold provided
$\frac{\hbar}{|z_{0}|}\ll \gamma_p \ll\hbar /L$. From these
expressions we may conclude that the centroid  transmission and
reflection time delays contain the formation times $t_{\rm
form,R(T)}$, as the largest times in the limit of small
$\gamma_p$, which arise because of the averaging over the spatial
packet distribution. The next-to-leading term (on the right-hand
side in Eq.~(\ref{tformT})) not depending on $\gamma_p$ coincides
with $t_{\rm T}$ given by Eq.~(\ref{Hart-time}). The Hartman
effect discussed above is described by this quantum term.
Corrections to the group times (\ref{gt}), (\ref{gr})), appeared
due to the finite packet width, invoke dependence on the  length
$L$ of the region of non-zero potential. Dependence on $L$ may
indicate  that the passage time of the barrier is proportional to
its length. The concepts of the group times introduced in the
previous section can be reliably used, if $\gamma_p \ll\hbar /L$.
Also, if these inequalities are fulfilled and $\frac{\hbar}{|z_{0}|}\ll \gamma_p $, we can exploit asymptotic centroids. The quantity
\begin{eqnarray}
\delta t_{\rm f}^{\gamma}=t^{\rm (cen)}_{\rm T}-t_{\rm form,I}\simeq \frac{2\hbar}{v_{\rm I}\varkappa} -\frac{\gamma_p}{\sqrt{2\pi}
v_{\rm I}} \frac{L}{\hbar\varkappa}\Big(1-\frac{p^2}{\varkappa^2}\Big)
\end{eqnarray}
has a meaning of the forward delay time, compare with Eqs.
(\ref{forwgamma}). The second (correction) term in the second
equality  is positive for $E>U/2$ and negative for $E<U/2$.

A more complete information about temporal behavior of the packets can be extracted from explicit forms of spatial distributions. To elucidate these aspects further, in the next section we consider a specific example of the propagation of the Gaussian momentum packet.

\subsection{Tunneling of the Gaussian wave packet}\label{Wpexample}

We consider now in details the tunneling of the packet with the
Gaussian envelop in the momentum space, see
Eq.~(\ref{Gauss-phi-prof}). To be sure that we operate really in
the  tunneling regime we have to keep $\gamma_p\ll \varkappa$,
$\varkappa >0$. Moreover we assume  that $\gamma_p\ll p$. Thus,
the integration over $k$ in Eqs.~(\ref{wp-free}),  (\ref{twp}) and
(\ref{rwp}) can be extended to $-\infty$. As in the previous
section we assume that $\hbar\xi'(k)=z_0-k\, t_0/m$ and we choose
the initial position of the packet $z_0$ and time $t_0$ such that
$\langle\hbar\xi'(k)\rangle_k=0$.

The probability densities to find a particle in the point $z$ at
the moment of time $t$ is given by $|\Psi_{>}(z,t)|^2=|\Psi_{\rm
T}(z,t)|^2$ for $z\ge L/2$ and by $|\Psi_{<}(z,t)|^2=|\Psi_{\rm
I}(z,t)|^2 + |\Psi_{\rm R}(z,t)|^2 +2\Re\big(\Psi^*_{\rm
I}(z,t)\Psi_{\rm R}(z,t)\big)$ for $z\le -L/2$, where the wave
functions are given by Eqs.~(\ref{wp-free}),~(\ref{twp}) and
(\ref{rwp}). The interference term in $\Re\big(\Psi^*_{\rm
I}(z,t)\Psi_{\rm R}(z,t)\big)$ is small, if $z$ is sufficiently
far from the left border of the potential,
$|z+L/2|\gg\hbar/\gamma_p$. Then the first term in
$|\Psi_{<}(z,t)|^2$ describes the free motion of a wave packet
with a Gaussian envelop and  equals to
\begin{eqnarray}
|\Psi_{\rm I}(z,t)|^2 &=&
\sqrt{\frac{2\gamma_{p,\rm I}^2(t)}{\pi\hbar^2}}
\exp\Big(-2\, \gamma_{p,\rm I}^2(t)\, \big(z-\widetilde{z}_{\rm I}(t)\big)^2/\hbar^2\Big)\,,\quad
\label{PsiI2}
\end{eqnarray}
where the time evolution of the packet centrum and the width are
determined  by
\begin{eqnarray}
\widetilde{z}_{\rm I}(t) &=& v_{\rm I}\, t\,,
\quad
\gamma_{p,\rm I}^2(t)={\gamma_p^2}/{\Big(1+4\gamma_p^4\frac{(t-t_0)^2}{m^2\hbar^2}\Big)}.
\end{eqnarray}
The packet becomes smeared on the time scale $t-t_0 \gsim t_{\rm
sm}=\hbar m/\gamma_p^2$. This corresponds to $z-z_0 \gsim v_{\rm
I}t_{\rm sm}$.  Further, to simplify expressions we will restrict
ourselves to the times $t-t_0\ll t_{\rm sm}$, and to the distances
$z-z_0\ll v_{\rm I}t_{\rm sm}$. For a particle moving not too far
from the barrier, the typical values of times and space
coordinates  are $t-t_0\sim |t_0|$ and  $z-z_0\sim |z_0|$. Then
assuming $|z_0|\ll \hbar p/\gamma_p^2$ we can neglect the smearing
of the wave packet and put further $\gamma_{p,\rm I}\simeq
{\gamma_p}$.

It is important to realize that for the packet described by
Eq.~(\ref{PsiI2}) freely moving through the spatial segment
$[-L/2,L/2]$, even for large $L$ there is small but finite
probability $|\Psi_{\rm I}\big(L/2,-L/(2v_{\rm
I})\big)|^2=\exp({-2\gamma_p^2 L^2/\hbar^2})$ to find the particle
at $z=L/2$, while the centre of the packet is still at the point
$z=-L/2$ and reaches the point $z=L/2$ only after the time
$L/v_{\rm I}$.

For the transmitted wave keeping the second derivatives of the
transmission amplitude  we find
\begin{eqnarray}
|\Psi_{\rm T}(z,t)|^2
&=& |T(p)|^2\frac{\gamma_{p,\rm T}}{\gamma_p}
\sqrt{\frac{2\gamma_{p,\rm T}^2(t)}{\pi\hbar^2}}
\exp\Big(-2\, \gamma_{p,\rm T}^2(t)\,\Big[ \big(z-\widetilde{z}_{\rm T}(t)\big)^2-\frac{\gamma^2_{p,\rm T}}{\gamma_{p,\rm T}^2(t)}l_{\rm T}^2\Big]/\hbar^2\Big)\,,
\label{PsiT2}
\end{eqnarray}
where the motion of the packet's centrum is described by equation
\begin{eqnarray}
\widetilde{z}_{\rm T}(t) &=& L + v_{\rm T}\,\big[t-({v_{\rm I}}/{v_{\rm T}})\,t_{\rm T}\big]
\, .
\label{tildeZT}
\end{eqnarray}
The speed of the transmitted packet $v_{\rm T}$ given here by
Eq.~(\ref{vRvT-approx}) is larger than the speed of the incident
packet.

 The
length $l_{\rm T}$ in Eq.~(\ref{PsiT2}) is
\begin{eqnarray}
l_{\rm T}=  \hbar \frac{\rmd}{\rmd p} \log|T(p)|. \label{lA}
\end{eqnarray}
 The width is time dependent at the order $\gamma_p^4$,
\begin{eqnarray}
\frac{1}{\gamma^2_{p,\rm T}(t)} &=& \frac{1}{\gamma_{p,\rm
T}^2}+4\, \gamma_{p,\rm T}^2\frac{(t-t_0-t_{\gamma,\rm
T})^2}{m^2\hbar^2}\,. \label{gammapT}
\end{eqnarray}
Here $\gamma_{p,\rm T}$ is given by the Eq.~(\ref{width-g2}), and
the time is delayed by the quantity
\begin{eqnarray}
t_{\gamma,\rm T}=m\,\hbar \frac{\rmd^2 }{\rmd p^2}\phi_{\rm T}(p)
= m\,\hbar \frac{\rmd^2 }{\rmd p^2}\delta_{s}(p)\, .
\end{eqnarray}
The time-dependent term follows directly from the last two terms
in the first equation in (\ref{centroid-width}), if we apply
Eq.~(\ref{fg-Taver}).

At the points $z\simeq \widetilde{z}_{\rm T}(t)\pm l_{\rm T}$
the exponent in Eq.~(\ref{PsiT2}) equals unity. For a broad
barrier  the tunneling amplitude $|T|$ can be presented as
\begin{eqnarray}
T(p)=t(p)\exp\Big(-\int_{z_1}^{z_2} \varkappa(z,E)\rmd z/\hbar\Big)\,,
\nonumber
\end{eqnarray}
cf. the semiclassical expression Eq.~(\ref{scl-wf-scatt}),
$z_{1,2}$ are the classical turning points and $t(p)$ is a
prefactor, which depends on $p$ rather slowly. The value  $l_{\rm
T}$ contains two terms
\begin{eqnarray}
l_{\rm T}=v_{\rm I}t_{\rm trav}^{\rm (tun)}+\delta l_{\rm T}\,,
 \quad t_{\rm trav}^{\rm (tun)}
=\int_{z_1}^{z_2}\frac{m\rmd z}{\varkappa(z,E)},\quad \delta
l_{\rm T}=\hbar t^{'}(p)/t(p) \,.\label{lA-decomp}
\end{eqnarray}
The quantity $t_{\rm trav}^{\rm tun}$  has the meaning of a
traversal time between the turning points provided $\varkappa >0$
, cf. Eq. (\ref{trav}). This is exactly the scale, which is
missing in the Hartman effect. The second term, $\delta l_{\rm
T}$, in (\ref{lA-decomp}) is of the order of a quantum length
scale, being much shorter for the thick barrier than the first
term.

For the
rectangular barrier in the limit $\varkappa L/\hbar\gg 1$ [see,
Eq.~(\ref{broad-rect})] we have explicitly
\begin{eqnarray}
l_{\rm T}=v_{\rm I} t_{\rm trav}^{\rm (tun)}+ \frac{\hbar}{p}
\frac{\varkappa^2-p^2}{\varkappa^2}\,. \label{lA-rect}
\end{eqnarray}
Here $t_{\rm trav}^{\rm (tun)} =mL/\varkappa$. Using
Eqs.~(\ref{width-g2}), (\ref{lA}) and (\ref{lA-rect}) we find how
the packet width changes after tunneling through the broad
rectangular barrier
\begin{eqnarray}
\frac{1}{\gamma^2_{p,\rm T}}\approx\frac{1}{\gamma^2_{p}} +
\frac{2}{\hbar} \frac{\rmd l_T}{\rmd p} \approx
\frac{1}{\gamma^2_{p}}+ 2
\Big(\frac{L}{\hbar\varkappa}+\frac{1}{p^2}\Big)
\Big(1+\frac{p^2}{\varkappa^2}\Big)\,. \label{gammaT-broad-rect}
\end{eqnarray}
Thus, the longer is the barrier, the broader becomes the
transmitted wave packet forming for $z\geq L/2$. Further we
continue to assume that  $|z_0|\ll \hbar p/\gamma_p^2$, and we
assume  that $L\ll \hbar \varkappa/\gamma_p^2$.

For the  reflected wave packet Eqs. (\ref{PsiT2}),
(\ref{tildeZT}), (\ref{gammapT}) and (\ref{lA}) can be applied
after the formal replacement of indices "T"$\to$"R" and the
scattering amplitude $T(p)\to R(p)$; the reflected-packet centrum
moves according to $ \widetilde{z}_{\rm R}(t) = -L - |v_{\rm
R}|\,\big[t-({v_{\rm I}}/{|v_{\rm R}|})\,t_{\rm R}\big],$ where we
take into account that $v_{\rm R}<0$. For the broad barrier we can
put $|v_{\rm R}|\simeq v_{\rm I}$ and $|R|\simeq 1$, so that
$\gamma_{p,\rm R} =\gamma_p$ and $l_{\rm R}=0$, and we write
\begin{eqnarray}
|\Psi_{\rm R}(z,t)|^2 &=& \sqrt{\frac{2\gamma_{p}^2}{\pi\hbar^2}}
\exp\Big(-2\, \gamma_{p}^2\big(z-\widetilde{z}_{\rm
R}(t)\big)^2/\hbar^2\Big) \,,\quad \widetilde{z}_{\rm R}(t)=-L -
v_{\rm I}\,(t-t_{\rm R})\,.
\label{PsiR2}
\end{eqnarray}
The peak of the reflected wave packet is formed at $z=-L/2$ at the
time moment $t(-L/2)=-L/2v_{\rm I}+t_{\rm R}$, $t_{\rm R}=t_{\rm
T}$, i.e. with a delay $t_{\rm T}$ compared to the time moment
when the incident packet reached $z=-L/2$. For the thick
rectangular barrier $t(-L/2)=-L/2v_{\rm I}+2\hbar/\varkappa v_{\rm
T}$.

Comparing  Eqs.~(\ref{PsiI2}) and (\ref{PsiT2}) we observe that
the tail of the transmitted wave packet begins to be formed for
$z\geq L/2$ already at time, when the maximum of the incident wave
packet has not yet reached the point $z=-L/2$. For example, if the
incident packet is at some coordinate $z'<-L/2$ at the time
$z'/v_{\rm I}$, the relative probability of the transmitted packet
to be at $z=+L/2$ is equal to
\begin{eqnarray}
\frac{|\Psi_{\rm T}(L/2,z'/v_{\rm I})|^2} {|\Psi_{\rm
I}(z',z'/v_{\rm I})|^2} \approx |T(p)|^2 \frac{\gamma_{p,\rm
T}^2}{\gamma_p^2} \exp\Big(2\,\gamma_{p,\rm T}^2 l_{\rm
T}^2/\hbar^2\Big) \exp\Big(-2\frac{\gamma_{p,\rm
T}^2}{\hbar^2}\Big[\half L + \frac{v_{\rm T}}{v_{\rm I}}\,z'-
v_{\rm I} t_{\rm T}\Big]^2\Big)\,.
\end{eqnarray}
However, the maximum of the transmitted packet does not emerge
from under the barrier even at the moment, when the
incident-packet's maximum is at the left border of the potential.
Indeed, from Eq.~(\ref{tildeZT}) we see that it happens, when
$t(L/2)=-(L/2 -v_{\rm I}t_{\rm T})/ v_{\rm T}$. At free
propagation at this moment the incident packet would be at $z=
-(L/2 -v_{\rm I}t_{\rm T})\, v_{\rm I}/v_{\rm T}>-L/2$ since
$v_{\rm I}<v_{\rm T}$.
 For the thick rectangular barrier
$t(L/2)=-L/2v_{\rm T}+2\hbar/\varkappa v_{\rm T}$. For
$L\gamma_p\ll \hbar$  the formation of the transmitted wave packet
peak at $z=L/2$ is delayed compared to the moment, when the
incident one arrives at $z=-L/2$, by the quantum time $\Delta
t\simeq 2\hbar/\varkappa v_{\rm I}$ and transmitted peak at
$z=L/2$ is formed approximately at the same time (at negligible
$\gamma_p$), as the reflected wave packet peak at $z=-L/2$ (the
Hartman effect). However for $L\gamma_p\gg \hbar$ the same
difference of times is approximately $\Delta t\simeq
L^2\gamma_p^2/\hbar \varkappa v_{\rm I}$, i.e. it depends on $L$.

The finite width of a packet describing a moving particle alters
the notion of the particle being at some spatial point. The
probability to find particle at given point becomes essentially
non-zero already before the centre of the packet has reached it,
with an advancement
\begin{eqnarray}
\label{gammadec} t_{\rm dec}^{\gamma} =\frac{\hbar}{v_{\rm I}\,
\gamma_p}= \frac{\hbar}{\gamma}\,,
\end{eqnarray}
where $\gamma_p$ and $\gamma$ are the momentum and energy
dispersions in the Gaussian packet given by Eqs.~(\ref{p0-gammap})
and (\ref{disp-E}), respectively. This is in accord with the
uncertainty principle derived by Mandelstam and Tamm in
Ref.~\cite{Tam-Tamm},
\begin{eqnarray}
\label{uncert}
 \Delta E\Delta {\cal T} \geq \frac{\hbar}{2}|\overline{\dot{\cal T}}|, \quad
 \Delta E=\left(\overline{(E-\bar{E})^2}\right)^{1/2}, \quad
 \Delta {\cal T}=\left(\overline{({\cal T}-\bar{\cal T})^2}\right)^{1/2}\,,
\end{eqnarray}
where ${\cal T}$ is a physical quantity not dependent on time
explicitly; the bar means quantum-mechanical averaging. The value
$\delta t_{\rm var}=|\Delta {\cal T}/\overline{\dot{{\cal T}}}|$
is the {\it variation time} during which the observable ${\cal T}$
changes its value more than its dispersion. Taking the coordinate
as the observable ${\cal T}$,  $\Delta{\cal T}$ as average spatial
width and $\Delta E\sim \gamma$ as energy width, for the Gaussian
wave packet under consideration we conclude that minimal time the
packet needs to pass a certain space point is $\sim \delta t_{\rm
var}$. Thus, the minimal duration of the emerging of the
transmitted wave packet on the right side of the barrier is
$\delta t_{\rm var} \sim \hbar/\gamma =t^{\gamma}_{\rm dec}$.
 The same time the incident packet needs
to enter the barrier on the left side. Since the information may
reach the given point with an advancement $\sim \delta t_{\rm
var}$ the real (forward) delay/advance time for the particles of
the transmitted packet is not $t_{\rm T}=t_{\rm bs}$ but should be
counted from $2\,t_{\rm dec}^{\gamma}$:
\begin{eqnarray}
\delta t_{\rm f}^{\rm (tun)}\equiv t_{\rm bs} -2\, t_{\rm
dec}^{\gamma}. \label{forward}
\end{eqnarray}

{\em Now consider propagation of the wave packet inside a
rectangular barrier.} The internal wave function
(\ref{internal-wf}) with the coefficients (\ref{rectan-C^2})
contains the growing and evanescent parts
\begin{eqnarray}
\psi_U(z,E)&=&  \psi_{\rm grow}+\psi_{\rm evan}= D_{+} (k)\,
e^{\varkappa\, z/\hbar} + D_{-}(k) \, e^{-\varkappa\, z/\hbar}\,,
\nonumber\\ D_{\rm \pm}(k) &=& \half (C_+ \pm C_-) = \half (1\pm i
k/\varkappa)\, T(k)\, e^{(ik\mp \varkappa)L/2\hbar} \nonumber\\
&&=\frac12\sqrt{1+\frac{k^2}{\varkappa^2}}\, T(k) \,e^{\pm
i\beta_{\varkappa}(k)+(ik\mp\varkappa)L/2\hbar}\,,\qquad
\beta_{\varkappa}(k)=\arctan(k/\varkappa).
\label{inter-WF-broad-rect}
\end{eqnarray}
Hence the internal wave packet
\begin{eqnarray}
\Psi_{U}(z,t)=\intop_0^\infty \frac{\rmd k}{2\pi\hbar}\, \varphi(k)\,
\psi_U(z,E)e^{-iEt/\hbar}
\end{eqnarray}
can be written as the sum of growing and evanescent parts and their interference
\begin{eqnarray}
|\Psi_{U}(z,t)|^2 =|\Psi_{\rm grow}(z,t)|^2 + |\Psi_{\rm evan}(z,t)|^2 +
2\Re\{\Psi_{\rm grow}(z,t)\Psi^*_{\rm evan}(z,t)\}.
\label{PsiU2}
\end{eqnarray}
First two terms do not contribute to the current density.
The current conservation for the particle motion  under the barrier is due to
the presence of the interference term.

Introducing the phase of the amplitudes for the growing and
evanescent parts of the wave function
\begin{eqnarray}
\phi_{D,\pm}(k)=\arg D_\pm(k)=\phi_{\rm T}(k)+\frac{kL}{2\hbar}\pm
\beta_{\varkappa}(k) \label{phiD}
\end{eqnarray}
we can cast the growing part, $|\Psi_{\rm grow}(z,t)|^2\equiv
|\Psi_{+}(z,t)|^2$, and the evanescent part, $|\Psi_{\rm
evan}(z,t)|^2\equiv |\Psi_{-}(z,t)|^2$, in the form
\begin{eqnarray}
|\Psi_{\pm}(z,t)|^2&=&\frac14\Big(1+\frac{p^2}{\varkappa^2}\Big)|T(p)|^2
e^{\pm 2\varkappa\, (z-L/2)/\hbar}
\frac{\gamma^{(\pm)}_{p,D}}{\gamma_p}\sqrt{\frac{2\gamma_{p,D}^{(\pm)2}(t)}{\pi\hbar^2}}
\exp\Big[2\, \frac{\gamma_{p,D}^{(\pm)2}}{\hbar^2}
\Big(l^{(\pm)}_D \mp z\, \frac{p}{\varkappa}\Big)^2\Big]
\nonumber\\ &&\times \exp\Big[ -
2\frac{\gamma_{p,D}^{(\pm)2}(t)}{\hbar^2}  \,\Big( v_{\rm I}\, (t- t^{(\pm)}_D) + 2\, \gamma_{p,D}^{(\pm)2}
\frac{t-t^{(\pm)}_{\gamma,D} }{m\, \hbar}\Big(l^{(\pm)}_D \mp z\,\frac{p}{\varkappa}\Big)
 \Big)^2 \Big]\,.
\label{PsiGD2}
\end{eqnarray}
The spatial widths of the internal packets are given by
\begin{eqnarray}
\frac{1}{\gamma^{(\pm)2}_{p,D}} &=& \frac{1}{\gamma_p^2}-2\left[\frac{\rmd^2}{\rmd p^2}\log|D_\pm(p)|
-\frac{z}{\varkappa\hbar}\Big(1+\frac{p^2}{\varkappa^2}\Big)
\right]
\nonumber\\
&=& \frac{1}{\gamma_{p,\rm T}^2}
-\frac{2}{\varkappa^2}\Big(1+2\frac{p^2}{\varkappa^2}\Big)-
2\frac{z\pm L/2}{\varkappa\hbar}\Big(1+\frac{p^2}{\varkappa^2}\Big)\,,
\label{gammaDpm}
\end{eqnarray}
thus acquiring  a weak time dependence
\begin{eqnarray}
&&\frac{1}{\gamma^{(\pm)2}_{p,D}(t)}=\frac{1}{\gamma_{p,D}^{(\pm) 2}}+4\gamma_{p,D}^{(\pm)2}
\frac{(t-t_0-t^{(\pm)}_{\gamma,D})^2}{m^2\, \hbar^2}\,, \quad
t^{(\pm)}_{\gamma,D}=m\hbar\frac{\rmd^2\phi_{D,\pm}(p)}{\rmd p^2}=
m\hbar\frac{\rmd^2\delta_{s}(p)}{\rmd p^2}\pm \frac{m\hbar p}{\varkappa^3}
\,,
\label{gammaDpm-t}
\end{eqnarray}
operating on the large time scales for $t-t_0\gg t_{\rm sm}$, where the smearing time $t_{\rm sm}$ is given by Eq.~(\ref{tsmear}). On the shorter time scales and for $\gamma_p^2
L/\varkappa \hbar\ll 1$ we can approximate $\gamma^{(\pm)}_{p,D}(t)\approx \gamma_{p,\rm T}$.
Then Eq.~(\ref{PsiGD2}) simplifies as follows
\begin{eqnarray}
|\Psi_{\pm}(z,t)|^2&\approx&\frac14\Big(1+\frac{p^2}{\varkappa^2}\Big)|T(p)|^2 e^{\pm 2\varkappa\, (z-L/2)/\hbar}\sqrt{\frac{2\gamma_{p,\rm T}^{4}}{\pi\hbar^2\gamma_p^2}}
\nonumber\\
&&\times
\exp\Big(2\frac{\gamma_{p,\rm T}^{2}}{\hbar^2} \,
\Big[ \big(l^{(\pm)}_D \mp z\, p/\varkappa\big)^2-v_{\rm                                                                                                                                                 I}^2\, \big(t- t^{(\pm)}_D\big)^2\Big] \Big)\,.
\label{PsiGD2-short}
\end{eqnarray}
We see that in this approximation the time dependence decouples completely from the spatial dependence. As time elapses starting from negative values the profile of the probability density increases as a whole, reaches the maximum at $t=t_D^{(\pm)}$ and then decreases for $t>t_D^{(\pm)}$ on the time scale $t_{\rm dec}^\gamma$, see Eq.~(\ref{gammadec}). Hence the probability to find a particle inside the barrier decreases with passage of time
on a typical time scale $t_{\rm dec}^{\gamma}$. The increase follows the approach of the incident packet with the time delay $t_D^{(\pm)}=\hbar \frac{\rmd \phi_{D,\pm}(E_p)}{\rmd E_p}$. Using the definitions of the phases $\phi_{D,\pm}$ and $\phi_{\rm T}$
from Eqs.~(\ref{phiD}) and (\ref{ampl-phase}) and
$\beta_{\varkappa}$ from Eq.~(\ref{inter-WF-broad-rect}) we can
write the time delay through the transmitted group
time~(\ref{t_RT-sym}) as
\begin{eqnarray}
t^{(\pm)}_D =
\hbar \frac{\rmd \phi_{\rm T}(E_p)}{\rmd E_p} +\frac{L}{2\,v_{\rm I}}\pm \frac{\hbar}{v_{\rm I}\varkappa}
=t_{\rm T} - \frac{L}{2\,v_{\rm I}}\pm \frac{\hbar}{v_{\rm I}\varkappa}\,.
\end{eqnarray}
We see that for the thick barriers, $\varkappa L/\hbar\gg 1$, we
deal with the time advance $t^{(\pm)}_D \simeq  - L/2v_{\rm I}$.
For the thick rectangular barrier $t^{(+)}_D\simeq  - L/2v_{\rm
I}+3\hbar/v_{\rm I}\varkappa$, $t^{(-)}_D\simeq  - L/2v_{\rm
I}+\hbar/v_{\rm I}\varkappa$. Thus  the maximum of the probability
for evanescent and growing waves delays compared to the peak of
the incident wave reached $z=-L/2$ respectively by the time steps
$\hbar/v_{\rm I}\varkappa$ and $3\hbar/v_{\rm I}\varkappa$.

As follows from Eq.~(\ref{PsiGD2-short}), because of the finite
width of the momentum distribution, the probability density is
modulated by the factor $\exp\big(2\gamma_{p}^{2}\big(l^{(\pm)}_D
\mp z\, p/\varkappa\big)^2/\hbar^2\big)$ with the characteristic
length
\begin{eqnarray}
l^{(\pm)}_{D}=\hbar \frac{\rmd}{\rmd p}\log|D_\pm(p)|= l_{\rm T}+\frac{p}{\varkappa}
\Big(\frac{\hbar }{\varkappa}\mp\frac{L}{2}\Big) \,.
\label{lD}
\end{eqnarray}
For the growing wave this factor is maximal at $z=-L/2$, and for
the evanescent wave, at $z=+L/2$. For the broad barrier the
characteristic length is equal to
\begin{eqnarray}
l^{(\pm)}_{D}=\big(1\mp\half)\, v_{\rm I}\, t_{\rm trav}^{(\rm
tun)} + \frac{\hbar}{p}  \,. \label{lD-broad}
\end{eqnarray}
Thus, the length of the barrier enters the internal wave through
the time delay $t_{D,\pm}$ and the length $l^{(\pm)}_{D}$.

For completeness we give also the expression for the last interference term in Eq.~(\ref{PsiU2}) in the limit $\gamma_p^2 L/\varkappa\hbar\ll 1$
\begin{eqnarray}
&&\Re\{\Psi_{\rm grow}(z,t)\Psi_{\rm evan}^*(z,t)\}
\approx \Re\{D_+(p)D_-^*(p)\} \, \sqrt{\frac{2 \,\gamma_p^2}{\pi\,\hbar^2}}
\nonumber\\
&&\qquad
\times\exp\Big(2\frac{\gamma_p^2}{\hbar^2}\Big[
\quart\big(l^{(+)}_D + l^{(-)}_D\big)^2- v^2_{\rm I} \big(t -\half(t^{(+)}_D + t^{(-)}_D)\big)^2
+\frac{p^2}{\varkappa^2}(z+L/2)^2
-\frac{\hbar^2}{4\varkappa^2}
\Big]
\Big)\,.
\end{eqnarray}

From Eqs.~(\ref{PsiT2}) and (\ref{PsiGD2-short})  we see that the probability to find particle inside the barrier ($-L/2\le z\le L/2$) and tunneled through it (at $L/2<z$) is enhanced compared to the case of the monochromatic wave with $E=E_p$.
To quantify these enhancements we introduce the following factors for the transmitted, growing and evanescent wave packets
\begin{eqnarray}
C_{\rm T}(z,t) =& \frac{|\Psi_{\rm T}(z,t)|^2}
{|T(p)|^2\,|\Psi_{\rm I}(\widetilde{z}_{\rm I}(t),t)|^2}
&=\frac{\gamma^2_{p,\rm T}}{\gamma_p^2}
\exp\big(2\, \frac{\gamma_{p,\rm T}^2}{\hbar^2}\,
\big[l_{\rm T}^2- \big(z-\widetilde{z}_{\rm T}(t)\big)^2\big]\big)\,,
\label{CT}\\
C_{\rm grow}(z,t) =& \frac{e^{-2\varkappa z/\hbar}|\Psi_{+}(z,t)|^2}
{|D_+(p)|^2\,|\Psi_{\rm I}(\widetilde{z}_{\rm I}(t),t)|^2}
&=\frac{\gamma^2_{p,\rm T}}{\gamma_p^2}
\exp\Big(2\frac{\gamma_{p,\rm T}^{2}}{\hbar^2} \,
\Big[ \big(l^{(+)}_D - z\frac{p}{\varkappa}\big)^2-v_{\rm I}^2\,
\big(t- t^{(+)}_D\big)^2\Big] \Big)\,,
\label{Cgrow}\\
C_{\rm evan}(z,t) =& \frac{e^{+2\varkappa z/\hbar}|\Psi_{ -}(z,t)|^2}
{|D_-(p)|^2\,|\Psi_{\rm I}(\widetilde{z}_{\rm I}(t),t)|^2}
&=\frac{\gamma^2_{p,\rm T}}{\gamma_p^2}
\exp\Big(2\frac{\gamma_{p,\rm T}^{2}}{\hbar^2} \,
\Big[ \big(l^{(-)}_D + z\frac{p}{\varkappa}\big)^2-v_{\rm I}^2\, \big(t-
t^{(-)}_D\big)^2\Big] \Big)\,.
\label{Cevn}
\end{eqnarray}
These enhancements occur owing to the fact that for the waves with
$E>E_p$ entering the packet the probability of penetration of the
barrier is larger than for the single wave with $E=E_p$.
Thus, analyzing the temporal aspects of the tunneling problem, we have to make a benchmark on the tunneling probability for the monochromatic wave. As follows from (\ref{CT}), the probability to meet the particle at $z=L/2$ becomes the same, as it were in  case of the monochromatic wave with $E=E_p$, for the first time on the right wing of the Gaussian at the time moment
\begin{eqnarray}
t_{\rm mon}^{(\rm r.w.)} =- \frac{l_{\rm T}}{v_{\rm
T}}-\frac{L}{2\, v_{\rm T}} + \frac{v_{\rm I}}{v_{\rm T}}\,t_{\rm
T} \,,
\end{eqnarray}
when the maximum of the incident wave packet is yet at
$z=\widetilde{z}_{\rm I}( t_{\rm mon}^{(\rm r.w.)})=v_{\rm I}
t_{\rm mon}^{(\rm r.w.)}<-L/2$ and the maximum of the transmitted
wave packet did not yet appear at $z=L/2$. Recall the traversal
time $t_{\rm trav}^{(\rm tun)} = m\,L/\varkappa$ is determined, as
in (\ref{lA-decomp}).
At the later time, the probability again becomes
the same on the left wing of the Gaussian at the time
\begin{eqnarray}
t_{\rm mon}^{(\rm l.w.)} = + \frac{l_{\rm T}}{v_{\rm
T}}-\frac{L}{2\, v_{\rm T}} + \frac{v_{\rm I}}{v_{\rm T}}\,t_{\rm
T} \,,
\end{eqnarray}
when the maximum of the transmitted wave packet achieves the point
$z=\widetilde{z}_{\rm T}( t_{\rm mon}^{(\rm l.w.)})=L/2+l_{\rm T}
$. Thus $\frac{1}{2}(t_{\rm mon}^{(\rm l.w.)}-t_{\rm mon}^{(\rm
r.w.)})=\frac{l_{\rm T}}{v_{\rm T}}$.

For the thick barrier:
\be
&&t_{\rm mon}^{(\rm r.w.)} =\frac{v_{\rm I}}{v_{\rm T}} \Big[-
t_{\rm trav}^{(\rm tun)} -\frac{L}{2\, v_{\rm I}}
-\frac{m\hbar}{p^2} + \frac{m\hbar}{\varkappa^2} +
\frac{2m\hbar}{p\varkappa}\Big],\nonumber\\ &&t_{\rm mon}^{(\rm
l.w.)} =\frac{v_{\rm I}}{v_{\rm T}} \Big[ t_{\rm trav}^{(\rm tun)}
-\frac{L}{2\, v_{\rm I}} +\frac{m\hbar}{p^2} -
\frac{m\hbar}{\varkappa^2} +\frac{2m\hbar}{p\varkappa}\Big]\,,
 \ee
$z=\widetilde{z}_{\rm T}( t_{\rm mon}^{(\rm l.w.)})=L/2+v_{\rm I}
t_{\rm trav}^{\rm (tun)}+ \hbar
(\varkappa^2-p^2)/(p\,\varkappa^2)$.

Note that working within the assumptions $|z_0|\ll \hbar
p/\gamma_p^2$ and  $L\ll \hbar \varkappa/\gamma_p^2$  we can use
$\gamma_{p,\rm T}(t)\simeq \gamma_{p,\rm D}(t)\simeq \gamma_p $
and $v_{\rm T}\simeq v_{\rm I}$ up to $1+O(\gamma_p^2)$
corrections. Finally for a thick barrier
 \be
\frac{1}{2}(t_{\rm mon}^{(\rm r.w.)}-t_{\rm mon}^{(\rm
l.w.)})\simeq t_{\rm trav}^{(\rm tun)}\simeq t_{\rm mon}^{(\rm
l.w.)}-t(\widetilde{z}_{\rm I}=-\frac{L}{2})\simeq  t_{\rm
mon}^{(\rm l.w.)}-t(\widetilde{z}_{\rm T}=\frac{L}{2}) .
 \ee
Up to small correction terms, this is the  difference of the time,
when the wave with $E\simeq E_p$ has passed the barrier and the
time, when the incident packet peak has reached it. On the other
hand it can be treated as the  difference of the time, when the
wave with $E\simeq E_p$ has passed the barrier and the time, when
the transmitted packet peak has been formed at the same point (on
its right boarder).

The above analysis allows us to reconsider the definition of the
transmission time through the broad barrier. If we are interested
in the time, the waves with $E\simeq E_p$ travel through the
barrier, we have to wait at least the time $\sim t_{\rm mon}^{(\rm
l.w. )}$ after the maximum of the transmitted packet appears to
the right from the barrier. Before this,  mainly the modes with
energies $E>E_p$ pass through the barrier. Thus we are able to
associate the time $t_{\rm trav}^{(\rm tun)} $ with the time of
penetration of the thick barrier by the peak of the wave packet.
The time $t_{\rm trav}^{(\rm tun)} \propto L$ naturally appears as
the time of propagation of approximately monochromatic waves
through a thick barrier. This can be considered as a resolution of
the Hartman paradox.

Summarizing, the physical picture of the tunneling of the wave
packet sharply peaked in the momentum space at $E=E_p <\max U$ incident on the very thick barrier (from large distances
to the left from the barrier) is as follows. The probability to
observe the particles  have passed the barrier reaches the same
value as it were in the stationary problem for $E=E_p$ at the
moment, when the peak of the incident wave packet did not yet
reach the barrier.  The peak of the transmitted wave packet is
formed at the right boarder of the barrier,  after a quantum time
delay (not dependent on the barrier depth) from the moment, when
the peak of the incident wave packet reached the left boarder of
the barrier (the Hartman effect). Then the peak of the transmitted
wave packet propagates to the right away from the barrier.  The
peak of the reflected wave packet is formed at the left boarder of
the barrier with approximately the same time delay. Then it
propagates back to the left from the barrier. The evanescent and
growing waves inside the barrier have no peaks. They increase with
time till the moment when the incident wave packet reaches the
left edge of the barrier with two different delays both of the
quantum time order and then decrease.  The modes with higher
energies pass through the barrier more rapidly than the less
energetic modes. The modes with $E\simeq E_p$ pass the barrier
during the time $t_{\rm trav}^{(\rm tun)} \propto L$, that
resolves the Hartman paradox.

\subsection{Resonance states and their time evolution }\label{quasistationary}

\begin{figure}
  \centering
  \includegraphics[width=7cm]{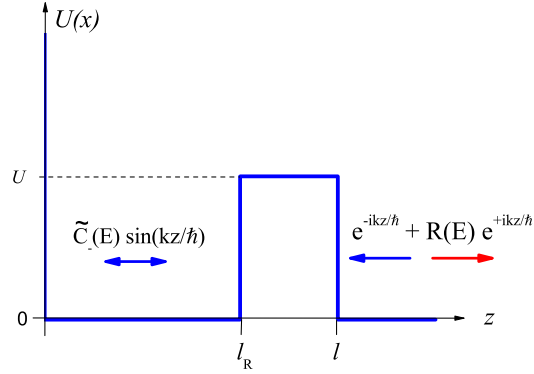}\\
  \caption{
Sketch for the problem of the 1D scattering on a potential
given by Eq.~(\ref{U-quasi}) with the incident wave
coming from the right. The internal wave function is given by Eq.~(\ref{QS-internal-wf}).
 }\label{fig:res-state}
\end{figure}

We turn now to the question of the temporal evolution of a quantum
system, which exhibits a resonance behaviour. Consider an example
of the particle motion restricted to a right half-space ($z>0$)
with a rectangular barrier of the height $U$ between $z=l_{\rm R}$
and $z=l$.
\begin{eqnarray}
U(z)=\left\{
\begin{array}{lll}
\infty &,& z\le 0\\ 0 &,& 0<z<l_{\rm R} \\ U &,& l_{\rm R}\le z\le
l\\ 0 &,& l<z\,.
\end{array}
\right.
\label{U-quasi}
\end{eqnarray}
We start this section assuming that $E<U$, so that the
classical motion is possible for $0<z<l_{\rm R}$ and $z>l$, and
for $l_{\rm R}<z<l$ we deal with the tunneling. Applying the
results of Sect.~\ref{sssec:stationary} we may use the wave
function (\ref{psi-f-2}) for $z\geq 0$ and identify $l=L/2$. The
internal wave function [Eq.~(\ref{internal-wf-hs})] contains only
the anti-symmetric part
\begin{eqnarray}
\psi_{U}(z,E) &=& \widetilde{C}_-(E)\,\chi_-(z,E)\,, \nonumber\\
\chi_-(z,E) &=& \left\{\begin{array}{lll} \sin(k\,z/\hbar) &,&
0\le z< l_{\rm R} \\ \sin(k\,l_{\rm
R}/\hbar)\cosh\big(\varkappa(z-l_{\rm R})/\hbar\big)
+\frac{k}{\varkappa}\cos(k\,l_{\rm
R}/\hbar)\sinh\big(\varkappa(z-l_{\rm R})/\hbar\big) &,& l_{\rm
R}\le z\le l
\end{array}\right. \,,
\label{QS-internal-wf}
\end{eqnarray}
where, as before, $k=\sqrt{2mE}$ and $\varkappa=\sqrt{2m(U-E)}$.
The logarithmic derivative is equal to
\begin{eqnarray}
d_-(E) &=& l\frac{\rmd}{\rmd z} \log\chi_-(z,E)\Big|_{z=l}=
\frac{l\varkappa}{\hbar} \frac{k + \varkappa \tan(k\,l_{\rm
R}/\hbar) \tanh\big(\varkappa(l-l_{\rm R})/\hbar\big) }
{k\,\tanh\big(\varkappa(l-l_{\rm
R})/\hbar\big)+\varkappa\tan(k\,l_{\rm R}/\hbar)} \nonumber\\ &=&
\frac{l\varkappa}{\hbar}\, \frac{\zeta(E,l_{\rm R}) +
\mathfrak{p}(E)} {\zeta(E,l_{\rm R}) - \mathfrak{p}(E)}\,,
\label{QS-dmin}
\end{eqnarray}
where we denoted
\begin{eqnarray}
\zeta(E,l_{\rm R}) =\frac{k\cot(k\,l_{\rm
R}/\hbar)+\varkappa}{k\cot(k\,l_{\rm R}/\hbar)-\varkappa} \,,
\quad \mathfrak{p}(E)=e^{-2\varkappa(l-l_{\rm R})/\hbar}\,.
\label{zeta-p}
\end{eqnarray}
Note that for $l_{\rm R}=0$ we recover from Eq.~(\ref{QS-dmin})
the result of Eq.~(\ref{rectang-ds}) for $d_-$. Working within a
half-space we have to put $d_+\equiv d_-$, then from
Eqs.~(\ref{RTtrhoughD}) and (\ref{Dfun}) we find the reflection
amplitude
\begin{eqnarray}
R(E) &=& e^{i\pi-i 2\,k l/\hbar} \frac{d_-(E) + i\, k\,
l/\hbar}{d_-(E) - i\, k\, l/\hbar}=e^{i\phi_{\rm R}(E)} \,,\quad
\phi_{\rm R}(E)=\pi- 2 k l/\hbar +\delta_{\rm s}(E)\,,
\label{QS-R-def}\end{eqnarray} with the scattering phase
$\delta_{\rm s}$ given by the relation
\begin{eqnarray}
e^{i\delta_{\rm s}(E)}= \frac{\varkappa+ i\, k}{\varkappa-i\, k}\,
\frac{\zeta(E,l_{\rm R})+\frac{\varkappa- i\, k}{\varkappa+i\,
k}\,\mathfrak{p}(E) } {\zeta(E,l_{\rm R})+ \frac{\varkappa + i\,
k}{\varkappa - i\, k}\,\mathfrak{p}(E) }
\,.
\label{QS-phase}
\end{eqnarray}
The coefficient $\widetilde{C}_-(E)$ of the internal wave function
defined in Eqs. (\ref{C-coeff-d}) and ~(\ref{internal-wf-hs}) can
be expressed with the help of Eq.~(\ref{QS-dmin}) and the relation
\begin{eqnarray}
\chi_-(l,E)=\frac{\sin(k l_{\rm R}/\hbar)}{\sqrt{\mathfrak{p}(E)}}
\frac{\zeta(E,l_{\rm R})-\mathfrak{p}(E)}{\zeta(E,l_{\rm R})-1}
\nonumber
\end{eqnarray}
following from Eq.~(\ref{QS-internal-wf}) as
\begin{eqnarray}
\widetilde{C}_-(E) &=& \frac{2\,i\sqrt{\mathfrak{p}(E)}}{\sin(k \,
l_{\rm R}/\hbar)} \frac{k\, e^{-ikl/\hbar}}{\varkappa - i k}
\frac{1-\zeta(E,l_{\rm R})} {\zeta(E,l_{\rm R}) + \frac{\varkappa
+ i\, k}{\varkappa - i\, k} \mathfrak{p}(E)} \nonumber\\&=& 2\,i
\sqrt{\mathfrak{p} \frac{k^2\,(\zeta-1)^2 + \varkappa^2\,
(\zeta+1)^2} {k^2\,(\zeta-\mathfrak{p})^2 + \varkappa^2\,
(\zeta+\mathfrak{p})^2} }\, e^{i\pi -ikl/\hbar + i\delta_{\rm
s}(E)/2}\,. \label{QS-Cmin}
\end{eqnarray}
In the last equation we used explicitly that $E<U$ and $\varkappa$
is real, and we suppressed the arguments of functions $\zeta$ and
$\mathfrak{p}$ for shortness. We also used that $1/\sin^2(k\,
l_{\rm R})=1+(\varkappa^2/k^2)(\zeta+1)^2/(\zeta-1)^2$.

If $U>U^{(n)}=\pi\, \hbar^2 (2\,n+1)/(4\,m\, l_{\rm R}^2)$,
equation $\zeta(E,l_{\rm R})=0$ has $n$ solutions,
$\{\varepsilon_i\}$, $i=1,\dots,n$, which constitute the spectrum
of bound states for the rectangular potential well
(\ref{U-quasi}), provided we put $l\to \infty$. For energies close
to $\varepsilon_i$ we can expand $\zeta(E,l_{\rm R})\approx
r_{\zeta,i}(E-\varepsilon_i)/4\, \varepsilon_i$, where
$r_{\zeta,i}= (l_{\rm R}\,
\varkappa_i/\hbar+1)(k_i^2/\varkappa^2_i +1)$, $k_i=\sqrt{2 m
\varepsilon_i}$, and $\varkappa_i=\sqrt{2 m (U-\varepsilon_i)}$.
Hence, the amplitude $R(E)$ possesses simple poles at energies
$E_i$.

Consider the case of a broad barrier. Then
$\mathfrak{p}(\epsilon_i)\ll 1$, and the poles are close to
$\varepsilon_i$,
\begin{eqnarray}
E_i &=& \varepsilon_i-\frac{4\,\varepsilon_i}{r_{\zeta,i}}
\frac{\varkappa_i + i\, k_i}{\varkappa_i - i\, k_i}\,
\mathfrak{p}(\varepsilon_i) = E_{{\rm R},i}
-\frac{i}{2}\Gamma_i\,, \label{polesR}
\end{eqnarray}
with the real, $E_{{\rm R},i}$ and imaginary, $-\Gamma_{i}/2$,
parts, given by
\begin{eqnarray}
E_{{\rm R},i} &=& \varepsilon_i -
\frac{4\,\varepsilon_i}{r_{\zeta,i}} \frac{\varkappa_i^2 -
k_i^2}{\varkappa_i^2 + k_i^2} \,
\mathfrak{p}(\varepsilon_i)\,,\quad \Gamma_i =
\frac{16\,\varkappa_i^2 k^2_i}{(\varkappa_i^2 + k_i^2)^2}
\mathfrak{p}(\varepsilon_i) \frac{\varkappa_i k_i}{2m(l_{\rm R}
\varkappa_i/\hbar+1)}\, . \label{ER-GR}
\end{eqnarray}
Expression for the width $\Gamma_i$ has a simple semiclassical
interpretation: $\Gamma_i=\hbar\,|T(E_{{\rm R},i})|^2/P(E_{{\rm
R},i})$ , where
 in the limit
$\varkappa(l-l_{\rm R})/\hbar\gg1 $, which we now consider,
$|T|^2$ is the transmission coefficient of the barrier,
 cf. Eq.~(\ref{rectang-R-T}),
and $P=2 (m/k) (l_{\rm R}+\hbar/\varkappa)$ is the period of the
particle motion within the potential well ($0<z<l_{\rm R}$). The
latter expression takes into account that the particle can enter a
depth $\hbar/\varkappa$ under the barrier.    In other words the
width is given by the product of the number of hits of the
particle off the barrier per unit of time and the probability of the barrier
penetration after each collision. This result survives for
arbitrary barrier within applicability of semiclassical
approximation \cite{Razavi}. Close to the resonance, $E\sim
E_{{\rm R},i}$, the amplitude can be written as
\begin{eqnarray}
R(E)\approx e^{i\pi- 2i\,k_il/\hbar
+2i\beta_\varkappa(\varepsilon_i)} \frac{E - E_{{\rm R},i} -
\frac{i}{2}\Gamma_i}{E - E_{{\rm R},i} + \frac{i}{2}\Gamma_i}
\approx e^{i\pi - 2\,i\,k_i l/\hbar + 2i\beta_\varkappa(E)+ 2\,
i\,\delta_{i}(E)}\,. \label{QS-R-res}
\end{eqnarray}
We see that the  phase shift can be approximately presented as
$\delta_{\rm s}(E)\approx 2 \beta_\varkappa(E) + 2\delta_{i}(E)$,
where   {\it the resonance scattering phase} is given by
\begin{eqnarray}
\delta_{i}(E) = \arctan\Big(\frac{\Gamma_i/2}{E_{{\rm R},i} -
E}\Big)\,,
\end{eqnarray}
and the non-resonant (potential) phase $\beta_\varkappa$ is
defined by  Eq.~(\ref{inter-WF-broad-rect}). Note that the values
$E_{{\rm R},i}/\hbar$ and $\Gamma_i/\hbar$ here have the same
meaning as the values $\omega_{\rm R}$ and  $\Gamma$, which we
used in Sect. \ref{Mech}, cf. poles of the Green's functions
(\ref{GF0-om}) and Eq.~(\ref{polesR}).

For the case $\Gamma_i\ll |E_{{\rm R},i+1}-E_{{\rm R},i}|$, which
we will further consider, we can write
\begin{eqnarray}
R(E)\approx \sum_{i=1}^n e^{-2\,i k_i l/\hbar +
i\beta_\varkappa(E)} \frac{i\Gamma_i} {E-E_{{\rm R},i}
+\frac{i}{2}\Gamma_i}\,.
\end{eqnarray}

Consider now the temporal aspects of this scattering problem.
Defining the dwell time in the same way as in
Sect.~\ref{sssec:time-stat}, after some manipulations we obtain
with the help of Eqs.~(\ref{QS-internal-wf}) and (\ref{QS-Cmin})
\begin{eqnarray}
t_{\rm d}(0,l,E) &=& \frac{1}{v}\intop_0^l |\psi_U(z,E)|^2 \rmd z
= \frac{m}{k}\mathfrak{p}\frac{k^2+\varkappa^2}
{k^2\,[\zeta-\mathfrak{p}]^2+\varkappa^2\,[\zeta +
\mathfrak{p}]^2} \nonumber\\ &&\times\Big( 2\,l_{\rm R}(\zeta+1)^2
-\frac{8l_{\rm R}\zeta
k^2}{k^2+\varkappa^2}+\frac{2\hbar}{\varkappa} \Big[(1-\zeta^2)
+\frac{k^2}{k^2+\varkappa^2}\, \big(2\zeta\log\mathfrak{p} +
\zeta^2/\mathfrak{p} -\mathfrak{p} \big)\Big] \Big)\,.
\label{QS-tdwell}
\end{eqnarray}
The value $t_{\rm d}(0,l,E)$ is the time needed by the incident
current $j_{\rm I}=v=k/m$ to fill the internal region $[0,l]$ with
the probability density $|\psi_U(z,E)|^2$. The quantity
(\ref{QS-tdwell}) is plotted in Fig.~\ref{fig:QS-tdwell} as a
function of the energy for different barrier penetrabilities
parameterized through the value
$\mathfrak{p}(E)=\exp(\sqrt{1-E/U}\log\mathfrak{p}(0))$ at the
zero energy. For a tiny barrier penetrability  the internal wave
function has a small amplitude for most energies $\propto
|\widetilde{C}_-|\propto \mathfrak{p}\ll 1$ and therefore the
dwell time is very short. Only for the energies close to the
resonance $\zeta\sim \mathfrak{p}$ the internal wave function can
acquire a large amplitude $|\widetilde{C}_-| \propto
1/\mathfrak{p}$ and the dwell time becomes very large. Exactly at
the resonance energy, $E=E_{{\rm R},i}$, we find that
\begin{eqnarray}
t_{\rm d}(0,l,E_{{\rm R},i})\approx \frac{4\hbar}{\Gamma_{i}}\,,
\label{QS-tdwell-res}
\end{eqnarray}
where we used that $\zeta(E_{{\rm R},i})=-\mathfrak{p}
(\varkappa_i^2-k_i^2) / (\varkappa_i^2+k_i^2)$ as follows from
Eq.~(\ref{ER-GR}). By varying the length of the resonator $l_{\rm
R}$ one can change the number of resonances in the potential, see
different panels in Fig.~\ref{fig:QS-tdwell} plotted for different
values of  $l_{\rm R}$.

\begin{figure}
\centering
\parbox{4.9cm}{\includegraphics[width=4.9cm]{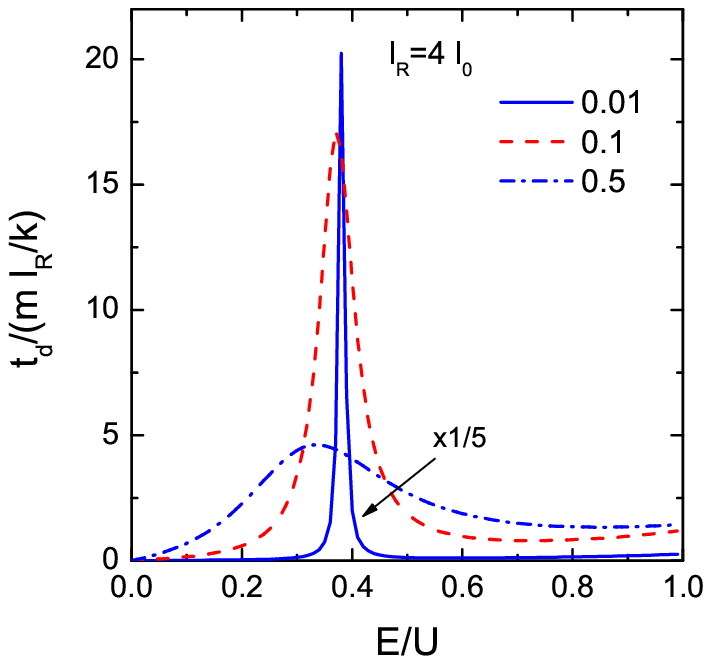}}\quad
\parbox{5cm}{\includegraphics[width=5cm]{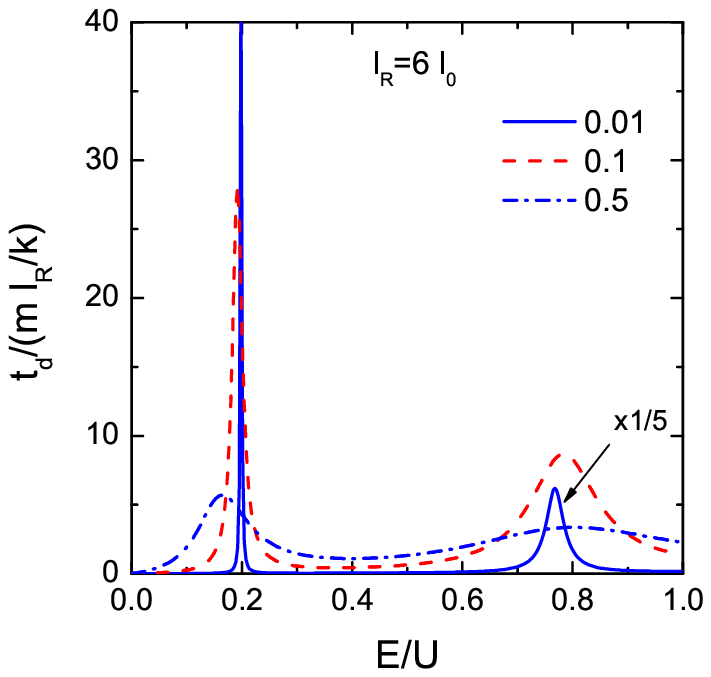}}\quad
\parbox{5cm}{\includegraphics[width=5cm]{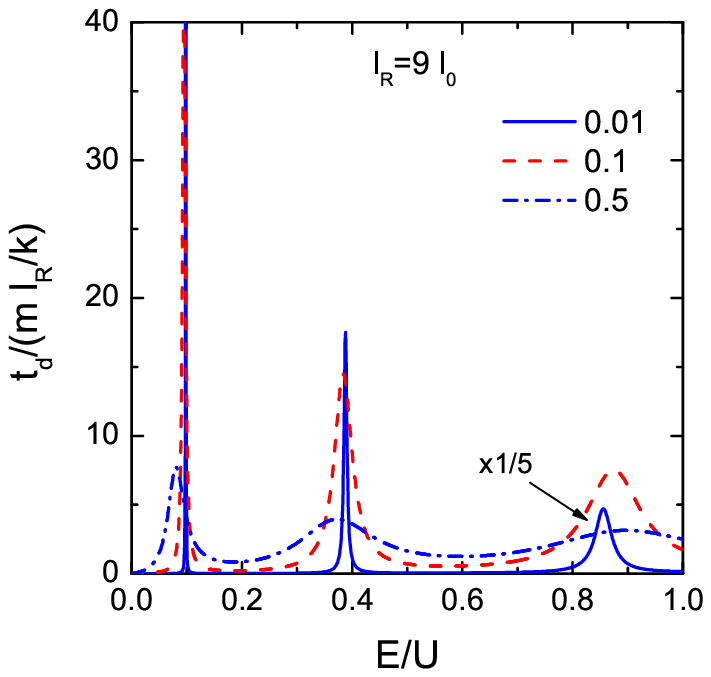}}\\
\caption{The dwell time (\ref{QS-tdwell}) for the potential
(\ref{U-quasi}) as a function of the energy for  various values of
the coefficient $\mathfrak{p}(0)$, see Eq. (\ref{zeta-p}), shown
by the line labels. Three panels demonstrate  results for various
values of $l_{\rm R}$ measured in units $l_0=\hbar/\sqrt{2mU}$.
The solid lines in all panels are multiplied by factor $1/5$.
}\label{fig:QS-tdwell}
\end{figure}

Describing the scattering problem in terms of the wave packet
(\ref{wp}) collected now with the wave function (\ref{psi-f-2})
with $T=0$, $R$ given by Eq.~(\ref{QS-R-def}) and the internal
function (\ref{QS-internal-wf}), we can define the reflection
group time in the same way as in Sect.~\ref{ssc:group-time}: This
is the time interval between the moment, when the maximum of the
incident wave packet moving towards the origin is at the position
$z=l$ and the moment, when the maximum of the reflected packet
moving away from the potential region is at the same position
$z=l$. Applying this definition to a wave-packet with the energy
distribution $\Phi(E)$ sharply peaked at the averaged energy
$\overline{E}$ with a small energy spread $\gamma$, $\gamma\ll
\Gamma_i$, we find
\begin{eqnarray}
t_{\rm R}(\overline{E}) = 2 \frac{l}{\widetilde{v}} +\hbar
\frac{\rmd \phi_{\rm R}(\overline{E})}{\rmd\overline{E}} =\hbar
\frac{\rmd \delta_{\rm s}(\overline{E})}{\rmd \overline{E}}\,,
\label{QS-tR}
\end{eqnarray}
where $\overline{E}\equiv m\widetilde{v}^2/2$. The physical
meaning of the quantity $t_{\rm R}$ is the following: If we send a
wave packet with the well-defined energy $\overline{E}$ and
observe the reflected packet at a fixed distance $z$ from the
scattering centre, $z\gg l$, then $t_{\rm R}(\overline{E})$ is the
time delay in the arrival of the emitted wave packet with respect
to the case without barrier. For energies close to the resonant
ones $\overline{E}\sim E_{{\rm R},i}$ and $\Gamma_i\ll |E_{{\rm R},i+1}-E_{{\rm R},i}|$ the reflection group time
can be written as
\begin{eqnarray}
t_{\rm R}(\overline{E})\approx - 2\frac{l_{\rm
R}}{\widetilde{v}}+\sum_i
2\hbar\frac{\rmd\delta_i(\overline{E})}{\rmd \overline{E}}= -
2\frac{l_{\rm R}}{\widetilde{v}}+
\sum_i\frac{\hbar\Gamma_i}{(\overline{E} - E_{{\rm R},i})^2+\quart
\Gamma_i^2}\,. \label{QS-tR-res}
\end{eqnarray}
We used that for  $\overline{E}$ close to $ E_{{\rm R},i}\simeq
\varepsilon_i$ we have
$\beta_\varkappa(\overline{E})\simeq\arctan(k_i/\varkappa_i)= -k_i
l_{\rm R}/\hbar\simeq -\widetilde{k} l_{\rm R}/\hbar$ with
$\widetilde{k}=\sqrt{2 m \overline{E}}$. We see that, if
$\overline{E}=E_{{\rm R},i}$, then the resonance time delay is as
large as
\begin{eqnarray}
t_{\rm R}(E_{{\rm R},i})\approx \frac{4\hbar}{\Gamma_i}\,.
\end{eqnarray}
Hence, the incident wave packet lingers in the interaction region
much longer than if it crossed  this distance with the mean
velocity. If the particle energy is de-tuned from any resonance
$|\overline{E}-E_{{\rm R},i}|\gg \Gamma_i$, then $t_{\rm R}$
changes the sign and we deal with the time advancement
\begin{eqnarray}
t_{\rm R}\approx - 2\frac{l_{\rm R}}{\widetilde{v}},
\label{QS-tR-off-res}
\end{eqnarray}
as for classical scattering on a hard sphere, cf. Eq.~(\ref{hsWigner}).  As we see, the internal part of the potential $0<z<l_{\rm R}$ is effectively excluded from the particle motion. Interestingly, in Eq.~(\ref{QS-tR-off-res}) there is no
contribution from the under-barrier region $[l_{\rm R},l]$. It
seems like the packet instantly passes under barrier but does not
enter the resonator $[0,l_{\rm R}]$. This is a manifestation of
the Hartman phenomenon discussed in Sect.~\ref{sssec:hartman}. In
this connection we have to emphasize that the group time $t_{\rm
R}$ is not a proper measure of the time the tunneling particle
spends under the barrier.

The dwell time (\ref{QS-tdwell}) and the reflection group time
(\ref{QS-tR}) are connected by the relation similar to
Eq.~(\ref{tD-tR-tT}) following from Eq.~(\ref{app:Int-rel}):
\begin{eqnarray}
t_{\rm d}(0,l,\overline{E}) = t_{\rm R}(\overline{E}) -\delta
t_{\rm i}(\overline{E}), \label{QS-tD-tR-tT}
\end{eqnarray}
where  the interference time delay is given by
\begin{eqnarray}
\delta t_{\rm i} (\overline{E})=-
\frac{\hbar}{\widetilde{k}\widetilde{v}} \sin\big(2\,k\,
l+\phi_{\rm R}(\overline{E})\big) = \frac{\hbar}{2\overline{E}}
\sin\delta_{\rm s}(\overline{E})\,.\label{interfexpl}
\end{eqnarray}
Close to the resonance energy the interference time is not
singular, vanishing at $\overline{E}=E_{{\rm R},i}$, and is much smaller than
the reflection group time $t_{\rm R}$ and the dwell time $t_{\rm
d}$.

Since, as depicted in Fig.~\ref{fig:res-state},  there is obvious
symmetry in the motion of a particle towards the origin and away
from it, it is convenient to define a measure of time for a
reflected wave only. Then we define the scattering group time,
\begin{eqnarray}
t_{\rm s}(\overline{E})=\frac12 t_{\rm R}(\overline{E}),
\label{QS-ts}
\end{eqnarray}
 as a half of the bidirectional
scattering time defined in Eq.~(\ref{gtil}), $ t_{\rm s}=t_{\rm
bs}/2$.   The time $t_{\rm s}$ corresponds to the group time
defined for the classical motion in Eq.~(\ref{grTime}). The
similar time quantity is introduced in Eq.~(\ref{hsScat}) for
classical particles undergoing the scattering on a hard sphere. In
view of the relation (\ref{QS-tD-tR-tT}) it is convenient to
introduce also the single-way dwell time
\begin{eqnarray}
{t}_{\rm d}^{\rm s.w.}(0,l,\overline{E})=\frac12 t_{\rm
d}(0,l,\overline{E}), \label{QS-tdwell-s}
\end{eqnarray}
so that close to the resonance energy for a narrow resonance we have
\begin{eqnarray}
 t_{\rm s}(\overline{E})\approx {t}_{\rm d}^{\rm
s.w.}(0,l,\overline{E})\approx \sum_{i=1}^n
\hbar\frac{\rmd\delta_i(\overline{E})}{\rmd \overline{E}}
=\sum_{i=1}^n\frac{\hbar\Gamma_i/2}{(\overline{E} - E_{{\rm
R},i})^2+\quart \Gamma_i^2} \equiv \sum_{i=1}^n \frac{\hbar}{2}
A_i(\overline{E})\,. \label{3-times-A}
\end{eqnarray}
Each of the functions $A_i(\overline{E})$  satisfies the sum rule
\begin{eqnarray}
\int_{-\infty}^{\infty}A_i(E)\frac{\rmd E}{2\pi}=1,
\label{A-res-sum rule}
\end{eqnarray}
cf. Eqs. (\ref{A1}), (\ref{A2}), (\ref{sumrulecl}) in classical
mechanics. Here the integral sits near each $i$-th pole and
thereby we are able to perform integration from $-\infty$ to
$\infty$ or from $0$ to $\infty$.
 Correspondingly, the integral over the energy of the
scattering group time or the single-way dwell time yield the
number of resonances in the system
\begin{eqnarray}
 \label{sumrulet}
\int_{0}^{\infty} t_{\rm s}(E) \frac{\rmd E}{\pi\hbar}\approx
\int_{0}^{\infty}{t}_{\rm d}^{\rm s.w.}(0,l,E) \frac{\rmd
E}{\pi\hbar} \approx n\,.
\end{eqnarray}
Thus the dwell time defined in Eq.~(\ref{QS-tdwell}) can be related to the number of
states per unit energy
\begin{eqnarray}
t_{\rm d}(0,l,E)=2{t}_{\rm d}^{\rm s.w.}(0,l,E) \approx 2\pi\hbar
\frac{\rmd n}{\rmd E}\,,
\end{eqnarray}
cf. the semiclassical relation (\ref{t-trav-scl}).

We turn now to a more detailed study of the wave function of the
scattering problem with the potential (\ref{U-quasi}).  For the
sake of  further applications let us now re-organize the wave
functions of the stationary problem shown in
Fig.~\ref{fig:res-state} to deal with the incident and reflected
currents equal to unity outside the barrier. For this we multiply
the wave function (\ref{psi-f-2}) with $R$ given by
Eq.~(\ref{QS-R-def}) and the internal wave function
(\ref{QS-internal-wf}) with the coefficient (\ref{QS-Cmin}) by the
factor  $i\sqrt{1/v} e^{ik\,l/\hbar-i\delta_s(E)/2}$ and obtain
\begin{eqnarray}
\psi(z;E)=\left\{
\begin{array}{lll}\frac{i}{\sqrt{v}} e^{ik\,l/\hbar-i\delta_{\rm s}(E)/2}
\widetilde{C}_-(E)\chi_-(z,E) &, & z \le l \\
 \sqrt{\frac{4}{v}}\sin(k\, (z-l)/\hbar+\delta_{\rm s}(E)/2) &, & z>l
\end{array}
\right.\,\,, \label{QS-wv-above}
\end{eqnarray}
Expressions for $\widetilde{C}_-(E)$ and $\delta_s(E)$ were
derived above for $E<U$. For $E>U$ the coefficient
$\widetilde{C}_-(E)$ is given by the first line in
Eq.~(\ref{QS-Cmin}) and the phase $\delta_{\rm s}(E)$ is defined
in Eq.~(\ref{QS-phase}) after the replacement $\varkappa\to
i|\varkappa|$, there and in Eq.~(\ref{zeta-p}). The wave functions
(\ref{QS-wv-above}) are normalized as
\begin{eqnarray}
\intop_0^{\infty}\psi^*(z,E)\, \psi(z,E') \,\rmd z = 2\pi \hbar \,\delta(E-E')\,.
\label{QS-wf-norm}
\end{eqnarray}
For our further study of the time evolution of a quantum system such a normalization is more convenient than that given by Eq.~(\ref{moment-norm}).

If we deal with a system with narrow and isolated
resonances, i.e. we assume that the potential barrier is broad,
$\mathfrak{p}\ll 1$, and  the resonator length $l_{\rm R}$  is
such that  $|E_{{\rm R},i+1}-E_{{\rm R},i}|\gg
\Gamma_i+\Gamma_{i+1}$, the internal wave function acquires for $E<U$ a
sizable magnitude only, if the energy $E$ is close to the
resonance one. In the vicinity of the $i$-th resonance $E\sim E_{{\rm R},i}$, we may put
$\sin(k_i l_{\rm R}/\hbar)=k_i/\sqrt{\varkappa_i^2+k_i^2}$
in Eq.~(\ref{QS-Cmin}) and it takes the form
\begin{eqnarray}
\widetilde{C}_-(E)&\approx&
2\,i\frac{4\varkappa_i^2\, \varepsilon_i \,
e^{-\varkappa(l-l_{\rm R})/\hbar}}{(k_i^2+\varkappa_i^2)(l_{\rm R} \varkappa_i/\hbar+1)}
\frac{e^{-i\,k_i\,l/\hbar+i\beta_\varkappa(E)}}{E-E_{{\rm R},i}+\frac{i}{2}\Gamma_i} \nonumber\\ &\approx& -i\sqrt{\frac{2\, v}{l_{\rm R}}} e^{-i\, k\,
l/\hbar+i\beta_\varkappa(E)+i\delta_i(E)} \sqrt{\frac{ \hbar\Gamma_i}{(E-E_{{\rm R},i})^2 + \Gamma_i^2/4}}.
\label{QS-Cmin-res}
\end{eqnarray}
Taking this into account the internal part of the wave function can be written for $E<U$ as follows
\begin{eqnarray}
\psi(z\le l;E<U)\approx \sum_{i=1}^n\sqrt{\frac{\hbar\,
\Gamma_i}{(E-E_{{\rm R},i})^2 + \Gamma_i^2/4}}\,
\sqrt{\frac{2}{l_{\rm R}}} \chi_-(z,E)\,,\quad
z \le l \,, \label{QS-wf-bound}
\end{eqnarray}
where each element of the sum contributes only fort
$|E-E_{{\rm R},i}|\ll |E_{{\rm R},i+1}-E_{{\rm R},i}|$\,.
This expression shows that only the particles with energies within the interval $E_{{\rm R},i} -\Gamma/2 <E<E_{{\rm R},i} +\Gamma/2$ penetrate inwards through the barrier and from the internal wave function.

The wave function in Eq.~(\ref{QS-wf-bound}) allows for the important generalization. The coordinate part of the internal wave function can be replaced by the stationary wave function of the closed quantum system, which is obtained from those shown in Fig.~\ref{fig:res-state} by extending the barrier to infinity, $l\to \infty$. Herewith the $n$ resonance states at energies $E_{{\rm R},i}$ (with the widths $\Gamma_i$) turn into $n$ bound states with energies $\varepsilon_i$ with the wave functions
\begin{eqnarray}
\psi^{\rm(bound)}_i(z)=C_i\chi_-(z,\varepsilon_i)\,.
\label{psi-bound}
\end{eqnarray}
Since the barrier was initially broad, $\Gamma_i$ being small, and
the difference between the energy of the bound state and the
energy of the resonance is small  $|{\varepsilon}_i-E_{\rm R,i}|= O\big(\mathfrak{p}(E_{\rm R,i})\big)$. Similarly the normalization coefficients $C_i$ differ
from $\sqrt{2/l_{\rm R}}$ by a small factor
$O\big(\mathfrak{p}(E_{\rm R,i})\big)$.
Then $\psi (z; E<U)$ in (\ref{QS-wf-bound}) is factorized as
$$\psi (z; E<U)\approx\sum_{i=1}^n\sqrt{A_i(E)}\psi^{\rm(bound)}_i(z)\,.$$
Such a factorization
of the internal wave function of a scattering problem into a wave
function of the corresponding bound state problem and an
enhancement factor $\sqrt{A_i(E)}$ is argued in
Refs.~\cite{Galitski-Cheltsov-64,MigPerelPopv71,Perelomov-Popov72,Migd}
to be possible for any finite-range potential and with some
modifications also for the Coulomb potential. These results have
found applications in the studies of nucleon-halo
nuclei~\cite{Migdal-72-2n} and di-proton
radioactivity~\cite{Galitski-Cheltsov-64,Pfutzner12}.

With the help of the wave functions (\ref{QS-wf-bound}) we can give a new interpretation for
the dwell time $t_{\rm d}^{\rm s.w.}(0,l,E)$. It can be presented
as the ratio of the density of states in the region of the
potential to the density of the free states in the same region
(i.e. the relative probability for the particle to be inside the
region of the potential  compared to the scattering in the absence
of the potential) multiplied by the time of the free motion inside
the potential region $l/v$:
\begin{eqnarray}\label{Baztime}
 t_{\rm d}^{\rm s.w.}(0,l,E) =\hbar \frac{l}{v}
 \frac{\int_{0}^l |
 \psi(z;E)|^2 \rmd z}
 {\int_{0}^l |
 \psi^{\rm (free)}(z;E)|^2 \rmd z}\,,
 \quad
 \psi^{\rm (free)}(z;E)=\sqrt{\frac4v}\sin(k\,z/\hbar)\,.
\end{eqnarray}
This is in accord with the ergodicity principle, see below Sect.
\ref{Ergodicity}.

Now consider the problem of the decay of  quasistationary states.
The problem can be formulated as follows. Assume that one sends a
stationary particle flux of the energy $E_1$ on the potential
shown in Fig. \ref{fig:res-state}, and we ask the question how
long particles from the beam will be delayed  inside the region of
the potential ($0<z<l_{\rm R}$) in dependence of the value $E_1$
of beam energy after the beam was suddenly switched off. Particles
from the beams having real energies $E_1$ within bands, $E_{{\rm
R}, i}\pm \alpha \Gamma_i$, $\alpha \sim 1$, form wave packets,
corresponding to initial (after switching off the beam)
quasistationary states, which leave long inside the potential well
(if the barrier is broad) till the particles describing by these
distributions penetrate through the barrier to infinity. The
particles with initial beam energies far from energies $ E_{{\rm
R}, i}\pm \alpha \Gamma_i$
enter the region of the potential well only with a tiny
probability. So, we may address the question how long a particle
corresponding the initial real energy $E_1$ from the band $E_{{\rm
R}, i}\pm \alpha \Gamma_i$ (or, better to say,  particles
corresponding to an energy distribution within the band provided
the beam had a finite energy dispersion))  stays in the resonance
quasistationary state till its decay? (The question how long does
it take for the particle to pass the barrier has been considered
above.) One can prepare, of course, a more complicated
quasistationary states by populating not one but several resonant
states using the incident wave packet with a broader energy
distribution.

A similar initial state can be prepared differently. The initial
localized state can be created right inside the potential well
(for $0<z<l_{\rm R}$), e.g. by reactions. If the barrier is broad,
at times much shorter than the decay times of the resulting
quasistationary states the produced particles with $E< {\rm
max}\,\, U$ are redistributed over the energy levels corresponding
to the stationary levels related to the same problem but with not
penetrable   barrier. On a longer time scale each of these levels
is actually a quasistationary level and our problem is   to find
the decay time.

Considering a general case, we assume that at the time $t=0$ our system is described by an arbitrary wave function $\Psi(z,0)$ localized inside the potential region,
i.e we assume $\Psi(z,0)= 0 $ for $z>l_{\rm R}$. The evolution of this state is determined by the unitary operator $\exp(-i\, \hat{H} t)$, so that at any later time $t>0$ the wave function of the system is equal to $\Psi(z,t)=\exp(-i\, \hat{H} t)\Psi(z,0)$. Expanding the initial wave function in terms of the eigenfunctions of the Hamiltonian $\hat{H}$, $ {\psi}(z;E)$ for $E>0$, normalized as in Eq.~(\ref{QS-wf-norm}) we can write
\begin{eqnarray}
\Psi(z,t)= \int_0^{\infty} \frac{\rmd E}{2\pi \hbar} \Phi(E)\,\psi(z;E) e^{-iEt/\hbar},
\label{QS-Psi-t}
\end{eqnarray}
where
\begin{eqnarray}
\Phi(E)= \int_0^{\infty} \rmd z \psi^*(z;E)\Psi(z,0)
=\int_0^{l} \rmd z \psi^*(z;E)\Psi(z,0)\,.
\label{QS-Phi}
\end{eqnarray}
The unitary evolution conserves the normalization of the wave
function  and since the initial wave function is normalized to
unity, then
\begin{eqnarray}
\int_{0}^{\infty} dz |\Psi(z,t)|^2= \int_{0}^{l} dz |\Psi(z,0)|^2=
\int_{0}^{\infty} \frac{\rmd E}{2\pi\hbar} |\Phi(E)|^2 =1
\label{QS-Phi-norm}
\end{eqnarray}
for any moment of time $t$. Note the difference in normalization
of the function $\Phi(E)$ in comparison with Eq.~(\ref{wp-norm}).

The overlap between the wave function at time $t$ and the initial wave function
gives the amplitude of the survival probability (also called integrity)
\begin{eqnarray}
\mathcal{G}(t)= \int_{0}^{\infty} \Psi^*(z,0) \Psi(z,t)\rmd z
= \int_0^{\infty} \frac{\rmd E}{2\pi\hbar} |\Phi(E)|^2 e^{-iEt/\hbar},
\label{QS-Gt}
\end{eqnarray}
so that $\mathcal{P}_{\rm surv}(t)=|\mathcal{G}(t)|^2$ is the
probability that the system remains in the same state after
passage of time $t$. Obviously at $t=0$ probability
$\mathcal{P}_{\rm surv}$ is equal to unity, $\mathcal{G}(0)=1$ in
view of Eq.~(\ref{QS-Phi-norm}). At any later time it becomes
smaller than unity, as follows from the Cauchy-Schwarz-Bunyakovsky
inequality
\begin{eqnarray}
|\mathcal{G}(t)|= \Big|\int_{0}^{\infty} \Psi^*(z,0) \Psi(z,t)\rmd
z\Big|\le \Big|\int_{0}^{\infty} |\Psi(z,0)|^2\rmd z\Big|^{1/2}
\Big|\int_{0}^{\infty} |\Psi(z,t)|^2\rmd z\Big|^{1/2}=1\,.
\end{eqnarray}
Since the function $|\Phi(E)|^2$ is integrable on the ray
$[0,+\infty)$, see Eq.~(\ref{QS-Phi-norm}), one can
prove~\cite{KrylovFock47} that
$\lim_{t\to \infty} \mathcal{G}(t)= 0.$
This means that an initial state will always decay at large times.
Under the assumption of a purely exponential decay one identifies
the lifetime of the system in the initial state, or its  decay
time, as $t_{\rm dec} = -\mathcal{P}_{\rm
surv}(t)/\dot{\mathcal{P}}_{\rm surv}(t)$, which in this case
would be a time-independent quantity.
However Khalfin in Ref.~\cite{Khalfin57} pointed out that
$\mathcal{P}_{\rm surv}(t)$ cannot be purely exponential. It
deviates from the exponent both for very large times  and for very
short times. This conclusion is obtained without any assumptions
about the quantum state and the system dynamics. For more
extensive discussion of  this issue we address the reader to the
review~\cite{Fonda78}. Possible manifestations of a
non-exponential decay in nuclear systems are discussed, e.g., in
\cite{Kelkar}. Peculiarities of a many-body quantum decay are
studied in \cite{Campo2011}, where important role of effects of
the quantum statistics is demonstrated.

Since, as argued above, the purely exponential decay is not possible for all times, it would be desirable to find such a definition of the decay time, which does not depend on the
assumption of a particular form of the survival probability amplitude. Following Fleming~\cite{Fleming73} let us define the decay time of the unstable state as
\begin{eqnarray}
t_{\rm dec}=\intop_0^\infty\rmd t |\mathcal{G}(t)|^2\,=
\frac{1}{2}\intop_0^\infty\frac{\rmd E}{2\pi\hbar}  |\Phi(E)|^4\,,
\label{QS-tdec-Phi}
\end{eqnarray}
where we used   Eq.~(\ref{QS-Gt}). The integral exists, if
$|\mathcal{G}(t)|^2\lsim 1/t^{1+\delta}$ for large $t$.
The sojourn time
\begin{eqnarray}
t_{\rm soj}(0,l)=\intop_0^\infty\rmd t\intop_0^l \rmd z
|\Psi(z,t)|^2\, \label{QS-tsoj}
\end{eqnarray}
is a characteristic  of how long the particle stays within
interval $[0,l]$ starting from initial moment $t=0$, see Eq.
(\ref{soj-quant}). By analogy to Eq.~(\ref{soj-dwell}) we can
express the sojourn time through the dwell time (\ref{QS-tdwell})
averaged with $|\Phi(E)|^2$ over the energy
\begin{eqnarray}
t_{\rm soj}(0,l)= \frac12 \int_0^{\infty} \frac{\rmd E}{2\pi
\hbar} |\Phi(E)|^2\, \intop_0^l \rmd z |
{\psi}(z;E)|^2
=\frac12\intop_0^\infty \frac{\rmd E}{2\pi\hbar} |\Phi(E)|^2
t_{\rm d}(0,l,E)\,. \label{QS-tsoj-tdwell}
\end{eqnarray}
Here in the last equation we take into account the different
normalization of the wave function $\psi_U$ used in
Eq.~(\ref{QS-tdwell}) and the wave function $\psi$, Eq. (\ref{QS-wf-bound}) used in the expansion (\ref{QS-Psi-t}), which produces the factor $1/v$ needed in
Eq.~(\ref{QS-tdwell}).

Let us now illustrate how the above formulae work for the case of
very narrow isolated resonances. The preparation of the initial
wave function $\Psi(z,0)$ for such a quasistationary states can be
done by putting infinite wall somewhere inside the barrier or one
can use the simple method of Refs.~\cite{Migd}:  in the potential
(\ref{U-quasi}) we extend the barrier to infinity by putting $l\to
\infty$. In the latter case we can expand initial localized wave
function in terms of the wave functions (\ref{psi-bound}) as
\begin{eqnarray}
\Psi(z,0)=\sum_i^{n} c_i \psi^{\rm(bound)}_i(z)+ \int_{U}^\infty \frac{\rmd E}{2\pi\hbar}
\tilde{c}(E) \, \psi(z,E)\,,
\quad \sum_{i=1}^n |c_i|^2 +\int_{U}^\infty \frac{\rmd E}{2\pi\hbar}
|\tilde{c}(E)|^2=1\, .
\label{Psi0-expand}
\end{eqnarray}
The wave function under the integral is given by Eq.~(\ref{QS-wv-above}) for $E>U$. If we now suddenly recover initial form of the potential (in the time $\ll \hbar/ \min(|\varepsilon_i -
\varepsilon_{i+1}|)$ for $i<n$) then the wave function does not change and we can substitute it in Eq.~(\ref{QS-Phi}), and using the wave functions (\ref{QS-wv-above}) and (\ref{QS-wf-bound}) obtain
\begin{eqnarray}
\Phi(E)\approx \sum_i^n c_i \, \sqrt{\hbar A_i(E)}+ \overline{\Phi}(E)\,\theta(E-U)\,,
\end{eqnarray}
where the part $\overline{\Phi}(E)$ corresponds to the modes over
the barrier, which do not contribute to the resonant scattering
and can be dropped thereby.

For simplicity let us now  assume that the initial wave function
corresponds to only one $j$-th bound state with $1<j<n$. We have
${c}_i =\delta_{ij}$ and $\tilde{c}(E>U)=0$, and
$|\Phi(E)|^2\approx\hbar A_j(E)$. Since close to the resonance
energy the dwell time is $t_{\rm d}(0,l,E)\approx \hbar A_j(E)$,
we find
\begin{eqnarray}
t_{\rm dec}&\approx& t_{\rm soj}(0,l)\approx \frac12\intop_0^\infty \frac{\rmd E}{2\pi}
\frac{\hbar\,\Gamma_j^2}{\big((E-E_{{\rm R},j})^2+\Gamma_j^2/4\big)^2}
\approx \frac12\intop_{-\infty}^\infty \frac{\rmd E}{2\pi}
\frac{\hbar\,\Gamma_j^2}{\big((E-E_{{\rm R},j})^2+\Gamma_j^2/4\big)^2} =\frac{\hbar}{\Gamma_j}.
\label{tdec-res}
\end{eqnarray}
Here we used that for $E_{{\rm R},j}\gg \Gamma_j$ the lower limit
of the integration can be extended to $-\infty$. Hence, the
particles occupied at $t=0$ a narrow quasistationary state with
the width $\Gamma$, will appear with the probability of the order
of one to the right from the barrier after passing of time
$\hbar/\Gamma$. Compare it with Eq.~(\ref{tdec}) introduced in
classical mechanics in Sect. \ref{Mech}.

The explicit form of the survival probability amplitude follows from Eq.~(\ref{QS-Gt}):
\begin{eqnarray}\label{AnalGa}
\mathcal{G}(t)=\int_0^{\infty}A_j(E) e^{-iEt/\hbar}\frac{\rmd E}{2\pi\hbar}
= e^{-iE_{{\rm R},j} t/\hbar-\Gamma_j\,t/2\hbar}
+ \frac{(1-i)}{\sqrt{2}}\int_{0}^{\infty}
\frac{\Gamma_j e^{-Et/\hbar}\sqrt{E_{{\rm R},j}/E} }{(iE+E_{{\rm R},j})^2+\Gamma_j^2/4} \frac{\rmd E}{2\pi}.
\end{eqnarray}
To get this expression we assumed the energy dependence of the
width $\Gamma (E)\simeq \Gamma \sqrt{E_{{\rm R},j}/E}$, as it
follows from analysis of the available phase space of the 1D
problem at small energies, and rotated the contour of integration
to coincide with the imaginary axis. For $t\gg \hbar/E_{{\rm R},
j}$ we find
\begin{eqnarray}
\label{AnalGab}
\mathcal{G}(t)\simeq e^{-iE_{{\rm R},j}
t/\hbar-\Gamma_j t/2\hbar} +
\frac{(1-i)}{\sqrt{8\pi}}
\frac{\Gamma_j}{E_{{\rm R},j}}
\Big(\frac{\hbar}{E_{{\rm R},j}\,t}\Big)^{1/2}
\,.
\end{eqnarray}
For times of $t\ll t_{\rm dec}\log\big(8\pi \,E^{3}_{{\rm
R},j}/\Gamma_j^{3}\big)$, i.e. almost in the whole time interval
of our interest, the second term can be neglected. If we extend
integration to $-\infty$ using that $A_j(E)$ is the sharp function
of $E$ near $E_{{\rm R},j} >0$, we get
\begin{eqnarray}
\label{AnalG}
\mathcal{G}(t)\simeq \int_{-\infty}^{\infty} A_j(E)
e^{-iEt/\hbar}\frac{\rmd E}{2\pi\hbar}\simeq e^{-iE_{{\rm R},j} t/\hbar-\Gamma_jt/2\hbar}.
\end{eqnarray}
Thus, the correction term in (\ref{AnalGa}) is fully compensated by the contribution
of negative energies.

We see that in case of a decaying system we deal with a packet
propagating outwards the resonator with an effective energy
distribution given by $A_j(E)$. As we argued in
Sect.~\ref{Wpexample}, the emerging of the packet on the outer
side of the barrier cannot last shorter then $\delta t_{\rm
var}\sim\hbar/\gamma=t_{\rm dec}^{\gamma}$, see the
Mandelstam-Tamm inequality (\ref{uncert}), where $\gamma$ is the
dispersion of the energy distribution. However, for the Lorentz
distribution $A_j(E)$ (i.e. for $\Gamma_j =const$) the dispersion
would be infinite. Here we have to remember that the resonance
wave functions (\ref{QS-wf-bound}) can be used only for energies
not far from the resonance one, i.e. for $E\in[E_{{\rm
R},j}-\alpha \Gamma_j,E_{{\rm R},j}+\alpha\Gamma_j]$, for
$\alpha\Gamma_j\ll |E_{{\rm R},j+1}-E_{{\rm R},j}|$. Taking these
energy limits into account we can estimate
\begin{eqnarray}
\gamma^2\simeq \int_{-\alpha\Gamma_j}^{+\alpha \Gamma_j}
\frac{\rmd E}{2\pi\hbar} E^2 A_j(E+E_{{\rm R},j}) \Bigg/
\int_{-\alpha\Gamma_j}^{+\alpha\Gamma_j} \frac{\rmd E}{2\pi\hbar}
A_j(E+E_{{\rm R},j}) =\Gamma_j^2\frac{ \left(2\alpha -\arctan
2\alpha\right)}{4 \arctan 2\alpha}.
\end{eqnarray}
For $\alpha \simeq 3.58$ we get $\gamma \simeq \Gamma_j$. The
condition $\gamma\simeq \Gamma$ looks rather natural for the
description of the wave packet in quasistationary state.

Thus, the probability to find particle outside the barrier becomes
essentially non-zero before the maximum of the wave packet
(according to the latter's position one defines the scattering
group time) has reached the point $z=l$ with advancement $\sim
t^\gamma_{\rm dec}\sim \hbar/\Gamma$ (if we put $\gamma \simeq
\Gamma$), and the real forward delay time is given by
\begin{eqnarray}
\label{forward-A}
\delta t_{\rm f}=t_{\rm s}-t^\gamma_{\rm dec}\,,
\end{eqnarray}
cf. Eq. (\ref{forward}) above. This quantity demonstrates an
advance of the formation and decay of the intermediate states
forming in the scattering on the potential (\ref{U-quasi}) at
energies $|E-E_{\rm R}|>\Gamma/2$ and a delay at energies in the
vicinity of the resonance, for $|E-E_{\rm R}|<\Gamma/2$.

Considering a scattering of a packet with an energy distribution $\Phi (E)$
on the potential well (\ref{U-quasi}), we may introduce the quantity
\begin{eqnarray}
\bar{t}_{\rm s}(\overline{E})= \int \frac{\rmd E}{2\pi\hbar} |\Phi (E)|^2 t_{\rm s}(E)
\approx \int \frac{\rmd E}{2\pi\hbar} |\Phi (E)|^2 \frac{\hbar}{2}\, A_i(E)\,.
\end{eqnarray}
Here $|\Phi (E)|^2$ is normalized  as in Eq.~(\ref{QS-Phi-norm}). For the Gaussian wave packet  $|\Phi (E)|^2 =\sqrt{2\pi \hbar^2/\gamma^2} \exp(-(E-\overline{E})^2/2\gamma^2)$ in the
limit of a narrow energy distribution $\gamma\ll \Gamma$ we derive
$\bar{t}_{\rm s}(\overline{E})={t}_{\rm s}(\overline{E})$ and in the
opposite limit case of a very broad distribution $\gamma \gg
\Gamma$ we get $\bar{t}_{\rm s}(\overline{E})=1/\gamma$. Thus $\delta
t_{\rm f}(\bar{E})=\bar{t}_{\rm s}(\overline{E})-1/\gamma$ for
$\gamma\ll \Gamma$ and $\delta t_{\rm f}(\overline{E})=0$ at $\gamma
\gg \Gamma$, and only for a very specific choice of the energy
distribution in the packet (e.g. for a Lorentzian distribution
with $\gamma\simeq \Gamma$) we arrive at $\delta t_{\rm
f}=\bar{t}_{\rm s}(\bar{E})-1/\Gamma$, as we obtained above for
the case of an initially localized state.

It is instructive to rewrite the survival probability amplitude (\ref{QS-Gt})
in the following form
\begin{eqnarray}
\mathcal{G}(t)=i\int_0^\infty\rmd z \int_0^\infty\rmd z'
\Psi^*(z,0) \, G^R(t,z,z')\, \Psi(z',0)\,,
\end{eqnarray}
where we introduce the retarded Green's function $G^R(t,z,z')$,
which describes the evolution of the wave function,
$\Psi(z,t)=\intop_0^\infty \rmd z' G^R(t,z,z')\Psi(z,0)$ forward
in time, i.e. $G^R(t,z,z')\equiv 0$ for $t<0$, cf. the same
quantity in classical mechanics (\ref{freesol}), (\ref{GF0-om}),
and (\ref{GF0-class}). The Green's function is expressed through
the eigenfunctions (\ref{QS-wf-bound})
 and
(\ref{QS-wv-above})
\begin{eqnarray}
G^R(t,z,z')=\intop_{-\infty}^\infty \frac{\rmd E}{2\pi} G_R(E,z,z') e^{-i E t/\hbar}
\,,\quad
G^R(E,z',z)= \intop_{0}^\infty \frac{\rmd E'}{2\pi\hbar}
\frac{\psi(z;E')\, \psi^*(z';E')}{E-E'+i0}\,.
\end{eqnarray}
The small shift of the pole in the last integral in the lower complex semi-plane assures that the Green's function vanishes for $t<0$, since
\begin{eqnarray}
\theta(t) e^{-i E t /\hbar}=i\int_{-\infty}^{+\infty}\frac{\rmd E'}{2\pi} \frac{e^{-i E' t/\hbar}}{E'-E+i 0}.
\end{eqnarray}
For energies $E<U$ and for the case of narrow resonances we can
use the wave function of the resonance states
(\ref{QS-wf-bound})
 with the replacement of their
coordinate parts by the wave-functions of the corresponding bound
states (\ref{psi-bound}). Then the coordinate and energy
dependence of the Green's function separate as follows
\begin{eqnarray}
G^R(E,z',z)\approx  \sum_{i=1}^n G_i^R(E) [\psi^{\rm
(bound)}_i(z')]^*\, \psi^{\rm (bound)}_i(z)\,, \nonumber\\
G_i^R(E)= \intop_{0}^\infty \frac{\rmd E'}{2\pi}
\frac{A_i(E)}{E-E'+i0} = \frac{1}{E-E_{{\rm
R},i}+\frac{i}{2}\Gamma_i}\,.
\end{eqnarray}
If the initial wave function (\ref{Psi0-expand}) contains only one state $j$,
the expression (\ref{AnalG}) reduces to the following one
\begin{eqnarray}\label{GRtime}
\mathcal{G}(t)=i\int_{-\infty}^{+\infty} \rmd z G^R(t,z,z)=i
G_j^R(t)\approx -\intop_{-\infty}^\infty \frac{\rmd E}{2\pi i}
\frac{ e^{-i E t/\hbar}}{E-E_{{\rm R},j}+\frac{i}{2}\Gamma_j}=
\theta(t)\,e^{-i E_{{\rm R},j} t/\hbar -\Gamma_j t/2\hbar}\,.
\end{eqnarray}
The function $A(E)$ plays the role of the spectral
density
and can be defined as
\begin{eqnarray}
A(E)=-2\int_{-\infty}^{+\infty} \rmd z \,\Im G^R(E,z,z)\,.
\end{eqnarray}

If we assume that initially we deal not with a pure quauntum
mechanical state (\ref{Psi0-expand}) but with a mixed state such
that $$ \Psi(z',0) \Psi^*(z,0) \approx \Big(\sum_{i=1}^{n}
n_T(\varepsilon_i)\Big)^{-1} \sum_{i=1}^n
n_T(\varepsilon_i)\psi^{\rm bound}(z')[\psi^{\rm bound}(z)]^*\,,
$$ which contains a number of quasistationary states characterized
by the thermal Fermi/Bose occupations $n_T$, we find that the
decay of such a system is described by
\begin{eqnarray}
\label{AnalG1-1}
\mathcal{G}_T(t)\simeq\int_{0}^{\infty}\frac{dE}{2\pi} A(E)\, n_T (E) e^{-iEt}
\Bigg/\int_{0}^{\infty}\frac{dE}{2\pi} A(E)\, n_T (E)\,.
\end{eqnarray}
This expression is to be compared with Eq.~(\ref{dwellF}) written
below with the help of the Wigner densities.

\subsection{Causality restriction}

From (\ref{QS-tD-tR-tT}), (\ref{interfexpl}) we arrive at
inequality
\begin{eqnarray}
\label{restr}
\delta t_{\rm s}\geq
-\frac{l}{v}-\frac{\hbar}{2kv}.
\end{eqnarray}
This restriction, cf. \cite{MugaEqu}, differs from the
corresponding condition, which we have derived  (for $R=l$) in
classical mechanics, see Eq. (\ref{hsWigner}), by the presence of
the second term in the r.h.s. of (\ref{restr}). The latter term is
of purely quantum origin. It shows the time, which the particle
needs in order to pass a half of the de Broglie wave length,
$\lambda =\hbar/k$. Following uncertainty principle free quantum
particles cannot distinguish distance $\xi <\hbar/2k$.

\section{Time shifts in non-relativistic quantum mechanics: 3D-scattering} \label{scattering}

\subsection{Scattering of the wave packet on the potential}

In the three dimensional scattering problem there appears a new specifics.
At large distances from  the interaction zone the wave packet is presented as
\begin{eqnarray}
\label{ampl}
\Psi (\vec{r}, t)= \int \frac{\rmd^3 k}{(2\pi\hbar)^3}
\mathcal{F}(\vec{k}-\vec{p}\,)\, \psi_{\vec{k}} e^{-iE_k t/\hbar},
\end{eqnarray}
where $\mathcal{F}(\vec{k}-\vec{p}\,)$ is the wave packet amplitude
peaked at $\vec{k}=\vec{p}$ and the stationary wave function
\begin{eqnarray}
\label{Psi-1}
\psi_{\vec{k}} \simeq e^{ikz/\hbar}+\frac{e^{ikr/\hbar}}{r} f(E,\theta_k)
\end{eqnarray}
is the sum of the incident and the scattered ($\propto f$) waves  \cite{LL}, normalized to
unit amplitude in the incident wave.
The cross section is determined through $f$ as
\begin{eqnarray}
\rmd\sigma =|f|^2 \rmd\Omega\,, \quad f(\theta)=\sum_{l=0}^\infty
 (2\,l+1)\,f_l\, P_l (\cos\theta_k)\,,
\end{eqnarray}
cf. Eq. (\ref{EM-scattampl}), (\ref{spham}). The scattering
amplitude is expressed through the phase shift:
\begin{eqnarray}
\label{fdelta}
 f_l =\frac{\hbar}{k}\sin\delta^l(k) e^{i\delta^l (k)}\,.
\end{eqnarray}
As follows from this expression, the amplitude $f_l$ is related to
the elements of the $S$ and $T$-matrices \cite{LL} as: $S_l -1=T_l
=2ik f_l/\hbar$, $S_l =e^{2i\delta_l}$. We would like to bring to reader's attention
that the partial phase $\delta_l$ vanishes identically if the scattering potential is put to zero. Thus the corresponding quantity in classical mechanics is $\delta^{\rm cl}-
\delta^{\rm cl}(U=0)$ in Eq.~(\ref{class-delta-delta}).

Presenting $E =E_p +\delta E$  and $k=p+\delta E/v_p +...$  in Eq.~(\ref{ampl}), we recognize in the scattered wave the factor
$\exp[i\delta E (r/v_p - t +\hbar\frac{\partial \ln f_l(E_p)}{\partial E_p})]$, $v_p =p/2E_p$.
Thus, using Eq.~(\ref{fdelta}) we find the time  delay/advance of the scattered wave
\begin{eqnarray}
\label{deltats}
\delta t_{\rm s}^l =\hbar\Im \frac{\partial \ln f_l(E_p)}{\partial
E_p}=\hbar\frac{\partial \delta^l (E_p)}{\partial E_p}.
\end{eqnarray}

On the other hand, expanding the plane wave part of the wave
function (\ref{Psi-1}), $e^{ikz/\hbar}$, in the Legendre
polynomials one gets series of converging and diverging waves:
\begin{eqnarray}\label{Psi}
\psi_{\vec{k}}
\simeq\frac{\hbar}{2ikr}\sum_{l=0}^{\infty}(2l+1)P_l(\mbox{cos}\theta_k)[(-1)^{l+1}e^{-ikr/\hbar}+e^{2i\delta^l
(k) } e^{ikr/\hbar}].
\end{eqnarray}
The first term in the squared brackets, the converging wave, arises as a part
of the incident wave. The second term is the result of the superposition of scattered waves ($\propto f$),
\begin{eqnarray}
 \label{psis}
\psi_{\rm s} \simeq \frac{\hbar}{kr}
\sum_{l}(2l+1)\,P_l(\cos\theta_k)\,\sin\delta^l(k)\, e^{i\delta^l(k)+ikr/\hbar},
\end{eqnarray}
 and the incident wave $e^{ikz/\hbar}$. Note that the optical
theorem ($\Im f(0) =\frac{k}{4\pi\hbar}\int |f|^2 d\Omega$) arises
as consequence of subtle interference that takes place in the
forward direction.

From the second term in the squared brackets (\ref{Psi}) one finds
the average exit time {\rm delay/advance} of the diverging waves
with the angular momentum $l$ (the $l$-th partial wave)
\cite{Wigner},
\begin{eqnarray}
\label{Wigner}
\delta t_{\rm W}^l= 2\hbar\frac{\partial \delta^l}{\partial E_p}.
\end{eqnarray}
This  result includes the interference with the incident wave and is twice larger than the time delay of the purely scattered wave (\ref{deltats}). The converging wave $\propto e^{i\delta E (r/v_p -t)/\hbar}$ propagates without any delay. Ref.~\cite{Smith}, see also \cite{WuOm}, introduced the collision life time for elastic collisions, as a difference between the time of the particle flight in presence of the potential and the free-flight time and found the relation
\begin{eqnarray}
\label{Qsmith}
Q_l =\lim_{R\to \infty}
\int^R \left(\widetilde{\psi}_l^* \widetilde{\psi}_l -\frac{2}{4\pi
vr^2}\right)\rmd^3 r =-i \hbar\frac{\rmd S_l}{\rmd E}\,S_l^*=\delta t_{\rm W}^l,
\end{eqnarray}
with $\widetilde{\psi}_l$ normalized here  to unit incident flux.
Note that Eq.~(\ref{Qsmith}) coincides with definition of the
dwell time, cf. Eq.~(\ref{dwell}), but now for the diverging waves
only and with other normalization.  As we will see in Sect.
\ref{sec:kin}, the meaning of the collision time is different. We
also stress that $\delta t_{\rm W}^l$ has the meaning of the group
delay/advance of the diverging wave occurring only at large
distances, cf. discussion of the wave zone in Sect. \ref{electro}.
The same wave at short distances near the scattering center is
disturbed and is delayed/advanced differently.

Besides time delays/advancements $\delta t_{\rm s}^l$ and $\delta t_{\rm W}^l$, the corresponding wave packets undergo a smearing since velocities of the particles depend on the energy. To be specific consider the diverging wave packet for given $l$, cf. second term in (\ref{Psi}):
\begin{eqnarray}
\label{psil}
\psi_l \simeq  \frac{\hbar}{2ikr} (2l+1)P_l (\cos\theta_k) \,e^{2i\delta^l(k)+ikr/\hbar}.
\end{eqnarray}
Let
\begin{eqnarray}
\label{calF} \mathcal{F}(\vec{k}-\vec{p}\,)=C\,\delta(\cos\theta_k
-\cos\theta_p) e^{-(k-p)^2/(4\gamma_p^2) } \,\quad C=const\,.
\end{eqnarray}
Expanding functions of $k$  in $k-p$ up to second order near the point $k=p$,  replacing these expressions in Eq.~(\ref{ampl}) and taking integral we find the diverging wave
packet, $\Psi_l$, cf. \cite{WuOm},
\begin{eqnarray}
\Psi_l &=& \hbar \,(2\,l+1)\,P_l(\cos\theta_p) \frac{ 2\pi^{3/2} C\, p\,
\widetilde{\gamma}_p}{ir (2\pi\hbar)^3}
\exp\left[-\frac{\widetilde{\gamma_p}^2}{\hbar^2} \left(v_p t
-r-2\,v_p\,\hbar\frac{\partial \delta^l}{\partial E_p}
+\frac{i\hbar}{p}\right)^2\right]\chi_{\vec{p}},
\\
\frac{1}{\widetilde{\gamma}^2} &=&
\frac{1}{\gamma^{2}}+\frac{1}{2p^2}+\frac{i}{2m\hbar}\left[t-\hbar\frac{\partial
\delta^l}{\partial E_p}-\hbar\frac{E_p}{2}\frac{\partial^2
\delta^l}{\partial E_p^2}\right],\quad \chi_{\vec{p}}=e^{ipr/\hbar
 -iE_p t/\hbar +2i\delta_p^l}\,.\nonumber
\end{eqnarray}
To get the law of the time propagation of maximum of the packet
one needs to keep only  linear terms in the expansion. The smearing of
the wave packet with passage of time appears due to the second
order terms kept in the expansion.  Because of the presence of the term
$\frac{\partial^2 \delta^l}{\partial
p^2}=\frac{1}{m}\frac{\partial \delta^l}{\partial
E_p}+v_p^2\frac{\partial^2 \delta^l}{\partial E_p^2}$
the smearing of the wave packet is advanced or delayed in dependence of the  sign of the term $\frac{\partial^2 \delta^l}{\partial p^2}$.

Similarly, we could consider the converging, the scattered and the
incident wave packets. We also could use the expansion (\ref{wp}) instead of Eq.~(\ref{ampl}) with a $\Phi (E,\theta_k)$ distribution, multiplied by $\delta(\cos\theta_k -\cos\theta_p)$, instead of Eq.~(\ref{calF}). Thus these results are similar to those derived above in Sect.~\ref{Wpexample} in one-dimensional case.

Refs.~\cite{GW,Nussenzweig,DP} defined  average time spent by the
wave packet within a chamber as
\begin{eqnarray}
\label{tvol}
t_{\rm vol}^N =\frac{1}{N} \int \rmd t\, t\int \rmd\Omega\,  r^2    \big(\vec{n}\,\vec{j}(r,\Omega,t)\big)
,\quad
\vec{n}=\vec{r}/r.
\end{eqnarray}
Here
\begin{eqnarray}
N=\int \rmd t \int \rmd\Omega\, r^2   |\vec{j}\,\vec{n}\,|
\end{eqnarray}
is the modulus of the integrated incident unit flux through the
chamber surface, cf. expression for the classical sojourn time
Eq.~(\ref{tsoj-q}). The flow density associated with the wave
function $\Psi$  is given by Eq.~(\ref{current}). The value
$\vec{j}\vec{n}$ is positive, when the particle exits the volume,
and negative, when it enters the volume. The incident current is
the sum of the scattering current and the interference term,
$\vec{j}=\vec{j}_{\rm s}+\vec{j}_{\rm i}$. Thereby we introduce
time delays: $ t_{\rm vol}^N = t_{\rm s}^N - \delta t_{\rm i}^N$,
all quantities being normalized by $N$.

Further, for simplicity, we consider only one $l$- partial wave. The scattering time   normalized by the scattered flux $N_{\rm s}$ is as follows
\begin{eqnarray}
&t_{\rm s} =\Big(\frac{\rmd N_{\rm s}} {\rmd\Omega}\Big)^{-1}
\int  \rmd t t \, r^2 \,  (\vec{j_s}\,\vec{n}\,) =t_{\rm free} +\delta t_{\rm s} ,
\label{tsc}\\
&N_{\rm s}/N =4\,\sin^2\delta^l\, ,
\nonumber
\end{eqnarray}
cf. Eqs. (\ref{psis}), (\ref{psil}), resulting in Eq.~(\ref{deltats}) for the scattering time delay/advancement. Here $ t_{\rm free}={r}/{v}$ is the time of the free flight in one
direction (at finite angles there is no interference).

The interference delay/advancement time for a one partial wave normalized by the scattered flux is, cf. Ref.~\cite{DP},
\begin{eqnarray}
\label{cossin}
-\delta t_{\rm i} =\Big(\frac{\rmd N_{\rm s}} {\rmd\Omega}\Big)^{-1}
\int \rmd t t\, r^2 \, (\vec{j}_{\rm i}\,\vec{n}\,)
=
\frac{\hbar^2}{4\,p^2\, |f_l|^2}
\frac{\partial }{\partial E_p} \big(p({f_l} + {f_l}^*)\big)=
\frac{\cos(2\delta^l)}{2\,\sin^2\delta^l}
\hbar\frac{\partial\delta^l}{\partial E_p}  .
\end{eqnarray}
 The total
delay/advancement in the diverging wave is
\begin{eqnarray}\label{tvolN}
\delta t_{\rm vol}^N =\delta t_{\rm W}=(\delta t_{\rm s}-\delta
t_{\rm i})\,4\,\sin^2\delta^l
=2\,\hbar\frac{\partial\delta^l}{\partial E_p}.
\end{eqnarray}
The factor $4\sin^2\delta^l$ arose due to different normalizations in Eq.~(\ref{tvolN}) and Eqs.~(\ref{tsc}), (\ref{cossin}). From here
\begin{eqnarray}
\label{dtvol}
\delta t_{\rm vol}=\frac{2\hbar}{4\sin^2 \delta^l}\frac{\partial\delta^l}{\partial E_p}
\end{eqnarray}
is the average time spent by the wave packet within the chamber normalized by the scattered flux $N_{\rm s}$.

\subsection{Resonance scattering}

For  one Breit-Wigner resonance
\begin{eqnarray}
\label{tan}
\tan\delta =-\frac{\Gamma}{2M}, \quad M=E-E_{\rm R} ,\quad
N_{\rm s} =N\,\Gamma\, A,
\end{eqnarray}
the forward delay/advance time
\begin{eqnarray}
\label{tfQ}
\delta t_{\rm f}\equiv \delta t_{\rm i}=\delta t_{\rm s} - t_{\rm
dec} =-\frac{\hbar(M^2-\Gamma^2/4)}{\Gamma(M^2 +\Gamma^2/4)},
\end{eqnarray}
being negative for $|M|> \Gamma/2$. Here the value
\begin{eqnarray}\label{tfQ1}
\delta t_{\rm vol}=\delta t_{\rm vol}^N/ (4\sin^2\delta)=t_{\rm dec}=\hbar/\Gamma
\end{eqnarray}
 has the meaning of the decay time of the quasistationary state with complex energy
$E_{\rm R}-i\Gamma/2$, cf. Eqs. (\ref{tdec}), (\ref{tdec-res}).

The probability for particle to enter the region of the resonance interaction is $P_{\Gamma}=\sin^2 \delta =\frac{\Gamma^2/4}{M^2 +\Gamma^2/4}$. Thereby the cross section of the resonance scattering can be presented as $\sigma \simeq 4\pi \lambda^2 P_{\Gamma}$, where $\lambda =\hbar/k$ is the de Broglie wave length. For $M=0$ (pure resonance) the cross section reaches its maximum $\sigma_{\rm max} = 4\pi \lambda^2$.

The probability for particle not to enter the region of the resonance interaction is $P_{M}=\mbox{cos}^2 \delta =\frac{M^2}{M^2 +\Gamma^2/4}$, $P_{\Gamma}+P_{M}=1$.  The
scattering time delay is
\begin{eqnarray}
\label{rstd}\delta
t_{\rm s}=\hbar\frac{\partial \delta}{\partial E}=\frac{\hbar}{2} A = 2\,t_{\rm dec} \, P_{\Gamma}\,,
\end{eqnarray}
$2t_{\rm dec}=\delta t_{\rm s}(E=E_{\rm R})=t_{\rm dec}^{(\rm
cl)}$.
The forward delay time,
 \be
 \delta t_{\rm f} =\hbar A/2
-t_{\rm dec}=t_{\rm dec}(P_{\Gamma}-P_{M})\,,
 \ee
is the time delay of the decay due to difference in the probability for the particle to enter the region of the resonance interaction and not to enter this region.

In Sect. \ref{sec:kin} we shall see that for a many-particle system the value $\delta t_{\rm dec}$ is the average time between collisions. {\em The forward delay/advance time, $\delta t_{\rm f}$, is then an average  delay/advance in the scattering counted from the collision time $\delta t_{\rm col}$.} Thus, this delay/advancement time characterizes delays and advancements of collisions in the quantum kinetic processes.


\subsection{Scattering on hard cores}

For the gas of hard core scatters~\cite{Mekjian} the scattering amplitude and its momentum derivative are
\begin{eqnarray}
\label{dhar}
\mbox{tan}\delta^l =-\frac{j_l (kR/\hbar)}{n_l (kR/\hbar)},\quad
\frac{\partial\delta^l}{\partial k}=-\frac{\hbar}{k^2R[j_l^2
(kR/\hbar)+n_l^2 (kR/\hbar)]}.
\end{eqnarray}
E.g., for $l=0$ from (\ref{dhar}) we find $\delta t_{\rm s}^0 =\hbar\frac{\partial\delta^0}{\partial E_k}=-R/v$ that agrees with Eq.~(\ref{hsScat}) for $b=0$, $\theta =\pi$. The same advancement, $\delta t_{\rm s}^l =-R/v$,  arises for rapid particles $kR/\hbar\gg l^2$ at $l\neq 0$. For slow particles, $kR/\hbar\ll l^{1/2}$, $\delta t_{\rm s}^l \propto (kR/\hbar)^{2l} R/v$, since the wave length $\lambda =\hbar/k\gg R$ in this case and the propagating wave almost does not feel presence of the sphere. For $l\gg 1$ the cross section becomes negligibly small.

\subsection{Semiclassical scattering}
Transition from the semiclassical expression for the phase shift~\cite{Demkov}
\begin{eqnarray}
 \delta t_{\rm W}^{\rm scl}=\frac{2\hbar\partial\delta^l}{\partial E} = \frac{1}{E}
\intop_{0}^{\infty} \Big(2\, U(r) +r\, U'(r)\Big)\frac{2R_l^2
(r)}{v_{\infty}} \rmd r\,,
\label{class-delay-3Da}
\end{eqnarray} where $R_l (r)\sim \sin (kr-l\pi/2 +\delta^l)$ is the radial wave function,
to the classical Eq.~(\ref{class-delay-3D}) occurs provided one exploits the semiclassical
expression for the wave function $R_M =\sqrt{v_{\infty}/v_r}\mbox{sin}(\int_{r_0}^r  v_r \rmd r+\pi/4)$. Substituting in Eq.~(\ref{class-delay-3Da}) $R_M$ instead of $R_l$ and
using that $R_M^2 \simeq v_{\infty}/(2v_r)$ we arrive at the result for the classical Wigner time delay (\ref{clW}).

The probability of the decay of a long living state  is determined
by the imaginary part of the  action: $W\simeq e^{2\Im S/\hbar}$.
In case of quasistationary level, the time scale
\begin{eqnarray}
\label{cldec}
 t^{\rm scl}_{\rm dec}=\hbar/(2|\Im E|)\, ,
\end{eqnarray}
characterizes decay of the state,  where $\Im E$ is the imaginary part of the energy.

The scattering delay counted from the decay time
in the given case is as follows
\begin{eqnarray}
\delta t_{\rm f}^{\rm scl}=\delta t_{\rm W}^{\rm scl}/2 -t^{\rm scl}_{\rm dec},
\end{eqnarray}
cf. Eq.~(\ref{tfQ}).

\subsubsection{Ergodicity, time shifts and  level density}\label{Ergodicity}

For the scattering on the potential, as well as for binary
collisions, in the virial limit the  energy level density (i.e.
the density of states) simply relates to the Wigner time delay as,
see \cite{DP,LLStat},
\begin{eqnarray}
\label{ergod1}
\frac{\rmd N^{\rm level}}{\rmd E_p} -
\frac{\rmd N^{\rm free}}{\rmd E_p}=\frac{1}{2\pi\hbar}\sum_{l} (2l+1)\delta t^l_{\rm W},
\end{eqnarray}
where $\frac{\rmd N^{\rm free}}{\rmd E_p}=\frac{4\pi
Vp^2}{(2\pi\hbar)^3 (\rmd E_p/\rmd p)}$, and $V$ is the system
volume, for binary collisions, $\rmd E_p/\rmd p$ is the relative
velocity of interacting particles and $\delta t^l_{\rm W}$ is
given by Eq.~(\ref{Wigner}) for given $l$. Since all thermodynamic
quantities such as entropy and pressure can be calculated, if one
knows the density of states, this condition allows to express
thermodynamical variables at low densities in terms of the phase
shifts and the time delays. It looks like an ergodic constraint:
deviation of the density of states from that for the ideal gas is
limited in time by the Wigner time delay, i.e. the  delay of
outgoing waves.

For the  scattering on the Breit-Wigner resonance the free term on the l.h.s. (\ref{ergod1}) should be dropped, since one should take into account that the phase additionally changes by $\pi$ when the energy passes  the resonance region, see Eq. (\ref{tan}). Thus one has
\begin{eqnarray}
\label{ergod}
\delta t_{\rm W}=2\hbar\frac{\partial\delta}{\partial E_p}={2\pi\hbar}\frac{\rmd N^{\rm level}}{\rmd E_p}.
\end{eqnarray}
Thereby $\delta t_{\rm W}$ can be interpreted as a time delay in
an elementary phase space cell.

\section{Time shifts in quantum field theory}\label{sec:qft}

\subsection{Time contour formulation}
From now on we use units $\hbar =c=1$. To be specific, we consider a multi-component system with different constituents "$a$" of non-relativistic particles and relativistic scalar bosonic field operators, $\hat{\phi}=\{\hat{\phi}_a(x)\}$, where from now on $x$ is a
4-coordinate. The free Lagrangian densities of these fields are
%
\begin{eqnarray}
\label{L0} \widehat{\mathcal{L}}^0_a=\left\{
\begin{array}{ll}
\frac{1}{2} \displaystyle\left( i \hat{\phi}_a^\dagger \partial_t
\hat{\phi}_a -i
\partial_t \hat{\phi}_a^\dagger\cdot\hat{\phi}_a  - \frac{1}{m_a} \nabla  \hat{\phi}_a^\dagger \nabla  \hat{\phi}_a\right)
\hspace*{-4mm}\quad& {\mbox{nonrel. particles,}}
\\[2mm]
\frac{1}{2}\displaystyle\vphantom{\frac{1}{m_a}}
 \left(\partial_\mu \hat{\phi}_a \cdot \partial^\mu \hat{\phi}_a
-  m^2_a  \hat{\phi}_a^2\right) \quad&{\mbox{neutral rel. bosons,}}\\[2mm] \displaystyle\vphantom{\frac{1}{m_a}}
\partial_\mu \hat{\phi}_a^\dagger \partial^\mu \hat{\phi}_a
- m^2_a  \hat{\phi}_a^\dagger \hat{\phi}_a 
\quad&{\mbox{charged rel. bosons.}}\\
\end{array}\hspace*{-1cm}\right.
\end{eqnarray}
%
We assume that these fields interact either via non-derivative coupling or via linear derivative coupling. In the latter case the interaction Lagrangian depends  not only on the fields but also on their derivatives $\widehat{\mathcal{L}}^{\rm int}=\widehat{\mathcal{L}}^{\rm int} \{
\hat{\phi}_a, \hat{\phi}_a^\dagger ,\partial^\mu \hat{\phi}_a ,
\partial^\mu\hat{\phi}_a^\dagger \}$. The variational principle of stationary action determines Euler--Lagrange equations of motion for the field operators $\hat{\phi}_a$
%
\begin{eqnarray}
\label{eqmotionL}
\partial_\mu\frac{\partial \widehat{\mathcal{L}}^0}{\partial\big(\partial_\mu\hat{\phi}_a^\dagger\big)}
-\frac{\partial \widehat{\mathcal{L}}^0}{\partial\big(\hat{\phi}_a^\dagger\big)}&=&
\frac{\partial \widehat{\mathcal{L}}^{\rm int}}{\partial\big(\hat{\phi}_a^\dagger\big)}
-
\partial_\mu\frac{\partial \widehat{\mathcal{L}}^{\rm int}} {\partial\big(\partial_\mu\hat{\phi}_a^\dagger\big)}
\equiv \frac{\delta \widehat{\mathcal{L}}^{\rm int}}{\delta\hat{\phi}_a^\dagger (x)},
\end{eqnarray}
%
and the corresponding adjoint equations, cf. Ref.~\cite{IKV3}. The ``variational'' $\delta$-derivative
%
\begin{eqnarray}
\label{delta-derivative}
  \frac{\delta}{\delta f(x)}\dots
  \equiv \frac{\partial}{\partial f(x)}\dots
  -\partial_\mu\left(\frac{\partial}
    {\partial(\partial_\mu f(x))}\dots\right)
\end{eqnarray}
%
of $\widehat{\mathcal{L}}^{\rm int}$ permits to include derivative couplings into the interaction Lagrangian $\widehat{\mathcal{L}}^{\rm int}$. In fact, the ``variational'' $\delta$-derivative means the {\em full} derivative over $f(x)$, implying that all derivatives acting on $f(x)$ in the action should be redirected to other terms by means of partial integration before taking derivative in $f(x)$.

Further we suppress particle index ``$a$''. Principle of stationary action leads to the Euler--Lagrange equations of motion for the field operators~\cite{IKV}
%
\begin{eqnarray}
\label{eqmotion} -\widehat{G}_0^{-1} \widehat{\phi}(x)&=&
\widehat{J}(x)=\frac{\delta \widehat{\mathcal{L}}^{\rm int}}{\delta\widehat{\phi}^\dagger},
\end{eqnarray}
%
\begin{eqnarray}
\label{G0}
\widehat{G}^{-1}_{0}=\left\{
\begin{array}{ll}
-\partial_\mu\partial^\mu -m^2 \quad&\mbox{for relativistic
bosons,}\\ i\partial_t
-\frac{1}{2m}\partial_{\vec{r}}^2\quad&\mbox{for non-rel.
particles,}
\end{array}\right.
\end{eqnarray}
cf. Eq.~(\ref{sourcecl}) of classical mechanics, which we introduced in Sect.~\ref{AnharmonicGeneral}. The $\widehat{J}(x)$ operator is a local source current of the field $\widehat{\phi}$, while $\widehat{G}_0^{-1}$ is the differential operator of the free evolution with the free propagator $G^0(y,x)$ as resolvent, $x,y$ are 4-time-space points.

In the non-equilibrium case, one assumes that the system has been prepared at some initial time $t_0$ described in terms of a given density operator $\widehat{\rho}_0=\sum_a P_a\left|a\right>\left<a\right|$, where the $\left|a\right>$ form a complete set of eigenstates of $\widehat{\rho}_0$. All observables can be expressed through $n$-point Wightman functions of Heisenberg operators $\widehat{A}(t_1),\dots ,\widehat{O}(t_n)$ at some later
times
%
\begin{eqnarray}\label{corrfct}
\left\langle\widehat{O}(t_n)\dots \widehat{B}(t_2)\widehat{A}(t_1) \right\rangle
=
\sum_a P_a\left\langle a\right| \widehat{O}(t_n) \dots \widehat{B}(t_2) \widehat{A}(t_1) \left|a\right\rangle .
\end{eqnarray}
%

\begin{figure}
\centerline{\includegraphics[width=7cm]{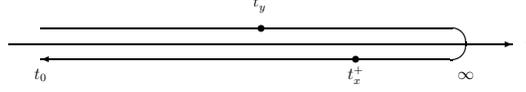}}
\caption{Schwinger-Keldysh time countour}
\label{fig:contour}
\end{figure}

The non-equilibrium theory is formulated on  closed real-time contour (see Fig.~\ref{fig:contour}) with the time argument running from $t_{0}$ to $\infty$ along the  time-ordered branch and back to $t_{0}$ along the  anti-time-ordered branch. Contour-ordered multi-point functions are defined as expectation values of contour ordered products of operators
%
\begin{eqnarray}\label{contex}
\left\langle \mathcal{T}_{\rm C} \widehat{A}(x_1)\, \widehat{B}(x_2)\dots\right\rangle
=\left\langle\mathcal{T}_{\rm C} \widehat{A}_{\rm I}(x_1)\, \widehat{B}_{\rm I}(x_2)
\dots e^{\left\{i\int_{\rm C} \widehat{\mathcal{L}}^{\rm int}_{\rm I}\rmd x\right\}}\right\rangle,
\end{eqnarray}
%
where $\mathcal{T}_{\rm C}$  orders the operators according to a time parameter running along the time contour ``C''. The left-hand side is written in the Heisenberg representation, whereas the right-hand side, in the interaction (${\rm I}$) representation. 

The contour ordering obtains its particular sense through the placement of external points on the contour.  One then has to distinguish between the physical space-time coordinates $x,\dots$ and the corresponding contour coordinates $x^{\rm C}$, which for a given $x$ take two values $x^-=(x^-_{\mu})$ and $x^+=(x^+_{\mu})$, $\mu\in\{0,1,2,3\}$, on the time ordered and anti-time ordered branches, respectively (see Fig.~\ref{fig:contour}). Closed real-time contour integrations are decomposed as
%
\begin{eqnarray}
\int_{\rm C} \rmd x^{\rm C} \dots
= \int_{t_0}^{\infty}\rmd x^-\dots
+ \int^{t_0}_{\infty}\rmd x^+\dots
=\int_{t_0}^{\infty}\rmd x^-\dots
-\int_{t_0}^{\infty}\rmd x^+\dots\, ,
\end{eqnarray}
%
where only the time limits are explicitly given. Thus, the anti-time-ordered branch acquires an extra minus sign, if integrated over physical times. For any two-point function $\mathcal{F}$, the contour values on the different branches define a $2\times 2$-matrix function
%
\begin{equation}\label{Fij}
\mathcal{F}^{ij}(x,y)=\mathcal{F}(x^i,y^j), \quad i,j\in\{-,+\},\,\,\,
\end{equation}
%
depending on the physical coordinates $(x,y)$. The contour $\delta$-function is
determined as
%
\begin{eqnarray}
\delta_{\rm C}^{ij} (x,y) = \delta_{\rm C} (x^i,y^j) = \sigma^{ij} \delta^4(x-y),
\quad
\sigma^{ij}=\left(\begin{array}{cc}1&0\\0&-1\end{array}\right)\,,\nonumber
\end{eqnarray}
%
where the matrix $\sigma^{ik}$ accounts for the integration sense on the two branches. For any multi-point function, the external point $x_{\rm max}$, which has the largest physical time, can be placed on either branch of the contour without changing the value, since the contour-time evolution from $x_{\rm max}^-$ to $x_{\rm max}^+$ provides unity. Therefore, one-point functions have the same value on both sides of the contour. Due to the change of operator ordering, genuine multi-point functions are discontinues in general, when two contour coordinates become identical.

Boson fields may take non-vanishing expectation values of the field operators $\phi(x)=\langle\widehat{\phi}(x) \rangle$. The corresponding equations of motion for these classical fields are provided by the ensemble average of the operator equations of motion (\ref{eqmotion})
\begin{eqnarray}
\label{eqmotion1}
-\widehat{G}_0^{-1} {\phi}(x)&=& J(x),\quad
\phi =\phi^0 (x) -\int_{{\cal C}}\rmd y\, G^0(x,y)\,J(y)\,,
\end{eqnarray}
now in a full analogy to Eq. (\ref{sourcecl}), which we used in classical mechanics.
Here $J(x)=\langle\widehat{J}(x)\rangle$, while $\phi^0(x)=\langle\widehat{\phi}_{\rm I}(x)\rangle$ is the freely evolving classical field, which starts from $\phi^0(t_0,{\vec x})$ at time $t_0$. Thereby, $G^{0}(x,y)$ is the free contour Green's function
%
\begin{eqnarray}\label{G0-int}
iG^{0} (x,y)= \big\langle \mathcal{T}_{\rm C}
\widehat{\phi}_{\rm I}(x)\,
\widehat{\phi}_{\rm I}^{\dagger} (y)\big\rangle
-\phi^0(x)\,(\phi^0(y))^*\,,
\end{eqnarray}
%
which resolves the equation
%
\begin{eqnarray}
\label{G0-eq.} \widehat{G}_{0}^{-1}G^{0}(x,y) &=&\delta_{\rm C}
(x,y)\,
\end{eqnarray}
%
on the contour. Graphically Eq.~(\ref{eqmotion1}) can be depicted as
\begin{eqnarray}
\includegraphics[width=6cm]{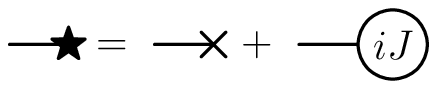}
\end{eqnarray}
with one-point function $iJ(x)$ as the driving term.

Performing replacements in Eq.~(\ref{eqmotion1}):
\begin{eqnarray}
\phi\to mz,\quad \widehat{G}_0^{-1}\to -\partial_t^2
-\Gamma\partial_t -E_R^2, \quad {J}\to-m\Lambda z^2 +F
\end{eqnarray}
we arrive at the results for unharmonic oscillator in external field, see Eqs.~(\ref{class-eom}), (\ref{sourcecl}), (\ref{freesol}), (\ref{class-eqG0}) Sect.~\ref{Mech}. Classical Maxwell equations follow with the help of replacements
\begin{eqnarray}
\phi\to A^\mu ,\quad \widehat{G}_0^{-1}\to -\partial_\nu\partial^\nu , \quad {J}\to \frac{4\pi}{c}j^\mu .
\end{eqnarray}

Subtracting the classical fields via $\widehat{\phi}(x) =\phi(x) +\widehat{\Psi}(x)$, we define the full propagator in terms of quantum-fluctuating parts $\widehat{\phi}(x)$ of the fields
%
\begin{eqnarray}\label{Ga1}
i\,G(x,y) = \big\langle{\cal T}_{\rm C} \widehat{\Psi} (x\,)\widehat{\Psi}^{\dagger}(y) \big\rangle
= \big\langle{\cal T}_{\rm C} \widehat{\phi} (x)\widehat{\phi}^{\dagger}(y)\big\rangle
 -\phi (x)\phi^{*} (y)\,.
\end{eqnarray}
%

Averaging the operator equations of motion (\ref{eqmotion}) multiplied by
$\widehat\phi^{\dagger}(y)$ and subtracting classical field parts one obtains the equation of
motion for the propagator
%
\begin{eqnarray}\label{Dyson0}
\widehat{G}_0^{-1}(x) G (x,y) = \delta_{\rm C}(x,y)
+ i\big\langle{\cal{T}}_{\rm C}\widehat{J}(x)\,\widehat\Psi^{\dagger}(y)\big\rangle_{\rm c} ,
\end{eqnarray}
%
which is still exact and accounts for the full set of initial correlations contained in $\widehat{\rho}_0$. The sub-label "c" indicates that uncorrelated parts are subtracted. Feynmann diagrammatic representation of the processes is not yet possible at this level. This description level should be still time reversible.

\subsection{Weakening of short-range correlations and Dyson equation}

In order to proceed further one suggests that the typical interaction time $t_{\rm int}$ for the change of the correlation functions is significantly shorter than the typical time,  which determines the system evolution. Then, describing the system at times $t-t_0 \gg t_{\rm int}$, one can neglect the short range correlations, which are supposed to die out beyond $t_{\rm int}$ in line with the  principle of weakening of initial and all short-range $\sim t_{\rm int}$ correlations \cite{Bogolyubov}. After dropping higher order correlations for
the driving terms on the right-hand side of the equation of motion (\ref{Dyson0}) one can apply the standard Wick decomposition. With the help of (\ref{contex}) the driving term can be expressed as functional of one-particle propagators rather than of higher order correlations
%
\begin{eqnarray}
\label{eqmotion2}
i\big\langle\mathcal{T}_{\rm C} {\widehat{J}}(x) \widehat\Psi^{\dagger}(y)\big\rangle_{\rm c}
&&= i\int_{\rm C} \rmd z\left\langle\mathcal{T}_{\rm C}
\frac{\partial }{\partial \widehat{\Psi}_{\rm I}(z)}
\left[ e^{\left\{i \int_{\rm C} \rmd z' \widehat{\mathcal{L}}^{\rm int}_{\rm I}\right\}}
\widehat{J}_{\rm I}(x)\right]\right\rangle_{\rm c1}
\,
\left<\mathcal{T}_{\rm C} \widehat{\Psi}(z) \widehat{\Psi}^{\dagger}(y) \right>
\nonumber\\
&&=\int_{\rm C} \rmd z\Sigma(x,z) \,G(z,y)\,.
\end{eqnarray}
%
Thus one recovers the Dyson equation in the differential form
%
\begin{eqnarray}\label{Dyson}
\widehat{G}^{-1}_0(x) G (x,y) = \delta_{\rm C} (x,y)+
\int_{\rm C} \rmd z \Sigma (x,z)\,G (z,y)\,.
\end{eqnarray}
%
Since we have separated the full propagator in Eq.~(\ref{eqmotion2}), the self-energy of the particle,
\begin{eqnarray}
-i\Sigma (x,y) =-\left<{\cal{T}}_{\rm C}\widehat{J}
(x)\widehat{J}^{\dagger}(y)\right>_{\rm c1},
\end{eqnarray}
here given in the Heisenberg picture, is one-particle irreducible (label $\rm c1$), i.e. the corresponding diagram cannot be split into two pieces, which separate $x$ from $y$ by cutting a single propagator line. In diagrams free and full propagators are usually given by thin and thick  lines, respectively. 
Therefore the Dyson equation (\ref{Dyson}) in a graphical form is depicted as
\begin{eqnarray}
\includegraphics[width=6cm]{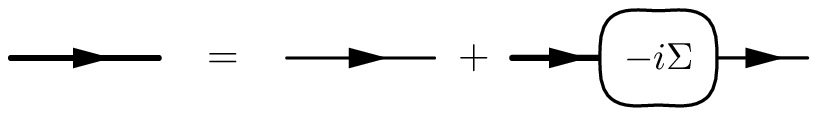}
\end{eqnarray}
with two-point function $-i\Sigma (x,y)$ as the driving term.

We would like to point out that in the derivation of the Dyson equation (\ref{Dyson}) with the application of Wick decomposition we have already lost the time reversibility. Any loss of information results in a growth of the entropy. Therefore, dropping of short range correlations on each time step would lead to a growth of the entropy, associated with the thus obtained Dyson equation, with time.

\subsection{ $\Phi$-Derivable Approximation Scheme}

For any practical calculation one has to apply some approximation scheme. In the weak-coupling limit, the perturbative expansion may be restricted to a certain order. Then no particular problems are encountered as far as conservation laws are concerned, since they are fulfilled order by order in perturbation theory. On the other hand, such perturbative expansion may not be adequate, as, for example, in the strong coupling limit, where re-summation concepts have to be applied. Such schemes sum up certain sub-series of diagrams to any order. Furthermore, with the aim to solve dynamical equations of motion, such as transport equations, one automatically re-sums all terms in the equations of motion to any order.

A $\Phi$-derivable approximation, first introduced by Baym~\cite{Baym} within the imaginary time formulation, is constructed by confining the infinite set of diagrams for $\Phi$ to either only a few of them or some sub-series of them. Note that $\Phi$ itself is constructed in terms of ``full'' Green's functions, where ``full'' now takes the sense of solving self-consistently the Dyson's equation with the driving terms derived from this $\Phi$ through relation $-i\Sigma = \mp  \delta \Phi/\delta G$. It means that even restricting  to a single diagram in $\Phi$, in fact, we deal with a whole sub-series of diagrams in terms of free Green's functions, and ``full'' takes the sense of the sum of this whole sub-series. Thus, a $\Phi$-derivable approximation offers a natural way of introducing closed, i.e. consistent approximation schemes based on summation of diagrammatic sub-series. In order to preserve the original symmetry of the exact $\Phi$, we postulate that the set of diagrams defining the $\Phi$-derivable approximation complies with all such symmetries. As a consequence,
approximate forms of $\Phi^{\rm (appr.)}$ define effective theories, where $\Phi^{\rm (appr.)}$ serves as a generating functional for approximate self-energies $\Sigma^{\rm
(appr.)}(x,y) = \mp  \delta \Phi^{\rm (appr.)}/\delta G$, which then enter as driving terms for the Dyson's equations (\ref{Dyson}). The propagators, solving this set of Dyson's equations, are still called ``full'' in the sense of the $\Phi^{\rm (appr.)}$-derivable scheme. For such re-summing schemes, the  conservation laws are preserved \cite{IKV,IKV2,KIV01,IKV3}.

\subsection{Typical time delays and advances}

For classical fields  Eq. (\ref{eqmotion1}) is similar to that we have considered in classical mechanics. Therefore  at least in some specific cases the field dynamics is characterized by the same typical time-scales as we considered above in Sect.~\ref{Mech}. Moreover, within the obtained quantum field description appear new typical time scales for delays and advances:

i) Since we neglected short range correlations we cannot anymore distinguish time effects on time scale $t\lsim t_{\rm int}$ and spatial effects on space scale $x\lsim x_{\rm int}\sim ct_{\rm int}$. Thus, we further consider a system at sufficiently large space-time scales only
\begin{equation}
t\gg t_{\rm int}.
\end{equation} 
On such a time scale we can describe the system in terms of the Feynman diagrams.

Note that the interaction time is of the order  $\Lambda_{\rm int}/c$, where $\Lambda_{\rm int}$ is the shortest distance at which we can use our interaction model and the chosen degrees of freedom.


ii) Some time scales  can be extracted right from expressions for the single particle
non-equilibrium Green's functions $G^{ij}$ for $i,j\in \{+,-\}$. Actually, only two quantities among four  $G^{ij}$ are independent \cite{IKV}. As these two quantities it is  convenient to use the Wigner density
\begin{eqnarray}
\label{Fdef}
F(t_1,\vec{r}_1; t_2,\vec{r}_2)=(\mp)i\,G^{-+}(x_1,x_2)
=\langle\Psi^{\dagger}(x_2)\Psi(x_1)\rangle\,,
\end{eqnarray}
and the retarded Green's function
\begin{eqnarray}\label{Gret}
i\,G^{R}(x_1,x_2)=\left\{
\begin{array}{ll}
\langle\Psi(x_1)\Psi^{\dagger}(x_2)\pm \Psi^{\dagger}(x_2)\Psi(x_1) >
\quad&\mbox{for}\,\, t_1 >t_2,\\
0\quad&\mbox{for}\,\, t_1 <t_2,
\end{array}\right.
\end{eqnarray}
where the upper sign is for fermions and the lower one is for bosons. The particle width in coordinate representation is determined as
\begin{eqnarray}
 \label{Gamma}
\Gamma(x_1,x_2) \equiv -2\Im \Sigma^R(x_1,x_2)\,.
\end{eqnarray}
The time scale related to this quantity is  the decay time $t_{\rm dec}=1/\Gamma$, where $\Gamma$ is the Fourier transform of $\Gamma(x_1,x_2)$ for the stationary system, cf. Eqs. (\ref{tdec}), (\ref{tdec-res}), (\ref{tfQ1}). The quantity
\begin{eqnarray}
\label{Adef}
 A(x_1,x_2) =-2\Im G^R(x_1,x_2)
\end{eqnarray}
is the spectral function in the coordinate representation,  cf. Eqs.~(\ref{3-times-A}), (\ref{AnalG}). With quantity $A$, where $A$ is the Fourier transform of $A_{12}$ for a stationary system, is associated a time delay $\delta t_{\rm s}=A/2$, as it has been
introduced above, cf. Eq. (\ref{Baztime}), which  in the virial limit is $\delta t_{\rm s}=\frac{\partial \delta}{\partial E_p} =2\pi \frac{dN^{\rm level}}{dE_p}$, cf. Eqs.~(\ref{A1}),~(\ref{A2}), (\ref{trad}), (\ref{rstd}), (\ref{ergod}). Note that
equation for the retarded Green's function decouples as
$$
G^R(x_1,x_2)=G^R_{0}(x_1,x_2)+\int_{\rm C}\rmd x_3 \int_{\rm C} \rmd x_4 G^R_{0}(x_1,x_3)\,\Sigma(x_3,x_4)\,G^R(x_4,x_2)
$$
and for stationary systems determines the spectrum of excitations.

As an example, consider a spatially uniform equilibrium  high temperature gas of non-relativistic Wigner resonances (i.e., when $\Gamma_{\rm eq}=-2\Im \Sigma^R_{\rm eq}$ and $\Re \Sigma^R_{\rm eq}$, being Fourier transformed to the energy-momentum space, do not depend on $p_0$ on relevant energy-momentum scales). In the mixed time-momentum representation,
\begin{eqnarray}
G(t_1 -t_2,\vec{p}\,)=\int \frac{\rmd p_0}{2\pi}\, G(p_0 ,\vec{p}\,)\,e^{-i\,p_0 (t_1 -t_2)},
\end{eqnarray}
and the retarded Green's function is
\begin{eqnarray}\label{Gret1}
G^R_{\rm eq} (t_1 -t_2,\vec{p}\,)= -i\,e^{-i\,E_p (t_1 -t_2)-\frac{1}{2}\Gamma_{\rm eq} (t_1 -t_2)}
\quad \mbox{for}\quad
t_{1}>t_2,
\end{eqnarray}
and $G^R =0$ for $t_{1}<t_2$, ${E}_p =E_p^0 +\Re \Sigma^R_{\rm eq}$, $E_p^0 =\frac{p^2}{2m}$, cf. Eq. (\ref{GRtime}). Using the Kubo-Schwinger-Martin relation \cite{Kad62} for the Fourier
transformed equilibrium Green's functions, $F_{\rm eq}(p_0, \vec{p}\,)=A_{\rm eq}(p_0, \vec{p}\,) \, f_{\rm eq}(p_0, \vec{p}\,)$ in the mixed representation we find
\begin{eqnarray}
\label{Mixed}
F_{\rm eq}(t_1 -t_2,\vec{p})=
\int \frac{\rmd p_0}{2\pi} A_{\rm eq}(p_0,\vec{p}\,)\, f_{\rm eq}^{\rm Bol}(p_0,\vec{p}\,)  e^{-i \, p_0\, (t_1 -t_2)}
= e^{\frac{\mu}{T}-i\,E_p\, (t_1 -t_2-\delta t_{\rm s}^T )
 - \frac{1}{2}\Gamma_{\rm eq}\, (t_1 -t_2-\delta t_{\rm col}^T)}.
\end{eqnarray}
Here $A_{\rm eq}$ is the spectral function for equilibrium system, $f_{\rm eq}^{\rm Bol}=e^{(\mu-p_0)/T}$ is the equilibrium Boltzmann distribution function, $\mu$ is the  chemical potential determined by the total number of particles $N$. New time scales are
\begin{eqnarray}\label{QFTshifts}
\delta t_{\rm s}^T =\frac{\Gamma_{\rm eq} }{2\,E_p\, T},\quad \delta
t_{\rm col}^T =-\frac{2\,E_p}{\Gamma_{\rm eq}\, T}.
\end{eqnarray}
The value $\delta t_{\rm s}^T$ shows a scattering delay time, and $\delta t_{\rm col}^T $ is the collision advancement time for equilibrium processes. For typical energies $E_p  \sim T$, $\delta t_{\rm col}^T\sim -1/{\Gamma}_{\rm eq}$. Thus the latter quantity demonstrates an
advance of particles of thermal energies compared to particles being at rest.
\begin{figure}
\centerline{\includegraphics[width=8cm]{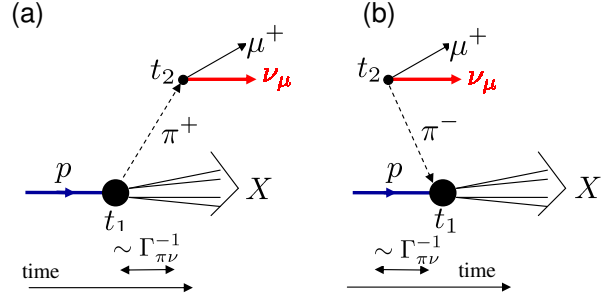}}
\caption{Graphical representation of the  two-step processes $p\to
n+X+\pi_{\rm virt}^+$, $\pi_{\rm virt}^+\to {\nu}+\mu^+$, as they
occur  in the neutrino experiment of the OPERA group
\cite{OPERA}.} \label{fig:opera}
\end{figure}

iii)~Since integration over $z$ in Eq. (\ref{Dyson}) in intermediate reaction states  includes all times $-\infty <t_z <\infty$, for $t_y <t_z <t_x$,  the process occurred at $t_z$ is delayed compared to that occurred at $t_y$, and for $t_z <t_y <t_x$ the process occurred at $t_z$ is advanced compared to that occurred at $t_y$. Both time processes should be incorporated as dictated by the Lorentz invariance. In Fig.~\ref{fig:opera} we demonstrate
example of a time  delay (a) and an advancement (b) in the specific two-step processes $p\to n+X+\pi_{\rm virt}^+$, $\pi_{\rm virt}^+\to {\nu}+\mu^+$. Such processes play important role in the recent neutrino OPERA experiment \cite{OPERA}. Let the life time of the off-mass-shell pion produced in the process $p\rightarrow n+X+\pi_{\rm virt}^+$ is $ t_{N\pi}^{\rm dec}= 1/\Gamma_{N\pi}$ and in the process $\pi_{\rm virt}^+\to {\nu}+\mu^+$ it is $t_{\pi\nu}^{\rm dec}= 1/\Gamma_{\pi\nu}$. The time $t_{\nu}^{\rm dec}= t_{N\pi}^{\rm dec}+t_{\pi\nu}^{\rm dec}$ characterizes duration of the full two-step process. This means that virtual pions, being produced in the process $p\to n+X+\pi_{\rm virt}^+$, undergo in the  subsequent process
$\pi_{\rm virt}^+\to {\nu}+\mu^+$ time delays and advances on a time scale $- t_{\pi\nu}^{\rm dec}\lsim t_2-t_1\lsim t_{\pi\nu}^{\rm dec}$, where $t_2$ characterizes act of the
production of $\nu$ and $t_1$, of the absorption of $p$. Note that
$t_{N\pi}^{\rm dec}\ll t_{\pi\nu}^{\rm dec}$.

The uncertainty in the production time reflects also the fact that a system undergoing some transition to a new state (decay) does not have a defined energy and is described in quantum mechanics be a wave packet possessing a finite spread in space and time of the order of $1/\Gamma$, cf Eq.~(\ref{uncert}).

iv)~One may introduce another time scales similarly  to that we introduced in classical and quantum mechanical descriptions. For example, for $1+1$ dimensional problem the quantity
\begin{equation}
\label{dwellF}
t_{\rm soj}^{({\rm 1D})}(a,b;\tau) =\int_0^{\tau}\rmd t\int_{a}^{b} \rmd z
F^{({\rm 1D})}(t,z; t,z)/\int \rmd z F^{({\rm 1D})}(t,z; t,z)
\end{equation}
is similar to that given by Eq. (\ref{sojt-class}) and (\ref{soj-quant}), and for $3+1$ theory,
\begin{equation}\label{dwellF3}
t_{\rm soj}(\Omega,\tau) =\int_0^{\tau}\rmd t\int \rmd^3 r F(t,\vec{r};
t,\vec{r})/\int \rmd^3 r F(t,\vec{r}; t,\vec{r})
\end{equation}
is similar to that given by Eqs. (\ref{sojt-class3}) and (\ref{tvol}).

\section{Time shifts in quantum kinetics}\label{sec:kin}

\subsection{Wigner transformation and gradient expansion}

Consider slightly inhomogeneous and slowly evolving systems. Then in the spirit of semiclassical approximation degrees of freedom can be subdivided into rapid and short-ranged and slow and long-ranged. For any two-point function $\mathcal{F}(x,y)$, one may introduce the variable $\xi =(t_1-t_2, \vec{r}_1-\vec{r}_2)$, which relates to rapid and short-ranged microscopic processes, and the variable $X= \frac{1}{2}(t_1+t_2,\vec{r}_1 +\vec{r}_2)$, which
refers to slow and long-ranged collective motions. The Wigner transformation~\cite{Wigner-trafo}, i.e., the Fourier transformation in  $\xi=x-y$ to 4-momentum $p$, leads to $\mathcal{F}(X,p)$ functions. Since the Wigner transformation is defined for physical time-space coordinates rather than for contour coordinates one has to decompose the contour integrations into the time-ordered $\{-\}$  and the anti-time ordered $\{+\}$ branches. Two-point functions then become matrices of the contour decomposed $\{-+\}$ components with physical space-time arguments. Thus
%
\begin{equation}
\label{W-transf}
\mathcal{F}^{ij}(X,p)=\int \rmd \xi e^{i p\xi}\,
\mathcal{F}^{ij}\left(X+\frac{\xi}{2},X-\frac{\xi}{2}\right),
\quad i,j\in\{-+\}
\end{equation}
%
leads to a four-phase-space representation of two-point functions.

The Wigner transformation of the Dyson equation (\ref{Dyson}) is straightforward. Taking the difference and half-sum of the Dyson equation (\ref{Dyson}) and the corresponding adjoint equation after the Wigner transformation we arrive at equations
%
\begin{eqnarray}
\label{Dyson-} i v_{\mu}  \partial^\mu_X G^{ij}(X,p) &=&
\int \rmd \xi e^{i p\xi} \int_{C} \rmd z \left( \Sigma (x^i,z)G(z,y^j) -
G(x^i,z)\Sigma (z,y^j) \right),
\\
\label{Dyson+} \widehat{Q}_X G^{ij}(X,p) &=& \sigma^{ij} +
\frac{1}{2} \int \rmd \xi e^{i p\xi} \int_{C} \rmd z \left( \Sigma
(x^i,z)G(z,y^j) + G(x^i,z)\Sigma (z,y^j) \right),
\end{eqnarray}
%
where $\sigma^{ij}$ accounts for the integration sense on the two contour branches, cf. Eq.~(\ref{Fij}). For non-relativistic kinematics $v^{\mu}=(1, \vec{p} /m)$, and $\widehat{Q}_X = p_0-{\vec p}^{\,2}/2m- \partial_{\vec{X}}^2/8m$. In this matrix notation, two of Eqs.~(\ref{Dyson-}) and (\ref{Dyson+}), involving $G^{-+}$ and $G^{+-}$ on the left-hand
side, are known as Kadanoff-Baym equations in the Wigner representation \cite{Kad62}. Particular combinations of these equations lead to the retarded and advanced equations, which
completely decouple and involve only integrations over physical times rather than contour times.

We will solely deal with the gradient approximation for slow collective motions by performing the gradient expansion of Eqs.~(\ref{Dyson-}) and (\ref{Dyson+}). This step preserves all the
invariances of the $\Phi$ functional in a $\Phi$-derivable approximation~\cite{IKV2}. Within the gradient expansion the Wigner transformation of a convolution of two two-point functions
entering the Dyson equations (\ref{Dyson+}), (\ref{Dyson-}) is given by
%
\begin{eqnarray}
\label{g-rule2} \int \rmd \xi e^{i p \xi} \left( \int \rmd z f(x,z)
\varphi(z,y) \right)\approx f(X,p) \varphi(X,p) + \frac{i\hbar}{2}
\{f(X,p), \varphi(X,p)\},
\end{eqnarray}
%
where
%
\begin{equation}
\label{[]}
\{f(X,p) , \varphi(X,p)\} =
\frac{\partial f}{\partial p^{\mu}}
\frac{\partial \varphi}{\partial X_{\mu}}
-
\frac{\partial f}{\partial X^{\mu}}
\frac{\partial \varphi}{\partial p_{\mu}}
\end{equation}
%
is the  Poisson bracket in covariant notation. Note that the smallness of the $\hbar\partial_X \cdot\partial_p$ comes solely from the smallness of space--time gradients $\partial_X$, while
momentum derivatives $\partial_p$ are not assumed to be small. Such a description is meaningful only if the typical time and space scales are large compared to microscopic ones
$|t_i -t_j|\sim t_{\rm mic}\sim (1/E_{\rm F}, 1/E_{\rm T})$,
$|\vec{r}_i -\vec{r}_j|\sim r_{\rm mic} \sim (1/p_{\rm F}, 1/p_{\rm T})$
(for the low energy excitations in the Fermi system and for the Boltzmann gas, respectively,  where index $\rm F$ labels the Fermi quantity and $\rm T$, the thermal one).

\subsection{Three  forms of quantum kinetic equation}

Only two real functions of all $G^{ij}(X,p)$ are required for a complete description of the system's evolution \cite{IKV2}. As these real functions, it is convenient to use the Wigner density $F(X,p)=(\mp)iG^{-+}(X,p)$ and the spectral function $A(X,p)$. The kinetic equation for off-shell particles is ordinary presented in the two different forms: in the  Kadanoff-Baym (KB) form, i.e. as it follows right after the first order gradient expansion in the Dyson equation for the Wigner density \cite{Kad62},
%
\begin{eqnarray}
\label{keqk1} \widehat{D}\, F(X,p) - \big\{\Gamma_{\rm in}(X,p) , \Re G^R(X,p)\big\} &=& C (X,p),
\end{eqnarray}
%
and in the Bottermans-Malfliet  (BM) form \cite{Bot90}, see discussion of these aspects in \cite{IKV2}. Similarly  kinetic equation can be written for $\widetilde{F}(X,p)\equiv iG^{+-}(X,p)$. The collision term
\begin{eqnarray}
C(X,p)=\Gamma^{\rm in}(X,p)\,\widetilde{F}(X,p) -
\Gamma^{\rm out}(X,p)\, F(X,p)
\label{Colterm}
\end{eqnarray}
is the difference of the gain and loss terms, $\Gamma^{\rm in}(X,p)=\mp i\Sigma^{-+}(X,p)$ (in the co-variant notations of \cite{IKV2}) is the reduced production (gain) source term, $\Gamma^{\rm out}(X,p)=i\Sigma^{+-}(X,p)$ is the reduced absorption (loss) term. The differential drift operator is defined as
%
\begin{eqnarray}\label{Drift-O}
\widehat{D}(\dots)= \big\{\Re [G^{-1}(X,p)], \dots\big\}
= Z^{-1}_{\mu}(X,p) \frac{\partial}{\partial X_\mu} +
\frac{\partial\Re\Sigma^R(X,p) }{\partial X^{\mu}} \frac{\partial }{\partial p_{\mu}},
\end{eqnarray}
with
\begin{eqnarray}\label{norm}
Z^{-1}_{\mu}(X,p) =\frac{\partial}{\partial p^\mu} \Re[G^{-1}(X,p)]= v_{\mu} - \frac{\partial \Re\Sigma^R(X,p)}{\partial p^{\mu}},\quad
v^\mu=\frac{\partial}{\partial p_{\mu}}G_0^{-1}(p).
\end{eqnarray}
Here $G^{-1}_{0}(p)$ is the Fourier transform of the inverse free Green's function (\ref{G0}):
\begin{eqnarray}
\label{G0p} G^{-1}_{0}=\left\{
\begin{array}{ll}
p^2-m^2\quad&\mbox{for relativistic bosons,}\\ p_0-{\vec
p}^{\,\,2}/(2m)\quad&\mbox{for non-rel. particles,}
\end{array}\right.
\end{eqnarray}
$m$ is the mass of the free particle. The BM form is obtained from the KB equation, if one puts $\Gamma^{\rm in}(X,p)=\Gamma(X,p) F(X,p)/A(X,p)$ in the Poisson bracket term. This replacement is legitimate, if one assumes only small (the first gradient order) deviations from the local or global equilibrium (in the local and global equilibrium the collision term  $C(X,p)=0$). Then the BM form differs from the KB form only in the second order of the gradient expansion (more precisely in the first order gradient times first order deviation from equilibrium), see~\cite{IKV2} for the details. At first glance these equations are equivalent in their common region of validity. However the KB equation has still one important advantage. It fulfills the conservation laws of the Noether 4-current and of the Noether energy-momentum exactly~\cite{KIV01,IKV3} provided self-consistent, i.e., $\Phi$ derivable approximations are used, whereas the BM form fulfills the conservation laws only approximately (up to zero gradients). Moreover the KB equation can be applied not only to description of the relaxation of the system towards the local equilibrium but in many other problems, in which the BM form is not applicable.

The third, non-local, form of the kinetic equation was introduced
in \cite{IV09}:
%
\begin{eqnarray}
&&\widehat{D} F(X,p) - \Big\{\Gamma(X,p)\frac{F(X,p)}{A(X,p)},\Re G^R(X,p)\Big\}=C^{\rm NL}(X,p)
\nonumber\\
&&\qquad\qquad\qquad
= \Big(1+\Big\{\frac{1}{A(X,p)},\Re G^R(X,p)\Big\}\Big)\, C^{\rm shift}(X,p)\,,
\label{new-kin} \\
&&
C^{\rm shift}(X,p)={C\left(X^\mu - \delta X^\mu , p^\mu-\delta p^\mu\right)},\quad
\nonumber\\
&&
\delta X^\mu =\frac{1}{A(X,p)}\frac{\partial \Re G^R(X,p) }{\partial{p_\mu}},
\quad
\delta p^\mu =-\frac{1}{A(X,p)}\frac{\partial \Re G^R(X,p) }{\partial {X_{\mu}}}\,.
\nonumber
\end{eqnarray}
If we replace $C^{\rm NL}(X,p)\to C(X,p)$ we obtain the BM form. If we expand $C^{\rm NL}(X,p)$ up to first gradient terms we arrive at the KB equation. Thus Eq. (\ref{new-kin})  coincides with the BM form up to the first order gradients and with the KB equation  up to second order gradients.

The retarded Green's function is
\begin{eqnarray}
\label{Asol}
G^R(X,p)=\frac{1}{M(X,p) + i \Gamma (X,p)/2}+O(\partial_X^2)
\end{eqnarray}
and
\begin{eqnarray}
\label{Ares}
A (X,p) &\equiv&-2\Im G^R(X,p)\simeq \frac{\Gamma (X,p)}{M^2 (X,p)
+\quart \Gamma^2(X,p)}\,
\end{eqnarray}
is the spectral function
%
with the  ``mass'' function and the width given by
%
\begin{eqnarray}\label{M}
M(X,p)=G^{-1}_{0}(p) -\Re\Sigma^R (X,p),\quad
\Gamma(X,p) \equiv -2\Im \Sigma^R(X,p) = \Gamma^{\rm out}(X,p)\pm \Gamma^{\rm in}(X,p)\,,
\end{eqnarray}
%
cf. the introduced above definitions (\ref{Gamma}) and (\ref{Adef}). We see that expression (\ref{Ares}) has the same resonance form, as Eqs.~(\ref{A1}),  (\ref{A2}) in mechanical example considered in Sect.~\ref{Mech}, as Eq.~(\ref{trad}) in classic-electrodynamical example, and as Eq. (\ref{AnalG}) in quantum mechanical example considered in Sect.~\ref{sec:quant}. Although the solution (\ref{Asol}), (\ref{Ares}) is simply
algebraic, it is valid up to first-order gradients.

To simplify the further consideration we imply that in relativistic case separation of particle and antiparticle degrees of freedom is performed. Therefore, below we deal with particle species. Antiparticle quantities can be introduced similarly.

In Eq. (\ref{new-kin}) the Wigner densities can be presented as
%
\begin{eqnarray}
\label{F}
F (X,p) =A(X,p)f(X,p), \quad \widetilde{F} (X,p) =A(X,p)(1\mp f(X,p)),
\end{eqnarray}
where $f(X,p)$ is a new generalized distribution function. In a local equilibrium this function takes the form
\begin{eqnarray}
f_{\rm l.eq} (X,p) =\frac{1}{e^{[p_\mu\, u^\mu(X) -\mu (X)]/T(X)}\pm 1},
\label{fleq}
\end{eqnarray}
where $u^\mu(X)=(1,\vec{u}(X))/\sqrt{1-\vec{u\,}^2(X)}$  and $\vec{u}(X)$ is the local velocity of a tiny element of the system. This distribution turns the collision term to zero.

Using Eq.~(\ref{F}) we are able to derive other form of the non-local kinetic equation (\ref{new-kin}):
\begin{eqnarray}\label{new-kin-eq}
&&\frac{1}{2}A^2(X,p)\Gamma(X,p)
\Big(\widehat{D}f(X,p)- \frac{M(X,p)}{\Gamma(X,p)}
\big\{\Gamma(X,p), f(X,p)\big\}\Big)=  A(X,p)\,\mathcal{C}^{\rm shift}(X,p),
\nonumber\\
&&\quad \mathcal{C}^{\rm shift}(X,p)=\frac{C^{\rm shift}(X,p)}{A^{\rm shift}(X,p)}\,,
\quad
A^{\rm shift}(X,p)= A( X^\mu -\delta X^\mu  , p^\mu -\delta p^\mu).
\end{eqnarray}
Equation~(\ref{new-kin-eq}) up to the second-order gradients coincides with the KB equation and up to the first-order gradients coincides with the BM form, cf. the kinetic equation in the BM form, Eq. (3.28) in \cite{IKV2}. Note that besides non-locality introduced by the shift of variables, $C ^{\rm shift}$ and $C$ may still include the memory effects, if the generating 2PI $\Phi$ functional contains diagrams with more than two vertices.

The key point which we want further to focus on is the variable shift in the collision term
\begin{eqnarray}
\delta X^\mu(X,p) &\equiv& (\delta t_f^{\rm kin} ,\delta\vec{r}_f)
=\frac{1}{A(X,p)}\frac{\partial \Re G^R(X,p) }{\partial{p_\mu}}
=\frac{1}{2}B^\mu(X,p) - \frac{Z^{-1,\mu}(X,p)}{\Gamma(X,p)},
\nonumber\\
\delta p^\mu(X,p)
&=&-\frac{1}{A(X,p)}\frac{\partial \Re G^R(X,p) }{\partial {X_{\mu}}},
\label{deltaX}
\end{eqnarray}
where
\begin{eqnarray}\label{B-j}
B^{\mu}(X,p)
&&= \big(B_0(X,p) , B_0(X,p) \,\vec{v}_{\rm gr}(X,p)\big)
=-2\Im \left[\left(v^{\mu} - \frac{\partial \Re\Sigma^R(X,p)}{\partial
p_\mu}\right)\,G^R(X,p)\right]\nonumber\\
&&=A(X,p)\left[
v^{\mu} - \frac{\partial \Re\Sigma^R(X,p) }{\partial p_\mu}
- \frac{M(X,p)}{\Gamma(X,p)}\frac{\partial \Gamma(X,p)}{\partial p_\mu}
\right]
\end{eqnarray}
is a flow spectral function and
\begin{eqnarray}\label{grv}
\vec{v}_{\rm gr}(X,p)=\frac{
\vec{v} + \frac{\partial \Re\Sigma^R(X,p) }{\partial \vec{p}}
+ \frac{M(X,p)}{\Gamma(X,p)}\frac{\partial \Gamma(X,p)}{\partial \vec{p}}}
{ v^{0} - \frac{\partial \Re\Sigma^R(X,p) }{\partial p_0}
- \frac{M(X,p)}{\Gamma(X,p)}\frac{\partial \Gamma(X,p)}{\partial p_0}}
\end{eqnarray}
has the meaning of a generalized group velocity  of off-mass-shell particles, see Ref.~\cite{V2011}. The latter quantity generalizes expression for the energy $p_0$-integrated transport velocity introduced in Ref.~\cite{MalflTr} and applied for localization phenomena and resonance scattering.

The non-local kinetic equation (\ref{new-kin-eq}) can be rewritten in a more convenient form
\begin{eqnarray}
A_S^\mu(X,p) \frac{\partial f(X,p)}{\partial {X^{\mu}}}  + A(X,p)\Big[
\frac{\partial\Re \Sigma^R(X,p)}{\partial {X_{\mu}}}
-\frac{\Gamma(X,p)}{A(X,p)}\frac{\partial \Re G^R(X,p)}{\partial{X_{\mu}}}\Big]
\frac{\partial f(X,p)}{\partial {p^{\mu}}} =A(X,p) \mathcal{C}^{\rm shift}(X,p).
\nonumber\\
\label{new-kin-eqS}
\end{eqnarray}
This will be the key equation for our further study. Here
\begin{eqnarray}
A^{\mu}_{S} (X,p)=\half A(X,p)\, \Gamma(X,p)\, B^{\mu}(X,p).
\end{eqnarray}
The first term on the left-hand side of Eq.~(\ref{new-kin-eqS}) is the entropy
drift term, the second term relates to the spatial changes of a mean field.
Dropping in  Eq.~(\ref{new-kin-eqS}) 4-phase space delays/advances in the collision term, i.e. replacing $\mathcal{C}^{\rm shift}(X,p)\to \mathcal{C}(X,p)=C(X,p)/A(X,p)$, we arrive at the BM form of the kinetic equation. In the latter case $A^{\mu}_{S}(X,p)$ is the BM Markovian (BMM) entropy flow  spectral function (memory effects are ignored)  relating to the entropy flow associated with the BM form of the kinetic equation~\cite{IKV2}:
\begin{eqnarray}
\label{entr-transp}
S^\mu_{\rm{BMM}}(X)= {\rm Tr}\int \frac{\rmd^4 p}{(2\pi)^4}
A^{\mu}_{S}(X,p)\sigma(X,p) ,\,\,\,
\end{eqnarray}
where
\begin{eqnarray}
\label{As}
\sigma (X,p)
=\mp \big(1\mp f(X,p)\big)\ln\big(1\mp f(X,p)\big)-f(X,p)\ln f(X,p) ,
\end{eqnarray}
satisfying the equation of motion
\begin{eqnarray}\label{SH}
\frac{\partial}{\partial{X^\mu}}S^\mu_{\rm{BMM}}(X) = -H(X)
={\rm Tr} \int \frac{\rmd^4 p}{(2\pi)^4}\ln \frac{1\mp f(X,p)}{f(X,p)}\,C(X,p).
\end{eqnarray}
Symbol ${\rm Tr}$ implies summation over internal degrees of freedom like spin, etc.

It is easy to demonstrate that the kinetic equation in the BM form conserves the BM effective current exactly~\cite{leupold,IKV3}
\begin{eqnarray}\label{curs}
j_{S}^\mu(X) =e{\rm Tr}\int \frac{\rmd^4 p}{(2\pi)^4}
A^{\mu}_{S}(X,p)\,f(X,p) \,,
\end{eqnarray}
provided we work within the $\Phi$-derivable approximation scheme, whereas the Noether
current
\begin{eqnarray}\label{Noether}
j_{\rm Noether}^\mu(X) =e{\rm Tr}\int \frac{\rmd^4 p}{(2\pi)^4} v^\mu A(X,p)\, f(X,p)
\end{eqnarray}
and the effective $B$-current
\begin{eqnarray}\label{curB}
 j_B^\mu(X) =e{\rm Tr}\int \frac{d^4 p}{(2\pi)^4} B^\mu(X,p) f(X,p)
\end{eqnarray}
are conserved  approximately (up to zero gradient order). Here $e$ is the (electric, baryon, or other) charge of the species and summation over the species, if necessary, is implied.

The maxima in the flow $A_S^\mu(X,p)$, the flow $B^\mu(X,p)$ and $v^\mu A(X,p)$ are shifted relatively to each other in the energy-momentum space. Difference of (\ref{curs}) and (\ref{curB}) with the Noether current (\ref{Noether}) is that the former two quantities
contain contributions of the drag and back flows. The drag flow is associated with the term $- \frac{\partial \Re\Sigma^R }{\partial p_\mu}A$ and the back flow, with $- \frac{M}{\Gamma}\frac{\partial \Gamma}{\partial p_\mu}A$ in the $B^\mu$ spectral function
(\ref{B-j}). Presence of these terms causes some additional delays and advances in the propagation of dressed off-shell particles.

The expressions for currents (\ref{curs}), (\ref{Noether}), and (\ref{curB}) are derived for the simple form of the 2PI-generating $\Phi$ functional. As soon as the $\Phi$ functional includes diagrams with more than two vertices, there appears additional term in the currents --- a so-called "memory" current $j^\mu_{\rm mem}$, see Eq.~(\ref{jmem}) below, and, if the interaction contains derivative couplings, there is yet another "derivative" current term $j^\mu_{\rm der}$. The same relates also to the entropy flow.

If we expand ${\cal C}^{\rm shift}$ in (\ref{new-kin-eq})
up to the first gradients as
\begin{eqnarray}
A(X,p)\, \mathcal{C}^{\rm shift}(X,p) &=& A(X,p)\left(\Gamma^{\rm in}_{\rm shift}(X,p)
-\Gamma_{\rm shift}(X,p) f^{\rm shift}(X,p)\right)
\nonumber\\
&\simeq& C(X,p) + \big\{\Gamma^{\rm in}(X,p) - \Gamma(X,p)\,f(X,p),\Re G^R(X,p) \big\}
\,,
\label{newcol}
\end{eqnarray}
we arrive at the generalized kinetic equation in the KB form for the distribution function $f(X,p)$:
\begin{eqnarray}\label{KB}
A(X,p)\widehat{D}f(X,p)+f(X,p)\big\{\Gamma(X,p) , \Re G(X,p)\big\}
- \big\{\Gamma^{\rm in}(X,p),\Re G(X,p)\}=C(X,p),
\end{eqnarray}
where the collision term
\begin{eqnarray}\label{colG}
C(X,p)=A(X,p)\,\Gamma^{\rm in}(X,p) -A(X,p)\,\Gamma(X,p)\, f(X,p)\,,
\end{eqnarray}
\begin{eqnarray}
A(X,p)\widehat{D} =\widetilde{B}^\mu(X,p) \frac{\partial}{\partial X_\mu} +
A(X,p)\frac{\partial \Re\Sigma^R(X,p)}{\partial X^{\mu}}
\frac{\partial  }{\partial p_{\mu}},\nonumber
\end{eqnarray}
\begin{eqnarray}
\widetilde{B}^\mu(X,p) =\Big(A(X,p)Z_0^{-1}(X,p), A(X,p)\,Z_0^{-1}(X,p)\vec{\widetilde{v}}_{\rm gr}(X,p)\Big),
\end{eqnarray}
and the generalized group velocity
\begin{eqnarray}\label{grKB}
{\widetilde{v}}_{\rm gr}^i(X,p)=
\frac{Z^{-1,i}(X,p)}{Z_0^{-1}(X,p)}=\frac{
{v}^i + \frac{\partial \Re\Sigma^R(X,p)}{\partial {p}^i} }{ v^{0} -
\frac{\partial \Re\Sigma^R(X,p) }{\partial p_0} },\quad i=1,2,3,
\end{eqnarray}
differ from the quantities introduced with the help of Eqs. (\ref{B-j}), (\ref{grv}). Recall that $v_0=1$ for non-relativistic particles and $2p_0$ for relativistic bosons. The
BM form is obtained after replacement $\Gamma^{\rm in}(X,p) =\Gamma(X,p) f(X,p) +O(\partial_x)$ in the second Poisson bracket in (\ref{KB}). The KB equation (\ref{KB}) exactly conserves the Noether current (\ref{Noether}), provided approximations  are $\Phi$ derivable \cite{KIV01,IKV3}, and it approximately (up to zero-gradient order) conserves the effective current (\ref{curs}) and the effective current (\ref{curB}).

Note that, since for off-mass-shell particles $p_0$ and $\vec{p}$ can vary independently, the value $\widetilde{v}_{\rm gr}(X,p)$ is not necessarily limited from the above by the velocity of light, it might be even not positive definite. Also, the factor $Z_0^{-1}(X,p)$ is not, in general, positive definite that leads to a non-trivial procedure of the particle-antiparticle separation for virtual particles. As known~\cite{Migdal}, in the quasiparticle limit
$Z^{\rm qp}_0 (X,p_0 (\vec{p}),\vec{p}) >0$ determines the quasiparticle spectrum branches and the branches with $Z^{\rm qp}_0 <0$ are related to the anti-quasiparticles after the replacement $p_0\to-p_0$ and $\vec{p}\to -\vec{p}$. To be specific,
we will further assume $Z_0^{-1}>0$, considering only the particle and not anti-particles.

The spectral functions $A$, $B_0$, $A_S^0$ fulfill sum-rules \footnote{Generally speaking, the sum-rules presented in such a form hold for non-relativistic particles and for relativistic neutral bosons. In the former case the integration goes from $-\infty$ to $\infty$, in the latter case from $0$ to $\infty$. Otherwise, antiparticle terms are not decoupled.}
\begin{eqnarray}\label{sumrule}
\int B_0(X,p) \frac{\rmd p_0}{2\pi} =\int v^0 A(X,p) \frac{\rmd p_0}{2\pi}= \int
A_S^0(X,p) \frac{\rmd p_0}{2\pi}=1\,,
\end{eqnarray}
cf. the sum-rules (\ref{sumrulemech}), (\ref{sumrulecl}) and (\ref{sumrulet}), (\ref{QS-Phi-norm}), which we obtained  in classical and quantum mechanics. Note that the sum-rule for $A(x,p)$ follows directly from the canonical equal-time (anti) commutator
relations for (fermionic) bosonic field operators.

Having at hand the kinetic equation for $F(X,p)$ (either in KB or in BM form) and the algebraic equation for $A$ is sufficient to recover all kinetic quantities. However there exists still one more, so called the mass-shell equation \cite{IKV2}, which, as well as the
KB equation (\ref{KB}), follows from the full Dyson equations expanded up to  the first gradient order:
\begin{eqnarray}
M(X,p)\,A(X,p)\,f(X,p)-\Re G^R(X,p)\Gamma^{\rm in}(X,p)&&=\frac{1}{4}\big\{\Gamma(X,p) ,f(X,p)A(X,p)\big\}
\nonumber\\
&&-\frac{1}{4}\big\{\Gamma^{\rm in}(X,p), A(X,p)\big\}.
\label{massShell}
\end{eqnarray}
Presenting $\Gamma^{\rm in}(X,p)=f(X,p)\,\Gamma(X,p) +\delta \Gamma^{\rm in}(X,p)$,
$\Gamma^{\rm out}(X,p)=\big(1\mp f(X,p)\big)\Gamma(X,p) +\delta \Gamma^{\rm out}(X,p)$, such
that in equilibrium $\delta \Gamma^{\rm in}_{\rm eq}(X,p)=\delta
\Gamma^{\rm out}_{\rm eq}(X,p)=0$, and using Eq. (\ref{M}),
from (\ref{massShell})  we find
\begin{eqnarray}\label{gamin}
\delta \Gamma^{\rm in}(X,p)=\frac{1}{4\Re G^R(X,p)}\big\{f(X,p),A(X,p)\,\Gamma(X,p)
\big\}+O(\partial_x^2)=\mp\delta \Gamma^{\rm out}(X,p).
\end{eqnarray}
Generally speaking the mass-shell equation should be considered on equal footing with the kinetic equation. Ref. \cite{IKV2} proved equivalence of this equation to the kinetic equation in the BM form up to first gradients. However in general case equivalence of (\ref{massShell}) and  the KB equation (\ref{KB}) is not proven.

\subsection{Memory effects}

A general treatment of the memory effects is a cumbersome task. Following Ref.~\cite{IKV2}, as an example, consider a system of non-relativistic fermions interacting via contact two-body
potential $V_0$. The  self-energy  up to three-vertex diagram becomes
%
\begin{eqnarray}
 -i \Sigma &=& -i\left(\Sigma^{(1)} + \Sigma^{(2)} + \Sigma^{(3)} \right)
= \includegraphics[width=7cm]{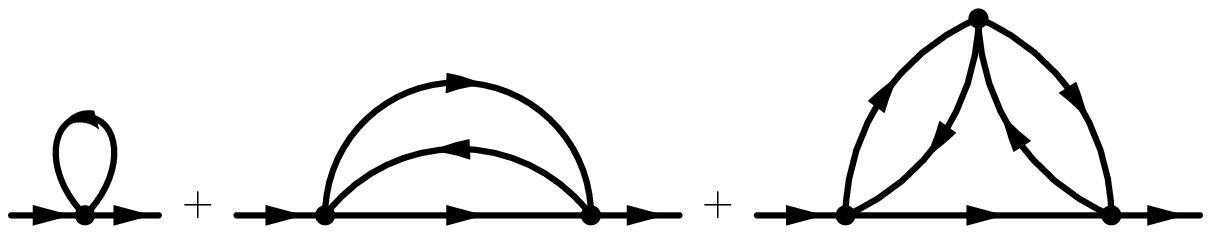}
\,\,.
\label{Pi-123}
\end{eqnarray}
The local part of the collision term is presented in the form
%
\begin{eqnarray}
C^{(2)} + C_{\rm loc}^{(3)} &=& d^2 \int \frac{\rmd^4 p_1}{(2\pi)^4} \frac{\rmd^4 p_2}{(2\pi)^4}
\frac{\rmd^4 p_3}{(2\pi)^4}
\left(
\left|
\parbox{7mm}{\includegraphics[width=7mm]{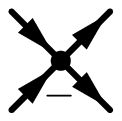}}
+
\parbox{8mm}{\includegraphics[width=8mm]{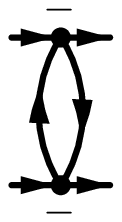}}
\right|^2 -
\left|\parbox{8mm}{\includegraphics[width=8mm]{Kin-mem-C-loop.eps}}\right|^2\right)
\nonumber\\
&&\times  (2\pi)^4\delta^4\left(p + p_1 - p_2 - p_3\right)
\left( F_2\, F_3\, \widetilde{F}\,\widetilde{F}_1 -
\widetilde{F}_2\,\widetilde{F}_3 \, F\, F_1 \right),
\label{C30}
\end{eqnarray}
%
where all the vertices in the off-shell scattering amplitudes are of the same sign, say $"-"$ for definiteness, i.e. there are no $"+-"$ and $"-+"$ Green's functions left, $d$ accounts summation over internal degrees of freedom, e.g. spin. Now the collision term contains a {\em nonlocal} part due to the last diagram (\ref{Pi-123}).

The current and the entropy flows are expressed in terms of the loop functions
\begin{eqnarray}
L^{jk}(x,y)=\parbox{2cm}{\includegraphics[width=2cm]{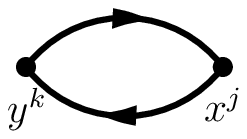}}
\nonumber
\end{eqnarray}
which in the Wigner representation takes the form
\begin{eqnarray}
L^{jk}(X,p)= \int \frac{\rmd^4 p}{(2\,\pi)^4}\widetilde{L}^{jk}(X,p' +p,p') \, ,
\end{eqnarray}
where
\begin{eqnarray}
\widetilde{L}^{jk}(X, p' +p, p')=d\, iV_0\, iG^{jk}(X,p' +p) \, iG^{kj}(X,p') \, .
\end{eqnarray}

The first-order-gradient memory correction to the collision term
induced by the third graph (\ref{Pi-123}) is
%
\begin{eqnarray}
\label{col-pi-delta} &&C_{\rm mem}^{(3)}(X,p) =
\left[\Sigma_{+-,{\rm mem}}^{(3)}(X,p)\, G^{-+}(X,p)
-
G^{+-}(X,p)\, \Sigma_{-+}^{(3),{\rm mem}}(X,p) \right]
\nonumber \\ &&= \frac{i}{2}
\int \frac{\rmd^4 p'}{(2\pi)^4} \frac{1}{d}
\left[ \widetilde{L}^{+-}(X,p'+p,p) - \widetilde{L}^{-+}(X,p'+p,p) \right]
\big\{L^{+-}(X,p'), L^{-+}(X,p')\big\}_{p' ,X}\, .
\end{eqnarray}
%

The memory current follows from integration of
(\ref{col-pi-delta}):
\begin{eqnarray}\label{jmem}
j^\mu_{\rm mem}(X)=e\,
\int \frac{\rmd^4 p}{(2\pi)^4}
\frac{i}{2}L^{+-}(X,p)L^{-+}(X,p)\frac{\partial }{\partial p_{\mu}}
\big(L^{+-}(X,p)+ L^{-+}(X,p)\big)\,.
\end{eqnarray}
Expression for the entropy flow is more cumbersome but it
simplifies in case of the local equilibrium:
\begin{eqnarray}\label{Smem}
[S^\mu_{\rm mem}(X)]_{\rm l.eq}=\int \frac{\rmd^4 p}{(2\pi)^4} \frac{i}{2}
L^{-+}(X,p)\,L^{+-}(X,p)
\left[\ln \frac{ L^{+-}(X,p)}{L^{-+}(X,p)} -1\right]
\frac{\partial L^{-+}(X,p)}{\partial p_{\mu}}.
\end{eqnarray}

\subsection{Time advances and delays}

It is rather natural to expect that the time and the  position characterizing collisions of propagating particles undergo some shifts, if one incorporates that collisions are not instant. In accordance with the uncertainty principle and the ergodicity the
energy and the momentum are shifted as well. This is taken into account in the non-local form of the kinetic equation (\ref{new-kin}), (\ref{new-kin-eq}), see also Eq. (\ref{deltaX}).
These effects are absent in  the kinetic equation written the BM form. In Ref. \cite{IKV2} the corresponding Poisson bracket terms in the KB kinetic equation (see Eq. (\ref{KB})) responsible for this phenomenon were associated with quantum fluctuations.

From (\ref{deltaX}) we find for the time delay/advance of collisions:
\begin{eqnarray}\label{sht}
\delta t_{\rm f}^{\rm kin}=\delta t_{\rm s}^{B} - t_{\rm col} \,,
\quad
\delta t_{\rm s}^{B}=B_0/2\,,\quad t_{\rm col}=Z_0^{-1}/\Gamma\,.
\end{eqnarray}
Here and further on in this subsection we will not write out the arguments $(X,p)$ of the quantities, unless it is explicitly needed. The value $\delta t_{\rm s}^{B}$ can be formally expressed through the  quantity having the meaning of an in-medium scattering phase shift~\cite{IV09}
\begin{eqnarray}\label{deltarho}
\delta t_{\rm s}^{B} =\frac{B_0}{2}\equiv  \frac{\partial \delta
(n) }{\partial p_0}, \quad \tan\delta (n) \equiv -\frac{\Gamma}{2M},
\end{eqnarray}
where $n=j^0_{\rm Noether}$ indicates dependence of $\delta$ on the particle density. Note that the quantity (\ref{deltarho}) describes the  delay/adwancement of the dressed particles (or the corresponding group of waves) at arbitrary distances in difference with a similar quantity $\delta t_{\rm s}$ (\ref{rstd}), which we exploited in description of the resonance quantum mechanical scattering, showing delay/advance of the scattered waves, as measured at large distances. The second relation (\ref{deltarho}) demonstrates measure of proximity of the virtual particle to the mass shell. In the virial limit, $\frac{B_0}{2}\to \frac{\partial
\delta}{\partial p_0}$, where $\delta$ has already the meaning of the real scattering phase shift. E.g. for the $\pi N\Delta$ system, the $B_0^{\Delta}$ spectral function  of the $\Delta(1232)$ isobar relates to the energy variation of the scattering phase shift $\delta_{33}^{\pi N}$ of the $P_{33}$ partial wave coupled to the $\Delta$ resonance. For the pion, $B_0^{\pi}$ in the virial limit relates to the the phase shift of the nucleon hole -- $\Delta$ scattering. For the nucleon, $B_0^{N}$ relates to $\delta_{\pi \Delta}$.
Since following Eq.~(\ref{B-j}) $B_0$ is expressed through $A$, the scattering time $\delta t_{\rm s}^{B}$ can be presented as the sum of the Noether scattering delay  and the drag and the back delay/advance terms:
\begin{eqnarray}
&&\delta t_{\rm s}^{B} =\delta t_{\rm s}^{A}+\delta t_{\rm s}^{\rm drag}
+ \delta t_{\rm s}^{\rm back},
\nonumber\\
&&\delta t_{\rm s}^{A} =\frac{v_0 A}{2}
>0, \quad
\delta t_{\rm s}^{\rm drag} =-\frac{A}{2}\frac{\partial \Re\Sigma^R}{\partial p_0},
\quad
\delta t_{\rm s}^{\rm back} =-\frac{AM}{2\Gamma}\frac{\partial
\Gamma}{\partial p_0}.
\end{eqnarray}
The collision time $t_{\rm col}$ in Eq.~(\ref{sht}) has the meaning of the average time between collisions. The value $t_{\rm col}$ does not contain factor 2 compared to the classical decay time $t^{\rm cl}_{\rm dec}$ given by Eq. (\ref{tdec}) of Sect.~\ref{Mech}, although in both cases $\Gamma$ enters the poles of the retarded Green's function similarly. This is because $t_{\rm col}$ describes dynamics of the Green's function (quadratic
form) rather than dynamics of the classical field ($z$-variable in
classical mechanics). Also, compared to the quantities:  $t^{\rm cl}_{\rm dec}$, see Eq. (\ref{tdec}) Sect. \ref{Mech}, $t_{\rm dec}$, $t^{\rm scl}_{\rm dec}$, see Eqs. (\ref{tfQ1}),
(\ref{cldec}) Sect. \ref{sec:quant}, the value $ t_{\rm col}$ contains additional renormalization factor, cf. also Eq.~(\ref{QFTshifts}) Sect. \ref{sec:qft}. Thereby the collision time can be presented as the sum of the decay time and the drag delay/advance terms:
\begin{eqnarray}
&&t_{\rm col}=t_{\rm dec}+ \delta t_{\rm col}^{\rm drag},
\\
&&t_{\rm dec}=\frac{v_0}{\Gamma}>0, \quad
\delta t_{\rm col}^{\rm drag}=-\frac{\partial \Re\Sigma^R}{\partial p_0}\frac{1}{\Gamma }
.\nonumber
\end{eqnarray}
The value $\delta t_{\rm f}^{\rm kin}$ given by Eq. (\ref{sht}) can be as positive as negative. The sum-rules (\ref{sumrule}) for $A$,  $B_0$ and $A_S^0$  can be
now presented as
\begin{eqnarray}
\int \frac{\rmd p_0}{2\pi} \delta t_{\rm s}^{A} =\int \frac{\rmd p_0}{2\pi} \delta t_{\rm s}^{B} &=
\int \frac{\rmd p_0}{2\pi} \delta t_{\rm s}^{A_S} =\frac{\hbar}{2} , \quad
\delta t_{\rm s}^{A_S} =\frac{A_S}{2}\,,
\end{eqnarray}
\begin{eqnarray}
\int \frac{\rmd p_0}{2\pi} \big(t_{\rm col}+\delta t_{\rm f}^{\rm kin}\big)
=\frac{\hbar}{2}.
\end{eqnarray}

Thus, in accord with the energy-time uncertainty principle, cf.
\cite{Tam-Tamm,Perelman}, $\delta t_{\rm s}^B$, $\delta t_{\rm s}^B$, $\delta t_{\rm s}^{A_S}$ are  minimal resolution times of the corresponding wave packets. The collision time $t_{\rm col}\sim t_{\rm dec}$ is  the time needed for the decay of unstable system.  The value $\delta t_{\rm f}^{\rm kin}$ is the minimal resolution time counted from the collision time, i.e. an average  time interval between two successive collisions. The corrected causality condition for collisions should now read as
\begin{eqnarray}
 t-r/v_{\rm gr}-\delta t_{\rm f}^{\rm kin} \geq 0.
\end{eqnarray}

In Sect.~\ref{sec:quant} we present the relation between the level density and the particle scattering phase shift. The density of states is often determined as \cite{Abrikosov}:
\begin{eqnarray}
\label{Noethlev}
\frac{\rmd N^{\rm level}_A}{\rmd p_0 /(2\pi)}=\int \frac{\rmd^3 X \rmd^3
p}{(2\pi)^3} v^0 A(X, p_0,\vec{p}).
\end{eqnarray}
Thereby in accord with Eq. (\ref{Noether}) the level density (\ref{Noethlev}) is related to the Noether particle density,
\begin{eqnarray}
\frac{\rmd N^{\rm Noether}}{\rmd^3X \rmd p_0 /(2\pi)}=
\frac{\rmd N^{\rm level}_A}{\rmd^3X \rmd p_0 /(2\pi)} f(X,p_0,\vec{p}).
\end{eqnarray}
Note that even in the limit of a small width the spectral density $A$ should include both the quasiparticle term and a regular terms, see Eq.~(\ref{SRS}) below, so that in the case of the conserved number of particles (e.g. baryons) $N^{\rm Noether}$ would coincide with the full number of particles: with the only quasiparticle Green's function Eq.~(\ref{Noethlev}) becomes incorrect.

We could introduce the level density of interacting particles differently relating it to the interacting particle density (\ref{curs}), cf. Ref.~\cite{Delano},
\begin{eqnarray}\label{NAS}
\frac{\rmd N^{\rm level}_{A_S}}{\rmd p_0 /(2\pi)}
=\int \frac{\rmd^3 X \rmd^3 p}{(2\pi)^3} A_S^0(X, p_0,\vec{p}),
\end{eqnarray}
or relating it to (\ref{curB}) with $B^0$ spectral function~\cite{Weinhold}
\begin{eqnarray}
\label{NB}
\frac{\rmd N^{\rm level}_{B}}{\rmd p_0 /(2\pi)}
=\int \frac{\rmd^3 X \rmd^3 p}{(2\pi)^3} B^0(X, p_0,\vec{p})\,.
\end{eqnarray}
Since following Eq.~(\ref{ergod}) in the virial limit the level density is related to the Wigner delay time, from Eqs.~(\ref{Noethlev}), (\ref{NAS}), and (\ref{NB}) we find relations
\begin{eqnarray}
\label{Bvir}
\delta t_{\rm W}^{\rm A}=v_0 A, \quad \delta t_{\rm
W}^{A_{S}}=A_S,\quad \delta t_{\rm W}^{B}=B_0.
\end{eqnarray}
For the non-relativistic particle scattering on a potential $B_0 =A$ and with $\delta t_{\rm W}^{B}$, as well as with $\delta t_{\rm W}^{A}$, we recover results (\ref{Wigner}), (\ref{rstd}) derived in Sect.~\ref{scattering}. For multi-component systems the total
Noether current can be presented as $j^{\rm tot} _\mu =\sum j^{\rm Noether}_\nu$, see Ref.~\cite{IKV}. On the other hand, the interaction between different species can be redistributed in many ways. For example, the interaction from some species can be redistributed to other ones, cf. Ref.~\cite{V08}. In the latter case the currents of some properly dressed species are described by Eq.~(\ref{NAS}) or by Eq.~(\ref{NB}), whereas some other species undergo free motion. For example, for the case of a resonance, like the $\Delta$ or $\rho$-meson resonances in hadron physics, the $B_0$-function relates to the energy variations of scattering phase shift of the scattering channel coupling to the resonance in the virial limit~\cite{Weinhold,IKHV}. Similarly, in a quantum mechanical scattering
divergent and scattered waves are relevant quantities only at large distances, being strongly distorted at short distances near the scatterer.

From Eq.~(\ref{jmem}) using Eqs.~(\ref{ergod}), (\ref{NB}), and (\ref{Bvir}) we may also recover the memory Wigner delay/advance time
\begin{eqnarray}\label{tmem}
\delta t^{\rm mem}_{\rm W}=\frac{1}{2}
\frac{\partial ({L}^{+-}+{L}^{-+})}{\partial p_{0}}{L}^{+-}{L}^{-+},
\end{eqnarray}
provided self-energies include diagrams with more than two vertices. The value $\delta t^{\rm mem}_{\rm W}\propto V_0^3$ and therefore $\delta t^{\rm mem}_{\rm W}$ disappears in the virial limit.

Finally, note that the quantities $t_{\rm soj}^{({\rm 1D})}(a,b;\tau)$ given by Eq. (\ref{dwellF}) and $t_{\rm soj}(\Omega,\tau)$ given by (\ref{dwellF3}), where $F$ is the exact non-equilibrium Green's function, have  the same form, being expressed through the Wigner density $F$ in the Wigner representation.

\subsubsection{Time advances and delays for Wigner resonances}

For the Wigner resonances  $\Re\Sigma^R$ and $\Gamma$ are assumed to be independent on $p_0$ and Eq. (\ref{sht}) is simplified as
\begin{eqnarray}\label{Wign}
\delta t_{\rm f}^{\rm kin} = \delta t_{\rm s}^B - t_{\rm col}
= -\frac{M^2-\Gamma^2/4}{\Gamma(M^2+\Gamma^2/4)}
= \delta t_{\rm f}
= \delta t_{\rm s}^A - t_{\rm dec}.
\end{eqnarray}
The energy weighted time shift $\int \frac{\rmd p_0}{2\pi}\Gamma A\delta t_{\rm f}^{\rm kin} =0$.

The value $\delta t_{\rm s}^A =\frac{\Gamma/2}{M^2 +\Gamma^2/4}>0$
coincides with the scattering time delay, $\delta t_{\rm s} =\frac{\partial\delta}{\partial p_0}$, introduced in quantum mechanics, see Eq. (\ref{deltats}). The collision time coincides with  the quantum mechanical time, $\delta t_{\rm vol}=\frac{1}{2\mbox{sin}^2 \delta}\frac{\partial \delta}{\partial p_0}$, when all the delays are put into scattering, see Eq. (\ref{tvol}), (\ref{tfQ1}). Thus, the value $-t_{\rm col}=-t_{\rm dec}=-1/\Gamma <0$ has the meaning of the collision time advance. The value $\delta t_{\rm f}^{\rm kin}$ given by (\ref{Wign}) coincides with the quantity $\delta t_{\rm f} =-\delta t_{\rm i} =
-\frac{\cos(2\delta)}{2\sin^2 \delta} \frac{\partial \delta}{\partial p_0}$, see Eq.~(\ref{tfQ}), being a forward delay/advance time. A propagating wave packet gets a
delay/advance $\delta t_{\rm i}$ due to interference of the incident and reflected waves. The collision term gets corresponding advance/delay $\delta t_{\rm f} =-\delta t_{\rm i}$, being the scattering time counted from the collision time, since  actual collision for the particles with the width may occur a time $\sim t_{\rm col}$ earlier than it happens in the case of the zero width. In the limit $|M|\ll\Gamma$ there is a time delay $\delta t_{\rm f}
=1/\Gamma$ and for $|M|\gg \Gamma$ there arises a time advance $\delta t_{\rm f} =-1/\Gamma$, and for $|M|=\Gamma/2$, $\delta t_{\rm f} =0$.

\subsection{Test particle method}

The conserving feature is especially important for devising numerical simulation codes based on the kinetic equation. If a test-particle method is used, one should be sure that the number of test particles is conserved  exactly rather than approximately. In the test-particle method the distribution function  $F$ (not $f$) satisfying Eq. (\ref{new-kin}) is represented by an ensemble of test particles \cite{IV09}
%
\begin{eqnarray}
\label{test-part} F (X,p) \sim \sum_i \delta^{(3)} \left({\vec{x}}
- {\vec{x}}_i (t)\right) \delta^{(4)} \left({p}-{p}_i (t)\right),
\end{eqnarray}
%
where the $i$-sum runs over test particles, $p=(p^0,\vec{p}\,)$. Then the drift term in KB equation for $F$ (derived from (\ref{new-kin}) if one expands the collision term in gradients
including the first gradient terms) just corresponds to the classical motion of these test particles subjected to forces inferred from $\Re \Sigma^{R}(X,p)$, while the collision term gives stochastic change of test-particle's momenta, when their trajectories ``cross''. For a direct application of this method, however, there is a particular problem with the kinetic equation in the KB form. Appearing additional Poisson-bracket term $\{\Gamma^{\rm in},\Re G^R\}$ spoils this simplistic picture, since derivatives acting on the distribution function $F$ arise here only indirectly and thus cannot be included in the collision-less propagation of test particles. This problem, of course, does not prevent a direct solution of the KB kinetic equation applying lattice methods, which are, however, much more complicated and time-consuming as compared to the test particle approach.

The effective BM-current  was used  in \cite{Leupold} as a basis for the construction of a test-particle ansatz for numerical solution of the nonrelativistic BM kinetic equation.
To fulfill the effective current conservation one introduces the test-particle ansatz for
%
\begin{eqnarray}
\label{test-part-L} \frac{1}{2} \Gamma B_0 F (X,p) \sim \sum_i
\delta^{(3)} \left({\vec{ x}} - {\vec{ x}}_i (t)\right)
\delta^{(4)} \left(p-p_i (t)\right),
\end{eqnarray}
%
rather than for the distribution function itself. Note that the
energy $p^0_i(t)\equiv E_i (t)$ of the test particle is an
independent coordinate, not restricted by a mass-shell condition.
Ref. \cite{Cass99} used this test-particle ansatz in the
relativistic case.

The BM kinetic equation   together with ansatz (\ref{test-part-L})
for the distribution function result in the  set of equations for
evolution of parameters of the test particles between collisions
\begin{eqnarray}
\label{x-dot}
\frac{ \rmd\vec{x}_i}{\rmd t}&=& \frac{1}{v_0
-\partial_{E_i}\Re\Sigma^R -(M/\Gamma)\partial_{E_i}\Gamma }
\left({\vec{v}}_i + \nabla_{p_i} \Re\Sigma^R
+(M/\Gamma)\nabla_{p_i}\Gamma\right),
\\%
\label{p-dot}
\frac{ \rmd\vec{p}_i}{\rmd t} &=& \frac{1}{v_0
-\partial_{E_i}\Re\Sigma^R -(M/\Gamma)\partial_{E_i}\Gamma}
\left(\nabla_{x_i} \Re\Sigma^R
+(M/\Gamma)\nabla_{x_i}\Gamma\right),
\\%
\label{e-dot} {\dot E}_i &=& \frac{1}{v_0
-\partial_{E_i}\Re\Sigma^R -(M/\Gamma)\partial_{E_i}\Gamma}
\left(\partial_{t} \Re\Sigma^R
+(M/\Gamma)\partial_{t}\Gamma\right).
\end{eqnarray}
All function on the right-hand side are evaluated in the point $(t,\vec{x}_i(t), E_i(t), \vec{p}_i(t))$. These equations of motion, in particular, yield the time evolution
of the mass term $M$, of a test particle \cite{Cass99,Leupold}
%
\begin{eqnarray}
\label{off-shellness}
\frac{\rmd M_i}{\rmd t}= \frac{M_i}{\Gamma_i}\frac{\rmd \Gamma_i}{\rmd t},
\end{eqnarray}
%
the origin of which can be traced back to the additional term
$\{\Gamma F/A,\Re G^R\}$ in the BM equation. Here $M_i(t)=
M[t,{\vec{ x}}_i(t);E_i(t),{\vec{ p}}_i(t)]$ measures an
``off-shellness'' of the test particle, and $\Gamma_i(t)=\Gamma[t,{\vec{
x}}_i(t);E_i(t),{\vec{ p}}_i(t)]$. Equation of motion (\ref{off-shellness}) yields $|M_i| =\alpha_i \Gamma_i$, where $\alpha_i>0 $ do not depend on time, and implies that once the width drops in time the particles are driven towards the mass on-shell, i.e. to $M=0$. This clarifies the meaning of the additional term $\{\Gamma F/A,\Re G^R\}$ in the off-shell BM kinetic equation (which follows from (\ref{new-kin}), if one suppresses variable
shifts in the collision term): it provides the time evolution of the off-shellness.

For the non-local form of the kinetic equation \cite{IV09} the set of equations for evolution of parameters of the test particles between collisions is the same as in the BM case.  The only difference with the BM case is that collisions of test particles occur with certain time (and space) delay (or advance) as compared with the instant of their closest approach to each other.

Following Eq. (\ref{deltaX})  the shift of the space variables is
\begin{eqnarray}\label{delX1}
\delta \vec{x} =\frac{1}{A}\frac{\partial \Re G^R } {\partial \vec{p}}
=\delta \vec{x}_{\rm drift}+\delta \vec{x}_{\rm col} =\frac{1}{2}\vec{B}+\vec{\widetilde{v}}_{\rm gr}t_{\rm col}.
\end{eqnarray}
We can further express the spatial shift in terms of time delays  and velocity $\frac{\rmd \vec{x}}{\rmd t}$ of a test particle on its trajectory, cf. Eq.~(\ref{x-dot}),
\begin{eqnarray}
\label{delX2}
\delta \vec{x}_i =\frac{\rmd \vec{x}_i}{\rmd t} \delta t_{i,{\rm s}}^B
+\vec{\widetilde{v}}_{i, {\rm gr}}t_{i,{\rm col}}.
\end{eqnarray}
Equation~(\ref{delX2}) demonstrates that before a delayed/advanced collision the test particle moves along its trajectory. Therefore, the scattering time delay $ \delta t_{i,{\rm s}}^B$ unambiguously results in a definite space shift. The collision itself is associated with an additional time delay $t_{i,{\rm col}}$, which implies that the collision is not instant, as it is treated in the BM kinetic equation, but requires certain time for complete
decoupling from intermediate states (e.g., the pion spends some time in the intermediate $\Delta$--nucleon-hole state, a soft photon requires certain time to be formed in multiple collisions of the proton with neutrons). Therefore, this additional delay gives rise to an additional shift of the particle with respect to its ``collisionless'' trajectory (\ref{x-dot}).

\subsection{Quasiparticle limit}\label{Quasiparticle}

The quasiparticle limit is understood, as the limit, when Green's functions are computed at $\Im \Sigma^R\rightarrow 0$. The quasiparticle width $\Gamma_{\rm qp}(X,\vec{p}) =-2\Im\Sigma^R(X,\vec{p},A\to A_{\rm qp})$ is then calculated with the quasiparticle Green's functions and the associated quasiparticle spectral functions $A_{\rm qp}$.
The letter reduces in this limit ($|M|=\alpha \Gamma$, $|\alpha | \gg 1$) to,
\begin{eqnarray}
A_{\rm qp}(X,p) =2\pi Z_0^{\rm qp}(X,\vec{p}\,)\,\delta \left(p_0 -E_p(X,\vec{p}\,)\right),
\quad
Z_0^{\rm qp}(X,\vec{p}\,)=\left(v^0 -\frac{\partial\Re \Sigma^R(X,p)}{\partial p_0}\right)_{p_0=E_p(X,\vec{p}\,)}^{-1}\,,
\end{eqnarray}
where $E_p(X,\vec{p})$ stands for the energy of a quasiparticle, being the root of the  dispersion relation
\begin{eqnarray}
\label{Mqp}
M(X,p_0 =E_p(X,\vec{p}\,), \vec{p})=0,
\end{eqnarray}
$Z_0^{\rm qp}(X,\vec{p}\,)>0$ on the quasiparticle branch $p_0=E_p(X,\vec{p}\,)$.\footnote{Here, we for simplicity assume that there is only one quasiparticle branch. Generalization is obvious.}

The quasiparticle spectral function does not fulfill the exact sum rule (\ref{sumrule}) but fulfills the corresponding quasiparticle sum-rule
\begin{eqnarray}
 \int Z_0^{{\rm qp}}(X,\vec{p}\,)\, A_{\rm qp}(X,p)\frac{\rmd p_0}{2\pi} =1\,.
\end{eqnarray}
Formally, within the quasiparticle picture this problem is avoided by a renormalization, after which one may already deal only with the quasiparticle degrees of freedom, being well separated from the degrees of freedom in the continuum. The reason for the difference between the exact and quasiparticle sum-rules is that the quasiparticle current $j_{\rm qp} ^\mu =(j_B^{\rm qp})^\mu = (j_S^{\rm qp})^\mu$ includes the quasiparticle drag flow and, thereby,  differs from the Noether current. This difference is compensated  due to the conservation of the Noether current by the presence of the back flow, as a back reaction of the whole energy sea to the particle drag flow. Thus, the exact sum-rule (\ref{sumrule}) for $A$ is recovered, provided one includes the width term. It is easily demonstrated in case of  a weak interaction~\cite{Schmidt}. Then
\begin{eqnarray}\label{SRS}
A(X,p)\simeq 2\pi Z_0^{\rm qp}(X,\vec{p}\,)\delta\big(M(X,p)\big)
+\mathcal{P}\frac{\Gamma(X,p)}{M^2(X,p)} \simeq 2\pi\,\delta (p_0-E_p(X,\vec{p}\,))\,,
\end{eqnarray}
where $\mathcal{P}$ indicates the principal value. The required sum-rule is recovered after using of the Kramers-Kronig relation between $\Gamma$ and $\Re\Sigma$.

The integration of all three forms of the kinetic equation,  the BM
form, the KB equation (\ref{KB}), and the non-local  Eq.~(\ref{new-kin-eqS})  over $\rmd p_0/ (2\pi)$ yields one and the same kinetic equation describing the  quasiparticle propagation in
matter:
\begin{eqnarray}\label{new-kin-eqhomQP}
\frac{\partial f_{\rm qp}(X,\vec{p}\,)}{\partial {t}} +
\vec{v}^{\,\rm qp}_{\rm gr}(X,\vec{p}\,)\, \nabla f_{\rm qp}(X,\vec{p}\,)
-\nabla E_p(X,\vec{p}\,) \frac{\partial f_{\rm qp}(X,\vec{p})}{\partial \vec{p}} = C_{\rm qp}(X,\vec{p}\,)\,,
\end{eqnarray}
here $E_p$ obeys mass-shell condition (\ref{Mqp}),
$$C_{\rm qp}^{\rm shift}(X,\vec{p}\,)= \int \frac{\rmd p_0}{2\pi}
A_{\rm qp}(X,p)\, \mathcal{C}^{\rm shift}(X,p;A\to A_{\rm qp})\to C_{\rm qp}(X,\vec{p}\,)$$
and the quasiparticle group velocity is
\begin{eqnarray}\label{grqp}
\vec{{v}}^{\,\rm qp}_{\rm gr}(X,\vec{p}\,) =
\frac{\partial E_p (X,\vec{p}\,)}{\partial \vec{p}}
=\frac{\left( \vec{v} + \frac{\partial \Re\Sigma^R (X,p) }{\partial
\vec{p}}\right)_{p_0=E_p(X,\vec{p}\,)} }
{ \left(v^{0} - \frac{\partial \Re\Sigma^R_{\rm qp}(X, p) }{\partial p_0}\right)_{p_0=E_p(X,\vec{p}\,)} }.
\end{eqnarray}
Memory effect corrections to the collision term derived in the extended quasiparticle approximation were considered in Ref.~\cite{Morawetz}.

In the quasiparticle limit the BMM entropy, Eq.~(\ref{entr-transp}), reads
\begin{eqnarray}
S^\mu_{\rm qp}=\int \frac{\rmd^3 p}{(2\pi)^3}
\sqrt{1- [\vec{v}\,_{\rm gr}^{\rm qp}(X,\vec{p}\,)]^2}
[u_{\rm gr}^{\rm qp}(X,\vec{p}\,)]^\mu\,\sigma_{\rm qp}(X,\vec{p}\,),
\end{eqnarray}
where we introduced notation
\begin{eqnarray}
[u_{\rm gr}^{\rm qp}(X,\vec{p}\,)]^{\mu} =
\left(
\frac{1}{\sqrt{1-[\vec{v}\,_{\rm gr}^{\rm qp}(X,\vec{p}\,)]^2}},
\frac{\vec{v}\,_{\rm gr}^{\rm qp}(X,\vec{p}\,)}
{\sqrt{1- [\vec{v}\,_{\rm gr}^{\rm qp}(X,\vec{p}\,)]^2}}\right).
\end{eqnarray}
The quasiparticle group velocity can significantly differ from the phase velocity.  In atomic Bose-Einstein condensates \cite{Hau} the light can be slowed down to the speed 17 m/s (a so-called slow light). The light  also becomes ultra-slow  in the hot atomic vapor of rubidium \cite{Kash}. On the other hand, the quasiparticle group velocity  can easily exceed the velocity of light in medium $c(n)<c$. It manifests as the Cherenkov radiation. An instability of quasiparticle modes also arises in a moving medium \cite{Pit84,V95,BaymPethick} (even if the modes are stable in the static medium), provided $|\vec{u}||\vec{p}|>E_p(p)$, where $\vec{u}$ is the speed of the medium and $E_p (p)$ is the quasiparticle spectrum branch. As the result, a
spatially inhomogeneous condensate of Bose excitations may be formed. This phenomenon is similar to the Cherenkov effect and the shock wave generation.

Obviously, the values of time delays are unlimited. The question about a possible limitation on the time advance and  possibility to have $v^{\rm gr}_{\rm qp}>c$ are subtle issues. As argued in Ref.~\cite{Chiao,Segev2000}, such a possibility does not contradict causality.
However the velocity of the wave front $v_{\rm fr}$  should always be limited by $c$, see Ref. \cite{Boyd} and references therein. An apparent superluminal propagation manifested in some laser experiments can be understood as consequence of a reshaping of the pulse envelope by
interaction within the medium, see discussion in \cite{Garrett,Faxvog,Boyd,Garrison98}. Although formally $v_{\rm gr}$ exceeds $c$, the forward wave front moves with velocity $\leq c$, as it should be. There are also arguments that the advance of the pulse of light in materials is generally limited only by few pulse widths (so called fast light). Thereby, the interpretation of a negative group delay as a superluminal propagation is according to Ref.~\cite{Garrison98} a semantic question, since the speed of the information transport never exceeds $c$, see Ref.~\cite{Stenner}.

Reference~\cite{MalflTr} used the $p_0$-averaged BM equation and averaged distributions,
\begin{eqnarray}
f^B(X,\vec{p}\,) =\int \frac{\rmd p_0}{2\pi} B_0(X,p) f(X,p)\,.
\end{eqnarray}
Averaged values
\begin{eqnarray}\label{groupav}
&&\vec{v}_{\rm gr}^{\,B}(X,\vec{p}\,) =\int \frac{\rmd p_0}{2\pi} B_0(X,p)  \,\vec{v}_{\rm gr}(X,p),
\quad
\vec{\widetilde{v}}_{\rm gr}^{\,B}(X,\vec{p}\,) =\int \frac{\rmd p_0}{2\pi}
\widetilde{B}_0(X,p)\,\vec{\widetilde{v}}_{\rm gr}(X,p)\,,
\nonumber\\
&&\vec{v}_{\rm gr}^{\,S}(X,\vec{p}\,)=\int \frac{\rmd p_0}{2\pi} A_S(X,p)\, \vec{v}_{\rm gr}(X,p)\,,
\end{eqnarray}
yield one and the same value $\vec{v}^{\,\rm gr}_{\rm qp}$ in the quasiparticle limit. It is worthwhile to mention that the quantity $\vec{v}_{\rm gr}B_0$ enters (through $\vec{B}$) the expression for  the entropy flow (\ref{entr-transp}) associated with the information transport. Therefore one may hope that the quantities (\ref{groupav}) behave reasonably in the whole $(p_0, \vec{p})$ plane.

Finally, to avoid possible confusion recall that there are no limitations on the values of the phase velocity and the group velocity $\widetilde{v}_{\rm gr}(M)$ for off-mass-shell particles  given by Eq.~(\ref{grKB}). However, the condition $v_{\rm gr}(M)<c$ might be satisfied.
For example, in the case of a Fermi liquid the width of particle-hole excitations
$\Gamma\propto p_0$ for $p_0\to 0$, cf. Ref.~\cite{V95}, and the restriction $v_{\rm
gr}(M)< c $ is then recovered, as it follows from Eq.~(\ref{grv}). Taking into account the dependence $\Gamma\propto p_0$ is also important in some other problems, e.g. in description of the growth of a static classical  pion condensation field in dense isospin symmetric nuclear matter~\cite{IKHV}. Also we arrived at such a dependence in description of damped oscillations in Sect.~\ref{Mech}. However, in general, it is not formally excluded that $v_{\rm gr}(M)>c$ for $p_0$ in some space-like region.

For moving media, an instability of off-mass-shell modes with $p_0 < |\vec{u}| \, |\vec{p}|$  may result in some measurable effects, like a heating of the medium, cf. \cite{Weldon,V95}.

\subsection{Kinetic entropy and time delays}

Any loss of information results in an increase of the entropy \cite{Wehrl}. Thus, it is important to find and compare values of the entropy related to the BM, KB and non-local forms of the kinetic equations.

First, let us for simplicity disregard memory effects. Above, see Eq.~(\ref{entr-transp}), we introduced expression for the BMM kinetic entropy flow. It also can be rewritten as follows
\cite{IKV2}
\begin{eqnarray}\label{entr-transp1}
&&S^\mu_{\rm{BMM}}(X) = {\rm Tr}\int \frac{\rmd^4 p}{(2\pi)^4} \left[
\left(v^\mu -\frac{\partial \Re \Sigma^R(X,p)}{\partial p_\mu}\right) A(X,p)\,\sigma(X,p)\right.\\
&&-\left.
\Re G^R(X,p) \left(\mp \ln\big(1\mp f(X,p)\big)
\frac{\partial}{\partial p_\mu}\Big[\Gamma(X,p)\, \big(1\mp f(X,p)\big)\Big]
- \ln f(X,p) \frac{\partial}{\partial p_\mu}\Big[ \Gamma(X,p)  f(X,p)\Big] \right)\right].\nonumber
\end{eqnarray}
This contribution is of the zero-gradient order. Similarly, one derives~\cite{IKV3} expression for the KB Markovian part of the entropy flow (KBM) relating to Eq.~(\ref{KB})
\begin{eqnarray}
&&S^\mu_{\rm{KBM}}(X) = {\rm Tr}\int \frac{\rmd^4 p}{(2\pi)^4}
\left(v^\mu -\frac{\partial \Re \Sigma^R(X,p)}{\partial p_\mu}\right)
\times\Bigg[ A(X,p)\,\sigma(X,p)
\nonumber\\
&&
-\Re G^R(X,p) \left(\mp \ln\big(1\mp f(X,p)\big)
\frac{\partial \Gamma^{\rm out}(X,p)}{\partial p_\mu}
-
\ln f(X,p)\frac{\partial \Gamma^{\rm in}(X,p)}{\partial p_\mu}
\right)\Bigg].
\label{entr-transp2}
\end{eqnarray}

For a discussion of the $H$ theorem it is necessary to get the entropy flow
including first order gradient terms.
The KB entropy flow contains an extra first-order gradient term compared to the KBM result~\cite{IKV3},
\begin{eqnarray}
\partial_\mu S^\mu_{\rm{KB}}(X)=\partial_\mu S^\mu_{\rm{KBM}}(X) +\partial_\mu \delta S^\mu_{\rm KB}(X)+\delta_{\rm cor}^\mu(X),
\end{eqnarray}
where
\begin{eqnarray}\label{delKB}
\delta S_{\rm KB}^\mu(X) = -{\rm Tr}\int \frac{\rmd^4
p}{(2\pi)^4}
\frac{M(X,p)}{\Gamma(X,p)}\, C(X,p)\,\frac{\partial}{\partial
p_\mu}\ln\frac{1\mp f(X,p)}{f(X,p)}.
\end{eqnarray}
In general, $C\propto O(\delta f, \partial_X )$, where $\delta f(X,p)=f(X,p)-f_{\rm l.eq}(X,p)$ with $f_{\rm l.eq.}$ given in Eq.~(\ref{fleq}), and the gradient expansion and the expansion in $\delta f$ near a local equilibrium are different, see discussion in the next subsection and in Ref.~\cite{IKV3}.
For near-local equilibrium configurations
\begin{eqnarray}
\delta_{\rm cor}^\mu(X) ={\rm Tr}\int \frac{\rmd^4 p}{(2\pi)^4}
\frac{C(X,p)}{A(X,p)}\Big\{\ln\frac{1\mp f(X,p)}{f(X,p)}, \Re G^R(X,p)\Big\}=
O(\delta f \partial_X\delta f).
\nonumber
\end{eqnarray}
Thus one may neglect $\delta_{\rm cor}^\mu(X)$ term, provided  the system is very close to the local equilibrium and gradients are small. Replacing in Eq.~(\ref{delKB}) the local equilibrium distributions everywhere except $C$ we obtain
\begin{eqnarray}
\delta S_{\rm KB}^0(X) = -{\rm Tr}\int \frac{\rmd^4 p}{(2\pi)^4}
\frac{M_{\rm l.eq.}(X,p)  }
{\Gamma_{\rm l.eq.}(X,p\,) }
\frac{C(X,p)}{T_{\rm l.eq.}(X)}.
\end{eqnarray}
This (the first  order in $\delta f$) correction is zero in the local equilibrium, where $C_{\rm l.eq.}=0$, and is sign-indefinite beyond the local equilibrium.

Counting the KB entropy flow from the BMM one, from  Eqs.~(\ref{entr-transp1}), (\ref{entr-transp2}), (\ref{delKB}) and (\ref{gamin}) we find
\begin{eqnarray}
S_{\rm KBM}^\mu(X) = S_{\rm BMM}^\mu(X) -
{\rm Tr}\int \frac{\rmd^4 p}{(2\pi)^4}
\Re G^R(X,p) \ln\frac{1\mp f(X,p)}{f(X,p)}\frac{\partial }{\partial
p_\mu}\Big[\frac{C(X,p)}{A(X,p)}\Big],
\end{eqnarray}
and finally
\begin{eqnarray}
S_{\rm KB}^\mu(X) =S_{\rm BMM}^\mu + {\rm Tr}\int \frac{\rmd^4 p}{(2\pi)^4}
\frac{C(X,p)}{A(X,p)}\frac{\partial \Re G^R(X,p) }{\partial
p_\mu}\ln\frac{1\mp f(X,p)}{f(X,p)}.
\end{eqnarray}
Thus we obtain
\begin{eqnarray}
&&S_{\rm KB}^\mu(X) =S_{\rm BMM}^\mu(X) +\delta S_{\rm KB}^\mu(X) ,
\nonumber\\
&&\delta S_{\rm KB}^\mu(X) = \int \frac{\rmd^4 p}{(2\pi)^4}
C(X,p) \delta X^\mu(X,p) \ln\frac{1\mp f(X,p)}{f(X,p)},
\label{skb}
\end{eqnarray}
with $\delta X^\mu(X,p)$ from (\ref{deltaX}). As we see, an additional purely non-equilibrium contribution $\delta S_{\rm KB}^\mu(X)$ is proportional to a weighted average space-time delay/advance. Expression (\ref{skb}) up to first gradients also holds for the non-local form of the kinetic equation.

Using (\ref{deltaX}) we obtain
\begin{eqnarray}
\delta S_{\rm KB}^0(X) =\int \frac{\rmd^4 p}{(2\pi)^4}
C(X,p)\,\delta t_{\rm f}^{\rm kin}(X,p)\, \ln\frac{1\mp f(X,p)}{f(X,p)}
\end{eqnarray}
that for  configurations close to the local
equilibrium produces an additional contribution to the specific heat
\begin{eqnarray}
\delta c_{\rm KB}(X) =-\int \frac{\rmd^4 p }{(2\pi)^4}
\frac{ p_0}{T_{\rm l.eq}(X)}\,C(X,p)\,
(\delta t_{\rm f}^{\rm kin})_{\rm l.eq} +
\int \frac{\rmd^4 p}{(2\pi)^4} p_0 C(X,p)\,
\frac{\partial \,(\delta t_{\rm f}^{\rm kin})_{\rm l.eq}}{\partial T}.
\end{eqnarray}
\emph{Presence or absence of an additional non-equilibrium correction $\propto C$ to the specific heat can be experimentally checked.}

We can also find the entropy flow directly for the non-local form of the kinetic equation. For that we  multiply the non-local kinetic  equation by $\mp \mbox{ln}[(1\mp f^{\rm shift})/f^{\rm shift}]$ and perform 4-momentum integration.  From the left-hand side of the thus-obtained equation we find instead of Eq.~(\ref{entr-transp1})
\begin{eqnarray}
&&S^\mu_{\rm{NL}}(X) = {\rm Tr}\int \frac{\rmd^4 p}{(2\pi)^4}
\left[\left(v^\mu
-\frac{\partial \Re \Sigma^R(X,p)}{\partial p_\mu}\right) A(X,p)\,
\sigma^{\rm shift}(X,p)\right.
\nonumber\\
&&-\left.
\Re G^R(X,p) \left(\mp\ln\big(1\mp f^{\rm shift}(X,p)\,)
\frac{\partial}{\partial p_\mu}\big[\Gamma(X,p) \big(1\mp f(X,p)\big)\big]-
\ln f^{\rm shift}(X,p)\,\frac{\partial}{\partial p_\mu}\big[\Gamma(X,p)\,  f(X,p)\big]
\right)\right] \nonumber\\
&&+O(\partial_{x}^2 )=S^\mu_{\rm{KB}}(X)+O(\partial_{x}^2 )
,\label{entr-transp3}
\end{eqnarray}
with
\begin{eqnarray}
\label{Asshift}
\sigma^{\rm shift}(X,p)
=\mp \big(1\mp f(X,p)\big)\ln\big(1\mp f^{\rm shift}(X,p)\big)
-f(X,p)\,\ln f^{\rm shift}(X,p)
\, .
\end{eqnarray}
Although formally Eq. (\ref{entr-transp3}) looks similar to the BMM term  (\ref{entr-transp1}), the former incorporates 4-space-time delays/adwancements.

To get the full result one should still add to $S^\mu$ the first-gradient-order memory correction $S_{\rm mem}^\mu$, which is non-zero if the generating $\Phi$ functional contains diagrams with more than two vertices, see Eq.~(\ref{Smem}). For example, in Ref.~\cite{IKV2}  it was shown that only with inclusion of the memory term the value
$S^0_{\rm BM}=S^0_{\rm BMM}+S^0_{\rm mem}$ yields appropriate thermodynamic expression for the equilibrium entropy. Since in the local equilibrium $\delta S^\mu =0$, the same relation holds in the local equilibrium for $S^0_{\rm KB}$, i.e.
\begin{eqnarray}
(S^0_{\rm KB})_{\rm l.eq}=(S^0_{\rm BM})_{\rm l.eq}
=(S^0_{\rm BMM})_{\rm l.eq}+(S^0_{\rm mem})_{\rm l.eq}.
\end{eqnarray}

$H$-theorem and the minimum of the entropy production are discussed in Appendix~\ref{minentr}.

\subsection{Examples of solutions of kinetic equations }

Consider a small portion of light resonances placed either in the uniform equilibrium medium consisting of heavy particles or in vacuum. To reduce the complexity of the problem we will exploit the following ansatz: assume that only the distribution $f$ depends on the Wigner variables $X=(t,\vec{r})$, whereas the dependence of $\Sigma^R (t,\vec{r})$  on $X$ is weaker and can be neglected.

Then the kinetic equation in the non-local form (\ref{new-kin}) simplifies as
\begin{eqnarray}\label{new-kin-eqhom}
A_S^\mu(p) \frac{\partial}{\partial{X^{\mu}}}f(X,p) = A(p)\, \mathcal{C}
\big(x_\mu- \half B_\mu(p) + Z^{-1}_\mu(p)/\Gamma(p),\, p_\mu\big).
\end{eqnarray}
The BM form of the kinetic equation (when ${\cal C}^{\rm shift}$ is replaced by $\mathcal{C}$) renders
\begin{eqnarray}
A_S^\mu(p) \frac{\partial}{\partial{X^{\mu}}}f(X,p) = A(p)
\,\Gamma^{\rm in}(p)-A(p)\,\Gamma(p) f(X,p)
\, .
\label{new-kin-eqhomBM}
\end{eqnarray}
The KB equation (when ${\cal C}^{\rm shift}$ is expanded up to the first-gradient order terms)  reads
\begin{eqnarray}\label{new-kin-eqhomKB}
\widetilde{B}^\mu(p) \frac{\partial}{\partial{X^{\mu}}}f(X,p)
+f(p)\big\{\Gamma(p),\Re G^R(p)\big\} - \big\{\Gamma^{\rm in}(p),\Re G^R(p)\}
= A(p)\, \Gamma^{\rm in}(p)-A(p)\,\Gamma(p)\, f(X,p)\, .
\end{eqnarray}

For uniformly distributed light resonances Eq. (\ref{new-kin-eqhom}) still simplifies as
\begin{eqnarray}\label{new-kin-eqhomHom}
A_S^0(p) \frac{\partial}{\partial{t}}f(t,p) = A(p)\, \mathcal{C}
\big(t- \half\,B_0(p) + Z^{-1}_0(p)/\Gamma(p),\, p \big)\,.
\end{eqnarray}
Equations~(\ref{new-kin-eqhomBM}) and (\ref{new-kin-eqhomKB}) are simplified accordingly.

\subsubsection{ Uniformly distributed light resonances in equilibrium medium of heavy particles}

Consider behavior of a dilute admixture of uniformly distributed light resonances in equilibrium medium consisting of heavy-particles. Thereby, we assume that  $\Sigma^R$ is determined by distribution of heavy particles, thus introducing ansatz $\Sigma^R\simeq \Sigma^R_{\rm eq}$. To further proceed we need to do an additional assumption:\\
(i)~Let us also \emph{assume} that the gain term
$\Gamma^{\rm in}(p)\simeq\Gamma^{\rm in}_{\rm eq}(p)=f_{\rm
eq}(p)\Gamma_{\rm eq}(p)$, this means it is a function of only equilibrium
quantities. Then only distribution of light resonances $f$ changes in time.
According to Eq. (\ref{colG}):
\begin{eqnarray}
C(p)
\simeq -A(p)\Gamma(p) \delta f(t,p), \quad
\delta f(t,p) =f(t,p)-f_{\rm eq}(p)
.\nonumber
\end{eqnarray}
Such an approximation (more accurately saying, an ansatz) is
called relaxation time approximation and it is often used in
Boltzmann kinetics without additional justification.

Replacing in the non-local  kinetic equation
(\ref{new-kin-eqhomHom})
\begin{eqnarray}
\label{KBgamNL}
\delta f_{\rm NL}(t,p)=\delta f(t=0,p) e^{- \alpha(p) t/\delta t_{\rm s}^B(p)},
\end{eqnarray}
cf. Eq.~(\ref{deltarho}), for  parameter $\alpha$ we find the equation
\begin{eqnarray}\label{alph}
\alpha(p)= e^{\alpha(p) -\alpha(p) t_{\rm col}(p)\,/\delta t_{\rm s}^B(p)},
\end{eqnarray}
that yields $\alpha \simeq 1$ for $\delta t_{\rm s}^B \simeq t_{\rm col}$
(then all three forms of kinetic equation coincide);
$\alpha\simeq  \frac{\delta t_{\rm s}^B}{ t_{\rm col}} \ln \frac{t_{\rm col}}{\delta t_{\rm s}^B}\ll 1$ for $\delta t_{\rm s}^B/ t_{\rm col}\ll 1$. In the case $\delta t_{\rm s}^B / t_{\rm col}\gg 1$  Eq.~(\ref{alph}) has no solutions. However, this case is not realized as it follows from explicit expressions for $\delta t_{\rm s}^B$ and $t_{\rm col}$.

From  the KB equation (\ref{new-kin-eqhomKB}) we find solution
\begin{eqnarray}\label{KBgam}
\delta f_{\rm KB}(t,p)=\delta f(t=0,p)  \, e^{- t/t_{\rm col}(p)} .
\end{eqnarray}
Contrary,  solving the BM equation (\ref{new-kin-eqhomBM}) we get a different solution
\begin{eqnarray}\label{BMgam}
\delta f_{\rm BM}(t,p)=\delta f(t=0,p) \, e^{-t/\delta t_{\rm s}^B(p)}.
\end{eqnarray}
These three solutions (\ref{KBgamNL}),  (\ref{KBgam}), and (\ref{BMgam}) coincide only for $|\delta t_{\rm s}^B(p) -t_{\rm col}(p)|\ll |\delta t_{\rm s}^B(p)|$. However this condition may hold only in very specific situations.
E.g., for Wigner resonances  it holds only for $|M- \Gamma/2|\ll \Gamma$.

The mass-shell equation (\ref{massShell}) produces another solution
\begin{eqnarray}
\label{BMmasshell}
\delta f_{\rm MS}(t,p)=\delta f(t=0,p) e^{-t/\delta t_{\rm MS}(p)},\quad
\delta t_{\rm MS}(p) =\frac{1}{4\,M(p)}\frac{\partial \Gamma(p)}{\partial p_0}.
\end{eqnarray}
Concluding, \emph{within the relaxation time approximation we arrive at somewhat contradictory results.}\\[2mm]

\noindent
(ii)~Using the BM replacement $\Gamma^{\rm in}=\Gamma f$ in the commutator term in the KB equation and in the mass-shell equation one proves that the latter two equations coincide with the BM equation. However, as we see from the non-local kinetic equation the parameter $\delta t_{\rm f}$ is not small compared to $\delta t_{\rm s}^B$ except for the case $|M- \Gamma/2|\ll  \Gamma$ (for Wigner resonances) that again puts in question correctness of the  gradient expansion for $t\sim \delta t_{\rm s}^B$.

\subsubsection{ Uniformly distributed  resonances in vacuum}

Now consider spatially uniform dilute gas of non-interacting resonances produced at $t<0$ and  placed in the vacuum at $t=0$. Then $\Sigma^R=\Sigma^R (p)$ and following Ref.~\cite{leupold} we put $\Gamma^{\rm in}(t>0)=0$ (the production of new resonances ceases). Using the latter ansatz we find
\begin{eqnarray}
C_{\rm vac.r.}= -A(p)\,\Gamma(p)\, f(t,p) .
\end{eqnarray}
From the BM equation we arrive at the distribution
\begin{eqnarray}
 f_{\rm BM}(t,p)= f(t=0,p) \,e^{-t/\delta t_{\rm s}^B(p)}.
\end{eqnarray}
However, the BM form of the kinetic equation does not hold for $\Gamma^{\rm in}=0$, since its derivation is based on the equation $\Gamma^{\rm in}=\Gamma f$.

On the contrary, from the KB equation we find another solution
\begin{eqnarray}\label{KBgam1}
f_{\rm KB}(t,p)= f(t=0,p)\, e^{- t/ t_{\rm col}(p)}.
\end{eqnarray}
Similarly, the solution of the kinetic equation in the non-local form is the same as  in Eq.~(\ref{KBgamNL}) with the replacement $\delta f\to f$.
From the mass-shell Eq.~(\ref{massShell}) we find solution (\ref{BMmasshell}), now for $f$ instead of $\delta f$. Thus we meet here  with the same problems as in previous example.

\subsubsection{ Collision-less dynamics of  propagating  resonances}

Let us find a class of spatially inhomogeneous  distributions of propagating virtual particles. We continue to assume  that $\Sigma^R$ does not depend on $X$.
As an ansatz, \emph{ let us use  the BM replacement $\Gamma^{\rm in}=\Gamma f$ both in the
commutator and in the collision terms in the KB equation}. Since in the shifted variables also $\Gamma^{\rm in}_{\rm shift}=\Gamma^{\rm shift} f^{\rm  shift}$, we obtain $C^{\rm
shift}=0$. Thus in all three cases we now deal with Eq.~(\ref{new-kin-eqhom}) with zero on the right-hand side  (the so-called Vlasov case). The solution of this equation is
\begin{eqnarray}\label{KBBMsol}
f(X,p)=f_0( t-\vec{r}\,\vec{v}_{\rm gr}/v^2_{\rm gr},p),
\end{eqnarray}
for an arbitrary function $f_0(\xi,p)$ and $\vec{v}_{\rm gr}$ being a function of $p$. Following Eq.~(\ref{gamin}), $\delta\Gamma^{\rm in}=0$  and the mass-shell Eq.~(\ref{massShell}) is also fulfilled.
Thus, in the given collision-less case solution (\ref{KBBMsol}) fulfills the KB, BM, non-local form and the mass-shell equations.

In the specific case of the one-dimensional propagation of a Gaussian distribution of off-shell particles, one gets
\begin{eqnarray}\label{disCol}
f(t,z,p)=f_0(p) \exp\left[-(z-{v}_{\rm gr}t)^2 (\Gamma_z /{v}_{\rm gr})^2\right],
\end{eqnarray}
where $f_0$ is an arbitrary function of $p$. The wave packet is propagating with the velocity $v_{\rm gr}$. Particles are distributed in $z$ near the maximum with the width $\delta z
\sim {v}_{\rm gr}/\Gamma_z$, $\Gamma_z$ is the energy width of the
initial distribution.

\subsubsection{ Collisional dynamics of  propagating resonances}\label{Collisional}

Let us continue to work with the assumption that $\Sigma^R$ does not depend on $X$.
The propagation of a resonance distribution in vacuum is described by the KB equation
(\ref{new-kin-eqhomKB}) with $C\neq 0$. The solution of the equation is
\begin{eqnarray}
\label{partsol}
f(X,p)= f_0(t-\vec{r}\,\vec{\widetilde{v}}_{\rm gr}/\widetilde{v}_{\rm gr}^2,p)
 \,e^{-t/t_{\rm col}(p)},
\end{eqnarray}
where $\vec{\widetilde{v}}_{\rm gr}$ is a function of $p$ and $f_0$ is an arbitrary function $f_0(\xi,p)$. The solution can be also presented as
\begin{eqnarray}\label{partsol1}
f(X,p)= \tilde{f}_0(t-\vec{r}\,\vec{\widetilde{v}}_{\rm gr}/\widetilde{v}_{\rm gr}^2,p)
\exp\left[-\frac{t-\vec{r}\,\vec{\widetilde{v}}_{\rm gr}/c^2}
{t_{\rm col}(p)\,(1-\widetilde{v}_{\rm gr}^2/c^2)}\right],
\end{eqnarray}
with another arbitrary function $\tilde{f}_0(\xi,p)$. The solution of the kinetic equation in the BM form is  obtained from (\ref{partsol}), (\ref{partsol1}) with the help of the replacement $t_{\rm col}\to \delta t_{\rm s}^B$, $\widetilde{v}_{\rm gr}\to {v}_{\rm gr}$. Note that the same solutions, but for $\delta f$ instead of $f$, exist also in case of propagation of light resonances in medium consisting of heavy particles, provided we exploit ansatz has been used above for the given case.

As follows from the solution (\ref{partsol1}),
if $\widetilde{v}_{\rm  gr}(p)>c$ and $v_{\rm gr}(p)>c$ in some $(p_0, \vec{p}\,)$ region,
there might occur an instability in respect to the growing of superluminal modes
that may result in some measurable effects, like a heating of the medium, cf. \cite{V95}.

\subsection{Validity of the gradient expansion}

Many works, e.g., the recent review \cite{Mosel}, use the BM form of the kinetic equation in practical simulations of the dynamics of resonances in heavy ion collisions, since it allows to apply a simplifying test-particle method. Thereby, they assume that the appropriate time for a relaxation of resonances is $\delta t_{\rm s}^B$ rather than $t_{\rm col}$. As we have demonstrated on examples, since $|\delta t_{\rm f} |\sim 1/\Gamma \gsim \delta
t_{\rm s}^B$, the gradient expansion may not hold on a typical time scale $t_{\rm ch}\lsim 1/\Gamma$ in the considered above problems at $C\neq 0$. Only if the typical time scale of the problem $t_{\rm ch}\gg 1/\Gamma$, the solutions of all three (BM, KB and non-local) forms of the kinetic equation and the mass-shell equation coincide.

Note that although the examples considered above show that the kinetic consideration might be not applicable for description of the system relaxation towards equilibrium at $t\lsim \delta t_{\rm s}^B ,  t_{\rm col}$, provided one considers propagation of off-mass shell particle
distributions, the quasiparticle limit proves to be the same for all three equations.
More generally, the kinetic approach holds at $t\lsim \delta t_{\rm s}^B , t_{\rm col}$ at least  for the wave packets with the energy integrated over a region near the maximum of the distribution, see Ref.~\cite{MalflTr}.

Another remark is in order. In spite of the formulated caution it might be practical to use one of the above kinetic equations for actual calculations even beyond its validity region, since all these kinetic equations reveal approximate or even exact (as for the KB form of the kinetic equation) conservation laws of the 4-current and the energy-momentum \cite{KIV01,IKV3}, thus, reasonably approximating the system evolution.

 \subsection{Hydrodynamical and thermodynamical limits}

Hydrodynamical limit \cite{V2011} is realized for $t\gg t_{\rm col}$, when  the distribution $f(X,p)$ gets the form of the local equilibrium distribution (\ref{fleq}). The hydrodynamical equations are derived from the conservation laws associated with the kinetic equation. The kinetic coefficients entering  hydrodynamical equations are derived from the BM equation (valid in this limit), see \cite{V2011}. They are expressed through the scattering delay time
$\delta t_{\rm s}^{A_S}$.

In the thermodynamical limit (global equilibrium, $\vec{u}=0, T,\mu =const$) all thermodynamical quantities can be expressed solely in terms of the spectral functions of the species~\cite{V08} and, thereby, they can be related to the above introduced Wigner time $\delta t_{\rm W}^A$. The memory term yields a contribution to thermodynamical entropy and specific heat and might be associated with the memory time.

In general all species are described with the help of the dressed Green's functions. However, since there are  relations between interaction and potential energies of the species, the interaction part  can be transported from some species to another ones. This procedure is nevertheless ambiguous \cite{V08}. In order to demonstrate how the interaction can be transported to one of the species consider the isospin-symmetric pion-nucleon-$\Delta$ isobar gas in the limit of very low density at finite temperature \cite{Weinhold}. It was assumed that pion and nucleon interact only via excitation of the intermediate  $\Delta$ resonances. In the virial limit the memory term disappears. Thermodynamical potential becomes as follows
\begin{eqnarray}
\Omega (T,\mu_{\rm bar})
&&=3 TV \int \frac{\rmd^3 q}{(2\pi)^3}\ln
[1-e^{-\omega_{\pi}^{\rm free}(\vec{q})/T}]
\nonumber\\
&&-4 TV \int \frac{\rmd^3 p}{(2\pi)^3}\ln
[1-e^{-(E_{N}^{\rm free}(\vec{p})-\mu_{\rm bar})/T}]
\nonumber\\
&&-16 TV \int \frac{\rmd^4 p }{(2\pi)^4}B_{0}^{\Delta}(p_0, \vec{p}) \ln
[1-e^{-(p_0 -\mu_{\rm bar})/T}],
\end{eqnarray}
where first two terms correspond to ideal gases of pions and nucleons and the third interaction term is expressed via the $B_0^{\Delta}$ function of the $\Delta$ resonance,
$\omega_{\pi}^{\rm free}= \sqrt{m_{\pi}^2+\vec{q}^{\,\,2}}$,
$E_{N}^{\rm free}\simeq m_N +\vec{p}^{\,\,2}/(2m_N)$. The baryon density is split in  the free  nucleon and dressed $\Delta$ contributions
\begin{eqnarray}
&&n_{\rm bar} =n_N +n_{\Delta}
=-\frac{1}{V}\left(\frac{\partial \Omega}{\partial \mu_{\rm bar}}\right)_{T,V},
\nonumber\\
&&n_N =n_N^{\rm free}
=4\int \frac{\rmd^3 p}{(2\pi)^3} \frac{1}{e^{(E_{N}^{\rm free}-\mu_{\rm bar})/T}+1},
\nonumber\\
&&n_{\Delta} =16\int \frac{\rmd^4 p}{(2\pi)^4}
\frac{B_0^{\Delta}}{e^{(p_0-\mu_{\rm bar})/T}+1}.
\end{eqnarray}
Decomposing $B_0^{\Delta}$ we may see that the first part of $n_{\Delta}=n_{\Delta}^{\rm Noether}$ is the proper contribution of the $\Delta$ to the baryon density, and the second part proportional to $\frac{\partial \Sigma_{\Delta}}{\partial p_0}$ is the contribution of $\pi N$ intermediate states to the dressed $\Delta$. Thus, the flow spectral function $B_0^{\Delta}$ is related to the density of the $\Delta$ states according to Eqs.~(\ref{NB}), (\ref{Bvir}). This result is in a line with the result (\ref{sht}) for the time shift of the $\Delta$. Moreover on example of a model with finite number of levels, Ref.~\cite{WeinholdDip} has demonstrated that namely expression (\ref{NB})  describes the level density in the given
interacting system.

In the  limit $n_B\to 0$ one has $B_0\to B_0^{\rm free}=A^{\rm free}$. In this limit the spectral function becomes the delta-function for stable particles but it remains a broad Lorentzian for "free" resonances. Thermodynamics  of free resonances was considered in \cite{V08}. Then all thermodynamic quantities are obtained from corresponding ideal gas expressions after replacements of the element of the 3-phase space $\frac{\rmd^3 p}{(2\pi)^3}\to \frac{d^3 p}{(2\pi)^3}\int \frac{\rmd p_0}{2\pi} A^{\rm free}(p_0,\vec{p}\,)$ and thereby they are expressed in terms of the Wigner time delay $\delta t_{\rm W}^{\rm free}=A^{\rm free}$.

\section{Space-time delays and measurements}\label{sec:measurement}

\subsection{Speed of the propagating wave packet}

Consider a propagation of an initial  distribution of off-mass-shell particles in a uniform medium or in vacuum. For a particle off mass shell there is, in general, no upper limit
on its speed. The distance of the order of the mean free path $\Delta z$ in the $z$ direction is passed by the maximum of the distribution at $(E_{\rm m},k_{\rm m})$ with
the velocity
\begin{eqnarray}
\label{vmsh}
v^{\,\rm shift}_{\rm m} =\frac{\Delta z}{\Delta t}
=\frac{\Delta z {\,'}+\delta z_{\rm f} }
  {\Delta t' +\delta t_{\rm f}^{\rm kin}},
\end{eqnarray}
where $\Delta z{\,'}$ and $\Delta t'$ would characterize two acts
of the collision/measurement (on average) without the variable shifts $\delta z_{\rm f}$ and $\delta t_{\rm f}^{\rm kin} $ in the collision integral, Eqs.~(\ref{deltaX}), (\ref{sht}). Thus, the arrival of the peak of the wave packet at point $\Delta z' +\delta z_{\rm f}$ (not $\Delta z'$) is delayed or advanced by  $\delta t_{\rm f}$. Under the assumption (done here for simplicity) that the variable shifts are small, the change of the speed of the propagating peak is
\begin{eqnarray}
\label{vshift}
&&\delta {v}_{\rm m} \simeq {v}_{\rm m}^{\rm shift}-{v}_{\rm ph}
=-\frac{B_0}{2\Delta t'} ({v}_{\rm ph}-{v}_{\rm gr})
 + \frac{Z^{-1}_0 }{\Gamma \Delta t'} ({v}_{\rm ph}
  -\widetilde{v}_{\rm gr})\,,
\end{eqnarray}
where
${v}_{\rm ph}=\Delta z{\,'}/\Delta t'.$ Although for $\Delta t'\to \infty$ the change of the velocity (\ref{vshift}) becomes negligibly small, it might be important for not too large
values of the time intervals $\Delta t'$. For the Wigner resonances ${v}_{\rm gr}={\widetilde{v}}_{\rm gr}=v_{\rm ph}$, and $\delta {v}_{\rm m} =0$. Nevertheless, even in this case actual the particles from the forward and backward tail of the distribution (for $E_{\rm m}\pm \Delta E$, $k_{\rm m}\pm \Delta k$) move between collisions with velocities slightly different from that of the peak, $\delta v_{\rm tail}\sim \pm \frac{\partial v_{\rm gr}}{\partial E}\Delta E \pm \frac{\partial v_{\rm gr}}{\partial k}\Delta k \pm v_{\rm gr}/(\Gamma_z \Delta t')$, where $1/\Gamma_z$ is the width of the wave packet, see Eq. (\ref{disCol}). This causes a smearing of the wave packet.

If the distance of the free flight $L$ is fixed by conditions of the measurement, then $\delta z_{\rm f}$ should be put zero in Eq.~(\ref{vmsh}), ${v}^{\,\rm shift}_{\rm m} = \frac{L}{\Delta t}$,
and
\begin{eqnarray}\label{vdelt}\delta {v}_{\rm m} =-\delta t_{\rm f}^{\rm kin} \frac{{v}_{\rm
ph}}{\Delta t'}= \left(-\frac{B_0}{2\Delta t'} + \frac{Z^{-1}_0
}{\Gamma \Delta t'}\right) {v}_{\rm ph}\,,
\end{eqnarray}
$\delta {v}_{\rm m}>0$ for $\delta t_{\rm f}^{\rm kin}<0$.

\subsection{Measurements and  resulting time delays and
advances}\label{Measurementsadvancements}

There are several sources of the time delays and advances.

(i)
Quantum mechanics, as well as quantum kinetics, says nothing about a motion of a single identifiable particle. Thus, the transmitted distribution of quantum particles is not the same entity as the incident distribution. To be sure that the particle beam propagating from $z=0$ to $z=L$ is described by a certain distribution, e.g. by Eq.~(\ref{disCol}), one should measure a small fraction of it (not disturbing a bulk) at $(t=0,z=0)$ and then at $(t=t_L, z=L)$. After the measurement at $(t=0, z=0)$, particles disturbed by the measurement are effectively taken out of the distribution. Thus, at $(t=t_L, z=L)$ we deal with other particles from the initial distribution which were not tagged at $(t=0, z=0)$. It could than happen that the first particles registered at $t=t_L,z=L$  may additionally advance those particles, which are almost identical to the particles registered at $t=0, z=0$, typically by a time step $$\delta t_{\rm tail} \sim \pm 1/\Gamma_z .$$ So, the typical time advance of the signal arriving at $z=L$  is  $\delta t_{\rm tail}$. Certainly this time advance can be diminished by performing precision measurements of the peaks of the distribution at $z=0$ and $z=L$ but this procedure needs a special care.

(ii) Another time delay/advance, $\sim \delta t_{\rm f}^{\rm kin} $, arises as it is seen from  the collision term in non-local form, that also manifests in appearance of the Poisson bracket fluctuation contributions in the KB equation (provided the kinetics is described by the KB equation or by its non-local form).  The result is given by Eq. (\ref{sht}). This time shift
characterizes (on average) the time delay/advance between two  successive collisions.

(iii) Other time delay/advance is associated with the memory effects (yielding $\delta t_{\rm mem}$) appearing in the processes of multiple interactions described by diagrams with more than two vertices. However, the value $\delta t_{\rm mem}$ diminishes in the case of a very dilute beam, since diagrams with three and more vertices bring additional powers of the  density.

Summing up these three delay/advance times for the total time shift  we finally obtain
\begin{eqnarray}
\delta t_{\rm tot} \sim \delta t_{\rm tail}+ \delta t_{\rm f}^{\rm
kin} +\delta t_{\rm mem}.
\end{eqnarray}

\subsection{Apparent superluminality in neutrino experiments as a time advance effect}

September 2011 the OPERA experiment \cite{OPERA} (see version 1 of the e-print) claimed  measurement of  muon neutrinos propagating with  superluminal  velocity, $(v-c)/c =[2.37\pm 0.32 ({\rm stat}) \pm 0.34({\rm sys}) ]\cdot 10^{-5}$, at average energies $<E>=17$ GeV. This data agreed with the  data  obtained earlier by the MINOS collaboration  \cite{MINOS}: $(v-c)/c =(5.1\pm 2.9) \cdot 10^{-5}$, $E$ peaking is at $\sim 3$ GeV with a tail extending to 100 GeV, $d=734$~km. An initial proton beam in OPERA experiment produces a bunch of pions. Neutrinos produced in the reactions $\pi\rightarrow\nu\bar{\mu}$ pass through the ground the distance
$L$ and reach a detector. For the distance $L=730$~km $\pm 20$~cm between the neutrino source in CERN and the detector in Gran Sasso the superluminality of the neutrino beam corresponds to the time advance $t_{\rm adv}=57.8\pm 7.8 (stat)+ 8.3-5.9 (sys)$~ns compared to that the neutrinos would move with the speed of light. In February 2012 the OPERA collaboration has informed \cite{OPERAfalse} that it has identified two possible effects that could have an influence on its neutrino timing measurement. The first possible effect concerns an oscillator used to provide the time stamps for a GPS synchronization. The second concerns an optical fibre connector that brings the external GPS signal to the OPERA master clock. At the 25th International Conference on Neutrino Physics and Astrophysics in Kyoto 8.06.2012 a final update on the OPERA time of flight measurement was reported $t_{\rm adv} =1.6\pm 1.1
(stat)+6.1-3.7 (sys)$ ns.

Not entering in details of the given experiment and its deficiencies we consider a principle possibility to get a time advance of the order of $\sim 10-10^2$ ns in the neutrino experiments. Although many different possibilities were discussed in the literature, the  effects, which we will consider, were not mentioned. Simplifying, we assume that the initial $z=0, t=0$ point is well fixed with the help of  heavy protons. The final point $z=L, t=t_L$ is fixed by reactions of the neutrinos on the forward front of the propagating packet  with
nuclei in the detector. We will argue that apparent superluminality can be associated with effects of the time advances considered in the given paper.

The maxima of the wave packets of protons and neutrinos produced in the two-step process $p\to \pi^{+}n\to n\mu^{+}\nu$ at CERN are separated by the time interval $\sim 1/\Gamma_{N\pi}+1/\Gamma_{\pi\nu}$, $N=p$ here. \emph{Thereby a time advance of the neutrinos arises owing to the advance  of pions compared to protons and neutrinos,}
$\delta t_{\rm adv}^\nu = \delta t_{\rm adv}^{N\pi}+\delta t_{\rm adv}^{\pi\nu}$, where
$\delta t_{\rm adv}^{N\pi}=-1/\Gamma_{N\pi}$ and $\delta t_{\rm adv}^{\pi\nu} = -1/\Gamma_{\pi\nu}$. The value $\delta t_{\rm adv}^{\pi\nu}= -1/\Gamma_{\pi\nu}=-26$ ns is due  to  the width $\Gamma_{\pi\nu}$ of the production of the neutrino in the process $\pi\rightarrow\nu\bar{\mu}$. The value $\delta t_{\rm adv}^{N\pi}\sim -10^{-23}$ s is much shorter and can be neglected. The origin of the resulting  time advance  $\delta t_{\rm
adv}^\nu$ arising in this process was illustrated by Fig. 2. So we believe  that the width $\Gamma_z\sim \Gamma_{\pi\nu}$ in the initial neutrino wave packet may yield $\delta t_{\rm tail}\sim \delta t_{\rm adv}^{\pi\nu}$ ($-26$ ns) of advance, see point (i) of  Sect.~\ref{Measurementsadvancements}.

Another contribution to the time advance $\delta t_{\rm f}^{{\rm kin},\nu} \sim -1/\Gamma_{\pi\nu}$ ($-26$ ns) arises if the virtual pion propagation is described by the KB kinetic equation or by its non-local form  for $|M_{\pi}| \gg \Gamma_{\pi\nu}/2$, as it follows from Eq. (\ref{sht}), see point (ii) of Sect.~\ref{Measurementsadvancements}.

Summing up two possible  time advances,  for the "most rapid" particles we find $\delta t^\nu = \delta t_{\rm adv}^\nu +\delta t_{\rm f}^\nu \sim -2/\Gamma_{\pi\nu}=-52$~ns. This agrees well with the firstly announced result of the OPERA experiment.

Note that  provided neutrino flux in the  ground corresponds to off-shell neutrinos, see Eq.~(\ref{vdelt}),  an additional time advance could occur. But this effect resulting in $\delta v_{\rm gr} \propto G_{\rm W}^2$, where $G_{\rm W}$ is the coupling constant of the weak interaction, is very small, since $c/\Gamma_{\nu} \gsim L$. Thus, most likely neutrinos produced at CERN undergo almost free flight to the detector in Gran Sasso. Likely, a smearing effect of the wave packet is also tiny at conditions under consideration.

Summarizing, one may expect few of $\delta t_{\rm adv}^{\pi\nu}$, as typical time advance in the neutrino experiments like those performed by the OPERA and MINOS. If our interpretation is correct, a possibly  measured in neutrino experiments, like the OPERA experiment, a time advance should not significantly depend on the distance $L$ between the source and the detector but its value is very sensitive to the conditions, by which the initial and final time moments are fixed in the measurements. From the description of the mentioned OPERA experiment it is not sufficiently clear how this fixation was performed.

\section{Conclusion}

The aim of the present paper is to give a coherent overview on how various measures used to quantify the durations of processes in classical and quantum physics appear and to explicate their interlinking.

\subsection{ Classical mechanics}

For time measurements in classical mechanics, besides ordinary
time characteristics, such as the oscillation period $P$, the
phase time shift $\delta t_{\rm ph}=\hbar\delta/E$ (here $E$
denotes the particle energy and $\hbar\delta=S_{\rm sh}$ stands
for the mechanical shortened action) and the decay time $t_{\rm
dec}$, we introduced another quantities such as the dwell time
$t_{\rm d}$, the sojourn time $t_{\rm soj}$, and the group delay
$t_{\rm gr}^{\rm 1D}=\partial\hbar\delta/\partial E$. We discussed
relations between these times. For example, we demonstrated that
the classical sojourn time delay is negative in case of attractive
one-dimensional potentials and positive for repulsive
one-dimensional potentials. In the three-dimensional case the
situation is more involved. For the spherically symmetric
potential there is no direct correspondence between the sign of
the potential and the sign of the classical sojourn time delay.
Also for the radial motion there appears extra factor of 2 in the
classical group time delay compared to the one-dimensional case,
because the coordinate integration is restricted by $r>0$ in the
former case, and goes from $-\infty$ to $\infty$ in the latter
case,  $t_{\rm gr}^{\rm 3D}=\delta t_{\rm W}= 2
\partial\hbar\delta/\partial E$. Exactly the latter quantity was
originally introduced by Wigner and Eisenbood for quantum
scattering~\cite{Wigner}.

Then we studied examples demonstrating time advances and delays of
a damped oscillator $z(t)$ under the action of different external
forces $F(t)$. We applied  the Green's function formalism
exploiting extensively in quantum field theory and quantum
kinetics. To establish a closer link to the formalism of the
quantum field theory, we introduced the dynamical variable -- a
"field" -- $\phi (t)=m\,z(t)$. The source term in the Lagrange
equation depends non-linearly on $\phi$ and linearly on the
external force. The formalism allows a natural diagrammatic
interpretation of the solution of the Lagrange equation.

First, we considered the response of the damped harmonic
oscillator with the resonance frequency $E_{\rm R}$ and of the
damping width $\Gamma$ to a sudden change of an external constant
force. The response is purely causal in this case. The larger is
the damping width $\Gamma$ of the oscillator and respectively the
shorter is the damping time $t_{\rm dec}= 1/\Gamma$, the longer is
the phase time $\delta t_{\rm ph}$, showing the time shift of the
oscillations. For $\Gamma \to 2E_{\rm R}$ the oscillation period
vanishes, $P \to 0$, and the phase shift $\delta t_{\rm ph}$
becomes infinite, but the ratio $\delta t_{\rm ph}/P$ remains
finite, $\delta t_{\rm ph}/P\to 1/4$.

Then we demonstrated that in the case, when an external force is acting
over a finite time interval, the damped harmonic oscillator can
exhibit apparently acausal reaction: the maximum of the oscillator
response may occur before the maximum of the external force. Thus,
if for the identification of a signal we would use a detector with
the threshold close to the pulse peak, such a detector would
register a peak of the response of the system before the input's
peak.

Next we considered a possibility of an advanced response also on the example of a periodic driving force with a constant frequency acting on a damped non-linear oscillator. In the linear approximation with respect to the anharmonicity parameter, there appears an overtone peaked at frequency $E_{\rm R}/2$.  Thus, there arises an extra phase time scale
characterizing dynamics of the overtone. When the frequency of the force is $E_p\sim \half E_{\rm R}$, the overtone can produce an additional maximum in $z(t)$, which would appear, as occurring before the actual action of the force. So, the system would seem
to ``react'' in advance.

In the case, when the external force acting on a damped anharmonic
oscillator with the resonance frequency $E_{\rm R}$ is a packet of
modes grouped near frequency $E_p$ with the width $\sim \gamma$,
typical time for which the envelop function fades away is $t_{\rm
dec}^{\gamma, {\rm (cl)} } = 1/\gamma$ for $\gamma \ll \Gamma$. In
linear approximation in anharmonicity parameter  there appear two
resonance group time delays, $t_{\rm gr}^{(1)}=A_1 /2
=\frac{\Gamma/2}{(E_p-E_{\rm R})^2 +\Gamma^2/4}$ and $t_{\rm
gr}^{(2)}=A_2 /2 =\frac{\Gamma/4}{(E_p-E_{\rm R}/2)^2
+(\Gamma/2)^2/4}$ one peaked at $E_{\rm R}$, and another one, at
$E_{\rm R}/2$. These group time delays appear because the system
responses slightly differently to various frequency modes
containing in the force envelop. Oscillations of the carrier wave
are delayed by the phase times, whereas the amplitude modulation
is delayed by the group times.
We introduced new quantity -- the forward time delay/advance,
$\delta t_{\rm f}^{\gamma}=t_{\rm gr}-t_{\rm dec}^{\gamma,{\rm
(cl)}}$ -- which takes into account that  the delays of the wave
groups are starting to accumulate before the external force
reaches its maximum with an advance determined by the width of the
force packet. When the external force frequency $E_p$  is near the
oscillator resonance frequency  $E_{\rm R}$ the forward time is
positive, that corresponds to a delayed response, but in the
off-resonance region the forward time changes its sign, that
corresponds to an advanced response. In the limit $\gamma \to 0$
we arrive at the case of a purely periodic force. An interesting
effect is that the frequency of the carrier wave is changed and
becomes time dependent. When $E_p$ approaches $E_{\rm R}$ not only
the amplitude of the system response grows but also the response
lasts much longer than the force acts. Thereby we demonstrate
effect of a smearing of the wave packet in classical mechanics.

Further on, time shifts appearing in the three-dimensional
classical scattering problem were considered on example of the
particle scattering on hard spheres of radius $R$.  We derived the
limits on the values of time advances: the sojourn time advance,
being equal to the Wigner time advance, is limited by $\delta
t_{\rm W}=2\frac{\partial\hbar\delta}{\partial E} >-2R/v$, where
$v$ is the particle velocity. We introduced another relevant
quantity, the scattering time delay, the difference of time, when
the particle touches the sphere surface, and the time, when the
particle freely reaches the center of the sphere.  This time
characterizing delay of scattered waves is twice smaller than the
Wigner time delay.

Next, we discussed time delays appearing in classical
electrodynamics. More specifically, we studied a problem of the
radiation of a damped charged oscillator induced by an external
electromagnetic plane wave. We introduced the scattering
amplitude, the cross section and related them to the phase shift.
As the cross section, the scattering group time delay has a
resonance shape. The damping is determined by the sum of the
oscillator and radiation damping widths. The scattering group time
delay (the scattering delay time), being twice as small compared
to the Wigner delay time.

For the scattering of light on a hard sphere of radius $R$, we
show that appearance of a temporal advance in the
signal propagation does not contradict causality.

\subsection{Quantum mechanics}

We studied time shifts arising in different quantum mechanical
problems. More specifically we considered  one dimensional
tunneling and three dimensional quantum scattering of
non-relativistic
 particles.

\subsubsection{ One dimensional quantum mechanical motion}

To be specific we assumed that the potential $U>0$ acts within a
finite segment $-L/2 \leq z\leq L/2$.  For the particle motion
above the barrier ($E>U$) both the dwell time and the traversal
time $t_{\rm trav}$ are relevant quantities depending on the
distance passed by the particle. High above the barrier they are
reduced to the free flight time $L/v$. However at energies below
the barrier the traversal time becomes imaginary. For the
rectangular barrier the dwell time is always smaller than the
classical transversal time for energies of the scattered particle
$E<U$ for a broad barrier and for $E<{\textstyle\frac{3}{4}}U$\,
for a thin barrier. For the case of the  tunneling through a very
broad barrier of an arbitrary form the dwell time in the region under the barrier is determined  by
the evanescent wave and describes, thereby, neither particle
transmission nor the dwell of transmitted waves. In this
particular case the dwell time is determined by the quantum time
scale, which  is   shorter than would be the classical traversal
time of the same region.

Then we considered propagation of wave packets. For a free moving
packet we recover the well-known result that the width of packet
increases with time. For a typical time of  smearing of the packet
we get $t_{\rm sm}\sim \hbar m/\gamma_p^2$, where $\gamma_p$ is
the width of the packet momentum distribution. Then we consider
scattering of wave packets with negligibly small momentum
uncertainty.  According to the method of the stationary phase, the
position of the maximum of an oscillatory integral is determined
by the stationarity of the complex phase of the integrand.
Eisenbood and Wigner used this method to introduce two measures of
time that could characterize the wave propagation within the
potential region: the difference of  time, when the maximum of the
incident packet is at the coordinate $z=-L/2$, and the time, when
the maximum of the transmitted packet is at $z=+L/2$, and the
difference of the time, when the maximum of the incident packet
and the maximum of the reflected packet are at the same spatial
point $z=-L/2$. We call these time intervals the transmission and
reflection group time delays, $t_{\rm T}$ and $t_{\rm R}$.
Moreover one introduces  bidirectional scattering group time
$t_{\rm bs}$ composed of the transmitted and reflected group times
weighted with probabilities of the transmission and the
reflection. We argue that in the case of tunneling  the group
times show time delays on the edges of the barrier  on the scale
of quantum length near the turning points. In case of the broad
barrier the bidirectional group time delay is mainly determined by
the reflection group time delay. The difference between the
bidirectional scattering time delay and the dwell time is now a
time delay/advance due to interference of waves $\delta t_{\rm
i}$. This interference time term is absent in the case of the
classical motion. The interference time proves to be negative
(advance) for under the barrier motion. Within semiclassical
approach for the tunneling  the bidirectional scattering time
equals zero. The difference with exact result is due to the fact
that in the region near the turning points, where the group time
delays are accumulated, semiclassical approximation is not
applicable. For the scattering of an arbitrary wave packet the
sojourn time appears as the dwell time averaged over the momentum
distribution in the packet. Therefore, from the definition of the
sojourn time one extracts basically the same information, as from
the definition of the dwell time. For the particle motion well
above the barrier the dwell and the group times are reduced to the
appropriate traversal time proportional to the length of the
distance passed by the particle, and the interference time
vanishes.

The phenomenon that for very broad barrier the dwell and the group
times are reduced to the quantum time not proportional to the
barrier length is known as the Hartman effect. We clarified the
reasons for the appearance of the Hartman effect and formulated
arguments,  why the group times and the dwell time are not
appropriate quantities to measure the tunneling time.

Operating with the group times and the related to them dwell time
one assumes that the position of a particle can be identified with
the position of the maximum of the wave packet. However, it is not
so easy to experimentally distinguish the peak position of a
spatially broad packet. Then, to specify the position of the
particle we studied the motion of the centroids (centers of mass)
of the incident, transmitted, and reflected wave packets. We
showed that the barrier acts as a filter letting  with higher
probability penetration for the modes with higher energies. As the
result, all three packets move with different velocities at large
distances from the potential region. Also the widths of
transmitted and reflected packets differ from the width of the
incident packet. We demonstrated that in case of $\gamma_p L\ll
\hbar$ the centroid transmission and the reflection time delays
are mainly determined by the wave packet formation times $t_{\rm
form, T}$, $t_{\rm form, R}$. These quantities show averaged
passage times by particles of the typical spatial packet length
$\hbar/\gamma_p$. The term $\propto L\gamma_p$ entering expression
for $t_{\rm form, T}$ but not entering $t_{\rm form, R}\simeq
t_{\rm form, I}$ corresponds to an advance, since the transmitted
wave packet moves with a higher velocity compared to the reflected
and  incident wave packets. Dependence on $L$ may indicate  that
the passage time of the barrier might be proportional to its
length.

A more complete information about temporal behavior of the packets
can be extracted from the explicit forms of the spatial
distributions. To elucidate these aspects further, we consider
explicitly Gaussian wave packet tunneling through a barrier. We
found that because of the smearing effect the longer is the
barrier, the broader is the transmitted wave packet, being formed
for $z\geq L/2$ with a delay depending on the length scale. In
further study, considering propagation of waves on time scales
shorter than $t_{\rm sm}$ we neglected the effects of smearing.
The probability (on the right wing of the packet) to meet the
particle at $z=L/2$ becomes the same, as it were in the case of
the monochromatic wave with $E=E_p$ (stationary problem), at the
time moment, when the maximum of the incident wave packet did not
yet reach the barrier and the maximum of the transmitted wave
packet did not yet emerge at $z=L/2$. For a very broad rectangular
barrier of the height $U$ placed at $-L/2\leq z\leq L/2$, the peak
of the transmitted wave packet is formed at the right boarder of
the barrier,  after a quantum time delay (not dependent on the
barrier depth) from the moment, when the peak of the incident wave
packet reached the left boarder of the barrier (the Hartman
effect). The probability to meet the particle at $z=L/2$ again
(now on the left wing of the packet) becomes the same, as it were
in case of the monochromatic wave with the energy $E_p$, when the
maximum of the transmitted wave packet achieves the point $z
\simeq L/2 +L v m/{\varkappa}+\hbar p
({p^{-2}}-{\varkappa_{p}^{-2}})$, where $\varkappa_p
=\sqrt{2mU-p^2}/\hbar$, the traversal time $t_{\rm trav}^{\rm tun}
=Lm/{\varkappa}$ is much larger than the (third) quantum time
term.  Thus, we are able to associate the time $t_{\rm trav}^{\rm
tun}$ with the passage time of the barrier for waves with $E\simeq
E_p$. Thereby, we believe that this observation can be
interpreted, as a resolution of the Hartman paradox.

Since the packet has a width, the probability to find particle at
a given point  becomes essentially non-zero already before the
center  of the packet with the energy dispersion $\gamma$ has
reached it, with an advance $ t_{\rm dec}^{\gamma}
=\hbar/\gamma\,$.  In accord with the Mandelstam-Tamm uncertainty
relation
it has meaning of the time, during which
the wave packet passes a given space point. Thus, the real
(forward)  delay/advance time at $z=L/2$ is not $t_{\rm bs}$ but
$\delta t_{\rm f}^{\rm tun}= t_{\rm bs}- t_{\rm dec}^{\gamma}$.

Next, we studied resonance states and their time evolution. We
considered a scattering on the potential well of the length
$l_{\rm R}$ (a resonator) with infinite wall at the origin
separated from the region $z>l$ by the rectangular barrier of the
height $U$ and thickness $l-l_{\rm R}$. The scattering amplitude
is shown to possess simple poles for complex energies $E=E_{{\rm
R},i}-i\Gamma_i/2$, $i=1,\dots n$.  The calculated dwell and
reflected group time exhibit strong resonance enhancement for
energies $E\sim E_{{\rm R}, i}$. Hence, the incident packet spends
in the interaction region much longer time than the typical
passage time of this region by a particle with the mean velocity
of the packet. For energies tuned from $E_{{\rm R},i}$ by more
than the resonance width the internal part of the potential
becomes inaccessible to the incident wave. In view of the symmetry
of the problem it is possible to introduce the single-way  dwell
time and the scattering time, which can be related to the number
of resonance states per unit  energy and are equal $t_{\rm d}^{\rm
s.w.}(0,l) \simeq  t_{\rm s}=\hbar A/2
=\frac{\hbar\Gamma/2}{(E-E_{\rm R})^2 +\Gamma^2/4}$. The sum-rule
for the scattering time delay is preserved. We constructed the set
of the eigenfunctions for the given scattering problem and used
them in the analysis of the evolution of some initially localized
state. The properties of the survival probability for this state
are discussed. The relations between  the survival probability and
the retarded Green's function are obtained. It is shown that the
decay time $t_{\rm dec}=\hbar/\Gamma$ of the resonance state can
be calculated as the sojourn time of the wave function within the
interval $(0,l)$. The forward scattering time $\delta t_{\rm f}=
t_{\rm s}-t_{\rm dec}$ is shown to correspond to some delay in the
scattering of particles with energies nearby the energy of the
quasistationary level and to an advance for $|E-E_{\rm
R}|>\Gamma/2$. The causality restriction becomes $\delta t_{\rm
s}=\frac{\partial\hbar\delta}{\partial E}
>-l/v -\hbar/2kv$. The  term $\hbar/2kv$ is
of purely quantum origin. It shows the time, which the particle
needs in order to pass the half of the de Broglie wave length of the particle,
$\lambda =\hbar/k$. Following the uncertainty principle free quantum
particles cannot distinguish distance $\xi <\hbar/2k$.

\subsubsection{ Three dimensional scattering problem}

We considered the three-dimensional scattering problem on a
central potential and discussed the difference compared with the
one-dimensional scattering. We introduced the sojourn time $\delta
t^N_{\rm vol}=\delta t_{\rm s}^N-\delta t_{\rm i}^N$ and the
corresponding scattering and interference times, all  normalized
by the incident flux. The Wigner group time delay $\delta t_{\rm
W}=\delta t_{\rm vol}^N =2\frac{\partial \hbar\delta^l}{\partial
E_p}$ appears, as the group time delay of the divergent wave
taking into account interference with the incident plane wave,
$\delta^l$ is the phase shift of the radial wave function. The
scattering group time $t_{\rm s}=t_{\rm free}+\delta t_{\rm s}$ is
normalized by the scattered flux. The scattering group time delay
$ \delta t_{\rm s}=\delta t_{\rm W}/2$ appears, as the delay of
the purely scattered wave. The decay time appears as $t_{\rm dec}=
\delta t^N_{\rm vol}/4\mbox{sin}^2\delta^l$.

Then we studied scattering on a Wigner resonance with a constant width $\Gamma$.
The probability for particle to enter the region of the
resonance interaction can be written as
$P_{\Gamma}=\sin^2 \delta
=\frac{\Gamma^2/4}{M^2 +\Gamma^2/4}
=A\Gamma/4$, where $M=E - E_{\rm R}$.
The cross section of the resonance scattering can be presented as
$\sigma \simeq 4\pi \lambda^2 P_{\Gamma}$ with $\lambda =\hbar/k$
standing for the de Broglie wave length. For $M=0$ (i.e. exactly
at the resonance) the cross section reaches its maximum
$\sigma_{\rm max} = 4\pi \lambda^2$. The scattering time delay
coincides with the single-way dwell time
$\delta t_{\rm s}= \hbar\frac{\partial \delta}{\partial E} =t_{\rm d}^{\rm s.w.}=\hbar A/2$.
The forward time delay coincides with the interference time. The
forward time delay
$\delta t_{\rm f}=\delta t_{\rm i}=\hbar A/2-t_{\rm dec}=t_{\rm
dec}\,(\sin^2\delta -\cos^2\delta)$
is the time delay of the decay because of the difference in the
probability for the particle to enter the region of the resonance
interaction ($\sin^2\delta$) and not to enter this region
($\cos^2\delta$). The forward delay/advance time, $\delta t_{\rm
f}$, is then an average delay/advance in the scattering counted
from the decay time $t_{\rm dec}$.  Explicitly, for the Wigner
resonances we derive expression $\delta t_{\rm f}=-\hbar\frac{M^2
-\Gamma^2/4}{\Gamma (M^2 +\Gamma^2/4)}$. Thus $\delta t_{\rm f}$
corresponds to a delay for $|M|<\Gamma/2$ and to an advance for
$|M|>\Gamma/2$.

We discussed quantum mechanical scattering on hard spheres of
radius $R$. For $l=0$ we find $\delta t_{\rm s}^0
=\hbar\frac{\partial\delta^0}{\partial E_k}=-R/v$. The same
advance, $\delta t_{\rm s}^l =-R/v$, arises for rapid particles
$kR/\hbar\gg l^2$ at angular momenta $l\neq 0$. For slow
particles, $kR/\hbar\ll l^{1/2}$, $\delta t_{\rm s}^l \propto
(kR/\hbar)^{2l} R/v$. When the de Broglie wave length $\lambda
=\hbar/k\gg R$ the propagating wave almost does not feel presence
of the sphere. Thus, for $l\gg 1$ the cross section becomes
negligibly small.

Then, using semiclassical expression for the phase shift we
considered similarity and difference  between semiclassical and
classical expressions for the time delays. Also, we discussed
ergodicity and related the scattering time shifts to the level
density. The Wigner delay can be interpreted as a time delay in an
elementary phase space cell: $\delta t_{\rm W}=2\delta t_{\rm
s}=2\pi\hbar dN^{\rm level}/dE_p$. Examples of the resonance
scattering and scattering on hard spheres were considered.

\subsection{ Quantum field theory}

We considered time shifts, as they appear in quantum field theory.
Knowing the Lagrangian one constructs the generating functional on
the Schwinger-Keldysh contour \cite{IKV}. Varying this functional
one reproduces the equation of motion for the mean field and four
Dyson equations for the non-equilibrium Green's functions
$G^{ij}$, $i,j=\{+,-\}$. These Green's functions can be expressed
in terms of Feynmann diagrams only, if the typical times in the
problem are longer than the typical time scale of the interaction
$t_{\rm int}$ (the principle of weakening of initial correlations)
and the typical spatial scale is longer than the interaction scale
$l_{\rm int}$. Further we assume that these conditions are
fulfilled.  Dropping of short-range correlations  on each time
step causes a growth of the entropy with time, which is associated
with the thus obtained Dyson equations. The scattering time delay
is expressed in terms of the imaginary part of the retarded
Green's function $A=-2\Im G^R$ and the decay time, in terms of the
imaginary part of the retarded self-energy, $\Gamma
=-2\Im\Sigma^R$. The equilibrium particle occupations (in the
Boltzmann limit) relate to a delay of the scattering time and an
advance of the collision time.

We discussed typical duration times for the reactions, which
occur via intermediate states. We showed that the reaction times
may cause an advance for some processes. Since integration over the 4-coordinate
$z$ in intermediate reaction states of Feynmann graphs includes all times $-\infty <t_z <\infty$, for $t_y <t_z <t_x$,  the process occurred at $t_z$ is delayed compared to that occurred at $t_y$, and for $t_z <t_y <t_x$ the process occurred at $t_z$ is advanced compared to that occurred at $t_y$. Both time processes must be incorporated as dictated
by the Lorentz invariance. If there occurs a  two-step process, e.g. $p\to n+X+\pi_{\rm virt}^+\to n+X+{\nu}+\mu^+$, its duration is characterized by the time
$t_{\nu}^{\rm dec}= t_{N\pi}^{\rm dec}+t_{\pi\nu}^{\rm dec}$.
Here $t_{N\pi}^{\rm dec}= \hbar/\Gamma_{N\pi}$ is the life time of the virtual pion produced in the process $p\rightarrow n+X+\pi_{\rm virt}^+$ and $t_{\pi\nu}^{\rm dec}= \hbar/\Gamma_{\pi\nu}$ is the life time of the virtual pion produced in the process
$\pi_{\rm virt}^+\to {\nu}+\mu^+$. This means that virtual pions, being
produced in the process $p\to n+X+\pi_{\rm virt}^+$, in the subsequent process
$\pi_{\rm virt}^+\to {\nu}+\mu^+$ undergo time delays and advances
on a time scale $- t_{\pi\nu}^{\rm dec}\lsim t_2-t_1\lsim t_{\pi\nu}^{\rm dec}$, where $t_2$ characterizes act of the production of $\nu$ and $t_1$, of the absorption of $p$.

Also, we expressed the sojourn time  in terms of the
non-equilibrium Green's function.

\subsection{ Quantum kinetics}

Assuming that time-space scales characterizing dynamics of
collective modes are larger than microscopic time-space scales one
exploits the Wigner transformation for the Green's functions (the Fourier
transformation in $\xi = (t_1-t_2, \vec{r}_1-\vec{r}_2)$). Within
the first-order space-time gradient approximation over
$\frac{1}{2}(t_1+t_2,\vec{r}_1+\vec{r}_2)$ one derives dynamical
equations for the four Green's functions.  It is important to
notice that although we derived four Dyson equations for four
complex Green's functions, only two real quantities are
independent. As independent real variables it is convenient to use
the spectral function $A=-2\Im G^R$ and the $iG^{-+}$ Green's
function (the Wigner density). It proves to be that the retarded
Green's function satisfies the algebraic equation (up to second-order space-time
gradients) and $iG^{-+}$ Green's function
fulfills the first-order gradient generalized kinetic equation.
The other, the mass-shell equation, describing propagation of the
off-mass-shell particles on equal footing with the generalized
kinetic equation, should coincide with the generalized kinetic
equation provided all approximations were done consistently.

We demonstrated that the generalized kinetic equation can be
presented in three forms, the proper Kaddanoff-Baym form, the
Botermans-Malfliet form and the non-local form. The
Botermans-Malfliet form follows from the Kaddanoff-Baym form
provided  space-time gradients are small and moreover the system
is close to the local equilibrium. The non-local form differs from
the Kaddanoff-Baym one only in the second-order gradient terms in
the expansion of the collision term and it coincides with the
Botermans-Malfliet  form, if one retains only zero gradient terms
in the gradient expansion of the collision term.
In the collision term of the non-local kinetic equation there
arise 4-coordinate-momentum shifts in space-time variables. They
appeared due to the Poisson bracket term that differs the
Kaddanoff-Baym form of the kinetic equation from the
Botermans-Malfliet one. We discussed the meaning of the
Botermans-Malfliet effective current, the Noether and the
effective $B$-current and the memory current. Then we analyzed
delays and advances, as they appear in the non-local form of the
kinetic equation for off-mass-shell particles. There appear
several time delays: the Wigner, scattering, collision, forward,
memory,  Noether, drag-flow and back-flow  delays. Thus, the
physical meaning of the Poisson bracket term in the Kaddanoff-Baym
equation is fully clarified. The forward time delay appears as the
time shift in the collision term: $\delta t_{\rm f}^{\rm
kin}=\delta t_{\rm s}^B -t_{\rm col}$, where the kinetic
scattering time delay is $\delta t_{\rm s}^B=\hbar B_0/2$ and the
collision delay is $ t_{\rm col}=\hbar Z_0^{-1}/\Gamma$,
$B_0=A(Z_0^{-1}-M\Gamma^{-1}\frac{\partial\Gamma}{\partial E})$,
and for non-relativistic particles $M=E-p^2/2m +\Re\Sigma^R$,
$Z_0^{-1}=1-\frac{\partial\Re\Sigma^R}{\partial E}$. In the
absence of the energy retardations in the response of the medium
(Wigner resonances) the kinetic times $\delta t_{\rm s}^B$,
$t_{\rm col}$ and $\delta t_{\rm f}^{\rm kin}$  appropriately
transform into the similar quantities introduced in quantum
mechanical scattering on potentials and in the case of the
resonance scattering.

Moreover we related time delays to the density of energy states
with and without interaction terms. In the low density limit the
time-shift appeared in the non-local collision term  is just the
forward time delay/advance discussed above for classical and
quantum mechanical motions. Then we discussed application of the
test-particle method to solve the Botermans-Malfliet and non-local
kinetic equations. The analysis of the test-particle trajectories
also sheds the light on the meaning of the space-time shifts. We
showed then how the appropriate quasiparticle limit can be
recovered. A superluminality problem is  briefly discussed.
Apparent superluminal propagation has been indeed manifested in
some laser experiments.
 This phenomenon can be understood as
consequence of a reshaping of the pulse envelope by interaction
within the medium. Although formally the group velocity  may in
some cases exceed $c$, the forward wave front moves with velocity
$\leq c$.

Next, we calculated entropy flow for the non-local form of the
kinetic equation and compared it with the flows for the
Bottermans-Malfliet and the Kaddanoff-Baym forms of the kinetic
equation. We related the forward time delay to the difference in the
expressions for the Kaddanoff-Baym and the Bottermans-Malfliet
kinetic entropies. Note that, in principle, the presence or absence of
an additional non-equilibrium correction to the specific heat proportional to the
collision term can be experimentally verified.

Then choosing some reduction ansatz for the initial
non-equilibrium configurations we found  specific solutions of the
kinetic equations in the Botermans-Malfliet,  Kaddanoff-Baym and
non-local forms and solutions of the mass-shell equation, which,
in general, differ from each other and coincide only, if the
typical time scale characterizing the system dynamics  is larger
than the forward time delay/advance. The latter is typically of
the order of the collision time. Thus, we uncover some problems
with simulations of heavy-ion collisions using the test-particle
method, if one deals with the kinetic equation in the
Botermans-Malfliet form for typical time of the order of the
mentioned time scales.  In specific energy-momentum regions (e.g.
for small $\Gamma$ and larger $|M|$) the typical scattering time,
which characterizes evolution of the Botermans-Malfliet equation
$\delta t_{\rm s}^B \sim \hbar A/2$ can be much shorter than
$t_{\rm col}\sim \hbar/\Gamma$ characterizing evolution of the
Kaddanoff-Baym equation. In spite of the mentioned problems it
might be practical to use one of the above kinetic equations for
actual calculations even beyond its validity region, since all
these kinetic equations reveal approximate or even exact (as the
Kaddanoff-Baym form of the kinetic equation) conservation laws of
the 4-current and the energy-momentum tensor, thus approximating
reasonably the system evolution. Hydrodynamical limit is realized
for $t\gg t_{\rm col}$, when particle distributions acquire the
form of the local-equilibrium distribution. For such a
distribution the collision term turns to zero. The hydrodynamical
equations are derived from the conservation laws associated with
the kinetic equation. The kinetic coefficients entering
hydrodynamical equations are derived from the Botermans-Malfliet
equation (valid in this limit). They can be expressed through the
scattering time delay $\delta t_{\rm s}^B$. We also demonstrated
possibility of appearance of an instability for superluminal
off-mass shell particles.

Finally we presented a possible interpretation of the apparent
superluminality, which may manifest in  experiments, like the
OPERA and MINOS neutrino experiments, and in similar experiments
expected to be settled in nearest future.
 The maxima of the wave packets of protons and neutrinos produced
in the two-step process $p\to \pi^{+}n\to n\mu^{+}\nu$ at CERN are
separated by the time interval $\sim
\hbar/\Gamma_{p\pi}+\hbar/\Gamma_{\pi\nu}$.
  Thereby a time advance of the neutrinos may arise owing to
 an advance  of pions compared to protons and
 neutrinos,
  $\delta t_{\rm adv}^\nu = \delta t_{\rm adv}^{N\pi}+\delta t_{\rm adv}^{\pi\nu}$, where
  $\delta t_{\rm adv}^{N\pi}=-\hbar/\Gamma_{N\pi}$ and $\delta
  t_{\rm adv}^{\pi\nu}=-\hbar/\Gamma_{\pi\nu}$.
  The value $\delta t_{\rm adv}^{\pi\nu}\sim -\hbar/\Gamma_{\pi\nu}=-26$ ns is due  to  the width $\Gamma_{\pi\nu}$
  of the production of the neutrino  in the process
$\pi\rightarrow\nu\bar{\mu}$. Thus, one may expect few of $\delta
t_{\rm adv}^{\pi\nu}$, as a typical time advance in the neutrino
experiments like those performed by the OPERA and MINOS.

Some details of calculations are deferred to Appendices
\ref{app:virial} - \ref{app:I-centroid1}.   In Appendix
\ref{minentr} we discuss $H$ theorem for three forms of the
generalized kinetic equation and argue for the minimum of the
entropy production related to the generalized kinetic equation.

Concluding,  we discussed similarities between description of time
delays and advances for various  systems, like classical
oscillating systems,  one dimensional quantum mechanical
tunneling, decay of quasistationary states, three dimensional
scattering,  reactions and quantum kinetical processes.

\section*{Acknowledgement}
We thank Yu.B.~Ivanov, B.M.~Karnakov and V.D.~Mur for fruitful discussions.
The work was partially supported by ESF Research Networking
Programme POLATOM and by Grants No. VEGA~1/0457/12 and No.
APVV-0050-11 (Slovakia).
\newpage
\appendix

\section{Virial theorem for infinite classical motion in central potential}\label{app:virial}

Here we give  a short recount of the derivation of the virial theorem (\ref{class-delay-3D}) by Demkov in
Ref.~\cite{Demkov}. The derivation is based on the suggestion of Fock~\cite{Fock30} to combine the variational
principle of mechanics and the scale transformation of coordinates.

Consider a particle of mass $m$ moving in a central field $U(r)$ diminishing sufficiently rapidly for $r\to
\infty$. The equation of motion of the particle between times $t_1$ and $t_2$ follows from the requirement of
the vanishing of the action variation around the true trajectory
\be
 \delta S=\intop_{t_1}^{t_2} \delta L(\vec{r},\vec{v}\,) \rmd t\,
 \label{minact}
 \ee
with the Lagrange function $L=\half m\, v^2-U(r)$\,.

Consider now a particular variation around the trajectory, $\delta \vec{r}=\epsilon
\vec{r}(t)$, with an infinitesimally small parameter $\epsilon$\,. The variation of the action is now not
zero, since the variation $\delta \vec{r}$ does not vanish at the ends of the time interval and is given
by $\delta S = \epsilon\, \Big( \frac{\prt L(t_2)}{\prt \vec{v}} \vec{r}(t_2)-\frac{\prt L(t_1)}{\prt \vec{v}}
\vec{r}(t_1)\Big) $. On the other hand we can calculate $\delta S$ by expanding the Lagrange function
in (\ref{minact}) directly. Equating terms linear in $\epsilon$ in both expressions, we obtain
\be
\delta S=\epsilon \intop_{t_1}^{t_2}\Big(\frac{\prt L}{\prt
\vec{r}}\vec{r} +\frac{\prt L}{\prt \vec{v}}\vec{v} \Big)\rmd t
=
\epsilon \Big( \frac{\prt L(t_2)}{\prt \vec{v}}
\vec{r}(t_2)-\frac{\prt L(t_1)}{\prt \vec{v}} \vec{r}(t_1)\Big)\,,
\label{app:virial1} \ee
and, substituting the Lagrange function, arrive at
\be
\intop_{t_1}^{t_2} \Big(m\, v^2(t) - r(t)\, \frac{\rmd  U
(r(t))}{\rmd r}\Big)\,\rmd t
=
m\,\big(\vec{v}(t_2)\vec{r}(t_2)-\vec{v}(t_1)\vec{r}(t_1)\big)\,.
\label{app:virial-2}
\ee
For  $t_1\to \pm\infty$  the particle speed is $v_{\infty}$\,. Let
assign the time $t=0$ to the position of the closest approach of
the particle to the center $r=r_0$. Then for large times (either
positive or negative) the distance from the origin is given by
$r=s+v_\infty \, |t|$\,, where $s$ is the difference of the
distance that the particle, moving in the potential, passed from
the moment $t=0$ to $t$ and the distance it would pass during the
same time interval, if it moved freely (for $U=0$)  with the
velocity $v_\infty$. The scattering time delay/advance of the
particle in the potential can be defined  as
 \be
 \delta t_s^{\rm cl} = -\,s/v_\infty\,.
  \ee
Since both sides of Eq.~(\ref{app:virial-2}) diverge in the limit $t_1\to \pm\infty$, we regularize them by
subtracting $v_{\infty}^2\, (t_2-t_1)$. Then on the left-hand side we can use the energy conservation $m\,
(v^2-v_{\infty}^2)=-2\, U$ and on the right-hand side we get $2\,m\,v_\infty\lim_{t\to\infty}(r-v_\infty\,
t)=2m\,v_\infty s$. We take here into account that at large distances from the center
$\vec{v}\uparrow\downarrow\vec{r}$ before collision ($t_1\to-\infty$) and $\vec{v}\uparrow\uparrow\vec{r}$
after collision ($t_2\to+\infty$)\,. Thus we  rewrite Eq.~(\ref{app:virial-2}) as
\be
\intop_{-\infty}^{+\infty} \Big(2\,U(r(t)) + r(t)\, \frac{\rmd
U(r(t))}{\rmd r}\Big)\,\rmd t =-2m\,v_\infty\, s
=2m\,v_{\infty}^2\, \delta t_s^{\rm cl}\,,
 \ee
and introducing the Wigner time delay $\delta t^{\rm cl}_{\rm W} = 2\delta t_s^{\rm cl}$ we recover
Eq.~(\ref{class-delay-3D}).

Another derivation of this relation from the point of view of hypervirial theorems, introduced by Hirschfelder for classical and quantum systems in Ref.~\cite{Hirschfeld60}, can be found in Ref.~\cite{Robinson-Hirsch63}.

\section{Relations for wave functions obeying  Schr\"odinger equation}
\label{app:schroed}


Consider two solutions of  Schr\"odinger equation with a potential
$U(x)$ and slightly different energies $E$ and $E'$:
 \begin{eqnarray}
&&-\frac{\hbar^2}{2\,m}\frac{\prt^2}{\prt z^2} \psi(z,E)+U(z)\, \psi(z,E) = E\, \psi(z,E)\,,
 \nonumber\\
&&-\frac{\hbar^2}{2\,m}\frac{\prt^2}{\prt z^2} \psi^*(z,E') + U(z)\, \psi^*(z,E') = E'^*\, \psi^*(z,E')\,.
 \end{eqnarray}
For the sake of generality we assume that the energy might be
complex. Let us multiply  first equation by $\psi^*(z,E')$ and
second one by $\psi(z,E)$, subtract one from another and integrate
from $a$ to $b$. Then we put $ E'\to E$. In this limit
$\Re(E'^*-E)=\delta E\to 0$ and  $\Im(E'^* -E)\to -2\Im E$.
Keeping  only the leading terms in $\delta E$ and $\Im E$ on the
right-hand side we obtain
 \begin{eqnarray}
\intop_{a}^{b}\rmd z\left[ E'^*\, \psi(z,E)\psi^*(z,E')-E\, \psi^*(z,E')\psi(z,E) \right] \approx (\delta E -
2\,i\, \Im E)\,\intop_{a}^{b}\rmd z |\psi(z,E)|^2 \,.
 \label{app:SchrEq-1}
 \end{eqnarray}
On the left-hand side
 \begin{eqnarray}
&&\frac{\hbar^2}{2\,m} \intop_{a}^{b} \rmd z \left[ \psi^*(z,E')\frac{\prt^2}{\prt z^2} \psi(z,E)
-\psi(z,E)\frac{\prt^2}{\prt z^2} \psi^*(z,E') \right]
\approx i\, \hbar\big( j(b;E)-j(a;E)\big)
 \nonumber\\
&&\qquad+ \delta E\frac{\hbar^2}{2\,m}
 \left[ \left(\frac{\prt}{\prt E}\psi^*(z,E)\right)\frac{\prt}{\prt z} \psi(z,E)
 -\psi(z,E)\frac{\prt}{\prt z}\left(\frac{\prt}{\prt E} \psi^*(z,E)\right) \right]\Bigg|_{a}^{b}\,,
 \label{app:SchrEq-2}
 \end{eqnarray}
where the currents at coordinates $a$ and $b$ are determined according to Eq.~(\ref{current}),
$j(z;E)=\mathcal{J}[\psi(z;E)]$\,. Equating real and imaginary parts of Eqs.~(\ref{app:SchrEq-1}) and (\ref{app:SchrEq-2}) we arrive at the relations
 \begin{eqnarray}
\intop_{a}^{b}\rmd z \, |\psi(z,E)|^2 =\frac{\hbar^2}{2\,m} \left[
\left(\frac{\prt}{\prt E}\psi^*(z,E)\right)\frac{\prt}{\prt z} \psi(z,E)
-\psi(z,E)\frac{\prt}{\prt z}\left(\frac{\prt}{\prt E} \psi^*(z,E)\right)
\right]\Bigg|_{a}^{b}\,,
 \label{app:Int-rel}
\end{eqnarray}
and
 \begin{eqnarray}
-2\Im E=\hbar {\big(j(b;E)-j(a;E)\big)}\Big/{\dsp\intop_{a}^{b}\rmd z \, |\psi(z,E)|^2} \,.
 \label{app:Gam-rel}
\end{eqnarray}
The first relation is used to get Eq. (\ref{phase-deriv-integ}).
The last relation demonstrates equivalence between the current
conservation and the vanishing of the imaginary part of the
energy. If the current is not conserved, $j(a;E)\neq j(b;E)$, and
we are dealing with exponentially increasing ($\Im E>0$) or
decreasing ($\Im E<0$) wave function in the interval $[a,b]$. For
a bound state the wave function can always be chosen real and
therefore  both $j(a;E)$ and $j(b;E)$ vanish and $\Im E=0$. In the
scattering problem (e.g., as given by Eqs.~(\ref{psi-f-1}) and
(\ref{psi-f-1})) the currents are independent of the coordinate
and $j(a;E)=j(b;E)$ thus yielding $\Im E =0$. Only the wave
functions satisfying the boundary conditions
 \be
&&a)~~\psi(z;E)\to e^{+i\,|z|\, \sqrt{2mE}}\,,\quad z\to
\pm\infty\,,
 \nonumber\\
&&b)~~\psi(z;E)\to e^{-i\,|z|\, \sqrt{2mE}}\,,\quad z\to
\pm\infty\,,
 \nonumber\\
&&c)~~\psi(z;E)\to e^{\pm i\,z\, \sqrt{2mE}}\,,\quad z\to \infty\,
;~~ \psi(z;E)=0\,,~~z<a\,,
 \nonumber\\
&&d)~~\psi(z;E)\to e^{\mp i\,z\, \sqrt{2mE}}\,,\quad z\to
-\infty\, ;~~ \psi(z;E)=0\,,~~z>b\,
 \ee
describe states with complex energies. The imaginary part of the energy is negative (decaying state) for cases $a)$ and $c),d)$ with upper signs,  and is positive (process of a state formation) for cases $b)$ and $c),d)$ with lower signs.

It is instructive to express Eqs.~(\ref{app:Int-rel})
and~(\ref{app:Gam-rel}) through the logarithmic derivatives
\be
d(z;E)=\frac{z}{\psi(z;E)}\frac{\prt}{\prt z}\psi(z;E)\,.
 \ee
After simple algebra we get from Eq.~(\ref{app:Int-rel})
 \begin{eqnarray}
\intop_{a}^{b}\rmd z \, |\psi(z,E)|^2 =
\frac{\hbar^2}{2\,m}\left[\frac{1}{z}
\big(d(z;E)-d^*(z;E)\big)\psi(z;E)\frac{\prt}{\prt E} \psi^*(z;E)-
\frac{1}{z}|\psi(z;E)|^2 \frac{\prt}{\prt E} d^*(z;E)
\right]\Bigg|_{a}^{b}\,.
 \end{eqnarray}
Since the integral on the left-hand side is real we can add to the
right-hand side its complex conjugated value and  halve it. Then
we obtain
\begin{eqnarray}
\intop_{a}^{b}\rmd z \, |\psi(z,E)|^2 &=&
\left[l_{\rm q}[\psi(z;E)]\, \Im d(z;E) -
\frac{\hbar^2}{2\,m\,z} |\psi(z;E)|^2 \frac{\prt}{\prt E} \Re d(z;E)
\right]\Bigg|_{a}^{b}
\nonumber\\
&=& l_{\rm q}[\psi(b;E)]\, \Im d(b;E) -l_{\rm q}[\psi(a;E)]\, \Im d(a;E)
\nonumber\\
&+&
\frac{\hbar^2}{2\,m}\left[ |\psi(a;E)|^2\,\frac{1}{a} \frac{\prt}{\prt E}\Re d(a;E)
-|\psi(b;E)|^2\,\frac{1}{b} \frac{\prt}{\prt E}\Re d(b;E)
\right]\,,
\end{eqnarray}
where we introduced the characteristic quantum length
characterizing a stationary wave function $\psi(z;E)$:
\begin{eqnarray}
l_{\rm q}[\psi(z;E)]=\frac{i\hbar^2}{2\,m\,z}\Big(\psi(z;E) \frac{\prt}{\prt E} \psi^*(z;E)-\psi^*(z;E)\frac{\prt}{\prt E} \psi(z;E) \Big)\,.
\end{eqnarray}
For example, for the plane wave this quantity is the de Broiglie
wave length $l_{\rm q}[\exp(i\, k\, z/\hbar)]=\hbar/k$. For the
wave function (\ref{psi-f-1}) and $z>L/2$ we find $l_{\rm
q}[T(E)\exp(i\, k\, z/\hbar)]=(\hbar/k)
|T(E)|^2\big(1+\frac{k}{z\, m}\hbar \frac{\rmd}{\rmd E}\phi_{\rm
T}(E)\big)$.

From Eq.~(\ref{app:Gam-rel}) straightforwardly follows
 \begin{eqnarray}
-2\Im E\intop_{a}^{b}\rmd z \, |\psi(z,E)|^2 = \frac{\hbar^2}{m\, b} |\psi(b;E)|^2\, \Im d(b;E)
-\frac{\hbar^2}{m\, a} |\psi(a;E)|^2\, \Im d(a;E)\,.
 \end{eqnarray}


\section{Asymptotic centroids of the wave packets}\label{app:I-centroid}

Let us perform derivation of Eq.~(\ref{xI-aver}). Substituting
Eq.~(\ref{wp-free}) in the standard  definition of the average
coordinate we have
\begin{eqnarray}
\bar{z}^{\rm (as)}_{\rm I}(t) &=& \intop_{-\infty}^{+\infty} \rmd z\, z |\Psi_{\rm I}(z,t)|^2 =
\intop_{0}^{+\infty} \frac{\rmd k}{2\pi\hbar}
\intop_{0}^{+\infty} \frac{\rmd k'}{2\pi\hbar}
 \varphi(k) \varphi^*(k')\, e^{i(E'-E)\, t/\hbar}
\intop_{-\infty}^{+\infty}\rmd z z e^{+i\,(k-k')\,z/\hbar}\,,
\label{app:zI-1}
\end{eqnarray}
 where
$E=k^2/2m$ and $E'=k'^2/2m$.
Changing variables to $Q=(k+k')/2$ and $q=k-k'$ with $\rmd k\rmd k'=\rmd Q\rmd q$ we write
\begin{eqnarray}
\bar{z}^{\rm (as)}_{\rm I}(t) &=&
\intop_{0}^{+\infty} \frac{\rmd Q}{2\pi\hbar}
\intop_{-\infty}^{+\infty} \frac{\rmd q}{2\pi\hbar}
 \varphi(Q+q/2) \varphi^*(Q-q/2)\, e^{-i\frac{Q\,q}{m} \frac{t}{\hbar}}
\intop_{-\infty}^{+\infty}\rmd z z e^{+i\,q\,z/\hbar}\,.
\label{app:zI-2}
\end{eqnarray}
 Using
\begin{eqnarray}
\intop_{-\infty}^{+\infty} \rmd z\, z^n
e^{+i\,q\,z/\hbar}=(-i\hbar)^n\, (2\pi\hbar) \frac{\rmd^n}{\rmd
q^n}\delta(q)
\label{app:zI-3}
\end{eqnarray}
after integration by parts we obtain
\begin{eqnarray}
\bar{z}^{\rm (as)}_{\rm I}(t)&=&i\intop_{0}^{+\infty} \frac{\rmd
Q}{2\pi} \Big\{ \half \varphi'(Q)\varphi^*(Q) -\half
\varphi(Q)\varphi'^*(Q)
-i\frac{Q}{m}\frac{t}{\hbar}|\varphi(Q)|^2\Big\}\,.
\label{app:zI-4}
\end{eqnarray}
Introducing the phase of the momentum profile function $\xi(k)$ as
$\varphi(k)= |\varphi(k)| e^{i\,\xi(k)}$ after the replacement
$Q\to k$ we  cast the integral in the form
\begin{eqnarray}
\bar{z}^{\rm (as)}_{\rm I}(t)&=&\intop_{0}^{+\infty} \frac{\rmd
k}{2\pi\hbar} \Big( -\hbar\,\xi'(k) + \frac{k}{m}
t\Big)|\varphi(k)|^2\, \label{xbar-1}
\end{eqnarray}
and recover Eq.~(\ref{xI-aver}).

Similarly to the above we derive
\begin{eqnarray}
\overline{[z^{\rm (as)}_{\rm I}(t)]^2} =
\intop_{-\infty}^{+\infty} \rmd z\, z^2 |\Psi_{\rm I}(z,t)|^2 =
-\hbar^2\intop_{0}^{+\infty} \frac{\rmd Q}{2\pi\hbar}
\intop_{-\infty}^{+\infty} {\rmd q}
 \varphi(Q+q/2) \varphi^*(Q-q/2)\, e^{-i\frac{Q\,q}{m} \frac{t}{\hbar}}
 \frac{\rmd^2}{\rmd q^2}\delta(q)\,.
\label{app:z2I-1}
\end{eqnarray}
The integration by parts over $q$  after the replacement $Q\to k$
yields
\begin{eqnarray}
&&\overline{[z^{\rm (as)}_{\rm I}(t)]^2} =
-\hbar^2\intop_{0}^{+\infty} \frac{\rmd k}{2\pi\hbar} \Big\{
\quart \varphi''(k)\varphi^*(k)+\quart
\varphi(k)\,\varphi''^*(k)-\half \varphi'(k)\varphi'^*(k)
\nonumber\\ &&\qquad - i\, \frac{k}{m}\frac{t}{\hbar}
\big(\varphi'(k)\varphi^*(k)-\varphi(k)\varphi'^*(k)\big)
-\frac{k^2}{m^2}\frac{t^2}{\hbar^2} |\varphi(k)|^2 \Big\}
\nonumber\\ &&\quad= \frac{\hbar}{4\pi}  |\varphi(0)|'|\varphi(0)|
+ \intop_{0}^{+\infty} \frac{\rmd k}{2\pi\hbar}
\Big\{\hbar^2\,(|\varphi(k)|')^2 +\Big(\hbar\xi'(k)
-\frac{k}{m}t\Big)^2  |\varphi(k)|^2 \Big\}\,,
\label{app:z2I-2}
\end{eqnarray}
and thereby Eq.~(\ref{packet-width-x}) is
recovered.

The results (\ref{app:zI-3}) and (\ref{app:z2I-1}) can be generalized as follows
\begin{eqnarray}
\intop_{-\infty}^{+\infty} \rmd z\, z^n |\Psi_{\rm I}(z,t)|^2
=
(-i\hbar)^n\intop_{0}^{+\infty} \frac{\rmd Q}{2\pi\hbar}
\intop_{-\infty}^{+\infty} \rmd q
 \varphi(Q+q/2) \varphi^*(Q-q/2)\, e^{-i\frac{Q\,q}{m} \frac{t}{\hbar}}
 \frac{\rmd^n}{\rmd q^n}\delta(q)\,.
\label{zIn}
\end{eqnarray}

Now we turn to the derivation of the asymptotic centroid evolution for the transmitted packets [Eq.~(\ref{zT-zR})].
To evaluate the integrals of the type
$
\intop_{-\infty}^{+\infty} \rmd z\, z^n |\Psi_{\rm T}(z,t)|^2
$
we can use Eq.~(\ref{zIn}) with the only replacement $\varphi\to
\varphi\, T$.  Then for the normalization integral we immediately
obtain
\begin{eqnarray}
\intop_{-\infty}^{+\infty} \rmd z\, |\Psi_{\rm T}(z,t)|^2=
\intop_{0}^{+\infty} \frac{\rmd k}{2\pi\hbar} |\varphi(k)|^2
|T(k)|^2 =\langle T(E) \rangle_k\,. \label{zT-norm}
\end{eqnarray}
Now we adopt Eq.~(\ref{app:zI-4}) and write
\begin{eqnarray}
\intop_{-\infty}^{+\infty} \rmd z\, z|\Psi_{\rm T}(z,t)|^2&=&
i\hbar\intop_{0}^{+\infty} \frac{\rmd k}{2\pi\hbar} \Big\{
\half\varphi^*(k) T^*(k) \frac{\rmd}{\rmd
k}\big(\varphi(k)T(k)\big) -\half \varphi(k) T(k)\frac{\rmd}{\rmd
k}\big(\varphi^*(k)T^*(k)\big) \nonumber\\
&-&i\frac{k}{m}\frac{t}{\hbar}|\varphi(k)|^2 |T(k)|^2\Big\}\,.
\label{zT-z-1}
\end{eqnarray}
Substituting  $\varphi(k)\, T(k)= |\varphi(k)|\,|T(k)|
e^{i\,\xi(k)+i\phi_{\rm T}(k)}$ we find
\begin{eqnarray}
\intop_{-\infty}^{+\infty} \rmd z\, z|\Psi_{\rm T}(z,t)|^2&=&
\intop_{0}^{+\infty} \frac{\rmd k}{2\pi\hbar} \Big\{
-\hbar\,\xi'(k)-\hbar\phi'_{\rm T}(k)
+\frac{k}{m}t\Big\}|\varphi(k)|^2 |T(k)|^2\,. \label{zT-z-2}
\end{eqnarray}
Dividing Eq.~(\ref{zT-z-2}) by Eq.~(\ref{zT-norm}) we recover the
first equation in (\ref{zT-zR}).

To get similar expressions for the reflected packet  we have to
replace $\varphi\to \varphi\, R$ in Eq.~(\ref{zIn}) and also
change $q\to -q$. The corresponding result in (\ref{zT-zR})
follows in the full analogy to Eq.~(\ref{zT-norm}), and
(\ref{zT-z-2}) with the change of the overall sign in the latter.

To calculate the width of the transmitted packet appeared in Eq.
(\ref{RT-width-def}) we need to calculate $\overline{[z^{\rm
(as)}_{\rm T}(t)]^2}$. Making the replacement $\varphi\to
\varphi\, T$ in Eq.~(\ref{app:z2I-2}) and taking into account that
$T(0)=0$ we can write
\begin{eqnarray}
&&\overline{[z^{\rm (as)}_{\rm T}(t)]^2} =\frac{1}{\langle |T(k)|^2\rangle_k}
\nonumber\\
&&\quad\times
\left\langle
\hbar^2\,(|T(k)|'+|T(k)||\varphi(k)|/|\varphi(k)|')^2 +\Big(\hbar\xi'(k)+ \hbar\phi_{\rm T}'(k) -\frac{k}{m}t\Big)^2 |T(k)|^2
\right\rangle_k
\nonumber\\
&&\quad=
\left\langle
\hbar^2\,\Big[\frac{\rmd}{\rmd k}\log(|\varphi(k)||T(k)|)\Big]^2\right\rangle_{k,\rm T} +\left\langle\Big(\hbar\xi'(k)+ \hbar\phi_{\rm T}'(k) -\frac{k}{m}t\Big)^2
\right\rangle_{k,\rm T}\,.
\label{app:z2T-1}
\end{eqnarray}
Here in the last equality we use the definition of the average (\ref{RT-aver}).
For the reflected packet we can write by analogy
 \begin{eqnarray}
&&\overline{[z^{\rm (as)}_{\rm R}(t)]^2} =
\left\langle
\hbar^2\,\Big[\frac{\rmd}{\rmd k}\log(|\varphi(k)||R(k)|)\Big]^2\right\rangle_{k,\rm R} +\left\langle\Big(\hbar\xi'(k)+ \hbar\phi_{\rm R}'(k) -\frac{k}{m}t\Big)^2
\right\rangle_{k,\rm R}\,.
\label{app:z2R-1}
\end{eqnarray}

\section{Relations for the sojourn time}\label{app:I-centroid1}
Let us perform derivation of the relation between the sojourn time
and the dwell time (\ref{soj-dwell}). Using Eq.~(\ref{wpp})
and
performing the integration over time we find
\be
t_{\rm soj}(a,b)&=&\intop_{-\infty}^{+\infty}\rmd t\intop_a^b\rmd z |\Psi(z,t)|^2
\nonumber\\
&=& \intop_{0}^{+\infty} \frac{\rmd
k}{2\pi\hbar}\,\intop_{0}^{+\infty} \frac{\rmd k'}{2\pi\hbar}\,
\varphi(k)\, \varphi^*(k')\, \intop_a^b\rmd z\, \psi(z,E)
\psi^*(z,E')\, (2\,\pi\, \hbar)\delta(E-E')\,,
 \nonumber\ee
 where we used that
$E=k^2/2m$ and $E'=k'^2/2m$. Taking the integral over momentum
$k'$ we obtain
\be
t_{\rm soj}(a,b)&=&\intop_{0}^{+\infty} \frac{\rmd k}{2\pi\hbar} |\varphi(k)|^2 \frac{m}{k} \intop_a^b\rmd x\, |\psi(x,E)|^2\,,
\ee
thus Eq.~(\ref{soj-dwell}) is recovered.

Now let us derive Eq. ~(\ref{int-cur-soj}).
 Using the definitions of the wave function on
the left and right sides of the barrier [Eqs.~(\ref{wp-free}),
(\ref{twp}) and (\ref{rwp})]  we can write the current as follows
\begin{eqnarray}
&&j(z\ge L/2,t)=\frac{i\hbar}{2\,m}\Big(\Psi_{\rm T}(z,t)\nabla_z
\Psi_{\rm T}^*(z,t)- \Psi_{\rm T}^*(z,t)\nabla_z \Psi_{\rm T}(\rm
T,t)\Big)\,, \nonumber\\ &&\quad=\frac{1}{2\,m} \intop_0^\infty
\frac{\rmd k}{2\pi\hbar}\intop_0^\infty \frac{\rmd k'}{2\pi\hbar}
\varphi(k)\,\varphi^*(k')\,T(E)\,T^*(E')
e^{i\,(E'-E)\,t/\hbar}e^{+i\, (k-k')\,z/\hbar}(k'+k), \nonumber\\
&&j(z\le -L/2,t)=\frac{i\hbar}{2\,m}\Big([\Psi_{\rm
I}(z,t)+\Psi_{\rm R}(z,t)]\nabla [\Psi_{\rm I}^*(z,t)+\Psi_{\rm
R}^*(z,t)] \nonumber\\ &&\qquad\qquad-[\Psi_{\rm
I}^*(z,t)+\Psi_{\rm R}^*(z,t)]\nabla [\Psi_{\rm I}(z,t)+\Psi_{\rm
R}(z,t)]\Big) \nonumber\\ &&\quad=\frac{1}{2\,m} \intop_0^{\infty}
\frac{\rmd k}{2\pi\hbar}\intop_0^\infty \frac{\rmd k'}{2\pi\hbar}
\varphi(k)\,\varphi^*(k') e^{i\,(E'-E)\,t/\hbar} \Big\{
(k'+k)\Big[e^{i\, (k-k')\,x/\hbar}- R^*(E')\,R(E)e^{-i\,
(k-k')\,k/\hbar}\Big] \nonumber\\ &&\qquad+(k-k') \Big[R^*(E')\,
e^{+i\, (k'+k)\,z/\hbar}-R(E)\, e^{-i\, (k+k')\,z/\hbar} \Big]
\Big\}\,.
\end{eqnarray}
Performing the replacement of momenta
$k=Q+\half q\,,\quad k'=Q-\half q\,,\quad \rmd k \rmd k' =\rmd
Q\rmd q$
and using that $E'-E=\frac{1}{2m}\big[(Q-\half q)^2-(Q+\half q)^2
\big]=Q\, q/m$,  we find
\begin{eqnarray}
j(z\ge L/2,t) &=& \frac{1}{2\,m} \intop_0^\infty \frac{\rmd
Q}{2\pi\hbar} \intop_{-\infty}^{\infty} \frac{\rmd q}{2\pi\hbar}
\varphi(Q+\half q)\,\varphi^*(Q-\half q)\, \nonumber\\ &\times&
T(E_{Q+q/2})\,T^*(E_{Q-q/2}) e^{-i\,Q\, q\,t/m\hbar}e^{+i\,
q\,z/\hbar}\, 2\,Q \,,\nonumber\\ j(z\le -L/2,t) &=&
\frac{1}{2\,m} \intop_0^\infty \frac{\rmd
Q}{2\pi\hbar}\intop_{-\infty}^\infty \frac{\rmd q}{2\pi\hbar}
\varphi(Q+\half q)\,\varphi^*(Q-\half q) e^{i\,Q\, q\,t/m\hbar}
\nonumber\\ &\times& \Big\{ 2Q\Big[e^{i\, q\,x/\hbar}-
R^*(E_{Q-q/2})\,R(E_{Q+q/2})e^{-i\, q\,z/\hbar}\Big] \nonumber\\
&+& q \Big[R^*(E_{Q-q/2})\, e^{+i\,2 Q\,x/\hbar}-R(E_{Q+q/2})\,
e^{-i\, 2Q\,x/\hbar} \Big] \Big\}\,.
\end{eqnarray}
Integrating over the time in Eq.~(\ref{tsoj-q}) with the help of expression
\begin{eqnarray}
\intop_{-\infty}^{+\infty}\rmd t\intop_{-\infty}^{t} \rmd t' e^{i\,Q\, q\,t/m\hbar}=
2\pi\frac{i\,m^2\hbar^2}{Q^2\,q}\delta(q)
\end{eqnarray}
we derive
\begin{eqnarray}
&&\intop_{-\infty}^{+\infty}\rmd t\intop_{-\infty}^{t} \rmd t'
\Big(j(L/2,t')-j(-L/2,t')\Big)
=
\intop_0^\infty \frac{\rmd Q}{2\pi\hbar}\intop_{-\infty}^{\infty}
\rmd q\, \varphi(Q+\half q)\,\varphi^*(Q-\half q)\,
\frac{i\,m\hbar}{Q}\delta(q) \nonumber\\ &&\times\Bigg[
\frac{1}{q}\Big(T(E_{Q+q/2})\,T^*(E_{Q-q/2}) + R(E_{Q+q/2}) \,
R^*(E_{Q-q/2})-e^{-i\, q\,L/\hbar}\Big)\, e^{+i\,
\frac{q\,L}{2\hbar}} \nonumber\\ &&\qquad - \frac{1}{2Q}
\Big(R^*(E_{Q-q/2})\, e^{-i\,2 Q\,L/2\hbar}-R(E_{Q+q/2})\, e^{+i\,
2Q\,L/2\hbar} \Big) \Bigg]\,. \nonumber
\end{eqnarray}
Taking into account that the expression in the squared bracket at
the term with $1/q$ vanishes for $q\to 0$, so that only the first
derivative of this expression contributes, we obtain
\begin{eqnarray}
&&\intop_{-\infty}^{+\infty}\rmd t\intop_{-\infty}^{t} \rmd t' \Big(j(L/2,t')-j(-L/2,t')\Big)
\nonumber\\
&&=-\intop_0^\infty \frac{\rmd Q}{2\pi} |\varphi(Q)|^2
\Bigg[
|T(E_Q)|^2\,\hbar\frac{\prt\phi_{\rm T}(E_Q)}{\prt E}
+|R(E_Q)|^2\,\hbar\frac{\prt\phi_{\rm R}(E_Q)}{\prt E}
 +\frac{m\,L}{Q}
+\frac{\hbar\,m}{Q^2}\Im\Big(R(E_Q)e^{iQ\, L/\hbar}\Big) \Bigg]\,.
\nonumber
\end{eqnarray}
Substituting here  definitions of the phases $\phi_{\rm R,T}$ from
Eq.~(\ref{ampl-phase}), we recover Eq.~(\ref{int-cur-soj}).

\section{$H$-theorem and minimum of the entropy
production}\label{minentr}


 In  \cite{IKV2} we presented arguments for the  $H$ theorem
and could prove it for some specific examples, e.g. for the $\Phi$ derivable theories for $\Phi$ diagrams with two vertices.
Equation for  the entropy flow for all three  forms of the
kinetic equation,
the BM, KB and non-local form, is as follows
\begin{eqnarray}\label{H}
\partial_{\mu}S^{\mu} =-H,
\end{eqnarray}
where now in the l.h.s. $S^{\mu}$ is either $S^{\mu}_{\rm BMM}$,
or $S^{\mu}_{\rm KB}$, or $S^{\mu}_{\rm NL}$, the latter quantity
is up to first gradients the  same  as  for the KB choice. The
memory contribution can be also incorporated as additional term
$\partial_{\mu}S^{\mu}_{\rm mem}$ in the l.h.s.

For the BM and the KB forms of the kinetic equation we multiply
the kinetic equation by $\mbox{ln} \frac{1\mp f}{f}$.
Using the multi-particle process decomposition \cite{IKV2} we arrive
at
the relation
%
\begin{eqnarray}
\label{s(coll)}
&&H=-\mbox{Tr}\int \frac{d^4{p}}{(2\pi)^4} \ln\frac{1\mp f}{f}\,C
= -\mbox{Tr}
\sum_{m,\tilde{m}}
\frac{1}{2}
 \int \frac{d^4{p}_1}{(2\pi)^4}
\cdot\cdot\cdot\frac{d^4{p}_m}{(2\pi)^4} \frac{d^4\tilde{p}_1}{(2\pi)^4}
\cdot\cdot\cdot\frac{d^4\tilde{p}_{\tilde{m}}}{(2\pi)^4}
\nonumber
\\
&&\times
\left[A_1 f_1\cdot\cdot\cdot A_m f_m A'_1 (1\mp f'_1) \cdot\cdot\cdot A'_{\tilde{m}}(1\mp
f'_{\tilde{m}})
-
A_1 (1\mp f_1) \cdot\cdot\cdot A_m (1\mp f_m)  A'_1
f'_1\cdot\cdot\cdot
A'_{\tilde{m}}f'_{\tilde{m}}\right]
\nonumber\\
&&\times
\ln\frac{f_1 \cdot\cdot\cdot f_m (1\mp f'_1)\cdot\cdot\cdot (1\mp f'_{\tilde{m}})}
        {(1\mp f_1)\cdot\cdot\cdot (1\mp f_m) f'_1 \cdot\cdot\cdot f'_{\tilde{m}}}
R_{m,\tilde{m}}\;
\delta^4\left(\sum_{i=1}^{m} p_i - \sum_{i=1}^{\tilde{m}} \tilde{p}_i \right) .
\end{eqnarray}
%
Here we assume different flavors and intrinsic quantum numbers to be absorbed
in the momenta $p_i$ and $\tilde{p}_i$.

In the case when all rates $R_{m,\tilde{m}}$ are
non-negative, i.e. $R_{m,\tilde{m}}\geq 0$, this expression is non-negative, since
$(x-y)\mbox{ln}(x/y) \geq 0$ for any positive $x$ and $y$.  In particular,
$R_{m,\tilde{m}}\geq 0$ takes place for all $\Phi$-functionals up to two vertices.
Then the divergence of $s^\mu$ is non-negative that proves the
$H$-theorem in this case.

For the non-local form of the kinetic equation we multiply the
latter by $\mbox{ln} \frac{1\mp f^{\rm shift}}{f^{\rm shift}}$ and
get $$H^{\rm shift}=-\mbox{Tr}\int \frac{d^4{p}}{(2\pi)^4}
\mbox{ln} \frac{1\mp f^{\rm shift}}{f^{\rm shift}} \frac{AC^{\rm
shift}}{A^{\rm shift}}\simeq -\mbox{Tr}\int
\frac{d^4{p}}{(2\pi)^4} \mbox{ln} \frac{1\mp f^{\rm shift}}{f^{\rm
shift}} C^{\rm shift}$$ instead of $H$.
 Thus Eq. (\ref{s(coll)}) continues to hold but now in
shifted variables.

Assume that the system is closed, i.e. there is no entropy flow through the volume
boundary.
Then
\begin{eqnarray}
\left[\frac{d \int S^{0} d^3 X}{d t}\right]_{\rm l.eq} =0
,\quad
\left[\frac{d^2 \int S^{0} d^3 X}{d t^2}\right]_{\rm l.eq}
=0,
\end{eqnarray}
 since both the
curle-bracket term and the $\ln$-term in   (\ref{s(coll)}) are zero in
the local equilibrium that results in zero of the function and its derivative.

Assuming the validity of the $H$-theorem (the entropy should be maximum  in the local equilibrium)
we have
\begin{eqnarray}\label{cond2}
\left(\frac{d^3 \int S^{0} d^3 X }{d t^3}\right)_{\rm l.eq} \leq 0
.
\end{eqnarray}
Thus we argued for the
principle of the minimum of the entropy production
(previously postulated by
Prigogine) now related to the generalized kinetic equation, provided
the $H$ theorem is satisfied.


\end{document}